\documentclass[prx,aps,amsfonts,showpacs,longbibliography,notitlepage,twocolumn,groupedaddress]{revtex4-1}
\usepackage[dvipsnames]{xcolor}

\usepackage{amsmath,amssymb,amsfonts,mathtools,dsfont,relsize}
\usepackage{microtype} 
\usepackage{graphicx}

\usepackage[colorlinks=true, urlcolor=violet, linkcolor=blue, citecolor=red, hyperindex=true, linktocpage=true]{hyperref}

\usepackage{amsthm}

\usepackage{soul} 

\usepackage{enumerate}
\usepackage{stmaryrd} 
\usepackage{suffix} 
\usepackage{comment}
\usepackage{xpatch}
\makeatletter
\patchcmd{\@ssect@ltx}
    {\addcontentsline{toc}{#1}{\protect\numberline{}#8}}
    {}
    {}
    {}
\makeatother

\makeatletter
\renewcommand{\p@subsection}{}
\def\l@subsubsection#1#2{}
\makeatother

\makeatletter
\g@addto@macro\bfseries{\boldmath}
\makeatother

\newcommand{\vpd}[0]{\vphantom{\dagger}}
\newcommand{\vps}[0]{\vphantom{*}}
\newcommand{\vpp}[0]{\vphantom{\prime}}

\DeclarePairedDelimiter{\set}{\lbrace}{\rbrace}

\DeclarePairedDelimiter{\norm}{\lVert}{\rVert}
\DeclarePairedDelimiter{\abs}{\lvert}{\rvert}
\DeclarePairedDelimiter{\expval}{\langle}{\rangle}

\DeclarePairedDelimiter{\ket}{\lvert}{\rangle}
\DeclarePairedDelimiter{\bra}{\langle}{\rvert}

\newcommand{\ii}[0]{\mathrm{i}}
\newcommand{\bvec}[1]{\boldsymbol{#1}}

\newcommand{\kron}[1]{\delta^{\,}_{#1}}

\newcommand{\ident}[0]{\mathds{1}}

\newcommand{\order}[1]{\mathsf{o} \hspace{-0.5mm} \left( #1 \right)}
\newcommand{\Order}[1]{\mathsf{O} \hspace{-0.5mm}  \left( #1 \right)}

\newcommand{\BKop}[2]{\ket{#1} \hspace{-0.4mm} \bra{#2}}

\newcommand{\inprod}[2]{ \left\langle #1 \middle| #2 \right\rangle}
\newcommand{\matel}[3]{\left\langle #1 \middle| #2 \middle| #3 \right\rangle}
\WithSuffix\newcommand\matel*[3]{\langle #1 | #2 | #3 \rangle}

\newcommand{\tr}[1]{{\rm tr} \left[  #1  \right]}
\DeclareMathOperator*{\trace}{tr}

\newcommand{\comm}[2]{\left[ #1, \, #2 \right]}
\newcommand{\acomm}[2]{\left\{ #1, \, #2 \right\} }

\newcommand{\chan}{\mathcal{W}}
\newcommand{\Chan}[1]{\mathcal{W}^{\vpd}_{#1}}
\newcommand{\Chandag}[1]{\mathcal{W}^{\dagger}_{#1}}

\newcommand{\observ}[0]{\mathcal{O}}

\newcommand{\Ham}[0]{H}
\newcommand{\Hadamard}[0]{H}

\newcommand{\MeasChannel}[0]{\mathsf{M}}
\newcommand{\QECChannel}[0]{\mathcal{R} }
\newcommand{\BellChannel}[0]{\mathcal{B}}

\def\DensMat{\rho}
\def\DensMatSS{\varrho}

\newcommand{\isometry}[0]{\mathbb{V}}

\newcommand{\umeas}[0]{\mathsf{V}} 
\newcommand{\Umeas}[1]{\umeas^{\vpd}_{\left[ #1 \right]}}
\newcommand{\Umeasdag}[1]{\umeas^{\dagger}_{\left[ #1 \right]}}
\newcommand{\wmeas}[0]{\mathsf{R}}
\newcommand{\Wmeas}[1]{\wmeas^{\vpd}_{\left[ #1 \right]}}

\newcommand{\proj}[2]{P^{#2}_{#1}}

\newcommand{\Pauli}[2]{\sigma^{#1}_{#2}}

\newcommand{\PX}[1]{X^{\vps}_{#1}}
\newcommand{\PY}[1]{Y^{\vps}_{#1}}
\newcommand{\PZ}[1]{Z^{\vps}_{#1}}

\newcommand{\LX}[0]{X^{\vps}_{{\rm L}}}
\newcommand{\LY}[0]{Y^{\vps}_{{\rm L}}}
\newcommand{\LZ}[0]{Z^{\vps}_{{\rm L}}}

\newcommand{\SSid}[1]{\widetilde{\ident}^{\vps}_{#1}}
\newcommand{\SSX}[1]{\widetilde{X}^{\vps}_{#1}}
\newcommand{\SSXp}[2]{\widetilde{X}^{#2}_{#1}}
\newcommand{\SSY}[1]{\widetilde{Y}^{\vps}_{#1}}
\newcommand{\SSZ}[1]{\widetilde{Z}^{\vps}_{#1}}
\newcommand{\SSProj}[1]{\widetilde{P}^{\,}_{#1}}
\newcommand{\SSproj}[2]{\widetilde{P}^{#2}_{#1}}

\newcommand{\measproj}[1]{\proj{#1}{\vpp}}

\newcommand{\Eig}[1]{\xi^{\vpp}_{#1}}

\newcommand{\Hilbert}[0]{\mathcal{H}}
\newcommand{\HilDim}[0]{\mathcal{D}}
\newcommand{\SpaceDim}[0]{D}
\newcommand{\Nspins}[0]{N}
\newcommand{\Nmeas}[0]{M}
\newcommand{\Nobs}[0]{\mathcal{M}} 
\newcommand{\Noutcome}[0]{\mathcal{N}}
\newcommand{\Dist}[0]{L}

\newcommand{\LRvel}[0]{v}
\newcommand{\Stab}[1]{\mathbf{S}^{\,}_{#1}}

\newcommand{\range}[0]{l}
\newcommand{\Msites}[0]{\Omega}

\usepackage{qcircuit}

\usepackage{pdfpages}
\makeatletter
\AtBeginDocument{\let\LS@rot\@undefined}
\makeatother

\begin{document}
\title{Locality and error correction in quantum dynamics with measurement}

\author{Aaron J. Friedman}
\email{aaron.friedman@colorado.edu}
\affiliation{Department of Physics and Center for Theory of Quantum Matter, University of Colorado, Boulder, CO 80309, USA}

\author{Chao Yin}
\email{chao.yin@colorado.edu}
\affiliation{Department of Physics and Center for Theory of Quantum Matter, University of Colorado, Boulder, CO 80309, USA}

\author{Yifan Hong}
\email{yifan.hong@colorado.edu}
\affiliation{Department of Physics and Center for Theory of Quantum Matter, University of Colorado, Boulder, CO 80309, USA}

\author{Andrew Lucas}
\email{andrew.j.lucas@colorado.edu}
\affiliation{Department of Physics and Center for Theory of Quantum Matter, University of Colorado, Boulder, CO 80309, USA}

\begin{abstract}
    The speed of light $c$ sets a strict upper bound on the speed of information transfer in both classical and quantum systems. In nonrelativistic quantum systems, the Lieb-Robinson Theorem imposes an \emph{emergent} speed limit $\LRvel \hspace{-0.2mm} \ll  \hspace{-0.2mm}  c$, establishing locality under unitary evolution and constraining the time needed to perform useful quantum tasks. We extend the Lieb-Robinson Theorem to quantum dynamics with \emph{measurements}. In contrast to the expectation that measurements can arbitrarily violate spatial locality, we find \emph{at most} an $(\Nmeas \hspace{-0.5mm} +\hspace{-0.5mm} 1)$-fold enhancement to the speed $\LRvel$ of quantum information, provided the outcomes of measurements in $\Nmeas$ local  regions are known. This holds even when classical communication is \emph{instantaneous}, and extends beyond projective measurements to weak measurements and other nonunitary channels. Our bound is asymptotically optimal, and saturated by existing measurement-based protocols. We tightly constrain the resource requirements for quantum computation, error correction, teleportation, and generating entangled resource states (Bell, GHZ, quantum-critical, Dicke, W, and spin-squeezed states) from short-range-entangled initial states. Our results impose limits on the use of measurements and active feedback to speed up quantum information processing, resolve fundamental questions about the nature of measurements in quantum dynamics, and constrain the scalability of a wide range of proposed quantum technologies.
\end{abstract}

\date{\today}
\maketitle
\tableofcontents

\section{Introduction} \label{sec:intro}
Information in nonrelativistic systems propagates at an \emph{emergent} speed $\LRvel$ that is much lower than the speed of light $c$ (much like the speed of sound in air). In quantum mechanical systems with local interactions, the Lieb-Robinson Theorem \cite{Lieb1972} establishes a finite speed of quantum information under unitary time evolution. In recent years, such quantum speed limits have been generalized to a wide range of 
physical systems, including  power-law interacting systems \cite{fossfeig,chen2019finite,Kuwahara:2019rlw,Tran:2020xpc}, interacting bosons \cite{gross,schuch,kuwahara2021liebrobinson,yin21}, spins interacting with cavity photons \cite{LRion}, local Lindblad dynamics \cite{Poulin}, and even toy models of holographic quantum gravity \cite{lucas2020non}. Although emergent locality seems generic to \emph{unitary} many-body quantum information dynamics in physically realizable systems, conventional wisdom is that there is no such emergent locality in the presence of \emph{measurements} and outcome-dependent feedback \cite{epr, li20, ippoliti, Bao:2021con, cirac, Verresen:2021wdv, Tantivasadakarn:2021vel}.

In their famous ``paradox,'' Einstein, Podolsky, and Rosen (EPR) \cite{epr} worried that measuring one qubit in the entangled pair $\ket{\text{Bell}^{\,}_{\pm}} = (\ket{00} \pm \ket{11})/\sqrt{2}$ \cite{Bell} \emph{instantly} affected the state of the other qubit, no matter their spatial separation. The paradox of EPR is that a single measurement of an entangled state could be used to send quantum information instantaneously over arbitrary distances, violating  the speed limit $c$ and relativistic causality. The resolution in the context of relativity is that any \emph{quantum} information ``teleported'' by a measurement can only be interpreted using an accompanying \emph{classical} communication, which travels no faster than $c$ \cite{Bell}. However, in nonrelativistic quantum systems, classical communication is effectively instantaneous---without a corresponding notion of locality, one might reasonably conclude that quantum information can be teleported (or entanglement generated) over arbitrary distances using local unitaries combined with a \emph{single} measurement.

Here, we present an asymptotically optimal bound \eqref{eq:main} on the extent to which the combination of measurements, local time evolution, and instantaneous classical communication can enhance useful quantum tasks. In particular, our bound limits (\emph{i}) the speed with which quantum information can be transferred and manipulated and (\emph{ii}) the preparation of resource states with long-range entanglement and/or correlations using  measurements. While our bound extends to weak measurements and generic quantum channels (beyond unitary time evolution), we find that only projective measurements provide optimal enhancements. In several cases, our bound \eqref{eq:main} is saturated by existing measurement-based protocols \cite{PhysRevLett.81.5932,1805.12570, 2204.04185,PhysRevLett.70.1895,Chung_2009}.

Importantly, we find \emph{at most} an $\Order{\Nmeas}$ enhancement to the speed $\LRvel$ of quantum information, provided that the outcomes of measurements in $\Nmeas$ local  regions are known \emph{and} utilized. Crucially, a single measurement does not destroy locality, nor can it 
teleport information over arbitrary distances, even in the limit of instantaneous classical communication. Our results elucidate the local nature of measurements and bound the most useful quantum tasks, which involve measurements. Moreover, our bound extends to other local quantum channels, thereby extending the Lieb-Robinson Theorem \cite{Lieb1972} to useful tasks implemented using arbitrary, local quantum channels.

Consider a short-range-entangled many-body state $\ket{\Psi}$ of qubits on a $\SpaceDim$-dimensional lattice that contains a localized logical qubit. We evolve $\ket{\Psi}$ under some spatially local, time-dependent Hamiltonian $\Ham(t)$ for total time $T$, during which we also perform measurements in $\Nmeas$ local regions, where both $\Ham(t)$ and the measurement protocol may be conditioned on the outcomes of prior measurements. Then, the maximal distance $\Dist$ that this protocol can teleport the logical qubit's state obeys the bound
\begin{equation}
    \label{eq:main}
    \Dist \, \leq \, (2 \, \Nmeas+\Nmeas^{\,}_0 ) \, \LRvel \, (T+T^{\,}_0) \, ,~~ 
\end{equation}
where $\Nmeas^{\,}_0$, $T^{\,}_0$, and the Lieb-Robinson velocity $\LRvel$ are all $\Order{1}$---i.e., independent of $\Dist$, $\Nmeas$, and $T$; for quantum circuits, $T$ is the circuit depth. Essentially, \eqref{eq:main} establishes that the extension of circuit depth (i.e., $\LRvel T$) \emph{is} $(2\Nmeas + \Nmeas^{\,}_0) \LRvel T$ in the presence of measurements and feedback. We also note that \eqref{eq:main} holds in the absence of measurements (where $\Nmeas^{\,}_0 =1$ or $2$), and is \emph{optimal} in numerous contexts.

The bound \eqref{eq:main} also constrains the preparation of many-body resource states with long-range entanglement and/or correlations, and also holds if the projective measurements are replaced by other nonunitary channels applied to $\Nmeas$ local regions. We also note that, in the most general case, there is an additive correction of $(\SpaceDim-1)\log_2 \Dist$ to $\LRvel (T+T^{\,}_0)$ in \eqref{eq:main}. However, as we discuss in Sec.~\ref{sec:Bounds}, we believe this term is not physical but simply an artifact of the proof strategy. That term does not appear in many examples of interest, including protocols with prefixed measurement locations, in $\SpaceDim=1$, and for discrete time evolution (generated by a quantum circuit). Proving that the term is absent in more general cases requires an alternate strategy, and is beyond the scope of this work.

The derivation of the bound \eqref{eq:main} is 
outlined in Sec.~\ref{sec:Bounds}, and 
rigorously proven in the Supplementary Material (SM) \cite{supp}. In contrast to the standard derivation of Lieb-Robinson bounds under unitary time evolution alone \cite{Lieb1972, fossfeig, chen2019finite, Kuwahara:2019rlw, Tran:2020xpc, gross, schuch, kuwahara2021liebrobinson, yin21}, \eqref{eq:main} does not recover from considering when commutators of the form $\comm{A^{\,}_x}{B^{\,}_y (t)}$ become nonzero. For example, a protocol consisting of a single measurement on site $x$, immediately followed by an outcome-dependent unitary operation on site $y$ (via instantaneous classical communication) leads to $\comm{A^{\,}_x}{B^{\,}_y (t)} \neq 0$ for $t=\Order{1}$ and \emph{arbitrary} distances $\Dist = \abs{x-y}$. However, 
such a protocol cannot transfer quantum information, nor generate correlations or entanglement between qubits $x$ and $y$. 

To extract a \emph{useful} bound \eqref{eq:main} in the presence of measurements (or other nonunitary channels), we instead show that, for short times $t \lesssim \Dist / 2 \Nmeas \, \LRvel$, the density matrix generated by applying \emph{any} measurement-assisted protocol to a short-range-entangled initial state is arbitrarily close in trace distance to a density matrix that cannot contain entanglement or correlations between sites $x$ and $y$. This is accomplished using a ``reference'' protocol, which compared to the ``true'' protocol does not act across some cut $C$ of the system separating sites $x$ and $y$. Thus, the true protocol cannot teleport quantum information nor generate correlations and/or entanglement in time $t$.

Importantly, our bound is furnished by the Heisenberg-Stinespring formalism \cite{Stinespring,  AaronMIPT, AaronDiegoFuture}, which provides for the unitary evolution of operators in the presence of  measurements and arbitrary outcome-dependent operations (facilitated by instantaneous classical communication). We stress that such operations cannot be captured by a local Lindblad master equation, which leads to the standard Lieb-Robinson bound $\Dist \leq \LRvel T$ \cite{Poulin}. In this sense, the enhancement to \eqref{eq:main} due to measurements requires feedback. We also stress that our bound is not simply a Lieb-Robinson bound on the unitary dynamics of the enlarged (``dilated'') system---due to instantaneous communication of outcomes, no such bound exists! Instead, our bound treats the effects of measurement and feedback separately to recover \eqref{eq:main}, as elucidated in the SM \cite{supp}.

We also prove bounds on the teleportation of \emph{multiple} qubits. Na\"ively, one might think that teleporting a single qubit requires a certain amount of entanglement in a resource state, and that the same entanglement can be used to send $Q$ qubits in succession, leading to a bound of the form $\Dist \lesssim 2 \, \Nmeas \, \LRvel \, (T-Q)$. However, as we establish in Sec.~\ref{subsec:multi qubit bound}, this is \emph{not} the case.  In addition to generating useful entanglement, error correction  via outcome-dependent operations is essential to successful teleportation. Importantly, one must correct for the errors accrued by each logical qubit \emph{individually} and in each repeating region \cite{AaronYifanFuture}.  The resulting bound for teleporting $Q$ qubits is instead $\Dist \lesssim \left( 1+ \Nobs / Q \right) \, \LRvel T$, where $\Nobs$ is the number of measurement \emph{outcomes} utilized for error correction. Note that $\Nobs \geq 2 \Nmeas$ (with $\Nobs = 2\Nmeas$ in familiar examples) \cite{AaronYifanFuture}.

Our main bound \eqref{eq:main} constrains \emph{arbitrary} local quantum dynamics in the presence of measurements and instantaneous classical communications, and applies to \emph{generic} useful quantum tasks. The bound \eqref{eq:main} also constrains protocols involving generic  nonunitary operations captured by local quantum channels (including, e.g., weak measurements). In addition to quantum teleportation (e.g., the optimal protocol presented in Sec.~\ref{sec:teleport}), the bound \eqref{eq:main} has deep  connections (and useful applications) to quantum error correction (QEC) and measurement-based quantum computation (MBQC) \cite{DominicMBQC_SPT, 0508124, AaronYifanFuture}, and also constrains the preparation of generic long-range entangled states \cite{Verresen:2021wdv, Tantivasadakarn:2021vel}, including the Bell state \cite{Bell},
\begin{equation}
\label{eq:Bell state}
    \ket{\mathrm{Bell}} \, = \, \frac{1}{\sqrt{2}} \left( \ket{00} + \ket{11} \right) \, , ~~
\end{equation}
the Greenberger–Horne–Zeilinger (GHZ) state \cite{GHZ89,GHZ07,Bell_GHZ},
\begin{equation}
    \label{eq:GHZ state}
    \ket{\text{GHZ}} \, = \, \frac{1}{\sqrt{2}} \left( \ket{00\dots 00} + \ket{11 \dots 11} \right) \, , ~~
\end{equation}
states $\ket{\Psi}$ corresponding to quantum critical points \cite{SubirQCP} and/or conformal field theories (CFTs) \cite{CFTbook} with algebraic (i.e., power-law) correlations \cite{TimShortcut}, for which
\begin{equation}
    \label{eq:power law corr}
    \matel{\Psi}{A^{\vpp}_x(t) B^{\vpp}_y (0)}{\Psi} \sim \abs{x-y}^{-\alpha} \, ,~~
\end{equation}
and the W state \cite{Dicke,W_state} of $\Nspins$ qubits,
\begin{equation}
    \label{eq:W state}
    \ket{W} \, = \, \frac{1}{\sqrt{\Nspins}} \sum\limits_{j=1}^{\Nspins} \, \ket{1}^{\,}_{j} \otimes \ket{\mathbf{0}}^{\,}_{j^c} \, , ~~
\end{equation}
where $\ket{\mathbf{0}}^{\,}_{j^c}$ denotes the $\ket{0}$ state of all qubits except $j$. The bound also applies to Dicke states \cite{Dicke} more generally, as well as  spin-squeezed states \cite{squeeze_rev,PhysRevA.47.5138, PhysRevA.46.R6797}
. Each of these states has various applications in quantum technologies \cite{Peres_1985, Shor_1995, Steane_1996_1, Steane_1996_2, PhysRevLett.77.3260, Gottesman_1998}.

In  Sec.~\ref{sec:outlook} we also discuss asymptotically optimal Clifford protocols for preparing the Bell \eqref{eq:Bell state} and GHZ \eqref{eq:GHZ state} states, whose existence establishes that our bound \eqref{eq:main} is optimal for these tasks.  We also present a protocol that prepares the W state \eqref{eq:W state} using depth $T \sim \log_2 \Nspins$ and $\Nmeas \sim \Nspins$ measurement regions, compared to depth $T \sim \Nspins$ in all known protocols \cite{cirac}. However, this protocol is not optimal with respect to \eqref{eq:main} for $\Nmeas>0$  \cite{ChaoWFuture}.

Several of the protocols we discuss are already known to the literature---at least in the $\Nmeas=0$ and $\Nmeas \sim \Dist$ limits. We reformulate these protocols to allow for straightforward interpolation between these two limits, revealing important resource tradeoffs (between $\Nmeas$ and $T$) and providing for optimization based on the details of particular quantum hardware, which we anticipate will be of considerable interest to the development of experimental protocols for quantum information processing. More importantly, we rigorously establish in these particular cases that better protocols simply do not exist.

The bound \eqref{eq:main} also provides 
insight into the EPR paradox \cite{epr}. First, the state-preparation process is crucial to understanding locality, as creating a well-separated Bell pair is itself a useful quantum task, which must obey \eqref{eq:main}. Even with instantaneous classical communication, locality ensures that unitarily separating two qubits by distance $\Dist$ takes time $T \sim \Dist/2 \, \LRvel \, \Nmeas$ using measurements in $\Nmeas$ regions. Second, the correct use of measurement outcomes is crucial, i.e., to determine which of the Bell states $\ket{00} \pm \ket{11}$ \cite{Bell,QC_book} has realized---otherwise, the resulting state is no better than a random \emph{classical} bit. Indeed, useful quantum tasks (such as QEC and MBQC) can \emph{only} be performed over distance $\Dist$ in $\Order{1}$ time if the outcomes of $\Order{\Dist}$ measurements are known \emph{and} utilized, regardless of how cleverly the task is performed. Locality then constrains the time needed both to generate entangled resource states and to perform useful quantum tasks, as captured by our main bound \eqref{eq:main}.

The rest of this paper is organized as follows. In Sec.~\ref{sec:teleport}, we introduce a teleportation protocol that is optimal with respect to \eqref{eq:main} and admits tradeoffs between circuit depth $T$ and the number of measurement regions $\Nmeas$. We demonstrate that quantum information is teleported only by using classical communication of the measurement outcome to determine an error-correction channel $\QECChannel$. 

In Sec.~\ref{sec:Stinespring} we review the Stinespring representation of generic quantum channels, focusing on projective measurements \cite{Stinespring, AaronDiegoFuture, AaronMIPT}. This formalism is crucial to the derivation of \eqref{eq:main}, as it implies a Heisenberg picture for operator dynamics in the presence of nonunitary quantum channels (e.g., measurements), which may be of broader use in the rapidly developing field of quantum dynamics in systems with measurement and feedback \cite{AaronMIPT}. 

In Sec.~\ref{sec:Bounds} we sketch the strategy for deriving \eqref{eq:main}. We first consider Clifford protocols such as the teleportation protocol of Sec.~\ref{sec:teleport}, showing precisely how protocols that violate \eqref{eq:main} fail to teleport quantum information. We then explain how the bound extends to the generation of entanglement and/or correlations, starting from a product state. Next, we present the general bound for continuous time dynamics, whose formal derivation is technical and relegated to the SM \cite{supp}. We then provide explicit bounds for the generation of Bell states \eqref{eq:Bell state}, the GHZ state \eqref{eq:GHZ state}, states with algebraic correlations \eqref{eq:power law corr}, the W state \eqref{eq:W state}, as well as Dicke and spin-squeezed states in Sec.~\ref{subsec:Corr bounds}. We further establish that the bound \eqref{eq:main} cannot be circumvented by teleporting multiple qubits (e.g., using only $\Nmeas$ measurement outcomes for all qubits together) in Sec.~\ref{subsec:multi qubit bound}. In Sec.~\ref{subsec:initial state}, we extend our bounds to protocols applied to two classes of initial states with short-range entanglement. In Sec.~\ref{subsec:other sys}, we explain how our bound applies to arbitrary local dynamics on systems of qubits, $d$-state qudits, fermions, and Majorana modes. Lastly, in Sec.~\ref{subsec:other channels}, we explain how \eqref{eq:main} extends to generic local nonunitary channels, including weak measurements.

In Sec.~\ref{sec:outlook}, we conclude with an outlook on the use of our formalism and bound \eqref{eq:main}. We revisit the EPR paradox \cite{epr} and review a number of practically relevant protocols (or codes) and quantum tasks, including Calderbank-Shor-Steane (CSS) codes, quantum routing in qubit arrays, and the preparation of various entangled (and/or correlated) resource states. A summary of results appears in Sec.~\ref{subsec:Summary}.

\section{Optimally fast teleportation}
\label{sec:teleport}
Before discussing the details our bounds, we begin by highlighting our bound's intriguing practical implication: A \emph{finite} number of local measurements can reduce the time $T$ needed to perform useful tasks (such as quantum teleportation or state preparation) by any constant factor! We showcase this established idea using a simple entanglement-swapping teleportation protocol (ESTP) \cite{PhysRevLett.81.5932,1805.12570,2204.04185,PhysRevLett.70.1895,Chung_2009}---or ``quantum repeater'' \cite{PhysRevLett.81.5932}---that teleports a state between two unentangled qubits separated by distance $\Dist$ in a time $T\sim \Dist/(2\Nmeas+1)$. What our work and the bound \eqref{eq:main} highlight (and which was not previously known) is that such protocols are provably optimal in that they saturate \eqref{eq:main} and use the fewest resources ($\Nmeas$ and $T$) possible to achieve successful teleportation. This remains true with arbitrarily complex adaptive protocols with continuous-time (but spatially local) dynamics. The ESTP circuit is depicted in Fig.~\ref{fig:SWAP circuit} for $\Dist = 15$, $T=4$.

Throughout, we use the convention $\PZ{} \ket{0} = \ket{0}$ and $\PZ{} \ket{1} = -\ket{1}$. The protocols of interest involve three distinct Clifford gates \cite{QC_book}: The single-qubit Hadamard gate $\Hadamard = \left( X + Z \right)/\sqrt{2}$ rotates between the $X$ and $Z$ eigenbases; the two-qubit controlled NOT (CNOT$(i\rightarrow j)$) gate applies $\PX{j}$ to the target qubit $j$ if the control qubit $i$ is in the state $\ket{1}$; and the SWAP gate acts on two-qubit states as $\ket{ab} \to \ket{ba}$ (see the SM \cite{supp} for further details).

\subsection{Standard teleportation protocol}
\label{subsec:STP}
As a warm up to the ESTP, we review the standard teleportation protocol (STP) on three qubits \cite{PhysRevLett.70.1895,QC_book}. The STP uses local operations and classical communication (LOCC) \cite{PhysRevLett.81.5932,cirac} to teleport an arbitrary logical state $\ket{\psi}$ of qubit $i$ to the target qubit $f$, where  $\ket{\psi}$ is given by
\begin{equation}
    \ket{\psi} = \alpha\ket{0}+\beta\ket{1} \, , ~~
    \label{eq:psi def}
\end{equation}
with $\abs{\alpha}^2+\abs{\beta}^2=1$ the only constraint. 

The three qubits are initialized in the state
\begin{align}
    \ket{\Psi}\, &= \, \ket{\psi} \otimes \ket{\mathrm{Bell}} 
    \notag \\ 
    &= \, \frac{\alpha}{\sqrt{2}} \left( \ket{000}+\ket{011} \right) + \frac{\beta}{\sqrt{2}} \left( \ket{100}+\ket{111} \right) \, ,~~
    \label{eq:STP initial}
\end{align}
where the Bell state $ \ket{\mathrm{Bell}} $ is defined in \eqref{eq:Bell state}. 

We next apply the Bell decoding channel $\BellChannel^{\dagger}_{i,a} =\Hadamard^{\,}_{i}\operatorname{CNOT}(i\rightarrow a) $ to the first two qubits. This channel is depicted in circuit form in the orange-shaded regions of Fig.~\ref{fig:SWAP circuit}. 
Applying $\BellChannel^{\dagger}_{i,a}$ to $\ket{\Psi}$ \eqref{eq:STP initial} gives
\begin{align}
    \ket{\Psi'} \, &= \, \frac{\alpha}{2} \left( \ket{000} + \ket{100} + \ket{011} + \ket{111} \right) \notag \\
    &+ \frac{\beta}{2} \left( \ket{010} - \ket{110} + \ket{001} - \ket{101} \right) \,,~~
\end{align}
and now, measuring $\PZ{1}$ and $\PZ{2}$ leads to four possible final states, distinguished by the outcomes $\bvec{m} = \left( m^{\,}_1 , m^{\,}_2 \right)$ of these two measurements (with $m^{\,}_{1,2} = \pm 1$), given by
\begin{align}\label{eq:4outcomes}
    \ket{\Psi^{\,}_t} \, = \, 
    \begin{cases}
        \, \ket{00} \otimes \ket{\psi} ~& \text{if } \bvec{m} = (1,1) \\
        \, \ket{01} \otimes X\ket{\psi} ~& \text{if } \bvec{m} = (1,-1) \\
        \, \ket{10} \otimes Z\ket{\psi} ~& \text{if } \bvec{m} = (-1,1) \\
        \, \ket{11} \otimes ZX\ket{\psi} ~& \text{if } \bvec{m} = (-1,-1) \, 
    \end{cases} \, , ~
\end{align}
meaning that $\ket{\psi}$ has been teleported to the third qubit up to a local rotation error $\QECChannel$ determined by the measurement outcomes. If $m^{\,}_1=-1$, then a $\PZ{}$ error occurred, so we apply $\PZ{}$ to $\ket{\Psi^{\,}_t}$; if $m^{\,}_2=-1$, then an $\PX{}$ error occurred, and we apply $\PX{}$ to $\ket{\Psi^{\,}_t}$. For all four outcomes, this error-correction step produces the desired state $\ket{\psi}$ \eqref{eq:psi def} on the final qubit (possibly up to a meaningless phase).

Importantly, if the  measurement outcomes are not used to perform error correction, then the outcome-averaged density matrix $\DensMat^{\,}_f$ for the target site corresponds to averaging over the four distinct final states in \eqref{eq:4outcomes}. Since each outcome is equally likely, the result is 
\begin{equation}
\label{eq:twirled density matrix}
    \overline{\DensMat}^{\vpd}_f \, = \, \frac{1}{4} \, \sum\limits_{n=0}^{3} \, \Pauli{n}{f}\,  \BKop{\psi}{\psi} \,\Pauli{n}{f} \, = \, \frac{1}{2} \, \ident \, ,~~
\end{equation}
which is also known as the ``twirl'' of $\BKop{\psi}{\psi}$ over the Pauli group. In other words, $\DensMat^{\,}_f = \ident/2$ is simply the maximally mixed state, which is equivalent to a random classical bit, and contains no quantum information! Moreover, the \emph{same} reduced density matrix \eqref{eq:twirled density matrix} recovers from \eqref{eq:STP initial}.

The only practical utility of teleporting the logical state $\ket{\psi}$ \eqref{eq:psi def} from the initial site $i$ to the final site $f$ is to reproduce the expectation values and statistics of $\ket{\psi}$ using operations on site $f$. This always requires repeating the experiment multiple times to extract statistics. However, if the measurement outcomes are not correctly used to determine the channel $\QECChannel$, then operations on the final qubit instead reproduce the statistics of the maximally mixed state \eqref{eq:twirled density matrix}. Thus, there is no practical sense in which a measurement-assisted protocol teleports the state $\ket{\psi}$ (or more generally, achieves a useful quantum task) without outcome-dependent error correction.

The STP also establishes that the tasks of separating a Bell pair and teleporting a logical state $\ket{\psi}$ over some distance $\Dist$ are equivalent up to $\Order{1}$ corrections to $\Nmeas$ and $T$. After all, the STP's initial state \eqref{eq:STP initial} presumes a Bell pair  shared by the ancilla qubit $a$ and final qubit $f$. For the bound \eqref{eq:main} to be meaningful, we must exclude initial states with long-range entanglement, the preparation of which is a useful task in and of itself. Supposing that the Bell sites $a$ and $f$ are separated by a distance $\Dist = d(a,f) \gg 1$, this leads to two important observations.

The first is that the full STP---including the separation of the Bell pair in \eqref{eq:STP initial}---obeys \eqref{eq:main}. More importantly, if the bound is violated (so that the state of qubits $a$ and $f$ is essentially unentangled), then teleportation fails. In this case, the combination of the measuring $\PZ{}$ on the $i,a$ qubits and applying the rotation channel $\QECChannel$ to qubit $f$ based on the measurement outcomes results in some unentangled state $\ket{\phi}$ of qubit $f$. Averaging over outcomes, the reduced density matrix of qubit $f$ is given by \eqref{eq:twirled density matrix} with $\ket{\psi} \to \ket{\phi}$, and thus no quantum information is transferred. The bound \eqref{eq:main} only constrains useful quantum tasks, which are equivalent to those that send quantum information. 

The second is that any \emph{unitary}  protocol that prepares the Bell state on $a,f$ in \eqref{eq:STP initial} obeys the bound \eqref{eq:main}, but does not \emph{saturate} it. In Sec.~\ref{subsec:ESTP}, we present a protocol that uses $2\Nmeas$ measurements to saturate \eqref{eq:main} starting from a product state by linking up $\Nmeas$ copies of the STP.

\subsection{Entanglement-swapping teleportation protocol}
\label{subsec:ESTP}
Consider a $1 \SpaceDim$ chain of $\Nspins$ qubits initialized in the  state
\begin{align}
\label{eq:Psi0}
    \ket{\Psi^{\,}_0} \, = \, \ket{\psi}^{\,}_1 \otimes \ket{\bvec{0}}_{2\cdots \Nspins} \, , ~~
\end{align}
in the computational ($Z$) basis, where $\ket{\psi}$ is the state of the initial logical qubit \eqref{eq:psi def}, and $\ket{\bvec{0}}$ denotes the $\ket{0}$ state on all other sites (the conventional initial state).

\begin{figure}[t]
\centering
\includegraphics[width=.48\textwidth]{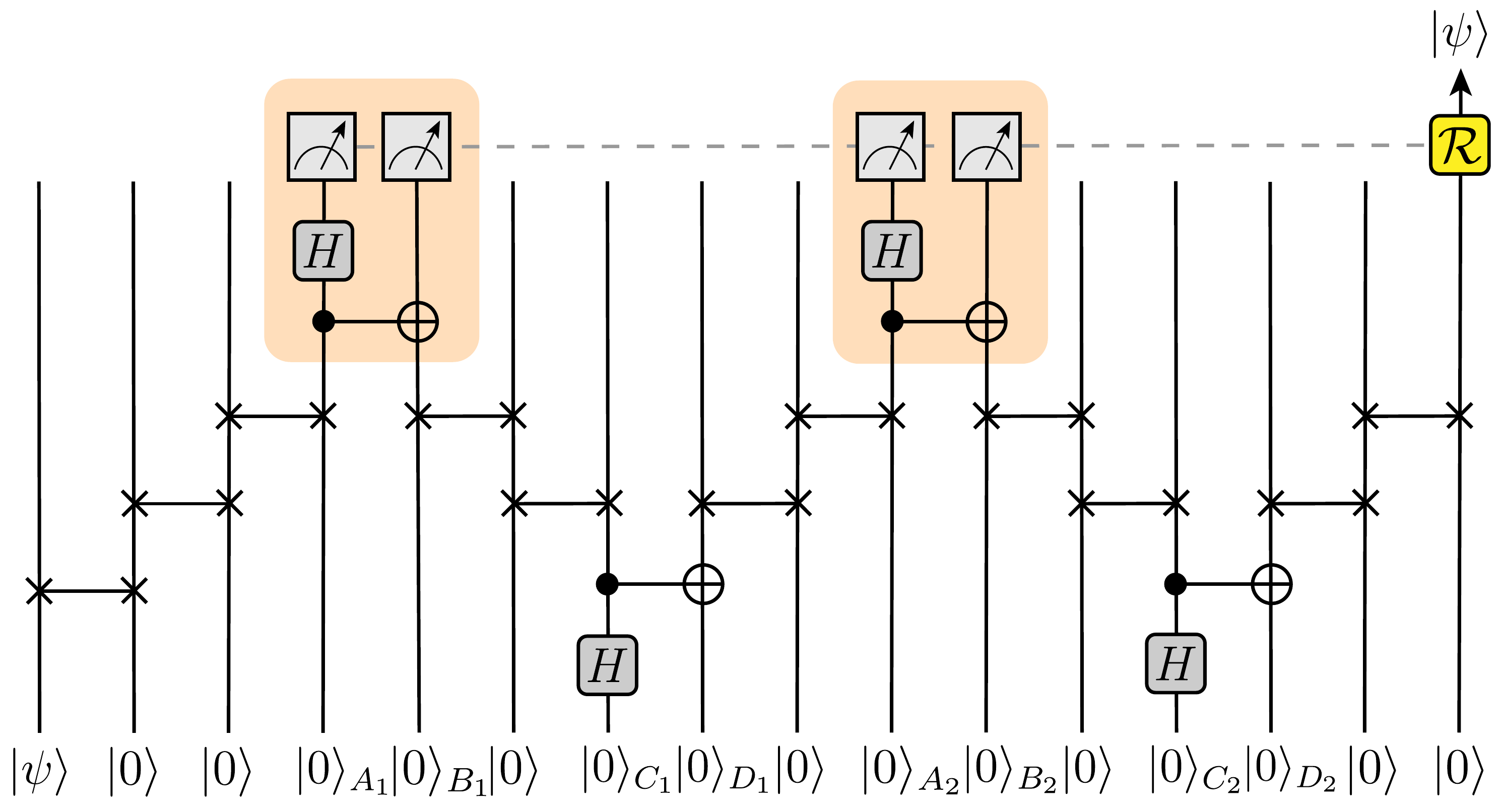}
\caption{The entanglement-swapping teleportation protocol (ESTP) is illustrated for $\Dist=15$, $T=4$, and $\Nmeas=2$ (gate notation \cite{GateNotation} is reviewed in the SM \cite{supp}). Bell pairs are generated on neighboring $C$ and $D$ qubits via a Hadamard--CNOT sequence ($\BellChannel$) and transported to $A$ and $B$ qubits via SWAP gates.  The shaded areas indicate the standard teleportation protocol (STP) and include two $Z$ measurements each; the dashed line denotes classical communication. The logical qubit $\ket{\psi}$ starts at $j=1$ and teleports to $j=16$ after applying the error-correction gate $\QECChannel$, which is determined by the measurement outcomes according to Tab.~\ref{tab:lookup_main}.
\label{fig:SWAP circuit}}
\end{figure}

The ESTP is realized by the computational-basis quantum circuit depicted in Fig.~\ref{fig:SWAP circuit} (for $\Dist=15$, $\Nmeas=2$, and $T=4$). Intuitively, the ESTP involves $\Nmeas$ copies of the STP, which are daisy chained together using SWAP gates to teleport the logical state $\ket{\psi}$ over greater distance. As an aside, we note that the measurements can instead be used to send more qubits over the same distance \cite{AaronYifanFuture}.

Heuristically, the ESTP begins with the entangling Clifford circuit $\mathcal{W}$, which encodes Bell pairs in $\Nmeas$ regions and separates them with Lieb-Robinson velocity $\LRvel=1$ in \eqref{eq:main}, realizing $\ket{\Psi^{\,}_t} = \mathcal{W} \ket{\Psi^{\,}_0}$  prior to the orange shaded regions in Fig.~\ref{fig:SWAP circuit}. Next, the STP \cite{PhysRevLett.70.1895} is applied in each of the orange boxes in Fig.~\ref{fig:SWAP circuit}, with $\PZ{}$ measurements indicated by pointer dials. The measurement outcomes are communicated classically (indicated by dashed lines) to determine the error-correction unitary $\QECChannel$, whose application to the final site completes the transfer of the state $\ket{\psi}$ \eqref{eq:psi def}.

We now consider the ESTP in detail. The chain is divvied into $\Nmeas$ repeating regions of size $\ell=2 (T-1)$, along with $T-1$ qubits  (including the initial logical qubit) to the left of the first region, and one or two qubits (including the target site) to the right of the last region. The first layers of $\mathcal{W}$ create Bell pairs on neighboring $C^{\,}_s$ and $D^{\,}_s$ qubits from the computational-basis state \eqref{eq:Psi0} via the ``Bell encoding'' channel  $\BellChannel^{\,}_s=\operatorname{CNOT}(C^{\,}_s\rightarrow D^{\,}_s) \Hadamard^{\,}_{\small C^{\,}_s} $ \cite{QC_book} on the neighboring $C$ and $D$ qubits in each region $s$. The SWAP gates then send $\ket{\psi}$ to $A^{\,}_1$, $C^{\,}_s$  to $B^{\,}_s$, and $D^{\,}_s$ to $A^{\,}_{s+1}$, with $A^{\,}_{\Nmeas+1} \equiv \Nspins$ the final site. The state is then
\begin{align}
\label{eq:psi bell pairs}
    \ket{\Psi^{\,}_t} \, = \, \ket{\psi}^{\,}_{A^{\,}_1} \otimes \left[ \bigotimes\limits_{s=1}^{\Nmeas}\, \ket{\mathrm{Bell}}^{\vpp}_{B^{\,}_s A^{\,}_{s+1}}\right] \otimes \ket{\Phi
}^{\,}_{\mathrm{rest}} ,
\end{align}
immediately before the orange regions in Fig.~\ref{fig:SWAP circuit}, where $\ket{\Phi}^{\,}_{\mathrm{rest}}$ is a decoupled state of all other qubits. 

The next step applies the STP to neighboring $A$ and $B$ qubits, indicated by the orange-shaded boxes of Fig.~\ref{fig:SWAP circuit}. First, the ``Bell decoding'' channel $\BellChannel^{\dagger}_s = \Hadamard^{\,}_{\small A^{\,}_s} \operatorname{CNOT}(A^{\,}_s\rightarrow B^{\,}_s)$ \cite{QC_book} acts on neighboring $A$ and $B$ qubits in each of the $\Nmeas \hspace{-1mm} \sim \hspace{-1mm} \Nspins/T$ regions $s$. Next, the outcomes of measuring $\PZ{A^{\,}_s}$ and $\PZ{B^{\,}_s}$ are recorded, which will determine the error-correcting unitary $\QECChannel$ to apply to the final site. 
Bell decoding ensures that the measurements in the nonunitary channel $\MeasChannel$ (indicated by pointer dials in Fig.~\ref{fig:SWAP circuit}) are in the computational ($\PZ{}$) basis, but can be omitted if one can measure in the Bell basis instead.

To be precise, after applying $\MeasChannel^{\,}_1$ to $\left( A^{\,}_1,B^{\,}_1 \right)$ we find
\begin{align}\label{eq:M1 psi}
    \MeasChannel^{\vpd}_{1} \ket{\Psi^{\,}_t}  =  \left( \mathcal{R}^{\,}_1 \ket{\psi} \right)^{\,}_{A^{\,}_2} \hspace{-0.5mm} \otimes \hspace{-0.5mm} \left[ \bigotimes\limits_{s=2}^{\Nmeas}\, \ket{\mathrm{Bell}}^{\vpp}_{B^{\,}_s A^{\,}_{s+1}}\right] \hspace{-0.5mm} \otimes \hspace{-0.5mm} \ket{\Phi}^{\,}_{\mathrm{rest}} \, , 
\end{align}
where the gate $\QECChannel^{\,}_1 \in \set{\ident, X, Y, Z}$ depends on the outcomes of $\MeasChannel^{\,}_1$. At this stage, the logical state $\ket{\psi}$ has been teleported a distance $2(T-1)$ from site $A^{\,}_1$ to site $A^{\,}_2$. 

This procedure is then repeated for the remaining regions $s = 2, \dots, \Nmeas$, collapsing the state  $\ket{\Psi^{\,}_t}$ onto one of the measurement outcomes,
\begin{align}
\label{eq:final state}
    \MeasChannel\BellChannel^{\dagger} \ket{\Psi^{\,}_t}  \, = \, \ket{\Phi}^{\,}_{1\cdots \Nspins-1} \otimes (\mathcal{R}\ket{\psi})^{\,}_{\Nspins} \, = \, \mathcal{R} \ket{\Psi^{\,}_f} \, ,~~
\end{align}
where $\ket{\Psi^{\,}_f}$ is the desired final state, $\ket{\Phi}$ is an arbitrary many-body state on sites $j \neq \Nspins$, and $\mathcal{R}$ is a single-site ``error-correction'' unitary, determined by the measurement outcomes (communicated instantaneously via the dashed lines in Fig.~\ref{fig:SWAP circuit}) according to Tab.~\ref{tab:lookup_main}.

\begin{table}[t]
\centering
\begin{tabular}{|c|c|c|}
\hline
$\;\;\QECChannel\;\;$ & \multicolumn{2}{c|}{\textbf{Measurement Outcomes}} \\ \hline\hline
& $\quad\quad m^{\vpp}_A \quad\quad$ & $ m^{\vpp}_B$ \\ \hline
$\ident$ & 1 & 1 \\ \hline
$X$ & 1 & -1 \\ \hline
$Y$ & -1 & -1 \\ \hline
$Z$ & -1 & 1 \\ \hline
\end{tabular}
\caption{A ``lookup'' table for the local rotation $\QECChannel$ is given for the various measurement outcomes. Here $m^{\,}_A,m^{\,}_B$ denote the product of $Z$-measurement outcomes on the qubits in sets $A$ and $B$, respectively.}
\label{tab:lookup_main}
\end{table}

Specifically, $\QECChannel = \prod_{s=1}^{\Nmeas} \QECChannel^{\,}_s$ is determined by the product of measurement outcomes for the $A$ and $B$ sites (see Tab.~\ref{tab:lookup_main}). Applying $\QECChannel$ to \eqref{eq:final state} on the final site ``undoes'' the error, giving $\ket{\Psi^{\,}_f}$ with the logical state $\ket{\psi}$ on the final site $j=\Nspins$. In this way, the ESTP enhances the standard teleportation protocol with SWAP gates and daisy-chained Bell pairs to cover more distance. 
The distance $\Dist$ over which the ESTP teleports the logical state $\ket{\psi}$ obeys \eqref{eq:main}, and more specifically, \eqref{eq:protocol dist}.

\subsection{Logical operator dynamics}
\label{subsec:ESTP Heisenberg}
\begin{figure}[t]
\centering
\includegraphics[width=.44\textwidth]{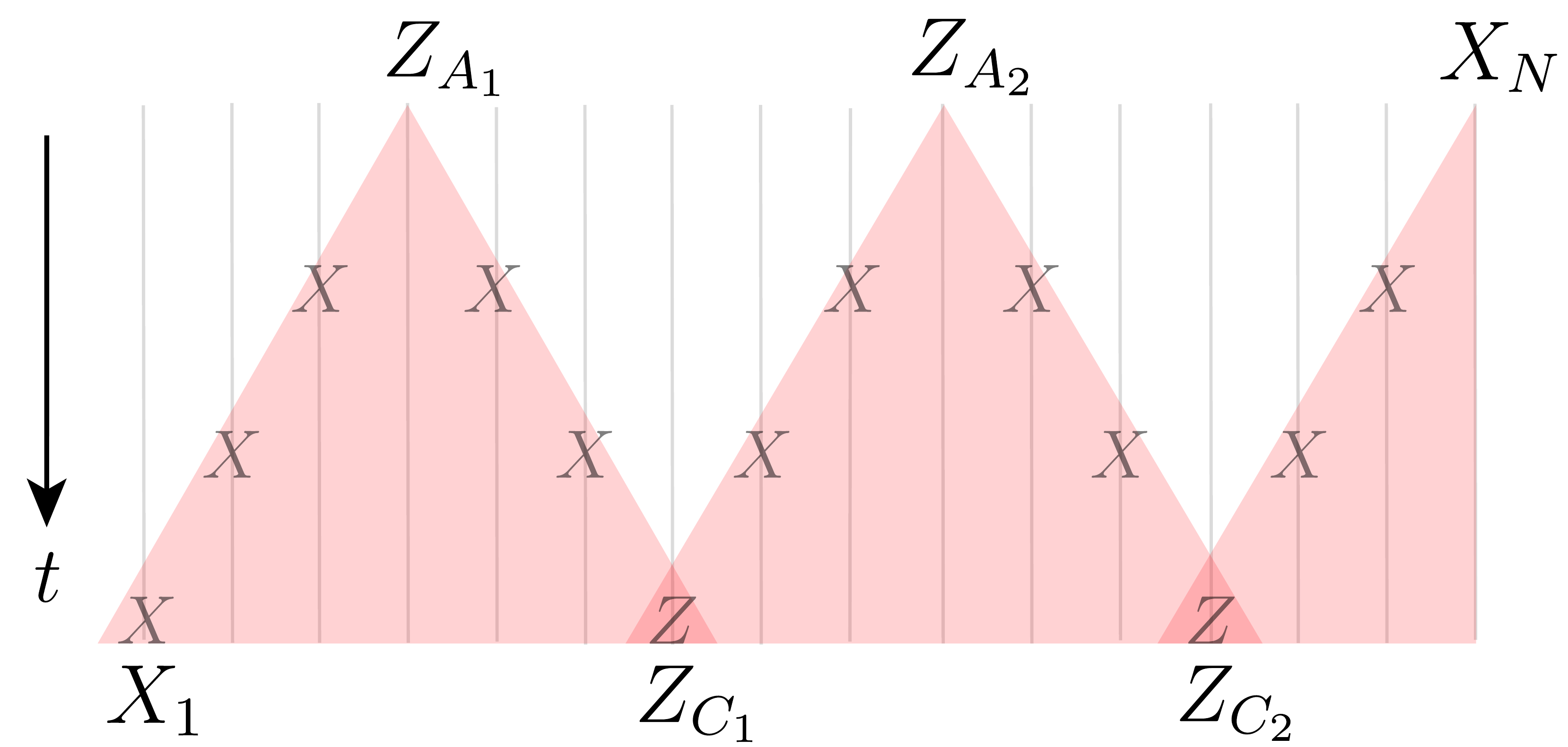}
\caption{Heisenberg picture of the 
logical operator $X^{\,}_{\rm L}$ for the ESTP depicted in Fig.~\ref{fig:SWAP circuit}. The local $Z$ operators grow with Lieb-Robinson velocity $v=1$, as depicted by the edges of the shaded cones. The check operators during the intermediate steps are shown inside their respective ``light cones.''  
\label{fig:logical growth}}
\end{figure}

An alternative perspective to the Schr\"odinger dynamics of states is afforded by the Heisenberg dynamics of the \emph{logical operators}---unitary operators that  reproduce the Pauli algebra acting on $\ket{\psi}$. One choice of logical operators for the final state $\ket{\Psi^{\,}_f}$ \eqref{eq:final state} corresponds to 
\begin{equation}
    \label{eq:nominal logicals}
    \LX \, = \, \PX{\Nspins} \, , ~~\LZ\,=\, \PZ{\Nspins} \, , ~ \, \text{with}\,~\LY \, = \, \ii \LX \LZ \, , ~~
\end{equation}
and we now evolve these logical operators forward in the Heisenberg picture---corresponding to \emph{backward} time evolution in the Schr\"odinger picture---to recover their action on the initial state \eqref{eq:Psi0},
\begin{align}
    \label{eq:Heisenberg}
    O (T) \, &= \, \mathcal{W}^{\dagger} \BellChannel \MeasChannel^{\dagger}  \QECChannel^{\dagger} \, O (0)\, \QECChannel \MeasChannel \BellChannel ^{\dagger} \mathcal{W} \, ,~~
\end{align}
where the channel $\MeasChannel = \MeasChannel^{\,}_A \MeasChannel^{\,}_B$ encodes the single-site measurements of $\PZ{}$ on the $A$ and $B$ qubits. The Heisenberg evolution of the logical operator $\LX = \PX{\Nspins}$ for the ESTP depicted in Fig.~\ref{fig:SWAP circuit} is illustrated in Fig.~\ref{fig:logical growth}.

Acting on states, $\MeasChannel^{\,}_A$ determines $m^{\,}_{A} = (-1)^{N^{\,}_A}$ with $N^{\,}_A$ the number of $A$ sites in the state $\ket{1}$, and likewise for $m^{\,}_B$; the error-correction channel $\QECChannel$ is determined by $m^{\,}_{A,B}$ according to Tab.~\ref{tab:lookup_main}. As a channel on the 
physical Hilbert space, $\MeasChannel$ is nonunitary, so some care is required to incorporate the effect of measurements.

Crucially, in the Heisenberg picture, the first channel applied to $O(0)$ in \eqref{eq:Heisenberg} corresponds to the error-correction channel $\QECChannel$ (i.e., conjugation of $O(0)$ by $\QECChannel=\QECChannel^{\,}_A \QECChannel^{\,}_B$). The unitary  $\QECChannel^{\,}_A$ acts as $\PZ{\Nspins}$ if $m^{\,}_A=-1$, and $\QECChannel^{\,}_B$ acts as $\PX{\Nspins}$ if $m^{\,}_B=-1$ (and both $\QECChannel^{\,}_{A,B}$ act trivially otherwise). Since the overall sign of the wavefunction is not physically significant, the ordering of $\QECChannel^{\,}_A$ and $\QECChannel^{\,}_B$ in $\QECChannel$ is inconsequential. Conjugating the logical operators by $\QECChannel$ gives
\begin{equation}
    \label{eq:logical R update}
    \QECChannel^{\dagger}  \PX{\Nspins}  \QECChannel \, = \, m^{\,}_A \, \PX{\Nspins} \, , ~\quad~\QECChannel^{\dagger}  \PZ{\Nspins}  \QECChannel \, = \, m^{\,}_B \, \PZ{\Nspins} \, .~~
\end{equation}
Next, conjugation by the measurement channel $\MeasChannel^{\,}_{A/B}$ acts nontrivially on the outcome $m^{\,}_{A/B}$ according to
\begin{equation}
    m^{\,}_{A/B} \, \to \, \PZ{A/B} = \prod_{j \in A/B} \, \PZ{j}
    \label{eq:Meas channel non Stinespring} \, ,~~
\end{equation}
and does nothing to $m^{\,}_{B/A}$. We derive this update in Sec.~\ref{sec:Stinespring} using the Stinespring Dilation Theorem \cite{Stinespring,ChoisThm,AaronMIPT} to represent measurements \emph{unitarily} by appending qubits to record their outcomes (see also SM  \cite{supp}). A similar update rule was derived in \cite{Chung_2009};  our perspective generalizes to arbitrary non-Clifford dynamics. We  find
\begin{subequations}
\label{eq:logicals post M}
\begin{align}
    \MeasChannel^{\dagger} \QECChannel^{\dagger} \, \PX{\Nspins} \, \QECChannel \MeasChannel \, &=  \,  \PX{\Nspins} \PZ{A} \, \label{eq:logical X post M} \\
    \MeasChannel^{\dagger} \QECChannel^{\dagger} \, \PZ{\Nspins} \, \QECChannel \MeasChannel \, &= \,  \PZ{\Nspins} \PZ{B}  \label{eq:logical Z post M} \, ,~~
\end{align}
\end{subequations}
and note that the generation of $\PZ{A,B}$ in the step above is crucial, and only occurs if $\QECChannel$ depends on $m^{\,}_{A,B}$.

We next conjugate by $\BellChannel$, giving
\begin{equation}
    \label{eq:logicals post H CNOT}
    \LX  \to  \PX{A} \PX{B}  \PX{\Nspins} ~ ,~~~~\LZ  \to  \PZ{A} \PZ{B}  \PZ{\Nspins} \, , ~~
\end{equation}
and applying the SWAP gates in $\mathcal{W}$ in reverse order moves the states of sites $A^{\,}_1 \to 1$, $A^{\,}_s \to D^{\,}_{s-1}$ for $s>1$,  $\Nspins \to D^{\,}_{\Nmeas}$, and $B^{\,}_s \to C^{\,}_s$ for all $1 \leq s \leq \Nmeas$. The result is
\begin{equation}
    \label{eq:logicals post SWAP}
    \LX \to \PX{1} \PX{C} \PX{D}  ~,~~~~\LZ \to \PZ{1}\PZ{C}\PZ{D} \, , ~~
\end{equation}
and applying $\BellChannel^{\dagger}$ to the $C,D$ sites gives
\begin{align}
    \LX (T) \, = \, \PX{1} \, \PZ{C} \, , ~\quad~ \LZ (T) \, = \, \PZ{1} \,  \PZ{D} \label{eq:final logicals} \, ,~~
\end{align}
for the \emph{initial} logical operators, as shown in Fig.~\ref{fig:logical growth}.  Thus, 
\begin{subequations}
\label{eq:XLTrelation}
\begin{align}
    \LX(T) \, \ket{\Psi^{\,}_0} \,  &= \, (\PX{} \ket{\psi})^{\vpp}_1\otimes \ket{\mathbf{0}}^{\vpp}_{2\cdots N}, \\
    \LZ(T) \, \ket{\Psi^{\,}_0} \,  &= \, (\PZ{} \ket{\psi})^{\vpp}_1\otimes \ket{\mathbf{0}}^{\vpp}_{2\cdots N} \, , ~~
\end{align}
\end{subequations}
as required for logical operators of the initial state \eqref{eq:Psi0}.

Crucially, if $\QECChannel$ did not depend on the measurement outcomes, Heisenberg evolution would not successfully transfer the logical operator from site $j=\Nspins$ to site $j=1$. Equivalently, in the Schr{\"o}dinger picture, no useful quantum information is teleported to qubit $j=\Nspins$ \emph{until} the measurement outcomes are communicated and  errors corrected (via $\QECChannel$). As in the STP, prior to application of $\QECChannel$ to site $j=\Nspins$, the outcome-averaged state is the maximally mixed state $\DensMat^{\,}_f = \ident/2$ (a random classical bit).

\section{Measurement channels}
\label{sec:Stinespring}
The protocols we consider combine  unitary time evolution and projective measurements to achieve useful quantum tasks. Note that the bound \eqref{eq:main} extends to \emph{generic} local quantum channels, as we explain in Sec.~\ref{subsec:other channels}. All quantum channels can be described using three equivalent representations: completely positive trace-preserving (CPTP) maps, Kraus operators (which are equivalent to CPTP maps), and isometries \cite{Stinespring, ChoisThm, AaronMIPT, AaronDiegoFuture}. The latter results from the Stinespring Dilation Theorem \cite{Stinespring}, and its equivalence to the other two representations follows from Choi's Theorem \cite{ChoisThm}. As we explain in Sec.~\ref{subsec:stinespring use}, we exclusively consider the outcome-averaged density matrix, since individual trajectories have no bearing on any useful quantum task. In the Stinespring picture, this corresponds to tracing out the detector degrees of freedom.

\subsection{Dilation Theorem and isometric measurement}
\label{subsec:iso meas}
The Stinespring Dilation Theorem \cite{Stinespring}  states that quantum channels can be represented using isometries and partial traces. An isometry is a length-preserving map from some physical Hilbert space $\Hilbert^{\,}_{\rm ph}$ (with dimension $\HilDim^{\,}_{\rm ph}$) to some \emph{dilated} Hilbert space $\Hilbert^{\,}_{\rm dil}$ (with dimension $\HilDim^{\,}_{\rm dil} \geq \HilDim^{\,}_{\rm ph}$). In particular, the isometries representing projective measurements have a unique form \cite{AaronDiegoFuture}, in which the extra degrees of freedom in  $\Hilbert^{\,}_{\rm dil}$ encode the \emph{outcome} of the measurement. For the Pauli measurements of interest, the binary outcomes $n=0,1$ are stored in qubits. 

We stress that the Stinespring Theorem \cite{Stinespring} establishes that \emph{all} nonunitary quantum channels are captured by partial traces and/or isometries (which involve ancillary ``Stinespring'' degrees of freedom). Weak and generalized measurements---as well as projective measurements with more than two outcomes---are all captured by this formalism. Surprisingly, even the measurement of unbounded operators (e.g., homodyne measurements and photon counting) are also captured via isometries \cite{AaronDiegoFuture}. 

While all of these cases are associated with distinct isometric channels, the \emph{locality} properties of these channels are always the same. In particular, these isometries all couple a local region of qubits to ancillary Stinespring qubits, where the latter may be nonlocally accessed via instantaneous classical communication. Hence, the results we derive for projective measurements apply to generic nonunitary quantum channels (e.g., weak measurements).

For convenience of presentation, we consider the measurement of Pauli-string operators $\observ$, which act on every site as one of the Pauli matrices or the identity (e.g., $\PX{i}$, $\PZ{i}$, $\PZ{i}\PZ{j}$, $\PZ{j-1}\PX{j}\PZ{j+1}$, etc.). The Pauli strings $\observ$ satisfy $\observ^2=\ident$, and their eigenvalues are $\pm 1$. It is convenient to write the \emph{spectral decomposition} of $\observ$, 
\begin{align}
    \observ  \equiv  \sum\limits_{\pm}  \, \pm \proj{\,}{(\pm)}  \, ,~~  \proj{\,}{(\pm)}  \equiv \frac{1}{2} \left( \ident \pm \observ \right) \,,~~
    \label{eq:ObservSpectralDecomp}
\end{align}
where the $\pm$ eigenvalues of $\observ$ have degeneracy $2^{\range-1}$ if $\observ$ acts nontrivially on $\range$ sites. We identify the label 0 with the $+1$ eigenvalue, and the label 1 with the $-1$ eigenvalue.

The eigenprojectors $\proj{\,}{(n)}$ for $n \in \set{0,1}$ satisfy 
\begin{equation}
    \observ \, \proj{\,}{(n) } =  (-1)^n \,   \proj{\,}{(n)} \, , ~~
\end{equation}
and are orthonormal and idempotent, satisfying
\begin{equation}
    \label{eq:Proj ortho idem complete}
    \proj{\,}{(m)} \proj{\,}{(n)} = \kron{m,n} \proj{\,}{(m)} ~~~\text{and}~~~ \sum\limits_{n=0,1} \, \proj{\,}{(n)} = \ident\, , ~~
\end{equation}
where the labels $m,n \in \set{0,1}$.

If the observable $\observ$ \eqref{eq:ObservSpectralDecomp} is measured in the state $\ket{\psi} \in \Hilbert^{\,}_{\rm ph}$ of the physical system, the post-measurement state $\ket{\psi'}$ in the Stinespring picture is given by
\begin{equation}
    \label{eq:SS iso measure}
    \ket{\psi} \to \ket{\psi'}  =  \sum\limits_{n=0,1} \, \left( \proj{\,}{(n)} \ket{\psi} \right)^{\vpd}_{\rm ph} \otimes \ket{n}^{\vpd}_{\rm ss} \, , ~~
\end{equation}
where $\ket{\psi'} \in \Hilbert^{\,}_{\rm dil}$ lies in the \emph{dilated} Hilbert space,
\begin{equation}
    \label{eq:Dilated Hilbert}
    \Hilbert^{\vpp}_{\rm dil} = \Hilbert^{\vpp}_{\rm ph} \otimes \Hilbert^{\vpp}_{\rm ss} \, , ~~
\end{equation}
where in \eqref{eq:SS iso measure}, the qubit in the state $\ket{n} \in \Hilbert^{\,}_{\rm ss}$ records the observed outcome $n$, and $\proj{\,}{(n)} \ket{\psi} $ is the post-measurement state of the physical system. The Stinespring states $\{ \ket{0},\ket{1}\}$ form a complete, orthonormal qubit basis, with $\inprod{m}{n} = \kron{m,n}$ and $\SSid{} = \BKop{0}{0} + \BKop{1}{1}$, where the tilde denotes an operator acting on $\Hilbert^{\,}_{\rm ss} \subset \Hilbert^{\,}_{\rm dil}$ \eqref{eq:Dilated Hilbert}.

The isometric channel $\isometry$ that represents the measurement process in \eqref{eq:SS iso measure} extends (or ``dilates'') the physical Hilbert space $\Hilbert^{\,}_{\rm ph}$ to $\Hilbert^{\,}_{\rm dil}$ according to
\begin{equation}
    \label{eq:SS isometry}
    \isometry^{\vpd}_{[\observ]} = \sum\limits_{n=0,1} \, \proj{\rm ph}{(n)} \otimes \ket{n}^{\vpd}_{\rm ss} \, , ~~
\end{equation}
where the outcome qubit (or ``Stinespring register'') does not exist prior to application of the channel $\isometry$. Isometries are length preserving (i.e., $\isometry^{\dagger} \isometry = \ident$), so the state $\ket{\psi'}$ \eqref{eq:SS iso measure} remains normalized as written \eqref{eq:SS iso measure}.

The probability to observe outcome $n \in \{0,1\}$ is
\begin{align}
    \label{eq:SS iso prob n}
    p^{\vpp}_n  =  \matel{\psi'}{\, \ident^{\vpp}_{\rm ph} \otimes \BKop{n}{n}^{\vpp}_{\rm ss} \, }{\psi'} = \matel{\psi}{\proj{\rm ph}{(n)}}{\psi}\, ,~~
\end{align}
and the expectation value of $\observ$ \eqref{eq:ObservSpectralDecomp} is given by
\begin{align}
    \label{eq:SS iso expval}
    \expval{\observ}^{\vpp}_{\psi} &= 
    \matel*{\psi'}{\,\sum\limits_{n=0,1} \, (-1)^n \BKop{n}{n}^{\vpp}_{\rm ss} \, }{\psi'}  \, , ~~
\end{align}
and we see that both the probability for outcome $n$ \eqref{eq:SS iso prob n} and expectation value of $\observ$ \eqref{eq:SS iso expval} are extracted via operations on the Stinespring register.

\subsection{Unitary measurement}
\label{subsec:unitary meas}

However, the Stinespring Theorem \cite{Stinespring} and isometric representation of channels on their own are not sufficient for the derivation of \eqref{eq:main}. Importantly, in the case where $\HilDim^{\,}_A=\HilDim^{\,}_B$, the corresponding isometric channel is \emph{unitary}. The crucial insight here---and in the accompanying works \cite{AaronDiegoFuture, AaronMIPT}---is the recognition that the extra Stinespring degrees of freedom in $\Hilbert^{\,}_{\rm dil}$ (but \emph{not} in $\Hilbert^{\,}_{\rm ph}$) are \emph{physical}. In the case of measurements, they reflect the  state of the measurement apparatus; for other quantum channels, they represent environmental degrees of freedom. In this sense, the Stinespring qubits are physical (and not merely a bookkeeping device). Importantly, because an isometry from a Hilbert space to itself is unitary, any isometry can be embedded in a unitary operator by extending the dimension of the initial Hilbert space. 

Thus, a \emph{unitary} representation of measurement recovers by including Stinespring degrees of freedom from the outset, which we initialize in some default state. In other words, we work at all times in the dilated Hilbert space $\Hilbert^{\,}_{\rm ph} \to \Hilbert^{\,}_{\rm dil}$, which includes the Stinespring (i.e., ``outcome'' or ``apparatus'') qubits for all \emph{possible} measurements (e.g., in adaptive protocols where the choice of measurements may be conditioned on past outcomes). The resulting representation of all generic quantum channels, measurements, and outcome-dependent operations is \emph{unitary} \cite{AaronMIPT, MW, AaronDiegoFuture, supp}, and corresponds to (discrete) time evolution of the system and measurement apparatus (or environment, more generally) under some particular entangling interaction. The fact that the unitary Stinespring representation of generic quantum channels corresponds to the physical time evolution of a larger, closed system is what allows for the Heisenberg evolution of operators in the presence of generic quantum channels.

As in Sec.~\ref{subsec:iso meas}, we restrict to systems of qubits and projective measurements of Pauli-string operators, so that the two outcomes are  labelled 0 and 1
. By convention, we initialize all Stinespring (or ``outcome'') qubits in the state $\ket{0}$, and denote the product of all Stinespring qubits in this state by $\ket{\mathbf{0}}^{\,}_{\rm ss}$. In Sec.~\ref{subsec:other channels} we explain how the results for projective measurements extend to generic, local quantum channels (e.g., weak measurements).

\begin{table}[t!]
    \centering
\begin{tabular}{|c|c|c|}
\hline 
~ & ~$\comm{\observ}{A}=0$~ & ~$\acomm{\observ}{A}=0$~ \\
\hline
$\widetilde{A} = \SSid{\,}$ &  ~$ A \otimes \SSid{\,}$~ & ~$ A \otimes \SSX{\,}$~ \\
\hline 
$\widetilde{A} = \SSX{\,}$ &  ~$A\otimes \SSX{\,}$~ & ~$A \otimes \SSid{\,}$~ \\
\hline
$\widetilde{A} = \SSY{\,}$ &  ~$A \, \observ  \otimes \SSY{\,}$~ & ~$\ii \,  A \, \observ \otimes \SSZ{\,}$ \\
\hline 
$\widetilde{A} = \SSZ{\,}$ &  ~$A \, \observ \otimes \SSZ{\,}$~ & ~$-\ii \,  A \, \observ \otimes \SSY{\,}$~\\
\hline 
$\widetilde{A}= \SSProj{\pm}$ &  $\, \frac{1}{2}  A\left( \ident \otimes \SSid{\,} \pm \observ \otimes \SSZ{\,}\right) \, $  & $\, \frac{1}{2} A \left( \ident \otimes \SSX{\,}  \mp \ii \observ \otimes \SSY{\,} \right) \,$ \\
\hline
\end{tabular}
    \caption{``Lookup'' table for the values of $\umeas^{\,}_{[\observ]} \, A \otimes \widetilde{A} \, \Umeasdag{\observ}$ 
    \eqref{eq:SSH Update Rule}, which corresponds to unitary measurement of observable $\observ$ in either the Heisenberg (observable) or Schr\"odinger picture (density matrix). The operator $A \otimes \widetilde{A}$ acts as $A$ on the physical Hilbert space and as $\widetilde{A}$ on the Stinespring register that stores the outcome of measuring $\observ$. The Pauli string $\observ$ acts on each site $j$ as \emph{either} $\ident$, $X^{\,}_{j}$,  $Y^{\,}_{j}$, $Z^{\,}_{j}$; we decompose $A$ in this basis, so that $\observ$ and $A$ either commute or anticommute.
    }
    \label{tab:Heisenberg Lookup}
\end{table}

The unitary operator describing the projective measurement of a Pauli-string operator $\observ^{\,}_n$ is given by
\begin{align}
    \Umeas{\observ^{\,}_n} &\equiv \frac{1}{2} \left( \ident +\observ^{\,}_n \right) \SSid{n} + \frac{1}{2} \left( \ident -\observ^{\,}_n \right) \SSX{n} \, , ~~
    \label{eq:Measurement Unitary}
\end{align}
where the parenthetical terms act on the physical Hilbert space and the twiddled operators act on the $n$th Stinespring qubit. The second term flips the default outcome $\ket{0}$ to $\ket{1}$ when the $-1$ eigenstate of $\observ^{\,}_n$ is observed \cite{AaronMIPT}.


The advantage of our Stinespring representation of measurement via unitary channels \eqref{eq:Measurement Unitary} is that it allows us to evolve operators. Note that the measurement unitary \eqref{eq:Measurement Unitary} is also Hermitian, so the evolution of density matrices and operators---in the Schr\"odinger and Heisenberg pictures, respectively---is equivalent, given by \cite{AaronMIPT}
\begin{align}
    A^{\prime}_{\rm ph} \otimes \widetilde{A}^{\prime}_{\rm ss} &= \Umeas{\observ} \, A^{\vpp}_{\rm ph} \otimes \widetilde{A}^{\vpp}_{\rm ss} \Umeasdag{\observ} \notag \\
    &= \Umeasdag{\observ} \, A^{\vpp}_{\rm ph} \otimes \widetilde{A}^{\vpp}_{\rm ss} \Umeas{\observ} \label{eq:Heisenberg Stinspring} \, , ~~
\end{align}
where the two pictures are only equivalent for qubits. Using \eqref{eq:Measurement Unitary}, the channel \eqref{eq:Heisenberg Stinspring} takes the form
\begin{align}
    &=  \frac{1}{4} \acomm{\observ}{A} \observ \otimes \acomm{\widetilde{A}}{\SSX{}} \SSX{} 
    + \frac{1}{4}  \acomm{\observ}{A} \otimes \comm{\widetilde{A}}{\SSX{}}  \SSX{} \notag \\
    &~- \frac{1}{4} \, \comm{\observ}{A} \, \observ \otimes \acomm{\widetilde{A}}{\SSX{}}  + \frac{1}{4} \comm{\observ}{A} \otimes \comm{\widetilde{A}}{\SSX{}} \, ,
    \label{eq:SSH Update Rule}
\end{align}
which we summarize in Tab.~\ref{tab:Heisenberg Lookup}. Note that for density matrices, $\widetilde{A}$ is generally a  projector $\BKop{0}{0}^{\,}_{\rm ss} = ( \SSid{} + \SSZ{})/2$, while for observables, $\widetilde{A}$ is generally the identity $\SSid{}$.

\subsection{Trajectories and expectation values}
\label{subsec:stinespring use}
By convention, $\DensMatSS$ denotes density matrices in the \emph{dilated} Hilbert space \eqref{eq:Dilated Hilbert}, with the initial state given by
\begin{equation}
    \label{eq:Initial State General}
    \DensMatSS (0)  \equiv  \DensMat \otimes \BKop{\bvec{0}}{\bvec{0}}^{\,}_{\rm ss} \, ,~~
\end{equation}
where $\BKop{\bvec{0}}{\bvec{0}}^{\,}_{\rm ss} = \bigotimes_{j=1}^{\Nmeas} \BKop{0}{0}^{\,}_{{\rm ss},j}$ initializes \emph{all} Stinespring registers in the default state $\ket{0}$ and $\DensMat$ is the initial density matrix for the \emph{physical} system. This state is then evolved under a hybrid protocol $\chan \in \operatorname{End}(\Hilbert^{\,}_{\rm dil})$ to produce
\begin{equation}
    \label{eq:Final State General}
    \DensMatSS (t) = \Chan{} \DensMatSS (0) \Chandag{}  \, ,~~
\end{equation}
and assuming $\chan$ contains a single measurement of some generic (many-body) observable $\observ$ \eqref{eq:ObservSpectralDecomp}, the probability to obtain outcome $n$ is given by
\begin{align}
    \label{eq:SS unitary prob n}
    p^{\vpp}_n  = \trace\limits_{\rm dil} \, \left[ \, \ident^{\vpd}_{\rm ph} \otimes \BKop{n}{n}^{\vpd}_{\rm ss} \, \DensMatSS (t) \,\right] \, ,~~ 
\end{align}
while the expectation value of $\observ$ in the state \eqref{eq:Initial State General} is
\begin{align}
    \label{eq:SS unitary expval}
    \expval{\observ}^{\vpd}_{\DensMatSS}  = \trace\limits_{\rm dil} \, \left[ \, \DensMatSS (t) \, \sum\limits_{n=1}^{\Noutcome} \, \Eig{n} \,\BKop{n}{n}^{\vpd}_{\rm ss}  \, \right] \, ,~~ 
\end{align}
where $\Noutcome$ is the number of unique eigenvalues (and thus, outcomes) of the measured observables $\observ$. 

More generally, the operator that projects onto the \emph{set}  
of outcomes $\bvec{n} \, = \, \set{ n^{\,}_1, \dots , n^{\,}_{\Nmeas} }$ is simply  
\begin{align}
    \label{eq:Outcome Projector}
    \measproj{\bvec{n}}  &\equiv \proj{\rm ss}{(\bvec{n})} = \bigotimes\limits_{j \in {\rm ss}} \, \BKop{n^{\,}_{j}}{n^{\,}_{j}}^{\vpp}_{{\rm ss},j}  =  \BKop{\bvec{n}}{\bvec{n}}^{\,}_{\rm ss} , ~~
\end{align}
which acts nontrivially only on $\Hilbert^{\,}_{\rm ss}$, projecting the outcome qubit $j$ onto $n^{\,}_j \in \set{0,1}$. Note that the density matrix \eqref{eq:Final State General} projected onto trajectory $\bvec{m}\subset \bvec{n}$ \eqref{eq:Outcome Projector} for a subset $\Msites \subset \Hilbert^{\,}_{\rm ss}$ requires renormalization: 
\begin{align}
    \label{eq:SS density matrix projected}
    \DensMatSS^{\vpp}_{\bvec{m}} (t) &\equiv  \trace\limits_{\Msites} \, \left[ \,\proj{\Msites}{(\bvec{m})} \, \DensMatSS (t) \, \right] / \trace\limits_{\rm dil} \, \left[ \, \proj{\Msites}{(\bvec{m})} \, \DensMatSS (t)\, \right] \, . ~~
\end{align}
The probability to realize a particular sequence of measurement outcomes along the protocol $\chan$ is defined in terms of the outcome projector \eqref{eq:Outcome Projector} via
\begin{align}
    p^{\vpp}_{\bvec{n}} = \tr{ \, \ident^{\vpd}_{\rm ph} \otimes \BKop{\bvec{n}}{\bvec{n}}^{\vpp}_{\rm ss} \, \DensMatSS (t) } \, , ~~
    \label{eq:Trajectory probability} 
\end{align}
which can be evaluated part way through the protocol $\chan$, provided that $n^{\,}_i =0$ for all $n^{\,}_i \in\bvec{n}$ corresponding to measurements that have \emph{not yet occurred}. Moreover, the joint and conditional expectation values of all observables are readily defined by implementing \eqref{eq:SS unitary expval} for each observable; conditional expectation values utilize \eqref{eq:SS density matrix projected}.

In deriving the bound \eqref{eq:main}, we exclusively consider the outcome-averaged density matrix (or evolution of operators). The physical rationale for this is simple: The bound \eqref{eq:main} constrains useful quantum tasks, all of which output (i.e., either prepare or manipulate) a particular quantum state (in the form of a density matrix $\DensMatSS$). Necessarily, that state is used to extract statistics and/or expectation values, which require numerous ``shots'' to resolve. Thus, in any experimental implementation, the statistics or expectation values always correspond to the state that the protocol prepares \emph{on average}, since the repeated shots will sample over histories of outcomes. In this sense, the outcome-averaged density matrix \emph{is} the effective output of any useful quantum task. In the case of a pure state $\DensMat$ of the physical system, the task must output $\DensMat$ for \emph{all} outcomes; in the case of a mixed state $\DensMat$, the different outcomes realize the various pure states that comprise $\DensMat$ with the required coefficients (i.e., probabilities). 

In the Stinespring formalism, for any protocol $\mathcal{W}$ involving time evolution, measurements, and outcome-dependent operations, the output density matrix
\begin{align}
\label{eq:outcome av}
    \DensMat^{\vpd}_{\rm av} (t) &= \trace\limits_{\rm ss} \left[ \mathcal{W} \, \DensMat^{\vpp}_0 \otimes \BKop{\bvec{0}}{\bvec{0}}^{\vpp}_{\rm ss} \, \mathcal{W}^{\dagger} \right]
\end{align}
is realized from the initial physical state $\DensMat^{\,}_0$ upon averaging over all outcomes (i.e., tracing out the Stinespring degrees of freedom is equivalent to averaging over outcomes).

\subsection{Operator dynamics}
\label{subsec:SS operator dyn}

Using the Stinespring formalism, we now consider the measurement-related aspects of the ESTP described in Sec.~\ref{subsec:ESTP}. The measurement channel $\MeasChannel=\MeasChannel^{\,}_A \, \MeasChannel^{\,}_B$ further factorizes over the individual measurements, e.g.,
\begin{align}
\label{eq:ESTP Meas Channel SS}
    \MeasChannel^{\vpd}_A \,  = \, \bigotimes\limits_{s=1}^{\Nmeas} \, \frac{1}{2} \sum\limits_{n_s=0,1} \, \left( \ident + (-1)^{n_s} \PZ{A^{\,}_s} \right) \otimes \widetilde{X}^{n_s}_{A^{\,}_s} \, ,~~
\end{align}
and likewise for $\MeasChannel^{\,}_B$ (with $A \to B$ above). The individual measurement unitaries are equivalent to \eqref{eq:Measurement Unitary}.

The error-correction channel $\QECChannel=\QECChannel^{\,}_A \, \QECChannel^{\,}_B$ is somewhat more subtle, and can be worked out from Tab.~\ref{tab:lookup_main} and the Stinespring encoding of outcomes,
\begin{subequations}
\label{eq:ESTP QEC SS}
\begin{align}
    \QECChannel^{\vpd}_A \, &= \,\ident \otimes \frac{1}{2} \left( \SSid{} + \SSZ{A} \right)  + \PZ{\Nspins} \otimes \frac{1}{2} \left( \SSid{} - \SSZ{A} \right) \label{eq:ESTP QEC SS A} \\
    \QECChannel^{\vpd}_B \, &= \,\ident \otimes \frac{1}{2} \left( \SSid{} + \SSZ{B} \right)  + \PX{\Nspins} \otimes \frac{1}{2} \left( \SSid{} - \SSZ{B} \right) \label{eq:ESTP QEC SS B} \, , ~~
\end{align}
\end{subequations}
where $\SSZ{A/B} = \bigotimes_{j \in A/B} \, \SSZ{j}$ is a shorthand.

For clarity, we briefly reconsider the first steps in the evolution of the logical operators in Sec.~\ref{subsec:ESTP Heisenberg}. The first step corresponds to conjugation of the logical operators $\LX (0)$ and $\LZ (0)$ for the final state $\ket{\Psi^{\,}_f}$  \eqref{eq:nominal logicals} by the error-correcting channel \eqref{eq:ESTP QEC SS}. We note that $\QECChannel^{\,}_A$ \eqref{eq:ESTP QEC SS A} acts trivially on $\LZ (0)$, while $\QECChannel^{\,}_B$ \eqref{eq:ESTP QEC SS B} acts trivially on $\LX (0)$. It is straightforward to verify the update
\begin{subequations}
\label{eq:ESTP QEC update SS}
\begin{align}
    \QECChannel^{\dagger}_A \, \LX (0) \, \QECChannel^{\vpd}_A\, &= \, \PX{\Nspins} \, \SSZ{A} \label{eq:ESTP QEC update SS X} \\
    \QECChannel^{\dagger}_B \, \LZ (0) \, \QECChannel^{\vpd}_B\, &= \, \PZ{\Nspins} \, \SSZ{B} \label{eq:ESTP QEC update SS Z} \, ,~~
\end{align}
\end{subequations}
where $\SSZ{A/B}$ replaces $m^{\,}_{A/B}$ in \eqref{eq:logical R update}. 

We next conjugate by the measurement channel, represented unitarily in the dilated Hilbert space. Similarly to the previous step, $\MeasChannel^{\,}_A$ acts trivially on $\SSZ{B}$ (and vice versa). Hence, we need only consider the following updates,
\begin{subequations}
\label{eq:ESTP Meas update SS}
\begin{align}
    \MeasChannel^{\dagger}_A \,  \PX{\Nspins} \, \SSZ{A}\, \MeasChannel^{\vpd}_A \, &= \, \PZ{A} \, \PX{\Nspins} \, \SSZ{A} \label{eq:ESTP Meas update SS X} \\
    \MeasChannel^{\dagger}_B \,  \PZ{\Nspins} \, \SSZ{B}\, \MeasChannel^{\vpd}_B \, &= \, \PZ{B} \, \PZ{\Nspins} \, \SSZ{B} \label{eq:ESTP Meas update SS Z} \, ,~~
\end{align}
\end{subequations}
as claimed in \eqref{eq:Meas channel non Stinespring}. Since all measurements accounted for, we next simply evaluate the Stinespring operators in the default state $\ket{\bvec{0}}$ on all Stinespring qubits, so that
\begin{equation}
    \matel*{\bvec{0}}{\SSZ{A/B}}{\bvec{0}} \, = \, 1 \, ,~~
\end{equation}
and the Stinespring operators in \eqref{eq:ESTP Meas update SS} vanish, reproducing \eqref{eq:logicals post M}. The remainder of the Heisenberg treatment of Sec.~\ref{subsec:ESTP Heisenberg} does not require the Stinespring formalism.

We also refer the reader to the SM \cite{supp} for the Stinespring treatment of another optimal teleportation protocol based on the transverse field Ising model \cite{Hong:2022tmu}.

\section{Lieb-Robinson Bounds}
\label{sec:Bounds}

We now prove the bound \eqref{eq:main}, extending Lieb-Robinson bounds \cite{Lieb1972} to nonrelativistic quantum dynamics involving \emph{arbitrary} local quantum channels (i.e., completely positive, trace-preserving maps) and instantaneous classical communication. In particular, we focus on the combination of unitary time evolution, projective measurements, and outcome-dependent local operations. We provide numerous application-specific bounds, and also prove that the bound \eqref{eq:main} is optimal in a number of settings.

Absent measurements, \eqref{eq:main} reduces to the usual bound $\Dist \leq \LRvel T$ \cite{Lieb1972}. In this sense, \eqref{eq:main} also captures standard Lieb-Robinson bounds---in such cases we have $\Nmeas^{\,}_0 \leq 2$, which is saturated by,  e.g., circuits with ``light-cone'' geometries. Importantly, \eqref{eq:main} extends the standard Lieb-Robinson Theorem \cite{Lieb1972} to protocols $\mathcal{W}$ involving measurements in $\Nmeas$ local regions. More formally, for an adaptive protocol $\mathcal{W}$, for each outcome ``trajectory'' $\bvec{n}$, $\mathcal{W}$ prescribes measurements in local regions $S^{\,}_k$, from which we construct the set of regions $\mathbb{M}^{\,}_{\bvec{n}}$ by appending the region of support $S^{\,}_k$ of the $k$th measurement to $\mathbb{M}^{\,}_{\bvec{n}}$, unless $S^{\,}_k$ already appears in $\mathbb{M}^{\,}_{\bvec{n}}$ or if $S^{\,}_k$ is a proper subset of some $S^{\,}_j \in \mathbb{M}^{\,}_{\bvec{n}}$. Then, if no  region $S^{\,}_k$ includes the initial task qubit $i$, we include a new region $S^{\,}_{-1}=\{i\}$; otherwise we relabel the region that includes $i$ as $S^{\,}_{-1}$. We similarly identify the region $S^{\,}_0$ with the task qubit $f$, and define $\Nmeas$ as the maximum over all trajectories $\bvec{n}$ of $\abs{\mathbb{M}^{\,}_{\bvec{n}}}-2$.

While in certain cases there is some freedom (i.e., ambiguity) in defining the measurement regions, especially in the limit $\Nmeas = \Order{\Dist}$, one should generally pick the minimum value of $\Nmeas$ possible. In the context of the ESTP, e.g., one can pick each $\PZ{}$ measurement to represent a region; however, absorbing the Bell decoding channel into the measurements shows that it is possible to identify \emph{pairs} of single-qubit measurements with a single region, in which case the task is optimal. This is also more transparent in the limit $\Nmeas = \Order{1}$. In general, any refinements to the definitions of $\Nmeas$, $\Nobs$, $\Nmeas^{\,}_0$, and $T^{\,}_0$ are $\Order{1}$ and depend on the particular protocol and/or task.

As noted in Sec.~\ref{subsec:stinespring use}, we need only consider the reduced dynamics of the physical system, which corresponds to averaging over outcomes \eqref{eq:outcome av}. In fact, \eqref{eq:main} derives from considering the reduced density matrix $\DensMat^{\,}_{if}$ for the task qubits $i$ and $f$. All useful quantum tasks either generate, transfer, or manipulate quantum information, entanglement, and/or correlations. Hence, the output of any such task is always a quantum state $\DensMatSS$, of which only the physical part is meaningful. Importantly, this state is subsequently utilized by extracting expectation values and/or statistics, which requires numerous experimental ``shots.'' As a result, the state that one samples in practice is the outcome-averaged output of the protocol, given by tracing over the Stinespring degrees of freedom \eqref{eq:outcome av}. In the case of pure states, the \emph{same} pure state must output for any sequence of outcomes; in the case of mixed states, the ratios in which the distinct pure states (that comprise the mixed state) appear is fixed by the mixed state itself. This also means that there is no reason to consider statistics over measurement outcomes, as they are trivial for pure states, and prescribed for mixed states.

We first consider Clifford circuits in Sec.~\ref{subsec:Clifford bound}, which are both simple and relevant to optimal protocols (e.g., the ESTP of Sec.~\ref{sec:teleport}). However, \eqref{eq:main} also applies to more general dynamics generated by the combination of arbitrary, local, time-dependent Hamiltonians $H(t)$, and local quantum channels (e.g., projective or weak measurements and outcome-dependent operations). We give a physical explanation of the derivation of this bound in Sec.~\ref{sec:General Bound}. 

The full, rigorous proof is both lengthy and technical, and further details appear in the SM \cite{supp}. Crucially, we need only assume that (\emph{i})~the physical system comprises qubits on some physical graph $G$ with vertex set $V$, edges $E$, and spatial dimension $\SpaceDim$; (\emph{ii})~the time-dependent Hamiltonian is a sum of local terms $H(t) = \sum_{j} H^{\,}_j (t)$ where $H^{\,}_j$ acts on finitely many sites neighboring $j$; (\emph{iii}) the nonunitary quantum channels are spatially local; and (\emph{iv}) that $T$ satisfies $\LRvel T \gtrsim 3$ or $H(t)$ generates a quantum circuit. In Secs.~\ref{subsec:other sys} and \ref{subsec:other channels}, we explain how this generalizes beyond qubits, nearest-neighbor Hamiltonians, and to generic local quantum channels.

In the SM \cite{supp} we establish the equivalence of several useful tasks: For example, preparing a Bell state of two qubits separated by distance $\Dist$ is equivalent to teleporting a state $\ket{\psi}$ \eqref{eq:psi def} by distance $\Dist$, up to $\Order{1}$ corrections to $T$, $\Nmeas$, and the number of qubits (as realized by the standard teleportation protocol of Sec.~\ref{subsec:STP}). We also provide in the SM \cite{supp} a straightforward proof that, e.g.,  teleportation of a state $\ket{\psi}$ \eqref{eq:psi def} from site $i$ to site $f$ is equivalent to moving the corresponding logical operators \eqref{eq:nominal logicals} from site $i$ to site $f$. This provides for the derivation of \eqref{eq:main} in terms of operator dynamics, for which the unitary representation of quantum channels \cite{AaronMIPT,AaronDiegoFuture} in Sec.~\ref{sec:Stinespring} is crucial. We also note that the possibility of outcome-dependent operations applied instantaneously at arbitrary distances is incompatible with a Lieb-Robinson bound \emph{except} using our unitary formalism---i.e., this would lead to completely nonlocal Lindblad operators, precluding the bound of \cite{Poulin}. Relatedly, a standard Lieb-Robinson bound does not apply to the dilated dynamics.

The bound \eqref{eq:main} applies to ``useful quantum tasks,'' which transfer quantum information or generate entanglement (and/or correlations) over distance $\Dist$. This includes, e.g., the preparation of generic many-body states in a region of size $\Dist^{\SpaceDim}$, including the GHZ 
\eqref{eq:GHZ state}, quantum critical \eqref{eq:power law corr}, W \eqref{eq:W state} \cite{W_state}, Dicke \cite{Dicke}, and spin-squeezed \cite{squeeze_rev} states (see SM \cite{supp}). These bounds are derived in Sec.~\ref{subsec:Corr bounds}, and closely resemble \eqref{eq:main}. The main caveat is a small restriction on the compatible initial states, which we discuss in Sec.~\ref{subsec:initial state}. We also prove that \eqref{eq:main} cannot be sidestepped by teleporting multiple qubits in parallel, and detail applications of the bound \eqref{eq:main} in Sec.~\ref{sec:outlook}. Note that protocols that do not teleport information or generate entanglement (and/or correlations) need not obey \eqref{eq:main}; conversely, protocols that violate \eqref{eq:main} cannot be useful, as we illustrate.

\subsection{Bounds for Clifford dynamics}
\label{subsec:Clifford bound}

We now prove \eqref{eq:main} for Clifford protocols, such as the ESTP. Importantly, the local gates in a Clifford circuit always map Pauli strings (i.e., operators that act on every site as $\ident$, $\PX{}$, $\PY{}$, \emph{or} $\PZ{}$) to Pauli strings. We restrict to single- and two-site Clifford gates, as is common practice. By convention, the time $T$ in \eqref{eq:main} is the circuit depth, equal to the minimum number of layers required to implement all (physical) two-site gates, parallelizing where possible. Since only two-site gates ``grow'' operators---and by at most one site per gate---this counting gives $\LRvel=1$  in \eqref{eq:main}. Allowing for three-site and larger gates and/or altering the convention for $T$ simply modifies $\LRvel$; in general, $T$ should be thought of as the actual run time of the protocol.

We now justify the bound \eqref{eq:main} in the presence of measurements, focusing on the teleportation of logical operators by the ESTP of Sec.~\ref{sec:teleport} for convenience of presentation. Although we refer to projective measurements below, all statements apply equally to generic quantum channels (i.e., other than physical time evolution); as we clarify in Sec.~\ref{subsec:other channels}. In general, we expect that only protocols combining time evolution, projective measurements, and outcome-dependent operations can saturate \eqref{eq:main}.

The proof of the bound \eqref{eq:main} for Clifford circuits follows from, e.g., the fact that all deviations from the ESTP of Fig.~\ref{fig:SWAP circuit} either (\emph{i}) continue to saturate the bound \eqref{eq:main} for a different task distance $\Dist$, and are thus equivalent to the ESTP; (\emph{ii}) achieve suboptimal teleportation and fail to saturate the bound \eqref{eq:main}; or (\emph{iii}) fail to teleport the logical state entirely \cite{AaronYifanFuture}. Additionally, protocols that violate \eqref{eq:main} cannot teleport information
. The same arguments apply to preparing entangled (and/or correlated) resources states, as we discuss in Sec.~\ref{subsec:Corr bounds}; in Sec.~\ref{sec:outlook}, we present several optimal Clifford  protocols that achieve other useful quantum tasks and saturate \eqref{eq:main}.

Note that successful teleportation requires that the logical operators $\LX (0)$ and $\LZ (0)$ \eqref{eq:nominal logicals}---which act as $\PX{\Nspins}$ and $\PZ{\Nspins}$ on the final state $\ket{\Psi^{\,}_f}$  \eqref{eq:final state}---trivially commute  except at $j=\Nspins$. The Heisenberg-evolved logical operators $\LX(T)$ and $\LZ(T)$ \eqref{eq:final logicals}---which act as $\PX{1}$ and $\PZ{1}$ on the initial state $\ket{\Psi^{\,}_0}$ \eqref{eq:Psi0}---must act as $\ident$ or $Z$ (and thus commute) on all sites \emph{except} $j=1$ to obey \eqref{eq:XLTrelation}. We denote by $j_* (t)$ the rightmost site on which the operators $\LX(t)$ and $\LZ(t)$ anticommute, and require that $j_*(T)=1$. Furthermore, useful quantum tasks other than teleportation are also generically captured by the evolution of operators, and may be analogously constrained.

In the absence of measurements, the two-site Clifford gates (with $\LRvel=1$) can only decrease $j_*$ by one site per time step. In this scenario, teleporting $\LX(0)$ and $\LZ(0)$ from site $j=\Nspins$ to site $j=1$ requires $T \geq \Nspins-1 = \Dist$ circuit layers. This bound is saturated by a ``staircase'' of SWAP gates and agrees with \eqref{eq:main} for $\Nmeas=0$. 

The ESTP depicted in Fig.~\ref{fig:SWAP circuit} generalizes to other choices of $\Dist,\Nmeas,T$ by identifying $\Nmeas$ repeating regions of size $\ell=2 (T-1)$, along with an additional $T-2$ qubits to the left (including the initial logical site $j=1$) and a final qubit (or two) $j=\Nspins$ to the right. Then, the distance $\Dist$ over which the ESTP teleports a qubit obeys
\begin{equation}
\label{eq:protocol dist}
    \Dist^{\vpp}_{\small \rm ESTP} \,\leq \, \left(2 \, \Nmeas + 1 \right) \left( T-1 \right) +1\, , ~~
\end{equation}
which saturates \eqref{eq:main} with $\Nmeas^{\,}_0=1$ and $T^{\,}_0=-1$. If one measures $\PX{j}\PX{j+1}$ and $\PZ{j}\PZ{j+1}$ instead of $\PZ{j}$ and $\PZ{j+1}$, then we find $\Dist \leq (2 \Nmeas + 1 )T - 1$, since the final layer of CNOT gates is no longer required. Importantly, this explains why the correct choice of ``measurement regions'' are pairs of neighboring qubits. Generally speaking, if a given protocol $\cal W$ obeys \eqref{eq:main} with either $\Nmeas^{\,}_0<0$ or $T^{\,}_0 <0$, then a more efficient implementation of $\mathcal{W}$ exists.

Importantly, the ESTP transfers information with effective speed $\LRvel = 2 \, \Nmeas + 1$, compared to $\LRvel= 1$ without measurements. Alternatively, one can view $(2\Nmeas+1)T$ as the correct extension of the depth $T$ for $\Nmeas>0$, compared to the standard Lieb-Robinson bound $\Dist \leq \LRvel T$  when $\Nmeas=0$ \cite{Lieb1972}. Including extra layers of SWAP gates prior to the measurements grows the region size $\ell=2(T-1)$, so that the ESTP remains optimal with increased $T,\Dist$; including additional measurement regions with the same $\ell$ also at best leaves the ESTP optimal with increased $\Nmeas,\Dist$. However, including other two-site Clifford gates generally leads to suboptimal teleportation with respect to \eqref{eq:main}. Any two-site gates applied \emph{after} the measurement channel $\MeasChannel$ have $\LRvel=1$ (compared to $\LRvel = 2 \, \Nmeas +1$ for the ESTP on its own). Given $\Dist^{\,}_{\small \rm ESTP}$ \eqref{eq:protocol dist}  for the ESTP, including $T'$ layers of two-site gates at any point after the measurement channel $\MeasChannel$ realizes a task distance
\begin{equation}
\label{eq:suboptimal post meas gates}
    \Dist^{\prime} \, \leq \, \Dist^{\vpp}_{\small \rm ESTP} + T' \, < \, 2 \, \Nmeas \, (T+T') \, ,~~
\end{equation}
which is suboptimal compared to \eqref{eq:main}. Hence, in optimal teleportation, the measurement channel $\MeasChannel$ is applied after \emph{all} two-site unitary channels to maximize $\Dist$ \eqref{eq:protocol dist}.

We now prove that it is not possible to realize a greater enhancement to $\LRvel$ than $\left(2 \, \Nmeas + 1 \right)$ using measurements in $\Nmeas$ regions. In step \eqref{eq:logicals post M} of the Heisenberg evolution of the logical operators \eqref{eq:nominal logicals}, the combination of the error-correction and measurement channels ($\QECChannel$ and $\MeasChannel$) attaches the measured Pauli operators to $\LX(0)$ and $\LZ(0)$ at the corresponding measurement locations (as depicted at the top of Fig.~\ref{fig:logical growth}), which may be \emph{arbitrarily far} from the final site $j=\Nspins$. Importantly, this process  \eqref{eq:ESTP QEC SS} attaches distinct operators to the two distinct logical operators---i.e.,  $\LX \to \PZ{A} \, \PX{\Nspins}$ while $\LZ \to \PZ{B} \, \PZ{\Nspins}$. There are no further measurements with which to contend in the Heisenberg evolution of $\LX(t)$ and $\LZ(t)$, since optimal teleportation requires that measurements and error-correction occur after \emph{all} unitaries to ensure that $\LRvel \sim 2 \Nmeas$ \eqref{eq:suboptimal post meas gates}.

Importantly, we still have $j_*=\Nspins$ after step \eqref{eq:logicals post M}, but we require $j_*=1$  \eqref{eq:XLTrelation}. Yet, there is no unitary operation that converts $\PX{\Nspins}$ to the identity, and converting $\PX{\Nspins}$ to $\PZ{\Nspins}$ also converts $\PZ{\Nspins}$ to $\PX{\Nspins}$, leaving $j_*$ unchanged. In fact, if the protocol does not utilize the Paulis seeded by the measurement channels in the step \eqref{eq:logicals post M}, the fastest way to realize $j_*=1$ \eqref{eq:XLTrelation} is to use $T= \Nspins-1 = \Dist$ layers of SWAP gates, in which case there is no enhancement due to the measurements! Thus, any advantage due to measurements must relate to the seeded Pauli operators. 

While there is no means of removing a \emph{single} $\PX{}$ operator, step \eqref{eq:final logicals} shows that it is possible to convert both $\PX{j}\PX{j+1}$ and $\PZ{j}\PZ{j+1}$ into stabilizer operators compatible with \eqref{eq:XLTrelation}. This is due to the fact that these two operators commute and share $\ket{\text{Bell}}$ as a common eigenstate. Specifically, the Bell encoding (decoding) channel $\BellChannel^{(\dagger)}_{j,j+1}$ simultaneously converts $\PX{j}\PX{j+1} \leftrightarrow \PZ{j}$ and $\PZ{j}\PZ{j+1} \leftrightarrow \PZ{j+1}$. Applying these channels leads to $\LX(1) = \PX{A}\PX{B} \PX{\Nspins}$ and $\LZ(1) = \PZ{A}\PZ{B}\PZ{\Nspins}$ \eqref{eq:logicals post H CNOT}. Then, achieving $j_*=1$ requires that the $2\Nmeas+1$ Pauli $\PX{}$ operators in $\LX(1)$ be relocated (unitarily) such that one appears on site $j=1$ and all others are grouped into pairs on neighboring sites, as depicted in Fig.~\ref{fig:logical growth} for $\LX(t)$. This is most efficiently accomplished via the SWAP gates of the ESTP (see Fig.~\ref{fig:SWAP circuit}): The leftmost $\PX{}$ and $\PZ{}$ operators move to site $j=1$; the operators on site $B^{\,}_s$ move to site $C^{\,}_s$ of the same region, while the operator from site $A^{\,}_{s+1}$ moves to site $D^{\,}_s$ (where $A^{\,}_{\Nmeas+1} = \Nspins$). Finally, the Bell encoding channel converts $\PX{C^{\,}_s} \PX{D^{\,}_s} \to \PZ{C^{\,}_s}$ and $\PZ{C^{\,}_s} \PZ{D^{\,}_s} \to \PZ{D^{\,}_s}$, both of which act trivially on (i.e., stabilize) the initial state \eqref{eq:Psi0}.  

This unitary reshuffling (the protocol $\mathcal{W}$) of the Paulis seeded by the measurement channel $\MeasChannel$ is optimal. Saturation of \eqref{eq:main} implies that $\mathcal{W}$ has depth $T$; hence, each Pauli moves a maximum distance of $\LRvel (T-1)$ under $\mathcal{W}$, saturated by $T-1$ SWAP gates for $\LRvel=1$ (i.e., two-site gates). The maximal distance between equivalent measurement sites (e.g., $A$ sites) is $\ell = 2 \, \LRvel \, (T-1)$; otherwise, $\mathcal{W}$ cannot restore commutation of the logical operators on sites $j > 1$, and hence teleportation fails. 

Note that no alternative protocol can do better: Restricting to two-site gates, the initial state \eqref{eq:Psi0}, and $\PZ{}$-basis measurements, any such protocol requires two-qubit channels to seed new Paulis (which cannot be faster than the Bell channels) and gates to move the Paulis (which are no faster than SWAP gates). In fact, an alternative teleportation protocol that is more easily realized in certain experimental platforms (based instead on \cite{Hong:2022tmu}) is detailed in the SM \cite{supp}, and also asymptotically saturates \eqref{eq:main} with the same optimal spacing $\ell=2(T-1)$. In this context, the separation between measurement sites obeys
\begin{equation}
\label{eq:Meas region separation}
    \ell^{*}_{\rm meas} \, = \, 2  \LRvel (T-1) \, ,~~
\end{equation}
for Clifford teleportation  with single-qubit measurements.

However, replacing the initial state \eqref{eq:Psi0} with a state that already contains Bell pairs on neighboring $C$ and $D$ sites, and/or  measuring $\PX{A}\PX{B}$ and $\PZ{A}\PZ{B}$ instead of $\PZ{A}$ and $\PZ{B}$ respectively obviates the need for Bell encoding and decoding in the ESTP (see Fig.~\ref{fig:SWAP circuit}). Together, these adjustments merely change $T^{\,}_0$ from $-1$ to $+1$, while also decreasing the righthand side of \eqref{eq:protocol dist} by one. For generality, we allow for $\Order{1}$ offsets $T^{\,}_0$ and $\Nmeas^{\,}_0$ in \eqref{eq:main}, which depend on details of the protocol and initial state that are unimportant in the large $\Dist,T$ limit from which Lieb-Robinson bounds are extracted \cite{supp, Lieb1972}. Thus, in general, the spacing \eqref{eq:Meas region separation} obeys the relation
\begin{equation}
\label{eq:Meas region separation general}
    \ell^{\vps}_{\rm meas} \, \leq 2 \, \LRvel \,\left( T + T^{\,}_0 \right)  \, ,~~
\end{equation}
where $T^{\,}_0$ is the depth required to prepare the initial state from a product state, and we allow for local measurements in any basis. If the spacing between equivalent measurements regions exceeds $\ell^{\,}_{\rm meas}$ \eqref{eq:Meas region separation general} then teleportation fails. 

In the Clifford setting, \eqref{eq:main} is justified by the arguments above in terms of the ESTP: No modification to the ESTP achieves a protocol distance $\Dist$ that violates \eqref{eq:main}, yet numerous alterations lead to less optimal protocols, or fail outright. Moreover, the distance $\ell=2 \, \LRvel \, (T+T^{\,}_0)$ \eqref{eq:Meas region separation general} between measurement regions is maximal
. Correspondingly, the generalization of \eqref{eq:protocol dist} to generic Clifford circuits is
\begin{align}
    \Dist \, \leq \, \left( 2 \, \Nmeas + 1 \right) \, \LRvel \, \left( T + T^{\,}_0 \right) \, , ~~
    \label{eq:Clifford teleport bound}
\end{align}
where $T^{\,}_0$ is the depth of the quantum circuit that prepares the initial state from some product state. The intuition for this bound follows from the usual Lieb-Robinson bound \cite{Lieb1972} and Fig.~\ref{fig:logical growth}: Essentially, measurements ``reflect'' operator light cones, allowing information to be transferred over a greater distance by daisy-chaining Bell pairs.

Importantly, the Clifford teleportation bound \eqref{eq:Clifford teleport bound} also extends to other useful  quantum tasks and to non-Clifford circuits involving time evolution and other quantum channels. For example, creating a Bell pair between qubits $i$ and $f$ with $d(i,f)=\Dist$,  is equivalent to teleporting a state from $i$ to $f$ up to $\Order{1}$ corrections to $\Dist,T,\Nmeas$ (as showcased by the STP of Sec.~\ref{subsec:STP}).  In Sec.~\ref{subsec:Corr bounds}, we derive similar bounds on the preparation of correlated resource states. In Sec.~\ref{sec:outlook}, we apply the bound \eqref{eq:main} to error-correcting stabilizer codes and present optimal protocols for preparing several resource states.

\subsection{Bounds for generic dynamics}
\label{sec:General Bound}

The generalized Lieb-Robinson bound \eqref{eq:main} applies not only to (Clifford) circuits but to \emph{generic} protocols involving evolution under some time-dependent, local Hamiltonian $\Ham (t)$ along with local quantum channels (e.g., measurements and outcome-dependent operations). In the case of interest involving measurements, we allow for both the Hamiltonian $\Ham (t)$ and all aspects of the measurement protocol at time $t$ to depend on the outcomes of prior measurements. We need only assume that $\Ham (t)$ is local: In the proofs in the SM \cite{supp}, we take $\Ham (t)=\sum_{X \in E} h^{\,}_{X} (t)$ to be a sum of terms acting on neighboring qubits $x,y$ connected by edges $X \in E$. We also generalize this in Sec.~\ref{subsec:other sys}---such details merely affect the Lieb-Robinson velocity $\LRvel$. Other details of the general proof are highly technical and relegated to the SM \cite{supp}; here we give a nontechnical explanation of how the bound \eqref{eq:main} recovers. The main assumption is merely that $\LRvel T$ is at least $\Order{\SpaceDim}$.

We first illustrate why the standard derivation of Lieb-Robinson bounds \cite{Lieb1972, fossfeig, chen2019finite, Kuwahara:2019rlw, Tran:2020xpc, gross, schuch, kuwahara2021liebrobinson, yin21} in terms of commutators
\begin{equation}
    \label{eq:Comm LR}
    \comm{A^{\,}_x}{B^{\,}_y (t)} \, \neq \, 0 \, , ~~
\end{equation}
for $\Dist=d(x,y)$ is not useful in the presence of, e.g., measurements and instantaneous classical communication. Consider a protocol that consists of first measuring $\PZ{x}$; if the measurement outcome is $m=1$, we apply $\PZ{y}$, and do nothing otherwise. This protocol has
\begin{align}
    \PX{y} (t) \, &= \, \MeasChannel^{\dagger} \, \QECChannel^{\dagger} \, \PX{y} \, \QECChannel \, \MeasChannel \notag \\
    &= \, \MeasChannel^{\dagger} \left(\PX{y} \SSproj{\rm ss}{(0)} + \PZ{y} \PX{y}\PZ{y} \SSproj{\rm ss}{(1)} \right) \MeasChannel \notag \\
    &= \, \PX{y} \proj{x}{(0)}\SSproj{\rm ss}{(0)} + \PX{y} \proj{x}{(1)}\SSproj{\rm ss}{(1)} \notag \\
    &~~~~- \PX{y} \proj{x}{(0)}\SSproj{\rm ss}{(1)} -\PX{y} \proj{x}{(1)}\SSproj{\rm ss}{(0)} \notag \\
\intertext{where we used the fact that $\PZ{y} \PX{y}\PZ{y}= - \PX{y}$. Projecting onto the $\ket{0}$ state of the Stinespring register leads to}
    \PX{y} (t) \, &\to \, \PX{y} \proj{x}{(0)} - \PX{y} \proj{x}{(1)} \, = \, \PX{y} \PZ{x} \, ,~~
    \label{eq:trivial commutator thing}
\end{align}
which implies that \eqref{eq:Comm LR} is given by
\begin{equation}
    \comm{\PX{x}}{\PX{y} (t)} \, = \, -2 \, \ii \, \PX{y} \, \PY{x} \, \neq \, 0 \, , ~~
\end{equation}
for $t= \Order{1}$ and \emph{arbitrary} separations $\Dist=d(x,y)$. However, this protocol cannot generate entanglement or correlations, nor can it be used to teleport. Thus, in the presence of measurements and instantaneous classical communications, a different strategy is required to derive a meaningful bound on \emph{useful} quantum tasks.

The proof of \eqref{eq:main} involves showing that it is not possible to teleport a qubit over distance $\Dist$ or generate  entanglement or correlations between qubits separated by distance $\Dist$ in time $T \ll \Dist / \Nmeas$. In the context of Clifford circuits, each discrete time step of the protocol either extends an operator's support by one site or leaves it in place.  In continuous time, by contrast, any operator $A(t)$ always has some nonvanishing support on $\Order{\Nspins}$ sites for any $t > 0$. The proof of \eqref{eq:main} in full generality essentially involves evolving operators using the unitary measurement formalism \cite{Stinespring, AaronMIPT, AaronDiegoFuture} described in Sec.~\ref{sec:Stinespring} and showing that for times $t \ll \Dist / \Nmeas$, the state prepared by \emph{any} local quantum channel $\cal W$ is arbitrarily close to one that cannot host entanglement or nontrivial correlations  between qubits separated by distance $\Dist$ (and thus, cannot teleport). While this proof strategy differs from that of the standard Lieb-Robinson Theorem \cite{Lieb1972} in numerous technical respects, the two proofs are similar in spirit.

The crucial component of our proof is the construction of a ``reference'' Hamiltonian $\widetilde{\Ham}(t)$ from the true Hamiltonian $\Ham(t)$. Compared to $\Ham(t)$,  $\widetilde{\Ham}(t)$ does not contain any of the terms that cross (i.e., act nontrivially on both sides of) some bipartition $C$ of the graph $G$ of physical qubits. Crucially, the ``task qubits'' $i$ and $f$ lie on opposite sides of the cut $C$, and thus cannot be entangled by the reference protocol $\widetilde{\Ham}(t)$. More generally, given an initial state that is separable with respect to the bipartition $C$ (e.g., the conventional product state $\ket{\bvec{0}}$), the state that results from the combination of measurements and feedback (or, more generally, local quantum channels) and evolution under the reference Hamiltonian $\widetilde{\Ham}(t)$ is also separable, and thus cannot have entanglement or correlations between the qubits $i$ and $f$. As we prove in the SM \cite{supp}, it is always possible to choose a cut $C$ that is sufficiently far from all measurement regions (in Fig. \ref{fig:logical growth}, the partition boundary lies at the intersection of two of the depicted light cones). This ensures separability of the reference density matrix $\widetilde{\DensMat}(t)$ prepared by the combination of the measurement protocol and evolution under $\widetilde{\Ham}(t)$.

We then prove that the true state of the reduced density matrix $\DensMat^{\,}_{if} (t)$ for the task qubits $i$ and $f$ is \emph{arbitrarily} close (in trace distance) to the reduced density matrix $\widetilde{\DensMat}^{\,}_{if} (t)$ at sufficiently short times. Importantly, the latter state is \emph{separable} with respect to $i$ and $f$ by construction; also note that the full density matrices $\DensMat(t)$ and $\widetilde{\DensMat}(t)$ may be quite distinct, especially for qubits near the cut $C$. 

In particular, the true correlations and/or entanglement between the qubits $i$ and $f$ in the state  $\DensMat (t)$---as well as the Heisenberg evolution of logical operators---are well approximated by $\widetilde{\DensMat} (t)$ at times $T \lesssim \Dist / 2\, \LRvel \,  (\Nmeas+1) $. Yet, by construction, $\widetilde{\Ham}$ cannot generate  entanglement or correlations between qubits  $i$ and $f$. Thus, when evolving logical operators such as \eqref{eq:logicals post M}, $\Ham$ only generates useful entanglement when the approximation $\widetilde{\Ham}(t) \approx \Ham(t)$ fails. The accuracy of this approximation is guaranteed by the Lieb-Robinson Theorem \cite{Lieb1972} for times $t < r/ \LRvel$, where $r$ is the distance of any measured site to the partition $C$. Then, the observation that a partition can always be chosen such that $r \gtrsim \Dist/2(\Nmeas+1)$ leads to the bound
\begin{equation}
\label{eq:General Bound init product state}
    \Dist \, \lesssim 2 \left( \Nmeas + 1 \right) \, \LRvel T \, , ~~
\end{equation}
where $\Nmeas$ is the number of measurement regions and $T$ is the total duration of Hamiltonian, or the depth of the quantum circuit. In this sense, $(\Nmeas+1) T $ captures the extension of depth $T$ to protocols $\mathcal{W}$ involving measurements in $\Nmeas$ local regions and feedback. As with the usual Lieb-Robinson Theorem \cite{Lieb1972}---which recovers in the measurement-free limit ($\Nmeas = 0$)---the bound \eqref{eq:General Bound init product state} derives in the asymptotic limit of large $\Dist,T$. For finite sizes (and depending on the particular task at hand), there may be small, $\Order{1}$ corrections, which are captured by $\Nmeas^{\,}_0$ and $T^{\,}_0$ in the more general bound \eqref{eq:main}. Moreover, because \eqref{eq:General Bound init product state} holds for $\Nmeas=0$, it captures standard Lieb-Robinson bounds as well (see also Secs.~\ref{subsec:other sys} and \ref{subsec:other channels}).

The Heisenberg-Stinespring picture also allows us to prove that adjusting the locations of measurements based on prior outcomes does not allow one to avoid the bound \eqref{eq:main}. The proof for such ``adaptive'' measurement protocols uses the same strategy, and appears in the SM \cite{supp}. The only caveat is that, for the most general adaptive protocols, we find the slightly modified bound
\begin{equation}
    \label{eq:General Bound init product state adaptive}
    \Dist\,  \leq\, 2 \left( \Nmeas + 1\right) \left( \LRvel T + \left( \SpaceDim-1 \right) \, \log^{\vpp}_2 \Dist \right) \, ,~~ 
\end{equation}
where we are confident that this $\log_2 \Dist$ enhancement is not physical, and merely an artifact of the proof strategy. 

Specifically, the $\log_2 \Dist$ correction to $\LRvel T$ is absent (\emph{i}) in $\SpaceDim=1$, (\emph{ii}) for discrete time evolution generated by a quantum circuit, and (\emph{iii}) for prefixed measurement locations. This term is also asymptotically unimportant in the limit $\LRvel T \gtrsim \log_2 \Dist$. While an alternate proof strategy that uses the fact that each of the $\Nmeas$ regions are only measured  $\lesssim T$ times likely avoids this spurious correction, such a proof would be quite different from the strategies that appear in the SM \cite{supp}, and is beyond the scope of this work. Crucially, \eqref{eq:General Bound init product state adaptive} imposes important limitation on the performance of adaptive measurement-based protocols, which have been shown to outperform their nonadaptive counterparts (in which measurement outcomes do not affect subsequent gate choices) \cite{teleport99, Hoyer_2005, 0508124, Browne_2010, AaronMIPT}.

Additionally, while the proofs of \eqref{eq:General Bound init product state} and \eqref{eq:General Bound init product state adaptive} assume an initial product state (e.g. $\ket{\bvec{0}}$), in Sec.~\ref{subsec:initial state} we extend these proofs to certain classes of entangled initial states. We also generalize \eqref{eq:General Bound init product state} and \eqref{eq:General Bound init product state adaptive} to other nonunitary (but local) quantum channels in Sec.~\ref{subsec:other channels}. The bound \eqref{eq:General Bound init product state} constrains quantum communication, information processing, teleportation, and the preparation of entangled resource states (e.g., Bell states)  \cite{supp, Bell}. 

\subsection{Bounds from correlations}
\label{subsec:Corr bounds}

The bound \eqref{eq:General Bound init product state} also applies to the preparation of \emph{correlated} resource states, as we now describe. The derivation of \eqref{eq:General Bound init product state}  in Sec.~\ref{sec:General Bound} establishes that no protocol can generate useful entanglement between two qubits $i$ and $f$ with $d(i,f) = \Dist$ unless $\Dist$ obeys \eqref{eq:General Bound init product state}. In general, we define
\begin{equation}
    \label{eq:corr dist}
    \Dist \, \equiv \,  \max\limits_{x,y \in V} \, d(x,y) \,= \, d(i,f) ,~~
\end{equation}
so that the task qubits $i$ and $f$ correspond to a pair of maximally separated vertices, and the task distance $\Dist$ is roughly the linear ``size'' of $V$. In the SM \cite{supp}, we also show that an asymptotically identical bound applies to protocols that generate correlations between qubits separated by distance $\Dist$. States whose preparation can be bounded in terms of correlations include the GHZ \eqref{eq:GHZ state} \cite{GHZ89}, Dicke \cite{Dicke}, W \eqref{eq:W state} \cite{W_state}, and spin-squeezed \cite{squeeze_rev} states, as well as states corresponding to conformal field theories (CFTs) and quantum critical points \cite{TimShortcut, SubirQCP, CFTbook}.

Correlations between qubits $i$ and $f$ along some measurement trajectory $\bvec{n}$ are captured by
\begin{align}
\label{eq:cor if}
    \mathrm{Cor}(i,f)^{\vpp}_{\bvec{n}} \, &\equiv  \, \expval{\observ^{\,}_i  \observ^{\,}_f  \measproj{\bvec{n}}} - \frac{1}{p^{\,}_{\bvec{n}}}  \expval{\observ^{\,}_i  \measproj{\bvec{n}} }  \expval{ \observ^{\,}_f \measproj{\bvec{n}} } \, , ~~
\end{align}
where the operators $\observ^{\,}_{i,f}$ (with norm $\norm{\observ} \leq 1$) are chosen to maximize the above expression and $\measproj{\bvec{n}}$ projects the  Stinespring (outcome) qubits onto the measurement trajectory $\bvec{n}$ with associated probability $p^{\,}_{\bvec{n}} = \expval{\measproj{\bvec{n}}}$.

We first consider the GHZ state \eqref{eq:GHZ state}, whose corresponding bound is proven rigorously in the SM \cite{supp}, and the strategy largely mirrors that presented in Sec.~\ref{sec:General Bound}. 
Again, we construct a reference protocol $\widetilde{\mathcal{W}}$ that, compared to the true dynamics generated by $\mathcal{W}$, does not couple qubits across some bipartition $C$ of the system. However, instead of comparing the reduced density matrices for qubits $i$ and $f$ produced by $\mathcal{W}$ versus $\widetilde{\mathcal{W}}$, we instead examine connected two-point correlators $\text{Cor}(i,f)$ \eqref{eq:cor if}. 

We then show that, just as $\DensMat (T)$ is arbitrarily close to a state with no entanglement between maximally separated qubits $i$ and $f$ at times $T \lesssim \Dist / 2 \, \LRvel \, \Nmeas$, $\DensMat (T)$ is also arbitrarily close to a state with no \emph{correlations} 
between $i$ and $f$ for the same times $T$. Thus, any protocol $\mathcal{W}$ that produces a GHZ state \eqref{eq:GHZ state} on a $\Nspins \sim \Dist^{\SpaceDim}$ qubits obeys
\begin{equation}
\label{eq:GHZ bound}
    \Dist\,  \leq \, 2 \, \left(\Nmeas + 1 \right) \, \LRvel  T \, ,~~
\end{equation}
where $\Dist$ is defined in \eqref{eq:corr dist}. Hence, the same bound \eqref{eq:General Bound init product state} applies to $\ket{\rm GHZ}$ \eqref{eq:GHZ state}, as well as GHZ-like states $\alpha \ket{\bvec{0}} + \beta \ket{\bvec{1}}$ for arbitrary $\alpha,\beta \in (0,1)$---essentially, generating \emph{any} amount of nonlocal entanglement and/or correlations over distance $\Dist$ is only possible if \eqref{eq:GHZ bound} is obeyed.

The bound \eqref{eq:GHZ bound} holds for all protocols $\mathcal{W}$ that prepare an $\Nspins$-qubit GHZ state from a product state (\emph{i}) using prefixed measurement locations, (\emph{ii}) in $\SpaceDim=1$, and/or (\emph{iii}) via a quantum circuit. For protocols in $\SpaceDim>1$ with continuous time evolution and adaptive measurement locations, we can only prove that $\Nspins$ obeys the bound \eqref{eq:General Bound init product state adaptive}. However, we expect that the extra $\log \Dist$ term is an unphysical artifact of the proof, and that \eqref{eq:GHZ bound} is generic. 


Another useful class of resource states are elements of the Dicke manifold \cite{Dicke}---a subspace of $\Hilbert^{\,}_{\rm ph}$ comprising $\Nspins$-qubit states that are symmetric under permutations. Examples include the GHZ \eqref{eq:GHZ state} and W \eqref{eq:W state} states; the latter is an equal-weight superposition of all states with a single spin in the state $\ket{1}$, with all others in the state $\ket{0}$. A generalization of the W state \eqref{eq:W state} is the $k$th Dicke state
\begin{equation}
    \label{eq:Dicke state}
    \ket{D^{\,}_k} \, \propto \, \sum\limits_{\substack{X \in V \\ \abs{X} = k}} \, \ket{\mathbf{1}}^{\vpd}_X \otimes \ket{\mathbf{0}}^{\vpd}_{X^c} \, , ~~
\end{equation}
where $V$ is the set of physical vertices and $0 < k < \abs{V}$. The Dicke state $\ket{D^{\,}_k}$ \eqref{eq:Dicke state} is the (unnormalized) sum over all states with exactly $k$ qubits (in the subset $X\subset V$) in the state $\ket{1}$, with all other qubits (the subset $X^c = V \setminus X$) in the state $\ket{0}$ \cite{Dicke}; the W state \eqref{eq:W state} corresponds to $k=1$.

We now state a bound on the preparation of the W  \eqref{eq:W state} and Dicke states \eqref{eq:Dicke state} from the product state $\ket{\mathbf{0}}$. A measurement-assisted protocol $\cal W$ in $\SpaceDim$ spatial dimensions prepares a Dicke state only if it satisfies the bound
\begin{equation}
    \label{eq:Dicke bound}
    \Dist \, \leq \, 2 \, \left( \Nmeas + 1 \right) \, \left( \LRvel T + (3  \SpaceDim - 1 ) \log^{\,}_2 \Dist + C \right) \, ,~~
\end{equation}
where $C$ is some finite constant. We note that the bound \eqref{eq:Dicke bound} mirrors \eqref{eq:General Bound init product state adaptive}, and is proven in the SM \cite{supp}. In the case of prefixed measurement locations, the factor $(3 \SpaceDim -1)$ is replaced by $2 \SpaceDim$---i.e., compared to the GHZ bound \eqref{eq:GHZ bound}, the bound on preparing \emph{arbitrary} Dicke states \eqref{eq:Dicke state} has this extra $\log_2 \Dist$ term; it is an open question whether or not the fully general bound \eqref{eq:Dicke bound} is optimal.

However, in the particular case of the $W$ state (and any Dicke state with $k \ll \abs{V}$ with finite correlations between any two regions), a tighter bound than \eqref{eq:Dicke bound} recovers for protocols with \emph{prefixed} measurement locations. By bounding the correlations between two \emph{regions} $I$ and $F$ (with $\abs{I}=\abs{F} \sim \Nspins/6$), we recover the bound
\begin{equation}
    \label{eq:W bound}
    \Nspins \, \leq \, 3 \left( \Nmeas + 1 \right)  \LRvel T  \, ,~~
\end{equation}
for a protocol $\chan$ that prepares $\ket{W}$ \eqref{eq:W state} from $\ket{\bvec{0}}$ \cite{supp}. The proof of the general bound \eqref{eq:Dicke bound}---and the bound \eqref{eq:W bound} for preparing the W state \eqref{eq:W state} using prefixed measurement locations---follow straightforwardly from the analogous proof bounding the preparation of GHZ states; a rigorous derivation appears in the SM \cite{supp}.


Another useful class of correlated resource states feature spin squeezing \cite{squeeze_rev, PhysRevA.47.5138, PhysRevA.46.R6797}. Letting $S^{\,}_{\alpha} = \sum_j \, \Pauli{\alpha}{j}/2$ be the collective spin operators for all physical qubits (with $\alpha=x,y,z$), a spin-squeezed state satisfies
\begin{equation}
\label{eq:Squeeze relation}
    \Delta S^{\,}_{\alpha} \, \Delta S^{\,}_{\beta} \, \geq \, \frac{1}{2} \abs{ \expval{\comm{S^{\,}_{\alpha}}{S^{\,}_{\beta}}}} \, = \, \frac{1}{2} \abs{\expval{S^{\,}_{\gamma}}} \, , ~~
\end{equation}
where $\alpha$, $\beta$, and $\gamma$ satisfy $\epsilon^{\alpha \beta \gamma}_{\,} \neq 0$ and the variances are given by $\Delta S^{\,}_{\alpha} = \sqrt{\expval{S^{2}_{\alpha}} - \expval{S^{\,}_{\alpha}}^2}$. The state $\ket{\mathbf{0}}$, e.g., has $\Delta S^{\,}_{x} = \Delta S^{\,}_{y} = \sqrt{\Nspins/4} = \expval{S^{\,}_z}/2$, saturating \eqref{eq:Squeeze relation}. 

For convenience, we choose a coordinate frame such that $\expval{\vec{S}} = \expval{S^{\,}_z} = J$, while the perpendicular components have $\expval{S^{\,}_x}=\expval{S^{\,}_y}=0$. The relation \eqref{eq:Squeeze relation} implies $\Delta S^{\,}_x \, \Delta S^{\,}_y \geq \abs{J}/2$, meaning that the variance $\Delta S^{\,}_x$ can only be made small if $\Delta S^{\,}_y$ is sufficiently large. This tradeoff is quantified via the \emph{squeezing parameter} \cite{squeeze_rev}, defined by
\begin{equation}
\label{eq:squeeze parameter}
    \xi^2  \, \equiv \, \Nspins \, \min\limits_{\alpha \perp z} \, \frac{\Delta S^2_{\alpha}}{\expval{\vec{S}}^2} \, = \, \Nspins^{-\nu} \, ,~~
\end{equation}
where the axis $\alpha$ is perpendicular to $\expval{\vec{S}}$ (which points in the $z$ direction), $\Nspins$ is the number of qubits, and the equality on the right holds for parametrically strong squeezing, with $0 < \nu \leq 1$. If the state is permutation-symmetric (like Dicke states), the squeezing parameter is related to correlation functions \eqref{eq:cor if} according to 
\begin{equation}
\label{eq:squeeze cor perm}
    \mathrm{Cor}(i,f) \, \geq \, \frac{1-\xi^2}{\Nspins-1} \, , ~~
\end{equation}
so that the preparation of permutation-symmetric spin-squeezed states with squeezing parameter $\xi$ \eqref{eq:squeeze parameter} obeys a bound analogous to \eqref{eq:Dicke bound}, due to the relation \eqref{eq:squeeze cor perm}.

However, in general, if the spin-squeezed state is \emph{not} in the Dicke manifold \eqref{eq:Dicke state}, then a given pair of sites $i$ and $f$ need not be correlated. Nevertheless, the \emph{average} correlation between sites remains large
\begin{equation}
\label{eq:squeeze cor relation}
    \overline{\mathrm{Cor}} \, \equiv \, \sum\limits_{u,v \in V} \, \mathrm{Cor}(u,v) \,\geq \, \Nspins^{1+\nu} \, , ~~
\end{equation}
for strongly squeezed states that obey \eqref{eq:squeeze parameter}, with $\Nspins = \abs{V}$.

Finally, using a slight modification of the aforementioned strategy used in the context of GHZ and Dicke states, the relation \eqref{eq:squeeze cor relation} implies a bound
\begin{equation}
    \Nmeas \, T^{\SpaceDim} \, \geq \, \Omega \left( \Nspins^{(1+\nu)/2} \right) \, , ~~
\end{equation}
where $\Nspins = \abs{V} \sim \Dist^\SpaceDim$ is the number of qubits,  $\SpaceDim$ is the spatial dimension, and we use the computer science ``big $\Omega$ notation'' in which the function $\Omega ( x )$ is at least linear in its argument (i.e., $\Omega ( x ) \geq a \, x + b$ for some $a,b$). Full details of the proof are provided in the SM \cite{supp}.

Finally, another class of states one might wish to prepare are ``critical'' states $\ket{\Psi}$ with algebraic correlations \eqref{eq:power law corr} \cite{TimShortcut}. These states may correspond, e.g., to a quantum critical point \cite{SubirQCP} or a conformal field theory (CFT) \cite{CFTbook}, and are characterized by correlations of the form
\begin{equation}
    \label{eq:power law cor 2}
    \overline{\text{Cor}}(x,y) \sim \frac{1}{\abs{x-y}^{\alpha}} \, , ~~
\end{equation}
for $\alpha > 0$ and generic two-point correlation functions.

It is straightforward to extend the proof of the bounds \eqref{eq:GHZ bound} and \eqref{eq:Dicke bound} for states with \emph{constant} correlations between maximally separated pairs of qubits $x,y \in V$ (with $d(x,y)=\Dist$) to account for algebraic dependence \eqref{eq:power law cor 2} of the correlations on $\Dist$. The resulting bound is
\begin{equation}
    \label{eq:critical bound}
    \Dist \,   \leq \,  2 \left( \Nmeas + 1 \right) \left( \LRvel T + \left( \alpha + \SpaceDim - 1 \right) \log^{\,}_2 \Dist  \right) \, ,
\end{equation}
where $\alpha$ is the correlation exponent and $(\alpha + \SpaceDim-1) \to \alpha$ in the case of prefixed measurement locations, in $\SpaceDim=1$, and if $H(t)$ generates a quantum circuit. Again, we expect that the $(\SpaceDim-1) \, \log_2 \Dist$ term is merely an artifact of the proof strategy (rather than physical); however, the term proportional to $\alpha \, \log_2 \Dist$ in \eqref{eq:critical bound} may be physical, implying an advantage to preparing such states.

\subsection{Multi-qubit bounds}
\label{subsec:multi qubit bound}


The bound \eqref{eq:main} implies that no better strategy than the ESTP exists for teleporting the logical state of a single qubit some distance $\Dist$. However, one might ask whether it is possible to teleport $Q>1$ qubits a distance $\Dist$ in time $T$ using only $\Nobs \sim  \Dist/T$ measurement \emph{outcomes} (in the multi-qubit case, the number of outcomes $\Nobs$ will prove more useful than the number of regions $\Nmeas$). We now prove by contradiction that this is not possible using Clifford circuits, focusing on a pedagogical example for concreteness. However, the resulting bound \eqref{eq:mainQ} applies to generic protocols $\cal W$ comprising local quantum channels---the fully general proof appears in the SM \cite{supp}.

Suppose that the Clifford protocol $\mathcal{W}$ teleports $Q=2$ logical qubits from sites $j=1,2$ to sites $j=\Nspins-1,\Nspins$ of a $1\SpaceDim$ lattice. As illustrated in Sec.~\ref{subsec:ESTP Heisenberg}, teleporting of a state $\ket{\psi}$ is equivalent to teleportation of the logical operators $\LX$ and $\LZ$ \eqref{eq:nominal logicals}. First consider $\LZ$ for each qubit, and suppose that $\mathcal{W}$ teleports these operators by $\Order{\Nspins}$ sites in time $T\sim \Nspins/3 \gg 1$ using a \emph{single} measurement,
\begin{subequations}
\label{eq:twofold logical operators}
\begin{eqnarray}
    Z_\mathrm{L}^{(1)} (0) \, &= \, \PZ{\Nspins-1} ~ ,~~~&Z_\mathrm{L}^{(1)} (T) \, = \, \PZ{1} ~~~\\
    Z_\mathrm{L}^{(2)} (0) \, &= \, \PZ{\Nspins} ~ ,~~~&Z_\mathrm{L}^{(2)} (T) \, = \, \PZ{2} \, , ~~
\end{eqnarray}
\end{subequations}
where 
$(i=1,2)$ labels the two logical qubits. 

The Heisenberg evolution of $Z_\mathrm{L}^{(i)}$ \eqref{eq:twofold logical operators} under $\mathcal{W}$ mirrors the discussion of Secs.~\ref{subsec:ESTP Heisenberg} and \ref{subsec:SS operator dyn}. Since $\mathcal{W}$ involves a local measurement of some Pauli-string operator $\mathcal{S}$ (where $\mathcal{S}^2=\ident$), by analogy to \eqref{eq:ESTP Meas update SS}, we expect that the combination of the measurement and error-correction channels ($\MeasChannel$ and $\QECChannel$) multiplies both logical operators $Z_\mathrm{L}^{(i)}$  \eqref{eq:twofold logical operators} by $\mathcal{S}$. Importantly, if neither logical operator $X_\mathrm{L}^{(i)}, Z_\mathrm{L}^{(i)} $ for the logical qubit $i$ were multiplied by $\mathcal{S}$, then the combination of $\MeasChannel$ and $\QECChannel$ would act trivially on the state and logical operators. This in turn implies that the bound \eqref{eq:main} is violated, and the arguments of Sec.~\ref{subsec:Clifford bound} then establish that teleportation of qubit $i$ must fail.

Without loss of generality, suppose that both logical $\PZ{}$s are multiplied by the Pauli string $\mathcal{S}$ under the combination of $\QECChannel$ and $\MeasChannel$ in the Heisenberg picture, so that
\begin{align}
\label{eq:twofold logical pick up S}
    Z_{\mathrm{L}}^{(1)} \to \mathcal{S} \PZ{\Nspins-1} ~~~\text{and}~~~Z_{\mathrm{L}}^{(2)} \to \mathcal{S} \PZ{\Nspins} \, , ~~
\end{align}
since the two logical operators are teleported using the \emph{same} local measurement of the Pauli string $\mathcal{S}$. 

We are free to define a new logical $\PZ{}$ operator for qubit $i=2$ by multiplying the two operators in \eqref{eq:twofold logical pick up S}, i.e., 
\begin{equation}
\label{eq:twofold new logical}
    Z_{\mathrm{L}}^{(2)} \to Z_{\mathrm{L}}^{(1)}Z_{\mathrm{L}}^{(2)} = \PZ{\Nspins-1}\PZ{\Nspins} \, , ~~
\end{equation}
which still anticommutes with $X_{\mathrm{L}}^{(2)} \sim \dots \PX{\Nspins}$, and is thus a valid logical operator. However, by construction, the redefined logical \eqref{eq:twofold new logical} does not contain a factor of the Pauli string $\mathcal{S}$, meaning that $\cal W$ teleports this logical operator was from site $j=\Nspins$ to site $j=2$ (in the Heisenberg picture) without being affected by the combination of the measurement and error-correction channels $\MeasChannel$ and $\QECChannel$. 

However, we note that the teleportation of any individual qubit obeys the single-qubit bound \eqref{eq:main}. If the combination of measuring $\cal S$ and any outcome-dependent operations act trivially on the logical operators for the qubit $i=2$, then that measurement cannot ``reflect'' the $i=2$ logical operator light cones (as in Fig.~\ref{fig:logical growth}), and the protocol $\cal W$ is equivalent to the same protocol without the measurement of $\cal S$, which obeys the measurement-free Lieb-Robinson bound $\Dist \leq \LRvel T$ \cite{Lieb1972}. Hence, the update \eqref{eq:twofold logical pick up S} is incompatible with the assumption that qubit $i=2$ is teleported a distance $\Dist= \Nspins-2$ sites in time $T \sim \Nspins/3$ (where $\LRvel=1$), which requires an enhancement due to measurements. Moreover, we could equally have chosen to modify $Z_{\mathrm{L}}^{(1)}(T)$  in \eqref{eq:twofold new logical}---thus, the above proof by contradiction  establishes that the teleportation of logical operators for two logical qubits a distance $\Dist> \LRvel T$ using a single measurement is impossible. 

Generally speaking, the ability to identify valid logical operators that do not acquire Pauli strings $\mathcal{S}$ under Heisenberg-Stinespring evolution implies the existence of one or more logical qubits that can be transmitted without knowledge of any measurement outcomes, in violation of \eqref{eq:main}. This implies that at least two measurement outcomes are required to teleport two logical qubits. It is straightforward to extend these arguments to see that, for any Clifford circuit, the bound \eqref{eq:main} generalizes to 
\begin{equation}
\label{eq:mainQ}
    \Dist \, \lesssim \, \left(1 + \frac{\Nobs}{Q} \right) \, \LRvel T \, ,~~
\end{equation}
where $\Nobs$ is the number of independent measurement \emph{outcomes} used for quantum error correction (QEC). 

The bound \eqref{eq:mainQ} applies to the teleportation of $Q$ qubits or the formation of $Q$ Bell states \eqref{eq:Bell state} between pairs of qubits separated by distance $\sim \Dist$ (these tasks are equivalent up to $\Order{1}$ corrections, as seen in the STP of Sec.~\ref{subsec:STP}). In the SM \cite{supp}, we prove \eqref{eq:mainQ} for arbitrary protocols $\cal W$ with continuous-time dynamics generated by some local Hamiltonian $H(t)$ and adaptive measurement locations. In the fully general case, the bound is
\begin{align}
    \label{eq:mainQ adaptive}
    \Dist \, \leq \, 2 \left( \left\lfloor \frac{\Nobs}{Q} \right\rfloor + 1 \right) \left( \LRvel T + \left( \SpaceDim-1 \right) \log^{\,}_2 \Dist \right) \, ,~~
\end{align}
where the  $\log_2 \Dist$ term is likely an artifact of the proof strategy (rather than physical), as before, and the overall factor of two may be suboptimal, since $\Nobs = 2 \Nmeas$ if the error-correction channels $\QECChannel$ are Clifford on $\Hilbert^{\,}_{\rm dil}$ \cite{AaronYifanFuture}. 

In general, \eqref{eq:mainQ adaptive} implies that error-correcting channels cannot be shared by distinct logical qubits---teleporting $Q$ logical qubits requires $Q$ times as many measurements. In the case where the outcome-dependent channels $\QECChannel$ are elements of the dilated Clifford group, it can be proven that exactly two measurement outcomes are required per teleported qubit (or Bell pair formed), per region of size $\ell^{*}_{\rm meas} \leq 2  \LRvel  (T-1)$ \eqref{eq:Meas region separation} \cite{AaronYifanFuture}. For more general outcome-dependent ``recovery'' operations, we expect that \emph{at least} two measurement outcomes per qubit per region are required \cite{AaronYifanFuture}. However, we relegate a more detailed consideration of multi-qubit teleportation to future work.

\subsection{Entangled initial states}
\label{subsec:initial state}

In deriving the bound \eqref{eq:General Bound init product state} we have thus far assumed the initial state $\DensMat^{\,}_0$ to be an unentangled product state. We now relax this assumption in two ways. We first sketch how the bound \eqref{eq:General Bound init product state} extends to entangled states that can be prepared from a product state using finite resources. We then identify a generalization of these states compatible with our entanglement bound
. Finally, we recover the most general form of the bound \eqref{eq:main}. Technical details and formal proofs appear in the SM \cite{supp}.


First, the bounds \eqref{eq:General Bound init product state} and \eqref{eq:General Bound init product state adaptive} readily extend to a class of short-range-entangled (SRE) initial states $\DensMat^{\,}_0$ given by
\begin{equation}
\label{eq:SRE state}
    \DensMat^{\vpd}_0 \, = \, \mathcal{W}^{\vpd}_0 \, \DensMat' \, \mathcal{W}^{\dagger}_0 \, , ~~
\end{equation}
where $\DensMat'$ is an unentangled product state and the  protocol $\mathcal{W}^{\,}_0$ uses finite resources $\Nmeas_0',T_0' \ll \Dist$ (where $T_0'$ is the duration of continuous time evolution or depth of the quantum circuit) and is compatible with \eqref{eq:General Bound init product state}, or more generally, \eqref{eq:General Bound init product state adaptive}.  As an aside, $\DensMat'$ need only be separable with respect to the optimal cut $C$, and $\mathcal{W}^{\,}_0$ extends the notion of a finite-depth circuit to allow for measurements, feedback, and continuous time evolution.

If $\mathcal{W}$ achieves a useful task starting from the SRE initial state $\DensMat^{\,}_0$ \eqref{eq:SRE state}, then $\mathcal{W}' = \mathcal{W} \, \mathcal{W}^{\,}_0$ achieves that same task starting from the unentangled initial state $\DensMat'$. Thus, $\mathcal{W}'$ obeys \eqref{eq:General Bound init product state} or \eqref{eq:General Bound init product state adaptive} without modification, from which we infer a bound on $\mathcal{W}$. While this bound holds for arbitrary $\Nmeas_0',T_0'$, the constraints on useful quantum tasks are most transparent when  $\Nmeas_0',T_0'$ are $\Order{1}$ (i.e., independent of $\Dist,\Nmeas,T$ for the protocol $\mathcal{W}$). The SRE initial state $\DensMat^{\,}_0$ \eqref{eq:SRE state} is maximally advantageous to a given quantum task if $\DensMat^{\,}_0$ realizes along the protocol $\mathcal{W}'$ that acts on $\DensMat'$. Then the protocol $\mathcal{W}'$ has $\Nmeas' = \Nmeas + \Nmeas_0'$ and $T' = T + T_0'$, where $\Nmeas$ and $T$ are the resource requirements for $\mathcal{W}$. 

Because $\mathcal{W}$ and $\mathcal{W}'$ achieve the same quantum task, they obey the same bound \eqref{eq:General Bound init product state}, which can be written
\begin{equation}
\label{eq:SRE bound}
    \Dist \, \leq \, 2   \left( \Nmeas + \Nmeas^{\prime}_0 + 1 \right) \LRvel   \left(  T + T_0^{\prime} \right) \, , ~
\end{equation}
and bounds the enhancement to \eqref{eq:General Bound init product state} and \eqref{eq:General Bound init product state adaptive} due to an SRE initial state $\DensMat^{\,}_0$ \eqref{eq:SRE state}. The modification \eqref{eq:SRE bound} to the bound \eqref{eq:General Bound init product state} extend to \eqref{eq:General Bound init product state adaptive} and all of the task-specific bounds in Sec.~\ref{subsec:Corr bounds}: In the case of an SRE initial state $\DensMat^{\,}_0$ \eqref{eq:SRE state} that can be prepared from a product state $\DensMat'$ using measurements in $\Nmeas_0'$ regions and evolution for time $T_0'$, any bound recovered for an initial product state is modified according to $\Nmeas \to \Nmeas + \Nmeas_0'$ and $T \to T+ T_0'$, which may also capture $\Order{1}$ corrections to $\Nmeas,T$ that are protocol dependent and asymptotically unimportant.

For bounds that derive from the entanglement between qubits $i$ and $f$ (e.g., Bell pair distillation, QEC, and teleportation), we can generalize to initial states with a ``finite range'' $\xi$ of entanglement. However, such initial states are \emph{not} compatible with correlation bounds (e.g., for the preparation of GHZ, Dicke, W, spin-squeezed, and critical states). The proofs of the bounds \eqref{eq:General Bound init product state} and \eqref{eq:General Bound init product state adaptive} do not require that the state $\DensMat'$ in \eqref{eq:SRE state} be unentangled per se, but only that its entanglement be fragile in the sense that some notion of separability applies to $\DensMat'$.

In particular, we define a state $\DensMat'$ to have entanglement range $\xi$ if any cut $C$ of the system with thickness $\xi$ or greater results in a state $\widetilde{\DensMat}' = {\trace}^{\,}_{\small C} \, [ \, \DensMat' \, ]$ that is \emph{separable} with respect to the partition $C$. For example, product states have entanglement range $\xi=0$, while the GHZ state \eqref{eq:GHZ state} has entanglement range $\xi=1$. 

In the context of entanglement-generating tasks, the bound \eqref{eq:SRE bound} extends to SRE initial states $\DensMat^{\,}_0$ \eqref{eq:SRE state} prepared via the protocol $\mathcal{W}^{\,}_0$ using finite $\Nmeas_0',T_0'$ from a state $\DensMat'$ with entanglement range $\xi$. Provided that $\Nmeas_0',T_0',\xi \ll \Dist$, we simply choose a cut $C$ with thickness $\xi$ or greater, and the same proof strategy articulated above holds for the protocol $\mathcal{W}' = \mathcal{W} \, \mathcal{W}^{\,}_0$. Intuitively, states with entanglement range $\xi$ are compatible with entanglement bounds but not correlation bounds because, e.g., it is difficult to prepare a well-separated Bell pair from the GHZ state, but trivial to prepare the GHZ state from itself. Note that $\xi \ll \Dist$ is asymptotically unimportant to \eqref{eq:SRE bound}.

Thus, the most general bound is given by \eqref{eq:SRE bound} for SRE initial states $\DensMat^{\,}_0$ \eqref{eq:SRE state} that can be prepared via the protocol $\mathcal{W}^{\,}_0$ from a state $\DensMat'$ with entanglement range $\xi$ using finite resources  $\Nmeas_0',T_0'$ (where $\mathcal{W}^{\,}_0$ is optimal with respect to the bound \eqref{eq:General Bound init product state}, so that $\Nmeas_0',T_0'$ reflect the \emph{minimum} resources required to prepare $\DensMat^{\,}_0$). When bounding the preparation of correlated states, we restrict  $\DensMat'$ to $\xi=0$ product states. 

We also allow for finite-size corrections $\delta \Nmeas$ and $\delta T$ to $\Nmeas,T$, which are unimportant in the limit $\Dist,T \gg 1$ from which \eqref{eq:SRE bound} derives; $\delta T$ also captures $\order{\Dist}$ corrections such as the (likely unphysical) $(\SpaceDim-1)\log_2 \Dist$ term for adaptive, continuous-time protocols. Defining the quantities 
\begin{subequations}
\label{eq:M0 T0 def}
\begin{align}
    \Nmeas^{\vpp}_0 \, &\equiv \, 2 \left( \Nmeas^{\prime}_0 + 1 +  \delta \Nmeas  \right) \label{eq:M0 def} \\
    T^{\vpp}_0 \, &\equiv \, T^{\prime}_0 + \delta T \, ,~~ \label{eq:T0 def}
\end{align}
\end{subequations}
the generalized bound \eqref{eq:SRE bound} takes the form
\begin{equation*}
    \tag{\ref{eq:main}}
    \Dist \, \leq \,  \left( 2 \,\Nmeas + \Nmeas^{\vpp}_0 \right) \, \LRvel   \left( T + T^{\vpp}_0 \right) \, , ~~
\end{equation*}
for both prefixed and adaptive measurement protocols, where the convention \eqref{eq:M0 T0 def} is chosen so that all of the asymptotically unimportant details of a given protocol are captured by $\Nmeas^{\,}_0$ and $T^{\,}_0$ in \eqref{eq:main}. Note that the generalized \emph{multi-qubit} bound will generally be of the form
\begin{equation}
    \label{eq:mainQ SRE}
    \Dist \, \leq  \, \left( 2\Nobs+\Nobs^{\vpp}_0 \right) \LRvel \left( T + T^{\vpp}_0 \right) / Q \, ,~~
\end{equation}
since the bound for a task $\mathcal{W}$ acting on an SRE initial state $\DensMat^{\,}_0$ \eqref{eq:SRE state} inherits from the \emph{combined} protocol $\mathcal{W}' = \mathcal{W} \, \mathcal{W}^{\,}_0$ that prepares the initial state $\DensMat^{\,}_0$ from a separable one $\DensMat'$ \emph{and} achieves a task on $Q$ qubits. 

\subsection{Models and Hamiltonians}
\label{subsec:other sys}

Thus far, our derivation of \eqref{eq:main}, \eqref{eq:mainQ SRE}, and the various specialized bounds have focused on systems of qubits undergoing projective measurements, outcome-dependent operations, and unitary time evolution generated by a Hamiltonian $H(t)$ acting on pairs of neighboring qubits. We now generalize these results to other  degrees of freedom (Hilbert spaces) and similar local Hamiltonians.

The mathematical proofs of \eqref{eq:main}, \eqref{eq:mainQ}, and all related bounds appear in the SM \cite{supp}, and rely on two key ingredients, making certain extensions straightforward. The first ingredient is that the protocol $\mathcal{W}$ is generated by some \emph{physical}, time-dependent Hamiltonian $H(t)$ that acts locally on the physical graph $G^{\,}_{\rm ph}$. However, note that $\mathcal{W}$ is \emph{nonlocal} acting on $G^{\,}_{\rm dil}$, and hence does not obey the standard Lieb-Robinson bound \cite{Lieb1972} applied to $G^{\,}_{\rm dil}$. Both $H(t)$ and all other aspects of $\mathcal{W}$ may be arbitrarily conditioned on \emph{any} prior measurement outcomes; crucially, however, such feedback is always captured by operations on $G^{\,}_{\rm ph}$ (as well as new Stinespring registers in the case of adaptive measurements) that are \emph{conditioned} on degrees of freedom in $G^{\,}_{\rm ss}$. This provides for a notion of locality, as the growth of physical operators via feedback only seeds nonlocal support in $G^{\,}_{\rm ss}$. The fact that measurements only couple a local set of vertices in $G^{\,}_{\rm ph}$ to a single vertex in $G^{\,}_{\rm ss}$ ensure that the combination of measurement and feedback is compatible with a sharp notion of locality. The second ingredient is the Lieb-Robinson bound for the physical Hamiltonian $H(t)$, which ensures that  operators only grow locally within $G^{\,}_{\rm ph}$, and from which $\LRvel$ in \eqref{eq:main} is extracted, independently of the dilated channels in $\mathcal{W}$, such as measurements and feedback.

We first note that the local (on-site) Hilbert space dimension $d$ does not affect \eqref{eq:main} or related bounds. While  $d$ may affect the constant $c$ that appears in all proofs in the SM \cite{supp} (and relates to details of the graph $G$), this constant is asymptotically unimportant to \eqref{eq:main}. We expect that any finite  $d$ is compatible with all of the bounds we derive---at most, $d$ may modify the Lieb-Robinson velocity $\LRvel$. Hence, we expect our bounds to apply equally to systems involving qubits, $d$-state qudits, fermions, and even Majorana modes. However, extending our bounds to bosonic systems requires greater care, both due to details that apply in the measurement-free case \cite{yin21} and due to the possibility of measuring unbounded operators. We defer a bosonic extension of \eqref{eq:main} to future work.

Relatedly, our bounds are not restricted to measurements of involutory (or even binary-valued) operators with eigenvalues $\pm 1$. Crucially, the proofs in the SM \cite{supp} make no such assumption. The protocol $\cal W$ may involve measurements of arbitrary bounded operators $\observ$, which have finitely many eigenvalues, all of which are finite. Such operators may require larger on-site dimensions of the Stinespring vertices $v \in V^{\,}_{\rm ss}$, which, like the on-site physical dimension $d$, does not affect any of our bounds. At most, measuring generic bounded operators modifies the constant $c$ that appears in the SM \cite{supp}, but is asymptotically unimportant to \eqref{eq:main} and all other bounds. 

Our bounds also extend beyond nearest-neighbor Hamiltonians $H(t)$ to those acting on local regions. In particular, an identical bound applies if $H(t)$ comprises local terms that act on at most $k$ qubits in some connected region $\Omega \subset V$, where the maximum distance between any two qubits in $\Omega$ is at most $\ell$, provided that both $k$ and $\ell$ are finite. As long as $k$ and $\ell$ do not scale with $\Dist$ or $T$, extending to $k>2$ and $\ell>1$ only affects the Lieb-Robinson velocity $\LRvel$ (which will be at most $\Order{\ell}$) and the constant $c$ (which appears in the proofs in the SM \cite{supp}, but is asymptotically unimportant to the bounds we derive). Hence, our bounds extend to the more general local Hamiltonians described above. However, we note that long-range Hamiltonians with two-body interactions that fall off as either $\exp(-\abs{x-y})$ or $\abs{x-y}^{-\alpha}$ require an alternate proof strategy. While we expect qualitatively similar bounds, we relegate such proofs to future work.

\subsection{Generic quantum channels}
\label{subsec:other channels}

As we have noted at various points 
in the discussion thus far, the main bound \eqref{eq:main} and the context-specific bounds in Secs.~\ref{subsec:Clifford bound}--\ref{subsec:multi qubit bound} constrain protocols $\cal W$ involving \emph{arbitrary} local quantum channels. Other than unitary time evolution, thus far we have only explicitly considered projective measurements and outcome-dependent operations; we now explain how all of our bounds extend to arbitrary quantum channels, provided that those channels are local in their action on the physical qubits.

First, consider the straightforward example of a quantum channel corresponding to weak measurements, which are captured by a dilated unitary channel of the form \cite{EhudMIPT}
\begin{align}
    \label{eq:weak meas}
    \Wmeas{\observ} \, &= \, \frac{1}{2} \sum\limits_{n=0,1}  \left( \ident + (-1)^n \observ \right) \otimes \exp\left(  \ii \alpha \SSXp{}{n} \right) \, , ~~
\end{align}
where $\alpha \in [0,\pi/2]$ interpolates between a trivial channel $\wmeas = \ident$ for $\alpha=0$ and a projective measurement of the form \eqref{eq:Measurement Unitary} for $\alpha = \pi/2$ (up to an overall phase of $\ii$). 

For intermediate values of $\alpha$, compared to a projective measurement, the state of the apparatus $\wmeas$ realizes the default state $\ket{0}$ if the system is in the $+1$ state of $\observ$, and a superposition of $\ket{0}$ and $\ket{1}$ otherwise. In this sense, one can still perform outcome-dependent operations $\QECChannel$ as with projective measurements. However, because the Stinespring register no longer reflects the measurement outcome with fidelity one, it is unlikely that the combination of weak measurements and feedback are useful to tasks such as teleportation, which require perfect fidelity.

Nonetheless, the bound \eqref{eq:main}---along with the various extensions and specialized version---apply equally to protocols $\cal W$ involving weak measurements \eqref{eq:weak meas}. The proof of \eqref{eq:main} only utilizes (\emph{i}) the structure of the \emph{dilated} graph $G^{\,}_{\rm dil} = G^{\,}_{\rm ph} \cup G^{\,}_{\rm ss}$ that includes all physical and Stinespring qubits and (\emph{ii}) the Lieb-Robinson bound $\Dist \leq \LRvel T$ for unitary time evolution on the physical qubits alone \cite{Lieb1972}. In particular, weak measurements of the form \eqref{eq:weak meas} do not affect the unitary Lieb-Robinson bound, and do not change the locality of the dilated graph $G^{\,}_{\rm dil}$ compared to projective measurements of the form \eqref{eq:Measurement Unitary}, because weak measurements do not introduce connectivity (i.e., edges of $G^{\,}_{\rm dil}$) not present for projective measurements. In other words, from a locality standpoint, weak measurements have precisely the same properties as projective measurements, and therefore obey exactly the same bounds, captured by \eqref{eq:main}. However, protocols involving weak measurements may not be able to saturate the bound \eqref{eq:main}.

As noted in Sec.~\ref{sec:Stinespring}, the Stinespring Dilation Theorem \cite{Stinespring} establishes that \emph{all} valid quantum channels---i.e., completely positive, trace-preserving maps---can be represented using isometries and partial traces. Channels that preserve the Hilbert space dimension $\HilDim$ realize unitary time evolution, which we treat separately; those that reduce $\HilDim$ correspond to partial traces, which generally destroy information, correlations, and/or entanglement. 

Hence, the quantum channels of interest are those that ``dilate'' the Hilbert space. Not only can these channels always be represented unitarily on $\Hilbert^{\,}_{\rm dil}$, but the associated unitary operator always reflects the actual time evolution of an enlarged, closed system that includes both the measurement apparatus as well as the original physical degrees of freedom \cite{AaronDiegoFuture}. Moreover, while the unitaries corresponding to these channels may differ from those corresponding to measurements in their action on the physical and/or Stinespring registers, they have the same local structure. Without loss of generality, we can always demand that any such channel couples the physical system to a single (or finite number) of Stinespring registers. As with weak measurements \eqref{eq:weak meas}, any quantum channel that acts on a finite number of physical degrees of freedom in some localized region has the same locality properties as a measurement (vis a vis the dilated graph $G^{\,}_{\rm dil}$), and therefore obeys the same bounds.

\section{Outlook}
\label{sec:outlook}
We extended the Lieb-Robinson Theorem \cite{Lieb1972} to quantum dynamics with projective measurements. Our bound \eqref{eq:main} implies a finite speed of quantum information and explicit notions of locality and causality even in the presence of measurements, tightening the existing resolution of the EPR paradox \cite{epr}. Additionally, \eqref{eq:main} reveals a limit on measurements as a resource for quantum information processing: Using error-correcting feedback, we show that the speed of information can be enhanced by \emph{at most}  $\Nmeas\hspace{-0.5mm}+\hspace{-0.5mm}1$. While adaptive protocols can provide a valuable speedup over nonadaptive ones \cite{teleport99,Hoyer_2005,0508124,Browne_2010}, \eqref{eq:main} fundamentally limits their performance in useful quantum tasks, such as the preparation of entangled resource states, including those that represent a novel phase of matter \cite{Tantivasadakarn:2021vel}. The multi-qubit bound \eqref{eq:mainQ} further limits the implementation of many-qubit entangling gates in future large-scale quantum devices 
(see the SM \cite{supp} for further examples).

In the remainder, we discuss applications of our formalism and main bound \eqref{eq:main} to quantum computation, error correction, routing, and the efficient preparation of entangled and correlated resource states, using the new insights and perspectives afforded by the Stinespring representation of measurements detailed in Sec.~\ref{sec:Stinespring} and the Lieb-Robinson bounds developed in Sec.~\ref{sec:Bounds}. We further expect that our bounds will provide valuable insight into quantum information dynamics in years to come, and reveal new routes for efficient quantum information processing, computation, metrology, and sensing.

\subsection{Error-correcting codes}
\label{subsec:CSS}

The main idea behind quantum error correction (QEC) is to encode quantum information (i.e., $Q$ logical qubits) nonlocally amongst $\Nspins$ physical qubits to protect against local sources of errors and decoherence. A particularly transparent QEC procedure involves the use of \emph{stabilizers} \cite{QC_book,Gottesman_1998} to reduce from a continuum of possible errors to a countable set of errors that one must then correct.

The $\Nspins$ Pauli $\PZ{}$ operators that define the computational basis for $\Nspins$ qubits are replaced by $\Nspins-Q$ unique stabilizer operators---which generate the Abelian stabilizer \emph{group} $\Stab{}$---along with $Q$ logical $\PZ{}$ operators $Z^{(k)}_{\rm L}$. The many-body Hilbert space contains two orthogonal, finite-dimensional subspaces relevant to QEC, known as the ``codespace'' and ``error space'' of the stabilizer group. 

The codespace is the simultaneous $+1$ eigenspace of all elements of the stabilizer group; the error space contains all possible states that are $-1$ eigenstates of at least one stabilizer element \cite{Gottesman_1998}. We then identify ``codewords'' with an orthogonal basis for the codespace; generic logical states $\ket{\Psi}$ are linear combination of the codewords. By measuring stabilizer generators projectively, one can detect local errors, and apply an operator to return to the codespace. However, it is not possible to detect \emph{logical} errors, which connect distinct codewords. 

Hence, it is advantageous to encode the logical state amongst many qubits, so that a conspiracy of numerous local errors is required to realize a logical error. Then, by measuring stabilizer generators faster than the local error rate, one can avoid the fatal buildup of a logical error.

An important characteristic of a stabilizer QEC code is its \emph{distance} \cite{Hamming,QC_book}. The (Hamming) distance is the minimum number of local errors required to convert between orthogonal codewords. It is thus the minimum size of any logical operator (over all $Q$ logical qubits). 

Thus, generating an arbitrary state $\ket{\Psi}$ in the codespace of a stabilizer code can be bounded in terms of the code distance $d$. Starting from the product state $\ket{\bvec{0}}$ in the computational ($\PZ{}$) basis, there exist logical operators with unit size. Growing the $Q$ pairs of logical operators $X^{(k)}_{\rm L}$ and $Z^{(k)}_{\rm L}$ (for $1 \leq k \leq Q$) can be no faster than teleporting a single pair of logical operators a distance $d-1$. Hence, the time $T$ required to generate $\ket{\Psi}$ obeys
\begin{equation}
    \label{eq:code dist bound}
    T \, \gtrsim \, d^{\, 1/\SpaceDim} \, / \, 2 \, \LRvel \, \Nmeas \, ,~~
\end{equation} 
where $\LRvel=1$ for circuits comprising two-site Clifford gates, and $\SpaceDim$ is the spatial dimension of the qubit array.

Note that there are numerous stabilizer QEC codes for which the code distance $d$ and associated bound \eqref{eq:code dist bound} are not optimal. The bound \eqref{eq:code dist bound} can only be saturated for stabilizer codes in which: (\emph{i}) all logical operations act on exactly $d$ qubits; (\emph{ii}) for each logical operator, the $d$ qubits lie in a contiguous region; (\emph{iii}) the two logical operators $X^{(k)}_{\rm L}$ and $Z^{(k)}_{\rm L}$ for the $k$th logical qubit have identical support; while (\emph{iv}) the logical operators for distinct logical qubits do not have identical support. However, it is unclear whether any stabilizer QEC codes meet all of the foregoing criteria.

We now consider a class of codes that do not saturate \eqref{eq:code dist bound}, known as Calderbank-Shor-Steane (CSS) codes \cite{Calderbank_1996, Steane_1996_2, QC_book}. In fact, CSS codes admit tighter bounds than \eqref{eq:code dist bound}. The logical operators and stabilizer group of a CSS code are both generated by two distinct sets of operators: One that acts nontrivially only as $\PX{}$ and one that acts nontrivially only as $\PZ{}$. Accordingly, error correction is performed independently within distinct $\PX{}$ and $\PZ{}$ sectors, and one defines \emph{two} distinct code distances $d^{\,}_x$ and $d^{\,}_z$ as the minimum size amongst $\PX{}$- and $\PZ{}$-type logical operators, respectively. The bound \eqref{eq:main}  then applies to preparing an arbitrary state $\ket{\Psi}$ in the codespace of a $\SpaceDim$-dimensional CSS code, with $\Dist$ given by $\Dist^{\SpaceDim} =\max{(d_X,d_Z)}$---the \emph{larger} of the two distances \cite{QC_book}, whereas \eqref{eq:code dist bound} reflects the \emph{smaller}.

A useful example to illustrate this distinction is the quantum repetition code (QRC). The QRC codespace is spanned by GHZ-like cat states \eqref{eq:GHZ state} of the form
\begin{align}
    \ket{\Psi^{\,}_\mathrm{QRC}} \, = \, \alpha\ket{000\dots} + \beta\ket{111\dots} \, ,
\end{align}
whose stabilizer group is generated by $\set{ \PZ{j} \PZ{j+1}}$. If the initial logical qubit is on site $j=1$, we take the final logical operators to be $\LX = \PX{1} \cdots \PX{\Nspins}$ and $\LZ = \PZ{\Nspins}$. This is a CSS code with $(d_X,d_Z) = (\Nspins,1)$, and its quantum code distance is $d=1$.  However, applying \eqref{eq:main} to this code, we find $\Nmeas T \gtrsim \Nspins = \Dist$ is required (when $\SpaceDim=1$), in agreement with the main bound \eqref{eq:main}. Thus, the bound \eqref{eq:code dist bound} underestimates the required $\Nmeas \, T$ by a factor of $\Nspins$.

A useful example of a $2\SpaceDim$ CSS code is the toric code \cite{toric_1998, Dennis_2002, Kitaev_2003, toric_2006}. The logical operators $\LX$ and $\LZ$ are Wilson loops \cite{WilsonLoop}---i.e., products of $\PX{}$ and $\PZ{}$ operators along any two paths that wrap around the two distinct ``legs'' of the torus. For an $\Dist^{\,}_x \times \Dist^{\,}_y$ torus, the code distances are given by, e.g., $d^{\,}_x = \Dist^{\,}_x$ and $d^{\,}_z = \Dist^{\,}_y$, and preparing a state $\ket{\Psi}$ in the toric codespace obeys \eqref{eq:main} with $\Dist = \max \left(\Dist^{\,}_x  , \Dist^{\,}_y \right) $.

Generally speaking, the application of the bound \eqref{eq:main} to quantum error-correcting codes can be quite sensitive to the code's details. Additionally, the parameters typically used to characterize such a code---such as the code distance $d$---do not lend themselves to optimal bounds. This is evident in CSS codes such as the QRC, where the bound predicted by the standard code distance $d = \min (d^{\,}_x, d^{\,}_z)$ is loose by a factor of $\Nspins$ compared to the bound predicted by $\Dist = \max (d^{\,}_x, d^{\,}_z)$. Schematically, given a set of logical operators with minimized support, we expect that a tighter bound recovers upon taking $\Dist$ to be roughly the diameter of the largest such logical operator. In general, further tightening may be possible by considering code-specific details, as with the toric code  \cite{toric_1998, Dennis_2002, Kitaev_2003, toric_2006}.

\begin{figure}[t]
\centering
\includegraphics[width=.35\textwidth]{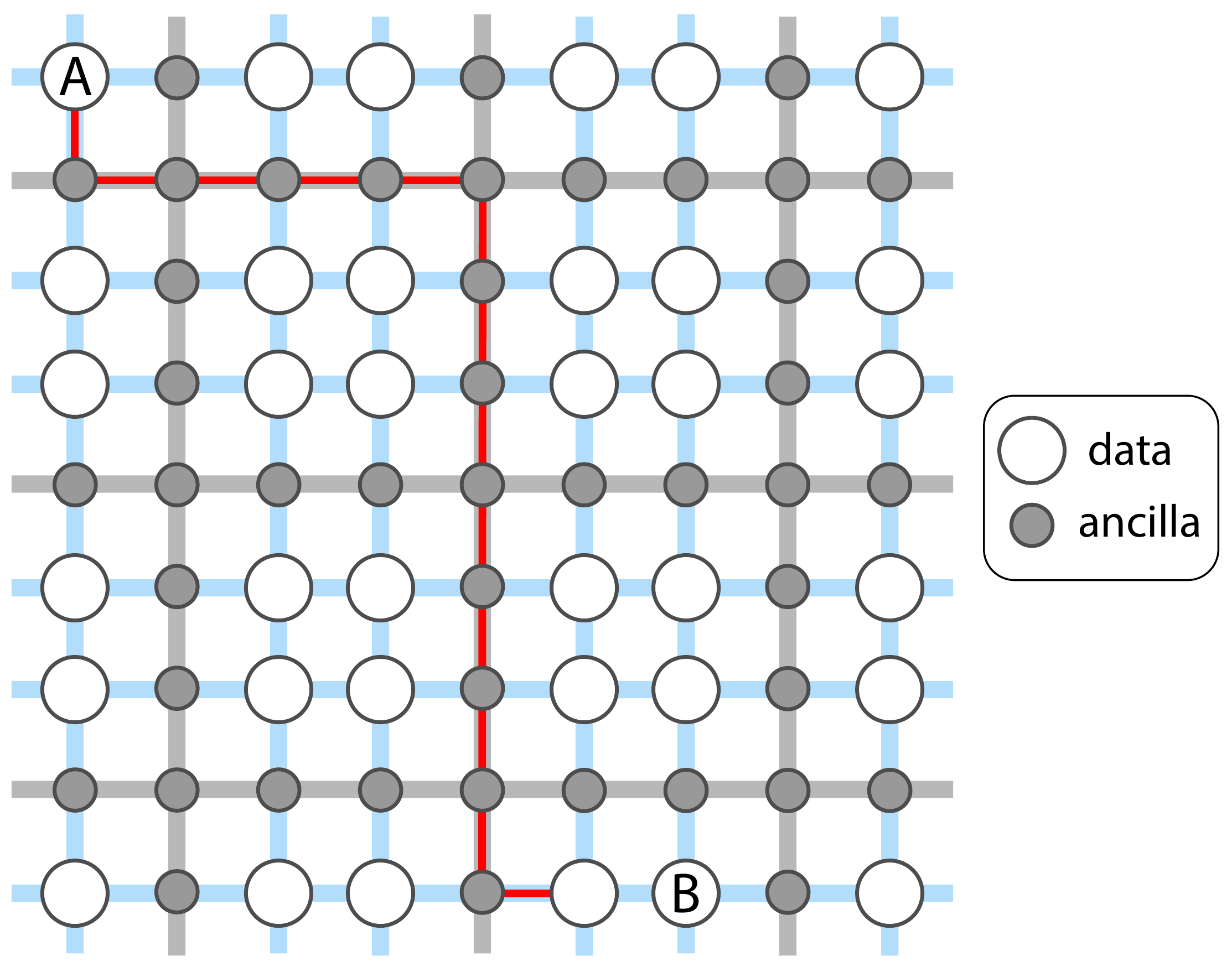}
\caption{A $2\SpaceDim$ lattice of qubits, where half of the qubits are ``data", containing information, and half are ``ancillas", facilitating quantum routing. The logical qubit, $A$, is first teleported next to $B$ by employing the ESTP along the red trajectory; subsequently, a local two-site gate is applied to entangle the pair.  Reversing the ESTP sends qubit $A$ back to its original location.  The run time is constrained by \eqref{eq:main}.}
\label{fig:2D lattice}
\end{figure}

\subsection{Routing in two-dimensional arrays}

Our bound \eqref{eq:main} also establishes that current quantum ``routing'' protocols utilizing teleportation are optimal, as we now explain. Numerous quantum tasks require operations applied to well separated qubits (e.g., several widely known quantum algorithms \cite{shor,grover} and simulating dynamics with all-to-all interactions). However, many experimental quantum devices that realize arrays of qubits have only limited connectivity and often use immobile qubits (with the notable exception of trapped-ion devices).

The saturating protocol for \eqref{eq:main} suggests that a quantum information processor with arbitrarily nonlocal connectivity can be efficiently built out of spatially localized (i.e., immobile) qubits using distributed quantum computing \cite{Beals_2013}. Essentially, one implements a two-site gate between arbitrarily distant qubits $A$ and $B$ via quantum routing \cite{1805.12570,2204.04185}, in which the state of qubit $A$ is teleported to an ancillary site adjacent qubit $B$, the two-body interaction is applied, and the state of the ancilla is then teleported back to the original location of qubit $A$. For example, see  Fig.~\ref{fig:2D lattice}: One can use roughly half of the qubits in a $2\SpaceDim$ lattice as ancilla registers to teleport logical qubits across the lattice's ancillary backbone for fast implementation of nonlocal gates. This entire process is optimized by using, e.g., the ESTP of Sec.~\ref{subsec:ESTP} with $\Nmeas$ measurement regions to decrease the time $T$ required.

\subsection{Bell pair distillation}
\label{subsec:Bell prep}

A simple Clifford protocol for preparing a well-separated Bell pair \eqref{eq:Bell state} follows from the entanglement-swapping teleportation protocol (ESTP) of Sec.~\ref{subsec:ESTP}. For Bell pair distillation, compared to Fig.~\ref{fig:SWAP circuit}, we remove qubit $A^{\,}_1$ along with all qubits to the left. The final Bell pair is recovered on sites $B^{\,}_1$ and $\Nspins=A^{\,}_{\Nmeas+1}$ using one fewer measurement region compared to the ESTP depicted in Fig.~\ref{fig:SWAP circuit}. All other aspects of the protocol are the same. 

Starting from the product state $\ket{\mathbf{0}}$, we apply Bell encoding channels $\BellChannel$ to the $C$ and $D$ sites in each of the $\Nmeas+1$ regions labelled $s=1,2,\dots,\Nmeas+1$. We then use $T-2$ layers of SWAP gates---as for the ESTP---to move each Bell qubit $C^{\,}_s$ to site $B^{\,}_s$ and each Bell qubit $D^{\,}_s$ to site $A^{\,}_{s+1}$ with $A^{\,}_{\Nmeas+2} = \Nspins$ the final site. We then perform Bell decoding $\BellChannel^{\dagger}$ on neighboring $A^{\,}_s$ and $B^{\,}_s$ qubits for $2 \leq s \leq \Nmeas+1$. For the case depicted in Fig.~\ref{fig:SWAP circuit}, there is only one measurement region (the first orange-shaded region of the ESTP is always excluded for Bell pair distillation). We then measure $\PZ{}$ on both the $A^{\,}_s$ and $B^{\,}_s$ qubits, which determines the error-correction channel $\QECChannel$ to apply to site $\Nspins$ according to Tab.~\ref{tab:lookup_main}. 

The default assignment recovers the Bell state \eqref{eq:Bell state} on qubits $B^{\,}_1$ and $\Nspins=A^{\,}_{\Nmeas+2}$. Importantly, $m^{\,}_A$ and $m^{\,}_B$ always correspond to $\PZ{}$ and $\PX{}$ errors, respectively. However, we can independently choose whether to assign the $+1$ or $-1$ value to an error for each $m^{\,}_A$ and $m^{\,}_B$. The four different assignments of $\pm$ values in Tab.~\ref{tab:lookup_main} to the QEC channel $\QECChannel$ corresponds to the four different Bell states.

This Bell-pair-producing modification of the ESTP is a Clifford protocol that distills a Bell state of two qubits separated by a distance $\Dist$ given by
\begin{equation}
    \label{eq:Bell distance}
    \Dist^{\vpd}_{\rm Bell} \, \leq \, 2 \, \left(\Nmeas+1 \right) \left(T-1 \right) + 1 \, , ~~
\end{equation}
for $T-2$ layers of SWAP gates, two layers of CNOT gates (for Bell encoding and decoding), and $\Nmeas$ measurement regions. The overall $+1$ compared to the circuit depicted in Fig.~\ref{fig:SWAP circuit} comes from including two extra SWAP gates on the far left and right in parallel with the CNOTs in the Bell decoding layer (prior to the measurements). Replacing the $\PZ{}$ measurements with Bell-basis measurements gives a task distance $\Dist^{\vpd}_{\rm Bell} \, < \, 2 \, \left(\Nmeas+1 \right) \, T$.

While similar protocols are certainly known in the literature, we note that (\emph{i}) the protocol above allows for arbitrary tradeoffs between $\Nmeas$ and $T$ and (\emph{ii}) the fact that the task distance \eqref{eq:Bell distance} saturates \eqref{eq:main} establishes that more efficient protocols do not exist.

\begin{figure}[t!]
\centering
\includegraphics[width=.46\textwidth]{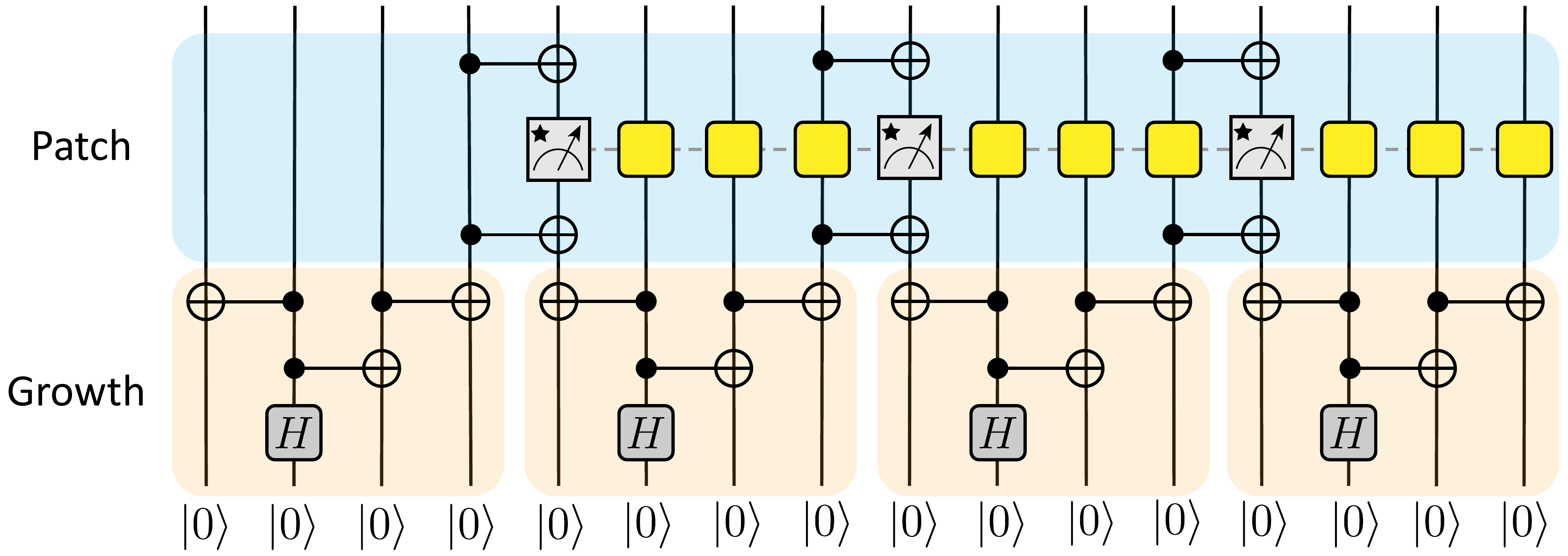}
\caption{A protocol for generating the GHZ state \eqref{eq:GHZ redef} is depicted diagrammatically for a $1\SpaceDim$ chain of $\Nspins=16$ qubits with $T=4$ and $\Nmeas=3$. The protocol creates and grows local GHZ states on each of the four-site patches shaded in orange. The blue-shaded portion of the protocol ``patches'' the local GHZ states into a single GHZ state using measurements (pointer dials), classical communication (the dashed line), and error correction (yellow boxes).}
\label{fig:GHZ protocol}
\end{figure}

\subsection{Preparing GHZ states}
\label{subsec:GHZ prep}

In Sec.~\ref{subsec:Corr bounds} we extended the bound \eqref{eq:main} to the preparation of GHZ states \eqref{eq:GHZ state}. We now discuss an optimal $1\SpaceDim$ protocol that prepares the GHZ state on $\Nspins$ qubits starting from the product state $\ket{\mathbf{0}}$ (on all physical and Stinespring qubits). While aspects of this protocol are well known to the literature, the fact that this protocol saturates \eqref{eq:main} implies (\emph{i}) that more efficient protocols do not exist and (\emph{ii}) that the bound \eqref{eq:main} is optimal with respect to the preparation of GHZ states \cite{GHZ89,GHZ07,Bell_GHZ,cirac}. 

As a reminder, the GHZ state is given by
\begin{equation}
    \label{eq:GHZ redef}
    \ket{\text{GHZ}} \, \equiv \, \frac{1}{\sqrt{2}} \left( \ket{\bvec{0}} + \ket{\bvec{1}} \right) \, , ~~
\end{equation}
and a protocol that prepares this state in $1\SpaceDim$ is depicted in Fig.~\ref{fig:GHZ protocol} for $\Nspins=16$, $\Nmeas=3$, $T=4$.  This protocol involves two distinct stages: the ``growth'' and ``patch'' stages, distinguished by orange and blue shading in Fig.~\ref{fig:GHZ protocol}. 

The $1\SpaceDim$ protocol follows from Fig.~\ref{fig:GHZ protocol}: For a given $\Nmeas \geq 0$ and $T \geq 3$, the $\Nspins$-qubit chain is divided into $\Nmeas+1$ regions of size $\ell = 2 (T-1)$. We apply Bell encoding $\BellChannel$ to form a local GHZ state of the central two qubits in each region. We then apply $T-3$ layers of CNOT gates to grow these local GHZ states by two qubits per layer. 

We then patch the local GHZ states together. We first apply a CNOT gate between each pair of regions, and then measure $\PZ{}$ on the target qubit and reset its state to $\ket{0}$. If the measurement outcome was 1, we apply $\PX{}$ to all other qubits in the same local patch as the measured state; otherwise, we do nothing. Finally, applying the same CNOT gate applied prior to measurement reincorporates the measured qubit into the GHZ state, which now contains all $\Nspins= \Dist +1 = \left( \Nmeas + 1 \right) \, \ell$ qubits.

This $1 \SpaceDim$ GHZ-preparation protocol with circuit depth $T$ and $\Nmeas=\Nobs$ single-qubit measurements satisfies
\begin{equation}
\label{eq:GHZ task distance}
    \Dist^{\vpd}_{\rm GHZ} \, \sim \, \Nspins \, = \, 2 \, \left( \Nmeas + 1 \right) \left( T - 1 \right)  \, , ~~
\end{equation}
which saturates \eqref{eq:main} with $\Nmeas^{\,}_0=2$ and $T^{\,}_0=-1$. 

In $\SpaceDim>1$, analogous protocols saturate \eqref{eq:main}, though a slight modification to the ``patch'' stage is required. A protocol in $\SpaceDim>1$ with circuit depth $T$ and $\Nmeas=\Nobs$ single-qubit measurements satisfies
\begin{equation}
\label{eq:GHZ task distance higher dim}
    \Dist^{\SpaceDim}_{\rm GHZ} \, \sim \, \Nspins \, = \, \Order{ \Nmeas \times \left( 2 \, T \right)^{\SpaceDim}} \, , ~~
\end{equation}
and further details of the protocol---including the extension to $\SpaceDim>1$---appear in the SM \cite{supp}. 

Most importantly, the fact that \eqref{eq:GHZ task distance} and \eqref{eq:GHZ task distance higher dim} saturate the bound \eqref{eq:main} implies that not only that the protocol depicted in Fig.~\ref{fig:GHZ protocol} optimal, but the bound \eqref{eq:main} itself is optimal with respect to the preparation of GHZ states.

\subsection{Preparing W states}
\label{subsec:W state}

\begin{figure}[t!]
\centering
\includegraphics[width=.46\textwidth]{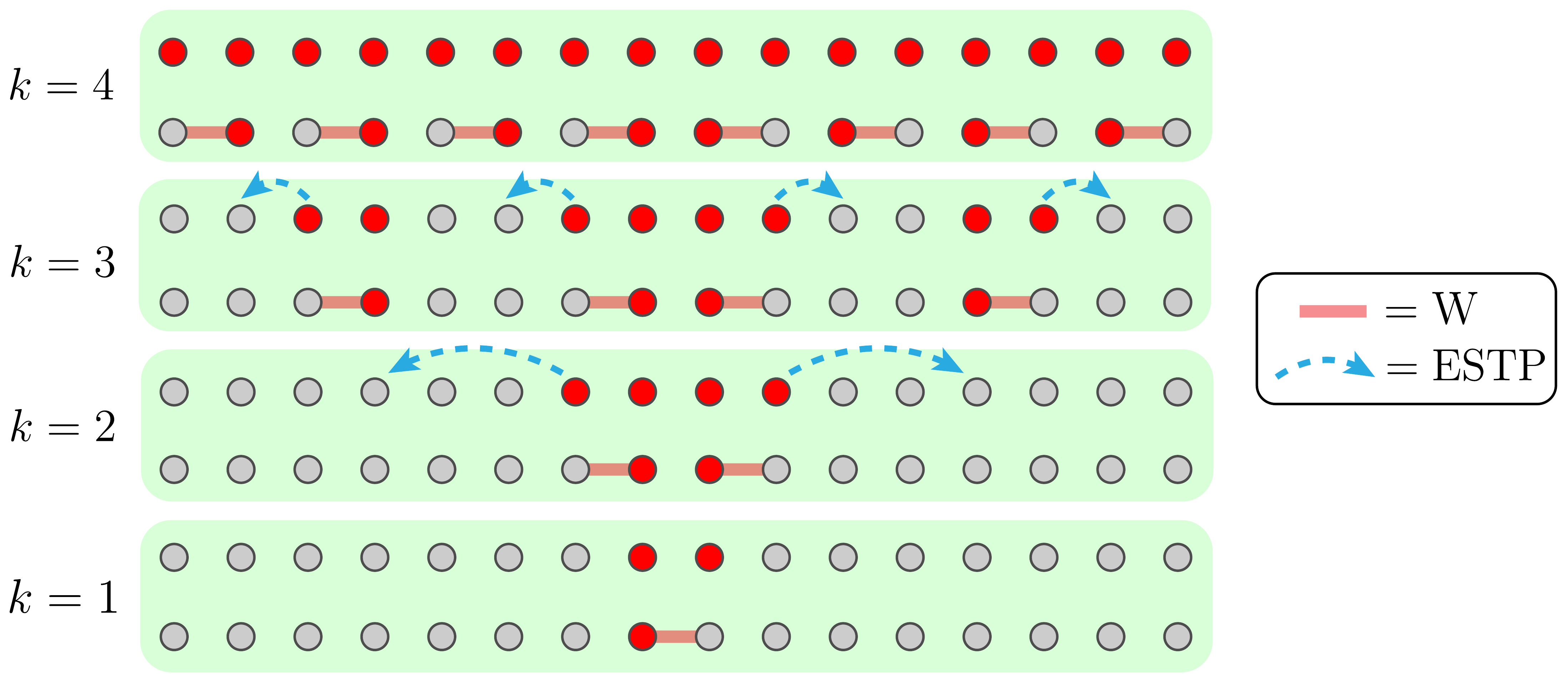}
\caption{A protocol that generates the W state on $\Nspins=2^n$ sites is depicted for a $1 \SpaceDim$ chain of $\Nspins = 16$ qubits ($n=4$). The gray circles denote qubits still in the initial $\ket{0}$ state; the red circles denote the set of qubits $A$ that participate in the W state of $2^k$ qubits in step $k$ of the protocol. Measurements are only used to expedite the equal spacing of participating qubits in each stage via the ESTP of Sec.~\ref{subsec:ESTP}.}
\label{fig:W protocol}
\end{figure}

We now present a protocol for preparing W states \eqref{eq:W state} on  a $1 \SpaceDim$ chain of $\Nspins$ qubits prepared in the state $\ket{\mathbf{0}}$ \cite{cirac,Cruz_2019}. The protocol is illustrated in Fig.~\ref{fig:W protocol}, and other technical details appear in the SM \cite{supp}. As a reminder, the W state for the $\Nspins$ qubits in the set $V$ is given by
\begin{equation}
    \label{eq:W redef}
    \ket{W} \, = \, \frac{1}{\sqrt{\Nspins}} \, \sum\limits_{v \in V} \, \ket{1}^{\vpp}_v \otimes \ket{\mathbf{0}}^{\vpp}_{V \setminus \{v\}} \, , ~~
\end{equation}
which is an equal-weight superposition over all configurations with a single qubit in the state $\ket{1}$ and all others in the state $\ket{0}$. Equivalently, the W state \eqref{eq:W redef} is the $k=0$ quantum Fourier transform \cite{coppersmith} of the state $\ket{1} \otimes \ket{\mathbf{0}}$. 

In contrast to the other protocols discussed thus far, the protocol that prepares the W state \eqref{eq:W redef}, depicted in Fig.~\ref{fig:W protocol}, does \emph{not} saturate the corresponding bound \eqref{eq:Dicke bound}, except in the $\Nmeas=0$ limit where $T = \Order{\Nspins}$. However, this protocol can be implemented in depth $T = \Order{\log \Nspins}$, compared to the $T=\Order{\Nspins}$ protocols described in \cite{coppersmith,cirac}.

The standard unitary ($\Nmeas=0$) protocols for preparing the W state \eqref{eq:W redef} are similar to the measurement-free protocols that prepare the GHZ state \eqref{eq:GHZ redef} \cite{coppersmith,cirac}. One first prepares the state $\ket{1000\dots}$ with a single flipped spin (or excitation) on the leftmost site; one then uses $\Order{\Nspins}$ SWAP gates to move this excitation to the right, applying local single-site rotation gates (which do not count toward $T$) so that a fraction of the excitation is ``left behind'' \cite{coppersmith,cirac}. Unlike the GHZ state \eqref{eq:GHZ redef}, there is no obvious means by which to use measurements to achieve an $\Order{2\Nmeas}$ enhancement to $\LRvel$. This is due in part to the fact that the W state \eqref{eq:W redef} is \emph{not} a stabilizer state. We also note that the protocol described in \cite{cirac} replaces these $T$ SWAP gates with $T$ copies of the ESTP using ancilla qubits, so that $\Nmeas,T = \Order{\Nspins}$; however, this is far more costly than the protocol we now describe.

The workhorse of this W-state-preparation protocol is the following non-Clifford two-qubit gate of \cite{Cruz_2019}:
\begin{align}
\label{eq:W gate}
    \operatorname{W}^{\vpd}_{i,j} \, \equiv \, 
    \quad 
    \begin{minipage}[h]{0.2\textwidth}
        \Qcircuit @C=0.8em @R=1.5em {
        \lstick{i} & \ctrl{1} & \targ & \qw \\
        \lstick{j} & \gate{H} & \ctrl{-1} & \qw
        }
    \end{minipage} \;\; ,
\end{align}
where the circuit above acts from left to right as
\begin{equation}
\label{eq:W gate operators}
    \operatorname{W}^{\vpd}_{i,j} \, = \, \text{CNOT} \left(j \to i \right) \, \text{CH} \left(i \to j \right) \, , ~~
\end{equation}
meaning that $\operatorname{W}^{\,}_{i,j}$ first applies a Hadamard gate $\Hadamard^{\,}_j = (\PX{j} + \PZ{j})/\sqrt{2}$ to qubit $j$ if qubit $i$ is in the state $\ket{1}$ (and does nothing otherwise), and then applies $\PX{i}$ to qubit $i$ if qubit $j$ is in the state $\ket{1}$ (and does nothing otherwise). 

The non-Clifford gate $\operatorname{W}^{\,}_{i,j}$ \eqref{eq:W gate} can be used to double the size of an existing W state. Suppose an $\ell$-qubit subset $A \subset V$ of the physical qubits participate in a W state, while the $\ell$-qubit subset $B \subset V$ realizes the state $\ket{\mathbf{0}}$
. For every qubit $a \in A$ we identify a partner $b(a) \in B$ (in Fig.~\ref{fig:W protocol}, these qubits are neighbors). We then have that
\begin{equation}
\label{eq:W state doubling}
    \ket{W}^{\vpp}_{A \cup B} \, = \, \prod\limits_{a \in A} \, \operatorname{W}^{\vpd}_{a,b(a)} \, \ket{W}^{\vpp}_A \otimes \ket{\mathbf{0}}^{\vpp}_B \, , ~~ 
\end{equation}
where the $\ell$ distinct gates $\operatorname{W}^{\,}_{a,b(a)}$ map the state $\ket{W}^{\,}_A$ on $\ell$ qubits to the state $\ket{W}^{\,}_{A \cup B}$ on $2\ell$ qubits. Since all $\ell$ gates can be applied in parallel, this costs $T=2$.

The following protocol creates a W state on $\Nspins = 2^n$ qubits arranged in a $1 \SpaceDim$ chain, and is illustrated in Fig.~\ref{fig:W protocol}. The qubits in the initial state $\ket{0}$ are depicted as gray circles in Fig.~\ref{fig:W protocol}, while those incorporated into the W state \eqref{eq:W redef} are depicted as red circles. The protocol involves $n$ rounds in which a two-stage subroutine---corresponding to ``growth'' and ``teleportation''---is applied.

We first seed a two-qubit W state \eqref{eq:W redef}---equivalent to one of the Bell states---on the two central qubits. This is achieved in $T=1$ using the Bell encoding channel $\BellChannel$ followed by $\PX{}$ on either qubit, and no teleportation is required. In each of the steps $1<k<n$, we first  double the size of the W state as in \eqref{eq:W state doubling} by applying $\operatorname{W}^{\,}_{a,b}$ gates, where $a$ runs over the incorporated qubits (the red circles in Fig.~\ref{fig:W protocol}) and the corresponding unincorporated qubit $b(a)$ is the outward neighbor to $a$. We then teleport the newly incorporated $b(a)$ qubits by $\Dist^{\,}_k = 2^{n-k}-1$ sites (indicated by blue arrows in Fig.~\ref{fig:W protocol}), which concludes the $k$th round. In the final round $k=n$, all $\Nspins/2$ remaining qubits are incorporated into the W state, without the need for teleportation, concluding the protocol.

As with the other protocols presented in this paper, this W-state-preparation protocol allows tradeoffs between $\Nmeas$ and $T$. However, measurements are only helpful in expediting the ``teleportation'' step of each round $k$. In the absence of measurements ($\Nmeas=0$), teleportation in round $k$ is achieved using $T = 2^{n-k}-1$ layers of SWAP gates, so that the total circuit depth is 
\begin{equation}
\label{eq:W prep M=0}
    T^{\vpd}_{\rm max} 
    \, = \, \frac{1}{2} \Nspins  + \log_2 \Nspins - 1\, < \, \Nspins , ~~
\end{equation}
which saturates the measurement-free bound $T \lesssim \Nspins$ \cite{Lieb1972}. 

Alternatively, we can use the ESTP to teleport the newly incorporated qubits in the steps $2 \leq k \leq n-2$. Full details appear in the SM \cite{supp}. Essentially, in round $k$ there are $2^{k-1}$ newly incorporated qubits, which we can teleport a distance $\Dist^{\,}_k = 2^{n-k}-1$ in parallel. In each of the $2^{k-1}$ teleportation regions, the ESTP first applies a SWAP gate to the qubit being teleported and its less central neighbor, and then generates $2^{n-k-1}-1$ Bell pairs per region. In each round $k$, only the outermost two regions utilize \emph{new} measurement locations, and thus the $k$th round with minimum depth involves $2^n-2^k$ total measurement outcomes, and $2^{n-k}-2$ new measurement regions. Note that we eschew the Bell-decoding channels of the ESTP as depicted in Fig.~\ref{fig:SWAP circuit}, replacing the measurements of $\PZ{A}$ and $\PZ{B}$ with $\Nmeas^{\,}_k$ Bell measurements of $\PX{A}\PX{B}$ and $\PZ{A}\PZ{B}$, respectively. The error-correction channel $\QECChannel$ is still determined according to Tab.~\ref{tab:lookup_main}.

Each round $k \in [2,n-2]$ requires circuit depth $T^{\,}_k = 3$ and measurements in $2^{n-k}-2$ \emph{new} regions, with $2^n-2^k$ new outcomes. The first round has $T=1$, the last round has $T=2$, and the penultimate round has $T=3$ (where teleportation is replaced by SWAP gates); these three rounds do not require measurements. The total resource costs are given by
\begin{subequations}
\label{eq:W prep ESTP resources}
\begin{align}
    T^{\vpd}_{\rm min} \, &= \, 3 \, \log_2 \Nspins - 3 \label{eq:W prep ESTP T} \\
    \Nmeas^{\vpd}_{\rm max} \, &= \, \Nspins/2 - 2 \, \log_2 \Nspins + 2 \label{eq:W prep ESTP M} \, ,~~
\end{align}
\end{subequations}
so that the depth scales as $T = \Order{\log \Nspins}$ while the number of independent regions obeys $\Nmeas \lesssim \Nspins$. Additionally, the number of measurement \emph{outcomes} utilized scales as $\Nobs=  \Order{\Nspins \, \log \Nspins}$.  While using the ESTP to expedite the teleportation steps leads to a significant speed up compared to the $\Nmeas=0$ case \eqref{eq:W prep M=0}, this protocol does not saturate the bound \eqref{eq:Dicke bound} for $\Nmeas > 0$. 

Importantly, although it does not saturate \eqref{eq:Dicke bound} or \eqref{eq:W bound}, this protocol generates the W state \eqref{eq:W state} faster than all protocols known to the literature---i.e., $T \sim \log_2 \Nspins$ instead of $T \sim \Nspins$ \cite{coppersmith, cirac}. Most likely, it is the protocol above---rather than the bound \eqref{eq:Dicke bound} or \eqref{eq:W bound}---that is suboptimal with respect to preparing $\ket{W}$ \eqref{eq:W state}. We also suspect that $T \gtrsim \log_2 \Nspins$ is a generic bound on the measurement-based preparation of $W$, Dicke, and other nonstabilizer states with long-range correlations. However, we leave the proof of these conjectures to future work \cite{ChaoWFuture}.

\subsection{Locality of measurements}
\label{subsec:local meas}

The unitary measurement formalism \cite{AaronMIPT} detailed in Sec.~\ref{sec:Stinespring} is a mathematical consequence of the Stinespring Dilation Theorem \cite{Stinespring}. Physically, the measurement channel \eqref{eq:Measurement Unitary} is unitary because---in an idealized limit---it describes the evolution of both the system and apparatus under measurement \cite{AaronDiegoFuture, AaronMIPT, MW}. A key advantage of this  unitary representation is the ability to evolve operators in the Heisenberg picture in the presence of generic quantum channels, leading to the main bound \eqref{eq:main}.

However, the locality of measurements is also quite transparent applying the Stinespring formalism to the evolution of \emph{states}. For example, consider the Bell state \eqref{eq:Bell state}, where one qubit is sent to Alice and the other to Bob, who each measure $\PZ{}$ on their respective qubits. The dilated state prior to measurement is
\begin{align}
    \ket{\text{Bell}} \, &= \, \frac{1}{\sqrt{2}} \left( \ket{00} + \ket{11} \right)^{\,}_{AB} \otimes \ket{00}^{\,}_{ab} \notag \\
    &= \, \frac{1}{\sqrt{2}} \left( \ket{00,00} + \ket{11,00} \right)\, , ~~\label{eq:Bell pre meas ZZ}
\end{align}
where, in the second line, the qubits correspond to $\ket{AB,ab}$ with $A$ and $B$ the physical qubits for Alice and Bob and $a$ and $b$ the corresponding Stinespring qubits (i.e., the states of the two measurement apparati). 

The measurement channels \eqref{eq:Measurement Unitary} corresponding to the two $\PZ{}$ measurements are given by $\umeas^{\,}_{A} = \text{CNOT}(A \to a)$ and $\umeas^{\,}_{B} = \text{CNOT}(B \to b)$. Following the measurements, the state of the physical and Stinespring qubits is
\begin{align}
    \ket{\Psi} \, &= \, \umeas^{\vpd}_{A} \, \umeas^{\vpd}_{B} \, \ket{\text{Bell}} \, = \, \umeas^{\vpd}_{B} \, \umeas^{\vpd}_{A} \, \ket{\text{Bell}} \notag \\
    &= \,  \frac{1}{\sqrt{2}} \left( \ket{00,00} + \ket{11,11} \right) \, , ~~
    \label{eq:Bell post meas ZZ}
\end{align}
i.e., each measurement merely entangles the state of the apparatus with the state of the physical qubit. 

More importantly, this entangling interaction is local. Note that the $a$ and $A$ qubits must be close by for the measurement to occur (and likewise for the $b$ and $B$ qubits). Additionally, the order of measurements is unimportant, and neither measurement ($A$/$B$) affects the state of the other physical qubit ($B$/$A$), suggesting that measurements alone cannot send quantum information nor generate entanglement or correlations.

In fact, this conclusion is even more generic. Provided that the state of the physical qubits is represented in the eigenbasis of the operator to be measured, measurements have \emph{no effect whatsoever} on the physical state. For example, suppose that Alice intends to measure $\PZ{}$ while Bob intends to measure $\PX{}$. In the $\PZ{A} \otimes \PX{B}$ basis, the same Bell state \eqref{eq:Bell pre meas ZZ} takes the form
\begin{align}
    \label{eq:Bell pre meas ZX}
    \ket{\text{Bell}} \, &= \, \frac{1}{2} \left( \ket{00} + \ket{01} + \ket{10} - \ket{11} \right) \otimes \ket{00}\, , ~~
\end{align}
and because we have written $\ket{\text{Bell}}$ \eqref{eq:Bell pre meas ZX} in the measurement basis, the measurement channels again act as CNOT gates. After the two measurements, we find
\begin{align}
    \ket{\Psi}  &=  \umeas^{\vpd}_{A} \, \umeas^{\vpd}_{B} \, \ket{\text{Bell}} \, = \, \umeas^{\vpd}_{B} \, \umeas^{\vpd}_{A} \, \ket{\text{Bell}} \notag \\
    &=  \frac{1}{2} \left( \ket{00,00} + \ket{01,01} + \ket{10,10} - \ket{11,11} \right) \, , 
    \label{eq:Bell post meas ZX}
\end{align}
so that the physical state is completely unaltered by the measurements. This is generic to any set of commuting measurement channels applied to qubits, when the state is expressed in the measurement basis. 

The unitary representation of measurements \cite{AaronMIPT,AaronDiegoFuture} applied to states elucidates why there is nothing paradoxical about measuring Bell states \cite{epr,Bell}. Additionally, this picture shows that information is not transferred---nor can entanglement or correlations be generated---using measurements of entangled states alone. 

However, as demonstrated by numerous protocols detailed herein, the inclusion of a classical communication channel indicating the outcomes of measurements, followed by a quantum error-correction channel conditioned on those outcomes, is not only compatible with---but can expedite by a factor of $\Nmeas+1$---the transfer of quantum information and generation of entanglement and/or correlations, provided that the bound \eqref{eq:main} is obeyed.

\subsection{Summary of results}
\label{subsec:Summary}

We prove an emergent, finite speed of quantum information in the presence of arbitrary local quantum dynamics, projective measurements, instantaneous classical communications, and outcome-dependent operations captured by the main bound \eqref{eq:main}. That bound extends to arbitrary, finite-dimensional degrees of freedom assigned to any graph $G$ in any spatial dimension $\SpaceDim$. The bound holds for both time-dependent Hamiltonians and quantum circuits acting on connected regions of bounded size, which  may depend arbitrarily on prior measurement outcomes. The bound extends the Lieb-Robinson Theorem~\cite{Lieb1972} to generic quantum dynamics with measurements, feedback, and instantaneous classical communication for the first time.

The bound also captures generic local quantum channels (e.g., weak measurements). Additionally, we derive the bound \eqref{eq:mainQ} for the generation of $Q$ Bell pairs and the teleportation of $Q$ logical qubits, and recover numerous bound on the preparation of various classes of correlated states in Sec.~\ref{subsec:Corr bounds}. In this sense, our bounds constrain \emph{generic quantum tasks}, which either generate or manipulate quantum information, entanglement, or correlations using local operations, which may depend on the outcomes of measurements, whose communication is instantaneous. 

Prior to this work---and the development of the Stinespring formalism (see also \cite{AaronMIPT,AaronDiegoFuture})---such a feat was not thought possible. Not only was the mathematical machinery required to evolve operators out of reach, but conventional wisdom held that measurements \emph{destroy} any emergent notion of spatial locality. Our bound \eqref{eq:main} proves that this is \emph{not} the case by establishing a finite, $\Order{\Nmeas}$ enhancement to the measurement-free Lieb-Robinson velocity $\LRvel$ \cite{Lieb1972, fossfeig, chen2019finite, Kuwahara:2019rlw, Tran:2020xpc, gross, schuch, kuwahara2021liebrobinson, yin21} provided that the outcomes of measurements in $\Nmeas$ local regions are known and utilized.

Our bound \eqref{eq:main} constrains the most efficient quantum protocols (which feature measurements) for transferring quantum information and generating long-range entanglement and/or correlations, while highlighting the essential role of error-correcting feedback in performing generic useful quantum tasks. Applications of \eqref{eq:main} include quantum error correction, measurement-based quantum computation, quantum routing, and the preparation of generic long-range-entangled many-body states of interest to condensed matter and atomic physics. We provide explicit, optimal protocols for numerous such tasks, establishing optimality of the bound \eqref{eq:main} in these contexts. 

As measurements both increase the speed of information \emph{and} the additional qubits and resources necessary for active error-correction,  our bound may elucidate fundamental limits on the operation of a large-scale quantum information processor built out of physical qubits. 

As with the Lieb-Robinson Theorem \cite{Lieb1972} itself, there is no telling what new applications of the bound \eqref{eq:main} the future may hold in store. In the near term, we expect that applications of \eqref{eq:main} and the Stinespring 
formalism \cite{AaronMIPT} 
will provide deep insight through more specialized consideration of various quantum tasks \cite{AaronYifanFuture, ChaoWFuture}. Additionally, \eqref{eq:main} provides for a more general classification of phases of matter, as it proves that states with short- and long-range entanglement cannot be connected by protocols $\mathcal{W}$ for which the \emph{combined} resources $\Nmeas \times T$ are finite, generalizing the concept of finite-depth circuits. More broadly, we hope that our results will lead to more efficient strategies for preparing useful resource states, achieving useful tasks, and optimizing compiling on near-term quantum devices.

\section*{Acknowledgements}
We thank Emanuel Knill and Rahul Nandkishore for feedback on this manuscript;  AJF also thanks Rahul Nandkishore for introducing Stinespring channels.  This work was supported by a Research Fellowship from the Alfred P. Sloan Foundation under Grant FG-2020-13795 (AL), by the U.S. Air Force Office of Scientific Research under Grant FA9550-21-1-0195 (CY, YH, AL), and by a Simons Investigator Award via Leo Radzihovsky (AJF).

\let\oldaddcontentsline\addcontentsline%
\renewcommand{\addcontentsline}[3]{}%
\bibliography{thebib}
\let\addcontentsline\oldaddcontentsline%

\bigskip

\onecolumngrid
\newpage



\section*{Supplementary material}

# Supplemental Material:
# Locality and error correction in quantum dynamics with measurement

Aaron J. Friedman, Chao Yin, Yifan Hong, and Andrew Lucas

*Department of Physics and Center for Theory of Quantum Matter,*
*University of Colorado, Boulder, CO 80309, USA*

(Dated: July 3, 2023)

**CONTENTS**

## 1. OVERVIEW

This Supplemental Material (SM) contains numerous technical details relevant to the results of the main text. A summary of the contents and main results of the SM appears below, for convenience. Importantly, we provide an overview of the derivation of the main bounds that is unencumbered by technical details. A summary of the main results—Theorem 5 and the corresponding main bound (S177)—appears in Sec. 6.9.

*Stinespring dilation, measurement, and error correction*

In Sec. 2, we introduce a unitary representation of measurement channels [S1–S4]. Using the Stinespring Dilation Theorem [S1], we first represent measurements using *isometries*. By introducing ancillary "Stinespring qubits" to store the measurement outcomes and identifying a "default" outcome, we can embed the isometries in *unitary* measurement channels that act on the *dilated* Hilbert space $\mathcal{H}_{\rm dil} = \mathcal{H}_{\rm ph} \otimes \mathcal{H}_{\rm ss}$. The Stinespring (or "outcome") qubits are contained in $\mathcal{H}_{\rm ss}$, and the default outcome is "0" (the +1 eigenvalue). This formulation allows for the evolution of observables under both time-evolution and measurement channels in the Heisenberg-Stinespring picture [S3].

In Sec. 3, we elucidate error correction in the three-qubit repetition code [S5] by using the Heisenberg-Stinespring formalism to evolve logical operators. This approach clarifies the precise manner in which the outcomes of measurements are used to determine the error correction unitary, and why this step is essential to the measurements being useful.

Importantly, while we explicitly consider *projective* measurements (which is the most useful quantum channel besides time evolution for executing useful tasks), our results extend to arbitrary local quantum channels (e.g., weak measurements). Essentially, any local quantum channel has a Stinespring representation [S1] with the same local structure as projective measurements, and thus obey the same bounds. We do not consider examples involving these other channels, as they are not generally useful to any quantum task; however, they are constrained by our bounds.

*TFIM teleportation protocol*

In Sec. 4 we present a state-transfer (or "quantum teleportation") protocol based on the Transverse Field Ising Model (TFIM) [S6]. This protocol is asymptotically optimal with respect to our main bound (S119) derived in Sec. 6, which is also the first equation of the main text. Like the entanglement-swapping teleportation protocol (ESTP) of the main text, the TFIM code allows for tradeoffs between the number of measurements $M$ and the circuit depth $T$, depending on the overhead and errors associated with unitary gates and measurements in a particular platform.

We provide a quick overview of Clifford circuits before detailing the TFIM code; we then use the Heisenberg-Stinespring picture for the logical operators to highlight the importance of error correction. The teleportation distance $L$ saturates the main bound (S119), and is given by (S62):

$$L_{\rm TFIM} \,=\, 2\,M\,(T-1)\,,$$

where $M$ is the number of two-site measurement *regions* and $T$ is the total time over which unitary gates are applied (see Fig. S1). For comparison, the ESTP presented in the main has teleportation distance

$$L_{\rm ESTP} \,=\, (2\,M+1)\,(T-1)+1\,,$$

which also saturates (S119). Using Bell-basis measurements instead of Bell decoding and pairs of $Z$ measurements, one finds that $L_{\rm ESTP} \leq (2M+1)T - 1$. By convention, $T$ is the minimum number of layers of multi-site unitary gates (acting in the *dilated* Hilbert space $\mathcal{H}_{\rm dil}$) needed to realize the protocol, parallelizing where possible.

*Equivalence of useful quantum tasks*

In Sec. 5, we move toward the establishment of general bounds on quantum information dynamics with measurements, setting the stage for the main proofs for continuous time dynamics in Sec. 6. In Sec. 5.1 we prove Proposition 1, establishing a precise connection between operator spreading and quantum teleportation. In Sec. 5.2 we prove Proposition 2, directly relating protocols that achieve teleportation over distance $L$ to those that create and separate a Bell state over distance $L$ up to $\mathsf{O}(1)$ corrections to $T$, $M$, and $N$. These Propositions provide for the derivation of bounds on generic useful quantum tasks—such as quantum teleportation and the generation of entangled resource states—by considering operator growth in a protocol that generates a well-separated Bell pair.

*Lieb-Robinson bounds with measurements*

The main conceptual achievement of this work is presented in Sec. 6, where we derive a general bound on the speed of quantum information in the presence of measurements (S119). In the main text, we explain how this bound extends to generic quantum channels as well. The main bound follows from Theorem 5, which we state in Sec. 6.4 and prove in Sec. 6.7. We frame Theorem 5 to be as general as possible, and numerous ingredients are therefore required in its proof—we divide the full proof into numerous steps, and prove various results along the way.

We state the main Theorem, extract the bound (S117), and provide some analysis thereof in Sec. 6.4. Theorem 5 constrains the time $T$ and number of measurement regions $M$ required to prepare a Bell state on qubits $i$ and $f$ separated by a distance $L = d(i,f)$—which is equivalent to teleporting a state between the qubits $i$ and $f$ by Proposition 2—starting from the initial state $|\mathbf{0}\rangle = \bigotimes_{v \in V} |0\rangle$, where $V$ is the set of vertices of an arbitrary graph $G$ defining the physical system. In either case, the "task distance" $L$ obeys the bound (S117)

$$L \leq 2\,(M+1)\,v\,T\,,$$

and subsequent Theorems and Corollaries extend this bound to more generic tasks and initial states, replacing (S117) above with the more general bound (S119) that appears in the main text.

The bound above is compatible with continuous-time dynamics generated by arbitrary local, time-dependent Hamiltonians $H$ (S86) with maximum coupling[1] $h$ (S87) and associated Lieb-Robinson velocity $v$ (S92), which follows from (S90) of the standard Lieb-Robinson Theorem [S7]. The local terms $H_j(t)$ may depend on the outcomes of prior measurements, which are instantaneously communicated. These details are discussed in Sec. 6.1, and we note that the bound (S117) depends on the details of the Hamiltonian $H(t)$ only through $v$ (S92); while we consider nearest-neighbor Hamiltonians in Sec. 6.1 for concreteness, our results generalize straightforwardly to arbitrarily local Hamiltonians, as these details only affect $v$ (S92). We expect a similar generalization to power-law interactions to hold.

In Sec. 6.2 we present formal details of Stinespring measurement trajectories (S15) and describe the construction of the "measurement set" $\mathbb{M}$, which contains the various local "measurement regions" $S$. The number of measurements $M$ is defined in (S94), and the range $l$ of measurements is defined in (S95). We allow for fully adaptive protocols in which the Hamiltonian dynamics and measurements at time $t$ may be conditioned on the outcomes of prior measurements made at times $t' < t$. The only other assumption required is that $T \gtrsim D$ (S111)[2].

In Sec. 6.3 we introduce the trace distance (S108) and the strategy for deriving the short-time bound (S112) in Theorem 5. Essentially, given some generic dilated quantum channel $\mathcal{W}$, we define a "reference" channel $\widetilde{\mathcal{W}}$ that, compared to $\mathcal{W}$, cannot generate entanglement (or correlations) across some cut $C$ of the system (where, e.g., the cut $C$ separates the two qubits $i$ and $f$ in the Bell pair, or the initial and final logical qubits in teleportation). This result relies on Lemma 4, which provides for the construction of reference protocols for particular trajectories, whether or not the desired outcomes obtain in the reference protocol. After stating the main Theorem in Sec. 6.4 and commenting on its interpretation, in Sec. 6.5 we prove (S112) for semi-adaptive protocols in Proposition 6 (where only the measurement locations are nonadaptive—i.e., the measurement set $\mathbb{M}$ is the same for all trajectories $\boldsymbol{n}$). That proof assumes that a cut can be chosen far from all measurement sets (S132), which we prove in Lemma 7 in Sec. 6.6.

We note that the most general version of the bound is of the form $L \leq 2\,(M+1)\,(v\,T + (D-1)\,\log_2 L)$, where we strongly suspect that the $\log_2 L$ term is not physical, but a spurious artifact of the proof strategy. Importantly, this term vanishes (*i*) in $D=1$, (*ii*) if the measurements are "prefixed" (rather than "adaptive"), and (*iii*) if the time evolution is discrete (i.e., generated by a quantum circuit). These three scenarios capture most of the cases of interest. Moreover, we expect that an alternate strategy using the fact that each measurement region is measured at most $\sim vT$ times (a reasonable assumption) could circumvention of this factor. However, such a complete reformulation of our already lengthy proof constitutes a technical refinement that is beyond the scope of this work.

Essentially at short times, Theorem 5 shows that the true state $\rho$ generated by the true protocol $\mathcal{W}$ is indistinguishable from the *separable* reference state $\widetilde{\rho}$ generated by $\widetilde{\mathcal{W}}$ (S112), and thus cannot contain useful entanglement or correlations. This holds for times $t \gtrsim r/v$, where $r$ is the distance from the cut $C$ to the nearest measurement region. Lemma 7 guarantees that there always exists a bipartition $C$ of the system that is sufficiently far from all measurement sets, with $r \sim L/2(M+1)$. This suggests that the task distance obeys $v\,t \gtrsim L/2(M+1)$, which resembles the main bound.

Finally, in Sec. 6.7 we prove Theorem 5 by showing that the generic case—in which all aspects of the protocol $\mathcal{W}$ at any time $t$ may depend on the outcomes of prior measurements—does not change the result of Proposition 6, proving Theorem 5. The saturation of (S117) by the ESTP and TFIM codes establish that the bound is asymptotically optimal, and rules out the possibility of meaningfully improving these protocols.

---

[1] The usual operator norm $\|H_j(t)\|$ is the maximal eigenvalue of the local term $H_j(t)$, which satisfies $\|H_j(t)\| \leq h$.

[2] The standard Lieb-Robinson Theorem [S7] also recovers in the asymptotic limit of $T, L \gg 1$.

*Extension to entangled initial states*

In Sec. 6.8 we extend the main bound (S119) to initial states with short-range entanglement (which is specified precisely in Definition 8), as well as initial states that can be prepared from some short-range-entangled (SRE) state $\rho_0$ (S159), where $\rho_0$ itself can be prepared from a separable (or product) state $\rho'$ using a finite-resource protocol $\mathcal{W}_0$ involving measurements in $M'_0$ local regions and evolution for duration $T'_0$.

Corollary 9 extends Theorem 5 to such entangled initial states. In the simple case of an SRE initial state (with entanglement range $\xi \ll L$), Theorem 5 applies directly, as long as the cut $C$ that defines the reference protocol has thickness $\xi$ or greater. In the more general case, the protocol $\mathcal{W}$ achieves a useful task (e.g., teleportation) starting from the initial state $\rho_0 = \mathcal{W}_0 \rho' \mathcal{W}_0^\dagger$, which can be prepared using the finite-resource protocol $\mathcal{W}_0$ from the SRE state $\rho$. Noting that the full protocol $\mathcal{W}' = \mathcal{W}\mathcal{W}_0$ obeys the main bound (S119) without modification, we infer a bound on $\mathcal{W}$. In the best case, the initial state $\rho_0$ would have been realized by $\mathcal{W}'$ starting from $\rho'$, and thus $\rho_0$ provides an advantage, in that $\mathcal{W}$ achieves the task distance of $\mathcal{W}'$. Hence, we infer

$$L \leq 2\,(M + M'_0 + 1)\,v\,(T + T'_0) \;=\; (2\,M + M_0)\,v\,(T + T_0)\,,$$

where the latter expression corresponds to the bound (S119) that appears in the main text. Note that the advantage provided by such entangled initial states $\rho_0$ in, e.g., teleporting quantum information or preparing entangled and/or correlated resource states (i.e., the offsets $T'_0$ and $M'_0$) cannot scale with $L, T, M$. We also absorb arbitrary other $\mathsf{O}(1)$ corrections $\delta M$ and $\delta T$ into the definitions of $M_0$ and $T_0$, respectively, via $T_0 = T'_0 + \delta T$ and $M_0 = 2M'_0 + 2$. We note that $M_0$ and $T_0$ may be positive if the entangled initial state $\rho_0$ is useful to the protocol $\mathcal{W}$ relative the state $\rho'$, and may be negative if the state $\rho_0$ is less useful than $\rho'$. In general, $M'_0 = T'_0 = 0$ for a product state, and are otherwise $\mathsf{O}(1)$ quantities. We also absorb the $\mathsf{o}(L)$ correction $\delta T = v^{-1}\,(D-1)\log_2 L$ into $T_0$ in (S119).

*Bounds on preparing other entangled resource states*

In Sec. 7 we consider the preparation of other useful entangled states via measurement-assisted protocols. In Sec. 7.1 we present Theorem 11, which proves a bound on the preparation of states with nonvanishing correlations (rather than entanglement) over distance $L$. That bound is asymptotically equivalent to the main bound (S119) and applies to Bell-pair distillation and the preparation of the Greenberger-Horne-Zeilinger (GHZ) state [S8] defined in (S166). In Sec. 7.2, we present a protocol for preparing the GHZ state (S166) that saturates (S119), meaning that the main bound (S119) is also optimal for GHZ. Importantly, our protocol admits tradeoffs between $M$ and $T$.

In Sec. 7.3 we present a bound on the preparation of permutation-symmetric states of qubits in the Dicke manifold (S208) [S9]. Among these states is the W state (S171) [S10], which for $N \geq 3$ has a distinct type of entanglement compared to Bell and GHZ states [S11]. The bound for generic Dicke states $|D_k\rangle$ (including the W state $|W\rangle = |D_1\rangle$) appears in Corollary 12. In contrast to the aforementioned bounds, the bound (S210) for preparing the $k$th Dicke state contains a $\log_2 L$ correction to $T$ even for prefixed measurements. However, for the case of the W state (S171) or Dicke states $|D_k\rangle$ with finite correlations between pairs of qubits, we instead recover the bound $L \leq 3MvT$ (S214).

In Sec. 7.4, we present a protocol that prepares the W state using projective measurements and non-Clifford gates. Without measurements, this protocol requires depth $T \lesssim N$ (S218), saturating the measurement-free Lieb-Robinson bound (S90). Using measurements to minimize circuit depth, we find $T = \mathsf{O}(\log N)$ and $M = \mathsf{O}(N)$ (S219), which does *not* saturate the bound (S210). We believe that $T \gtrsim \log N$ is a generically true bound, and we relegate the development of protocols with optimal $M \sim N/\log N$ to future work [S12]. Unlike the GHZ state (S166), the operators that stabilize the W state (S171) are not local Pauli strings. Hence, the W state cannot be prepared using Clifford gates, and it may not be possible to "patch" local W states into a single W state (unlike with GHZ).

In Sec. 7.5, we present Corollary 13, which bounds the preparation of spin-squeezed states [S13] in arbitrary dimensions. Finally, in Sec. 7.6, we present Corollary 14, which imposes the related bound (S232) on the preparation of quantum critical states [S14] and states of conformal field theories (CFTs) [S15] with power-law correlations.

*Multi-qubit bounds*

Lastly, we rule out any advantage to teleporting $Q > 1$ qubits in Theorem 15 in Sec. 8 via a proof by contradiction. We show that teleporting $Q > 1$ qubits with $MT \leq QL$ would imply that at least one qubit was teleported using $MT \leq L$, violating the main bound (S119) of Theorem 5. Hence, one can only teleport $Q$ logical qubits in parallel if

$$L \leq 2\,v\,T\left(\left\lfloor \frac{\mathcal{M}}{Q} \right\rfloor + 1\right),$$

where $\mathcal{M}$ is the number of measurement *outcomes* utilized, and $\lfloor \cdot \rfloor$ is the floor function.

## 2. STINESPRING FORMALISM

Measurements of quantum systems are captured by *quantum channels*, which can be described using three equivalent representations: completely positive trace-preserving (CPTP) maps, Kraus operators (which are equivalent to CPTP maps), and finally, isometries [S1–S4]. An isometry is simply a length-preserving map from a Hilbert space with dimension $\mathcal{D}_A$ to another with dimension $\mathcal{D}_B \geq \mathcal{D}_A$; when $\mathcal{D}_A = \mathcal{D}_B$, the isometry is also unitary. We now detail the Stinespring representation of quantum channels using isometries.

### 2.1. Projective measurements

An observable $A$ has a "spectral decomposition" given by

$$A \equiv \sum_{n=1}^{\mathcal{N}} a_n P^{(n)}, \tag{S1}$$

where $a_n$ is one of $\mathcal{N}$ unique eigenvalues of $A$ (which are the possible measurement outcomes) and $P^{(n)}$ satisfies $AP^{(n)} = a_n P^{(n)}$. The projectors are complete ($\sum_n P_n = \mathbb{1}$) and orthogonal ($P^{(m)} P^{(n)} = \delta_{mn} P^{(n)}$). If the eigenvalue $a_n$ is nondegenerate then $P^{(n)} = |n\rangle\langle n|$ simply projects onto the $n$th eigenstate of $A$ ($A|n\rangle = a_n|n\rangle$); if the $n$th eigenvalue is degenerate, then $P^{(n)}$ is a sum over projectors onto each individual eigenstate of $A$ with eigenvalue $a_n$.

Our interest lies in measurements of systems of qubits, where the operators of interest are always Kronecker products of Pauli matrices (e.g., $X_i$, $Z_i$, $Z_i Z_j$, and so on). All such Pauli operators have two eigenvalues $-1$ and $+1$: Single-site Pauli operators have nondegenerate spectra, while $\ell$-site Pauli *strings* have degeneracy $2^{\ell-1}$. For a Pauli-string operator $\mathcal{O}$ acting on $\ell$ sites, the projectors onto the $\pm 1$ eigenspaces are given by

$$P^{(\pm)} \equiv \frac{1}{2} \left( \mathbb{1} \pm \mathcal{O} \right), \tag{S2}$$

which, for the single-site operator $Z_j$, are simply the projectors onto the states $|0\rangle_j$ and $|1\rangle_j$. It is straightforward to verify that these projectors are complete, orthogonal, and idempotent.

### 2.2. Isometric measurements and the dilated Hilbert space

The intuition behind "dilating" the Hilbert space upon measurement is that one includes additional degrees of freedom to encode the quantumness of the measurement process itself (i.e., entanglement between the observer and the system) [S1]. The dilated Hilbert space then encodes both the post-measurement state of the system *and* the observer (whose state reflects the observed outcome); the dilated Hilbert space is given by $\mathcal{H}_{\text{dil}} = \mathcal{H}_{\text{ph}} \otimes \mathcal{H}_{\text{ss}}$, where the former is the physical Hilbert space and the latter is the "Stinespring" (or "outcome") Hilbert space, which stores the measurement outcome quantumly, entangling the measurement result itself with the system's post-measurement state.

The Stinespring formulation of quantum measurement follows from the Stinespring Dilation Theorem [S1], which guarantees that quantum channels can be represented isometrically. Measuring the $\ell$-site observable $\mathcal{O}$ in some many-body state $|\psi\rangle$ is captured by the isometric quantum channel

$$|\psi\rangle \to |\psi'\rangle = \sum_{\pm} \left( P^{(\pm)} |\psi\rangle \right) \otimes |\pm\rangle_{\text{ss}}, \tag{S3}$$

which entangles the post-measurement state of the system with the outcome $|\pm\rangle_{\text{ss}} \in \mathcal{H}_{\text{ss}}$, whose basis satisfies $\langle \mu | \nu \rangle_{\text{ss}} = \delta_{\mu,\nu}$ for $\mu, \nu = \pm 1$. This isometry maps the physical Hilbert space $\mathcal{H}_{\text{ph}}$ where $|\psi\rangle$ lives to the *dilated* Hilbert space $\mathcal{H}_{\text{dil}} = \mathcal{H}_{\text{ph}} \otimes \mathcal{H}_{\text{ss}}$. Note that there is no need to renormalize the post-measurement state of the system explicitly, in contrast to the more-ubiquitous Copenhagen picture, as isometries preserve norms.

The fact that this mapping is isometric can be verified by evaluating the post-measurement norm,

$$\langle \psi' | \psi' \rangle = \sum_{\mu,\nu=\pm} \left\langle \psi \middle| P^{(\mu)} P^{(\nu)} \middle| \psi \right\rangle \otimes \langle \mu | \nu \rangle_{\text{ss}} = \sum_{\pm} \left\langle \psi \middle| P^{(\pm)} \middle| \psi \right\rangle = \langle \psi | \psi \rangle = 1, \tag{S4}$$

by completeness of the projectors and orthonormality of the Stinespring kets. Writing $\mathbb{V} : |\psi\rangle \mapsto |\psi'\rangle$, the isometry $\mathbb{V}$ satisfies $\mathbb{V}^\dagger \mathbb{V} = \mathbb{1}_{\text{ph}}$, while $\mathbb{V}\mathbb{V}^\dagger = P_{\text{ph}}$ projects onto the physical subspace $\mathcal{H}_{\text{ph}} \subset \mathcal{H}_{\text{dil}}$.

## 2.3. Unitary measurement and dilated time evolution

To avoid extending (or "dilating") $\mathcal{H}$ explicitly with each measurement, we include outcome qubits for each anticipated measurement from the outset. In this case, $\mathbb{V}$ acts purely within $\mathcal{H}_{\text{dil}}$, and is therefore a *unitary* operator. Thus, by extending the Hilbert space to include the Stinespring qubits from the outset, the isometry implied in (S3) realizes a *unitary* operator acting on $\mathcal{H}_{\text{dil}}$. By convention, all Stinespring qubits are initialized in the "default" state $|0\rangle$. Going forward, we exclusively use 0 and 1 to denote the state of a qubit, where $Z$ acts as $Z|n\rangle = (-1)^n |n\rangle$.

The labelling convention for the Stinespring (or "outcome") degrees of freedom depends upon the number and nature of the planned measurements. For a single round of single-site measurements, e.g., we simply label the Stinespring qubit that encodes the result of measuring site $j$ as $|0\rangle_{\text{ss},j}$; the Pauli operators that act on this $j$th Stinespring qubit are denoted $\widetilde{X}_j$, $\widetilde{Y}_j$, and $\widetilde{Z}_j$, where the tildes distinguish these operators from those that act on the *physical* qubit $j$[3]. If the result of measuring the observable $Z_j$ is $+1$ (corresponding to "0" in the Stinespring register), we apply $\widetilde{\mathbb{1}}_j$ to the Stinespring qubit (i.e., leave the outcome qubit in its default state, $|0\rangle_{\text{ss},j}$); if the result is $-1$ (corresponding to "1" in the Stinespring register), we apply $\widetilde{X}_j$, sending the "default" state $|0\rangle_{\text{ss},j}$ to $|1\rangle_{\text{ss},j}$.

If we restrict to measurements of involutory Pauli-string operators $A$ (i.e., $A$ acts on every site $j$ as *either* $\mathbb{1}$, $X_j$, $Y_j$, or $Z_j$, so that $A^2 = \mathbb{1}$), then the unitary channel that respresents the measurement of $A$ is given by

$$\mathsf{V}_{[A]} = P^{(+)} \otimes \widetilde{\mathbb{1}}_n + P^{(-)} \otimes \widetilde{X}_{\text{ss}} = \frac{1}{2}(\mathbb{1} + A)\widetilde{\mathbb{1}}_{\text{ss}} + \frac{1}{2}(\mathbb{1} - A)\widetilde{X}_{\text{ss}} = \mathsf{V}_{[A]}^\dagger, \tag{S5}$$

which, in the case of Pauli measurements, is also Hermitian. The unitary operator $\mathsf{V}$ reproduces (S3) acting on the initial state $|\psi\rangle \otimes |0\rangle_{\text{ss},j}$. Note that $\mathsf{V}$ acts like a controlled NOT (CNOT) gate (S39), with the physical qubit the "control" qubit, and the Stinespring qubit the "target" qubit.

The Heisenberg update rule for operators $\mathcal{O} = B \otimes \widetilde{C}$ under measurement is given by

$$\mathcal{O}' = \mathsf{V}_{[A]}^\dagger \mathcal{O} \mathsf{V}_{[A]} = \mathsf{V}_{[A]}^\dagger B \otimes \widetilde{C} \mathsf{V}_{[A]} \tag{S6}$$

$$= \frac{1}{4}(B + AB + BA + ABA) \otimes \widetilde{C} + \frac{1}{4}(B - AB - BA + ABA) \otimes \widetilde{X}\widetilde{C}\widetilde{X}$$

$$+ \frac{1}{4}(B + BA - AB - ABA) \otimes \widetilde{X}\widetilde{C} + \frac{1}{4}(B - BA + AB - ABA) \otimes \widetilde{C}\widetilde{X} \tag{S7}$$

$$= \frac{1}{4}\{B, A\} A \otimes \{\widetilde{C}, \widetilde{X}\}\widetilde{X} + \frac{1}{4}\{B, A\} \otimes [\widetilde{C}, \widetilde{X}]\widetilde{X}$$

$$+ \frac{1}{4}[B, A] A \otimes \{\widetilde{C}, \widetilde{X}\} - \frac{1}{4}[B, A] \otimes [\widetilde{C}, \widetilde{X}], \tag{S8}$$

and because $\mathsf{V}$ is also Hermitian, the above update also applies to density matrices in the Schrödinger picture.

Generally speaking, operators evolved in the Heisenberg picture act trivially on $\mathcal{H}_{\text{ss}}$ (i.e, $\widetilde{C} = \widetilde{\mathbb{1}}_{\text{ss}}$), so that

$$\mathsf{V}_{[A]}^\dagger B \otimes \widetilde{\mathbb{1}} \mathsf{V}_{[A]} = \frac{1}{2}\{B, A\} A \otimes \widetilde{\mathbb{1}} + \frac{1}{2}[B, A] A \otimes \widetilde{X}, \tag{S9}$$

and decomposing $B$ as a sum over Pauli strings, each string in $B$ *either* commutes or anticommutes with $A$, and only one term above is nonzero. If $A$ and $B$ commute, the result is $B \otimes \widetilde{\mathbb{1}}$; if $A$ and $B$ *anti*commute, the result is $B \otimes \widetilde{X}$.

However, in the context of error correction (and also when evolving density matrices), we may also encounter terms that involve projectors onto particular Stinespring outcome states. In this case, the observable $\mathcal{O}$ (or density matrix $\varrho$), may act on the Stinespring register as $\widetilde{C} = \widetilde{Z}_{\text{ss}}$, in which case we have

$$\mathsf{V}_{[A]}^\dagger B \otimes \widetilde{Z} \mathsf{V}_{[A]} = \frac{1}{2}\{B, A\} \otimes \widetilde{Z} + \frac{1}{2}[B, A] \otimes \widetilde{X}\widetilde{Z}, \tag{S10}$$

and performing similar calculations for $\widetilde{C} = \widetilde{X}$ and $\widetilde{Y}$, we find the relations

$$\mathsf{V}_{[A]}^\dagger B \otimes \widetilde{X} \mathsf{V}_{[A]} = \mathsf{V}_{[A]}^\dagger B \otimes \widetilde{\mathbb{1}} \mathsf{V}_{[A]} \left(\mathbb{1} \otimes \widetilde{X}\right) \tag{S11a}$$

$$\mathsf{V}_{[A]}^\dagger B \otimes \widetilde{Y} \mathsf{V}_{[A]} = \mathsf{V}_{[A]}^\dagger B \otimes \widetilde{\mathbb{1}} \mathsf{V}_{[A]} \left(A \otimes \widetilde{Y}\right) \tag{S11b}$$

$$\mathsf{V}_{[A]}^\dagger B \otimes \widetilde{Z} \mathsf{V}_{[A]} = \mathsf{V}_{[A]}^\dagger B \otimes \widetilde{\mathbb{1}} \mathsf{V}_{[A]} \left(A \otimes \widetilde{Z}\right), \tag{S11c}$$

---

[3] In more general contexts in which a site is measured at multiple times, such as hybrid quantum circuits [S3, S16, S17], the Stinespring degrees of freedom carry spatiotemporal labels (essentially, one creates a "spacetime lattice" to store measurement outcomes) [S3].

|  | $[\mathcal{O}, A] = 0$ | $\{\mathcal{O}, A\} = 0$ |
|---|---|---|
| $\widetilde{\mathcal{O}} = \widetilde{\mathbb{1}}$ | $\mathcal{O} \otimes \widetilde{\mathbb{1}}$ | $\mathcal{O} \otimes \widetilde{X}$ |
| $\widetilde{\mathcal{O}} = \widetilde{X}$ | $\mathcal{O} \otimes \widetilde{X}$ | $\mathcal{O} \otimes \widetilde{\mathbb{1}}$ |
| $\widetilde{\mathcal{O}} = \widetilde{Y}$ | $\mathcal{O} A \otimes \widetilde{Y}$ | $\mathrm{i}\,\mathcal{O} A \otimes \widetilde{Z}$ |
| $\widetilde{\mathcal{O}} = \widetilde{Z}$ | $\mathcal{O} A \otimes \widetilde{Z}$ | $-\mathrm{i}\,\mathcal{O} A \otimes \widetilde{Y}$ |
| $\widetilde{\mathcal{O}} = \widetilde{P}^{(\pm)}$ | $\frac{1}{2}\mathcal{O}\left(\mathbb{1}\otimes\widetilde{\mathbb{1}} \pm A \otimes \widetilde{Z}\right)$ | $\frac{1}{2}\mathcal{O}\left(\mathbb{1}\otimes\widetilde{X} \mp \mathrm{i} A \otimes \widetilde{Y}\right)$ |

TABLE 1. "Lookup" table for the updates $\mathsf{V}_{[A]}\,\mathcal{O}\otimes\widetilde{\mathcal{O}}\,\mathsf{V}_{[A]}^{\dagger} = \mathsf{V}_{[A]}^{\dagger}\,\mathcal{O}\otimes\widetilde{\mathcal{O}}\,\mathsf{V}_{[A]}$ to density matrices or operators corresponding to unitary measurement of observable $A$. The operator $\mathcal{O}\otimes\widetilde{\mathcal{O}}$ acts as $\mathcal{O}$ on the physical Hilbert space and as $\widetilde{\mathcal{O}}$ on the Stinespring qubit that stores the outcome of measuring $A$, which acts on each site $j$ as *either* $\mathbb{1}$, $X_j$, $Y_j$, $Z_j$ (i.e., a Pauli string). We decompose $\mathcal{O}$ in this basis so that every $\mathcal{O}$ and $A$ either commute or anticommute.

and the general update rules corresponding to the measurement of Pauli string operator $A$ are given in Tab. 1.

As a final remark, the crucial observation regarding this formalism is that the Stinespring measurement unitary (S5) corresponds to the physical time evolution of the physical system and the detector (i.e., the measurement apparatus) during the projective measurement process [S3, S4]. For example, a $Z$-basis "readout" measurement of an atomic qubit is captured by a fluorescent measurement; in the limit where a large number of photons are shone on the atom, the measurement channel acts on the atom and detector as (S5) [S4]. This identification of the measurement unitary with the time evolution of the enlarged (i.e., "dilated") system is critical in that it allows for the evolution of operators in the Heisenberg picture even in the presence of projective measurements *and* outcome-dependent feedback.

### 2.4. Expectation values and measurement trajectories

The advantage of the Stinespring formulation of measurement is the physically intuitive manner in which information about the measurement outcomes is stored (although not having to renormalize wavefunctions explicitly is also appealing). The Stinespring measurement unitary (or isometry) entangles the system and the observer—with the latter codified by the "outcome" register, labelled "ss"—creating an entangled superposition over measurement "trajectories" in which outcome $n$ is recorded *and* the system finds itself in the $n$th eigenstate of the observable in question thereafter.

Information about the measurement outcomes is extracted from the Stinespring qubits. Owing to the entanglement between the outcome register and the physical system, one can easily check that applying the projector $|1\rangle\langle 1|_{\mathrm{ss},j}$ to the Stinespring register that encodes the outcome of measuring site $j$ projects the full system onto the subspace (or "measurement trajectory") in which outcome "1" (i.e., spin down) obtained, up to normalization.

In fact, we find that the probability $p_n$ to obtain outcome $n$ is given by

$$p_n \;=\; \langle\, \mathbb{1}_{\mathrm{ph}} \otimes |n\rangle\langle n|_{\mathrm{ss},j}\,\rangle_{\varrho} \;=\; \mathrm{tr}_{\mathrm{dil}}\left[\,\mathbb{1}_{\mathrm{ph}} \otimes |n\rangle\langle n|_{\mathrm{ss},j}\,\varrho(t)\,\right]\,, \tag{S12}$$

i.e., the expectation value of the projector $\widetilde{P}_{\mathrm{ss}}^{(n)}$ (which acts only on the outcome register) in the state $\varrho(t)$, which is evolved under the measurement unitary in question and time evolved as appropriate.

Expectation values of the measured observables can be extracted straightforwardly by summing over the outcomes $a_n$ (S1) weighted by their probabilities $p_n$ (S12). For an observable $A$ given by (S1), we have

$$\langle A \rangle_\rho \;=\; \langle\, \mathbb{1}_{\mathrm{ph}} \otimes \sum_n a_n\, |n\rangle\langle n|_{\mathrm{ss}}\,\rangle_{\varrho} \;=\; \sum_{n=1}^{\mathcal{N}} a_n \, \mathrm{tr}_{\mathrm{dil}}\left[\,\mathbb{1}_{\mathrm{ph}} \otimes |n\rangle\langle n|_{\mathrm{ss},j}\,\varrho(t)\,\right] \;=\; \sum_{n=1}^{\mathcal{N}} a_n\, p_n\,, \tag{S13}$$

which is equivalent to $\mathrm{tr}\left[A\rho(t')\right]$, where $\rho(t')$ is the *physical* density matrix immediately prior to the projective measurement of $A$. By convention, $\varrho$ denotes density matrices on $\mathcal{H}_{\mathrm{dil}}$, where the initial state is given by

$$\varrho(0) \;\equiv\; \rho \otimes |\mathbf{0}\rangle\langle\mathbf{0}|_{\mathrm{ss}}\,, \tag{S14}$$

where $|\mathbf{0}\rangle\langle\mathbf{0}|_{\mathrm{ss}} = \bigotimes_{j=1}^{\mathcal{M}} |0\rangle\langle 0|_{\mathrm{ss},j}$ represents the default initial state $|0\rangle$ of the Stinespring degrees of freedom, where $\mathcal{M}$ is the number of measured observables. More rigorously, we have $|\mathbf{0}\rangle\langle\mathbf{0}|_{\mathrm{ss}} \in \mathrm{End}\left(\mathcal{H}_{\mathrm{ss}}\right)$ and $\rho \in \mathrm{End}(\mathcal{H}_{\mathrm{ph}})$, where $\mathrm{End}(\mathcal{H})$ is the space of operators (i.e., endomorphisms) acting on a Hilbert space $\mathcal{H}$, which is itself a Hilbert space.

The formula (S12) is easily extended to give the probability for a sequence of measurement outcomes (a *measurement trajectory*). First, we define the object that projects onto the set of measurement outcomes $\boldsymbol{n} = \{n_1, \ldots, n_{\mathcal{M}}\}$,

$$P_{\boldsymbol{n}} \equiv \bigotimes_{j\mathrm{ss}} \widetilde{P}_{\mathrm{ss},j}^{(n_j)} = \bigotimes_{j\in\mathrm{ss}} |n_j\rangle\langle n_j|_{\mathrm{ss},j} = |\boldsymbol{n}\rangle\langle\boldsymbol{n}|_{\mathrm{ss}}, \tag{S15}$$

which acts nontrivially only on the Stinespring qubits, projecting each outcome register $j$ onto the outcome $n_j \in \{0, 1\}$. We then define the *joint probability* for a sequence of outcomes $\boldsymbol{n}$ according to

$$p_{n_1;n_2;\ldots;n_m} \equiv p_{\boldsymbol{n}} = \langle P_{\boldsymbol{n}} \rangle_\varrho = \mathrm{tr}\left[\, \mathbb{1}_{\mathrm{ph}} \otimes |\boldsymbol{n}\rangle\langle\boldsymbol{n}|_{\mathrm{ss},\boldsymbol{j}}\, \varrho(t)\,\right], \tag{S16}$$

where $\boldsymbol{n}$ is the length-$\mathcal{M}$ vector whose components denote which outcome obtained in each of the $\mathcal{M}$ measurements, and $\boldsymbol{j}$ is a length-$\mathcal{M}$ vector whose components label the corresponding Stinespring registers.

Thus, (S16) gives the probability to obtain outcomes $n_1$ *and* $\ldots$ *and* $n_{\mathcal{M}}$. The expectation value of these $\mathcal{M}$ measurements is given in the dilated Hilbert space by

$$\langle A^{(\mathcal{M})} \ldots A^{(1)} \rangle = \langle \mathbb{1}_{\mathrm{ph}} \times \bigotimes_{j=1}^{\mathcal{M}} \sum_{n_j} a_{n_j} |n_j\rangle\langle n_j|_{\mathrm{ss},j} \rangle_\varrho = \mathrm{tr}\left[\varrho(t) \bigotimes_{j=1}^{\mathcal{M}} \sum_{n_j} a_{n_j} |n_j\rangle\langle n_j|_{\mathrm{ss},j}\right], \tag{S17}$$

and finally, we can recover the physical density matrix corresponding to a particular measurement trajectory $\boldsymbol{n}$ for any subset $\Omega$ of the outcome registers, according to

$$\varrho_{\boldsymbol{n}}(t) \equiv \frac{\mathrm{tr}_\Omega\left[\varrho(t)\, \mathbb{1}_{\mathrm{ph}} \otimes \mathbb{1}_{\mathrm{ss},\Omega^{\mathrm{c}}} \otimes |\boldsymbol{n}\rangle\langle\boldsymbol{n}|_{\mathrm{ss},\Omega}\right]}{\mathrm{tr}\left[\varrho(t)\, \mathbb{1}_{\mathrm{ph}} \otimes \mathbb{1}_{\mathrm{ss},\Omega^{\mathrm{c}}} \otimes |\boldsymbol{n}\rangle\langle\boldsymbol{n}|_{\mathrm{ss},\Omega}\right]}, \tag{S18}$$

where $\Omega^{\mathrm{c}}$ includes all outcome registers not in $\Omega$. Note that the numerator involves a partial trace over the outcome registers in $\Omega$ only, while the trace in the denominator is over the full Hilbert space, and equal to $p_{\boldsymbol{n}}$ (S16). One can also recover the density matrix—averaged over outcomes of measurements in $\Omega$—by superposing the density matrices for individual trajectories $\varrho_{\boldsymbol{n}}(t)$ (S18) weighted by their probabilities $p_{\boldsymbol{n}}$ (S16), according to

$$\overline{\rho(t)} \equiv \sum_{\boldsymbol{n}} p_{\boldsymbol{n}}\, \varrho_{\boldsymbol{n}}(t) = \mathrm{tr}_\Omega[\varrho(t)], \tag{S19}$$

is given simply by tracing over all outcome registers in $\Omega$. In other words, at the end of a given calculation, one may consider specific measurement trajectories by applying projectors to the Stinespring registers, or average over all possible outcomes—weighted by their probabilities—by simply tracing out the Stinespring registers (equivalent to summing over outcomes, due to completeness of the outcome projectors). In this work, we will be interested in performing operations (such as quantum error correction) that perform equally well on all individual measurement trajectories. We will then not need to worry about the denominator in (S18), which makes the Stinespring formalism extremely useful as an analytical tool.

Importantly, we note that evaluating expectation values and correlation functions of observables—either averaged over measurement trajectories or along particular measurement trajectories—involves terms of the form $\mathrm{tr}\left[\mathcal{O} \otimes |\boldsymbol{n}\rangle\langle\boldsymbol{n}|_{\mathrm{ss}}\, \varrho\right]$. We can evaluate these quantities in either the Schrödinger or Heisenberg picture, and the latter is especially revealing. Considering operator evolution under measurement unitaries in the Heisenberg picture (S6), we note that if the observable $\mathcal{O}$ does not commute with any of the measurements, then a single $\widetilde{X}$ will appear on the corresponding Stinespring register. However, because $\langle n|X|n\rangle = 0$ for any outcome, $n$, such terms will *always* vanish, because no other operation will ever affect the Stinespring outcome register other than the measurement unitary that produced $\widetilde{X}$ in the first place. Hence, working in the Heisenberg picture, we can "ignore" any term that comes with a Stinespring $\widetilde{X}$ operator, as these terms cannot contribute to any physically observable quantity. More practically, if we wish to perform a quantum error-corrected computation that always succeeds, even a single $\widetilde{X}$ Pauli operator in a time-evolved operator acting on the dilated Hilbert space heralds the failure of our protocol with some nonzero probability. We elaborate on this point in subsequent sections.

### 3. SIMPLE EXAMPLE OF ERROR CORRECTION IN THE STINESPRING FORMALISM

Here we consider error correction in the Stinespring formalism using a minimal error-correcting code: the three-qubit bit-flip repetition code [S5]. The three-qubit code is a toy example of error correction where it is possible to correct for

arbitrary bit-flip ($X$) errors, but not sign ($Z$) errors. Note that the entanglement-swapping teleportation protocol (ESTP) of the main text—along with the transverse field Ising model (TFIM) code considered in Sec. 4—are robust to *both* bit-flip and sign errors, at the cost of requiring more qubits to implement. However, we consider the example of the three-qubit code to provide a quick introduction to quantum error correction and the codification—and subsequent corrective utilization—of measurement outcomes via the Stinespring formalism.

### 3.1. The three-qubit repetition code

We begin by setting up the three-qubit bit-flip error-correction protocol [S5]. The system is initialized in the state

$$|\psi(\alpha,\beta)\rangle = \alpha|000\rangle + \beta|111\rangle, \tag{S20}$$

where the logical operators are defined according to

$$X_{\rm L} = X_1 X_2 X_3 \cdot f_x(\mathbb{1}, Z_1 Z_2, Z_2 Z_3) \tag{S21a}$$
$$Z_{\rm L} = Z_1 \cdot f_z(\mathbb{1}, Z_1 Z_2, Z_2 Z_3), \tag{S21b}$$

where "L" is for "logical" and $f_\mu(\cdot)$ are arbitrary functions of their arguments (the functions labelled $\mu = x, z$ are independent); the operators $Z_1 Z_2$ and $Z_2 Z_3$ are the "syndrome" or "check" operators, which can be used to locate errors without changing the logical qubit. Acting on the state (S20), we find that

$$X_{\rm L} |\psi(\alpha,\beta)\rangle = f_x(\ldots) |\psi(\beta,\alpha)\rangle \tag{S22a}$$
$$Z_{\rm L} |\psi(\alpha,\beta)\rangle = f_z(\ldots) |\psi(\alpha,-\beta)\rangle, \tag{S22b}$$

in analogy to the action of the usual Pauli $X$ and $Z$ operators on the single-qubit state $|\psi(\alpha,\beta)\rangle = \alpha|0\rangle + \beta|1\rangle = (\alpha,\beta)^T$. The advantage of the logical operators (S21) over single-qubit operators is that it is possible to correct for errors while preserving the action (S22).

### 3.2. Correcting an $X$ error: The Schrödinger picture

Suppose that we accidentally apply $X_i$ without knowing the site $i$ on which the error occurred. We also identify $i = 0$ with no bit-flip error having occurred. The resulting state is then

$$|\psi'\rangle = \delta_{i,0}(\alpha|000\rangle + \beta|111\rangle) + \delta_{i,1}(\alpha|100\rangle + \beta|011\rangle) + \delta_{i,2}(\alpha|010\rangle + \beta|101\rangle) + \delta_{i,3}(\alpha|001\rangle + \beta|110\rangle), \tag{S23}$$

which we can correct by simply applying $X_i$ ourselves (or doing nothing if $i = 0$). However, doing so requires that we determine whether and where any errors occurred without spoiling the underlying state.

This is accomplished using *syndrome* measurements of the state $|\psi'\rangle$. In the original state (S20), measuring either syndrome operator $Z_1 Z_2$ or $Z_2 Z_3$ should return $+1$; if we find $-1$ instead, then a bit-flip error occurred. We now proceed using the unitary formulation of measurement, initializing the two syndrome outcome qubits labelled "12" and "23" in the state $|0\rangle$. In the dilated Hilbert space, (S23) becomes

$$|\varphi\rangle \equiv |\psi'\rangle \otimes |0\rangle_{12} \otimes |0\rangle_{23}, \tag{S24}$$

and performing the syndrome measurements sends $|\varphi\rangle \mapsto |\varphi'\rangle$ with

$$\begin{aligned}|\varphi'\rangle &= \frac{1}{2}\left\{(\mathbb{1}_{\rm ph} + Z_1 Z_2)\widetilde{\mathbb{1}}_{12} + (\mathbb{1}_{\rm ph} - Z_1 Z_2)\widetilde{X}_{12}\right\}\frac{1}{2}\left\{(\mathbb{1}_{\rm ph} + Z_2 Z_3)\widetilde{\mathbb{1}}_{23} + (\mathbb{1}_{\rm ph} - Z_2 Z_3)\widetilde{X}_{23}\right\}|\varphi'\rangle \\ &= \frac{1}{4}\left\{\widetilde{\mathbb{1}}_{12}\widetilde{\mathbb{1}}_{23}(\mathbb{1} + Z_1 Z_2 + Z_2 Z_3 + Z_1 Z_3) + \widetilde{X}_{12}\widetilde{\mathbb{1}}_{23}(\mathbb{1} - Z_1 Z_2 + Z_2 Z_3 - Z_1 Z_3) \right. \\ &\quad \left. + \widetilde{\mathbb{1}}_{12}\widetilde{X}_{23}(\mathbb{1} + Z_1 Z_2 - Z_2 Z_3 - Z_1 Z_3) + \widetilde{X}_{12}\widetilde{X}_{23}(\mathbb{1} - Z_1 Z_2 - Z_2 Z_3 + Z_1 Z_3)\right\}|\varphi\rangle, \end{aligned} \tag{S25}$$

and we note that the four terms above correspond to the four unique possible outcomes of the two syndrome measurements, so that exactly one of these terms is nonzero. If both measurements give $+1$, then only the $\widetilde{\mathbb{1}}_{12}\widetilde{\mathbb{1}}_{23}$ term survives; if both give $-1$, then only the $\widetilde{X}_{12}\widetilde{X}_{23}$ term survives; if only $Z_1 Z_2$ is negative, then only the $\widetilde{X}_{12}\widetilde{\mathbb{1}}_{23}$ terms survives; and finally, if only $Z_2 Z_3$ is negative, then only the $\widetilde{\mathbb{1}}_{12}\widetilde{X}_{23}$ terms survives. This is precisely how the Stinespring registers capture measurement outcomes.

The syndrome measurements allow us to determine where a bit-flip error occurred. If no qubits were flipped, the outcome registers will be in the state $|0\rangle_{12} \otimes |0\rangle_{23}$; if either syndrome measurement returns $-1$, the corresponding outcome register should be in the state $|1\rangle_{ij}$, and the Stinespring operator $\widetilde{X}_{ij}$ appears in the expression for $|\varphi'\rangle$ (S25). Due to the nature of the initial state (S20), having bit flip errors on *both* qubits two and three is equivalent to having a bit error on qubit one alone, up to an overall sign—which can be absorbed into the $f(\cdot)$ functions in definitions of the logical operators (S21)—since the syndromes cannot distinguish $|\psi\rangle$ from $X_1 X_2 X_3 |\psi\rangle$. Thus, the general rotation of $|\varphi'\rangle$ that corrects for arbitrary bit flip errors is

$$\mathcal{R} \equiv \mathbb{1} \otimes |0\rangle\langle 0|_{\text{ss},12} \otimes |0\rangle\langle 0|_{\text{ss},23} + X_1 \otimes |1\rangle\langle 1|_{\text{ss},12} \otimes |0\rangle\langle 0|_{\text{ss},23} \\ + X_3 \otimes |0\rangle\langle 0|_{\text{ss},12} \otimes |1\rangle\langle 1|_{\text{ss},23} + X_2 \otimes |1\rangle\langle 1|_{\text{ss},12} \otimes |1\rangle\langle 1|_{\text{ss},23} \,, \tag{S26}$$

which applies the required operator to the physical degrees of freedom based on the outcomes of the syndrome measurements, which are stored in the Stinespring registers. The terms in (S26) appear in the same order as the corresponding outcomes in (S25)—e.g., if the first term in (S25) is nonzero, then the appropriate choice for error correction is the first term in (S26). Also note that, in practice, one does not actually "apply" the operator $\mathcal{R}$ as written in (S26); rather, $\mathcal{R}$ codifies how to correct an error based on measurement outcomes, which are conveniently stored in Stinespring registers. In other words, in an actual experiment, the measurement results will be known, and one need only apply the corresponding term in (S26), as only one term will realize in any particular experiment. Finally, we note that the same procedure works if we treat the "application" of $X_i$ as a proper measurement and retain (but do not "use") the outcome.

### 3.3. Recovering the logical operators: The Heisenberg picture

We can also understand quantum error correction in the Stinespring formalism from the perspective of operators. Previously, we considered how a bit-flip ($X$) error affects the state (S20); however, the importance of that state lies in the action of the logic operators (S21) thereupon. Now, rather than evolve the state (S20) forward in time via the Schrödinger picture, we evolve the logic operators (S21) "backward" in time in the Heisenberg picture.

The final step is error correction, which depends on the outcomes of the syndrome measurements; in the Stinespring formalism, all measurements can be represented unitarily[4]. Hence, we consider quantities of the form

$$\mathcal{O} \to \mathsf{V}^\dagger_{[\text{err}]} \mathsf{V}^\dagger_{[12]} \mathsf{V}^\dagger_{[23]} \mathcal{R}^\dagger \mathcal{O} \mathcal{R} \mathsf{V}_{[23]} \mathsf{V}_{[12]} \mathsf{V}_{[\text{err}]} \tag{S27}$$

with $\mathcal{O}$ one of the logical operators, $X_\text{L}$ (S21a) or $Z_\text{L}$ (S21b).

For concreteness, let us consider $Z_\text{L}$ (S21b), and note that the function $f_z$ can always be written

$$f_z = A\,\mathbb{1} + B\,(Z_1 Z_2) + C\,(Z_1 Z_2)(Z_2 Z_3) + D\,(Z_2 Z_3) \,, \tag{S28}$$

which captures all unique combinations of the syndrome operators. The general form of $Z_\text{L}$ is then given by

$$Z_\text{L} = A\,Z_1 + B\,Z_2 + C\,Z_3 + D\,Z_1 Z_2 Z_3 \,, \tag{S29}$$

and $A + B + C + D = 1$ ensures that $Z_\text{L} |\psi(\alpha,\beta)\rangle = |\psi(\alpha,-\beta)\rangle$ (S20). We first apply the error-correction $\mathcal{R}$ finding

$$\begin{aligned} Z'_\text{L} &= \mathcal{R}^\dagger Z_\text{L} \mathcal{R} \\ &= \frac{1}{2} [A\,Z_1 + B\,Z_2 + C\,Z_3 - D\,Z_1 Z_2 Z_3] \otimes \widetilde{\mathbb{1}}_{12} \otimes \widetilde{\mathbb{1}}_{23} \\ &+ \frac{1}{2} [-A\,Z_1 + B\,Z_2 + C\,Z_3 + D\,Z_1 Z_2 Z_3] \otimes \widetilde{\mathbb{1}}_{12} \otimes \widetilde{Z}_{23} \\ &+ \frac{1}{2} [A\,Z_1 + B\,Z_2 - C\,Z_3 + D\,Z_1 Z_2 Z_3] \otimes \widetilde{Z}_{12} \otimes \widetilde{\mathbb{1}}_{23} \\ &+ \frac{1}{2} [A\,Z_1 - B\,Z_2 + C\,Z_3 + D\,Z_1 Z_2 Z_3] \otimes \widetilde{Z}_{12} \otimes \widetilde{Z}_{23} \,, \end{aligned} \tag{S30}$$

---

[4] Note that, in general, there will be entangling unitaries applied prior to the syndrome measurements to ensure that the state is correctable. However, for the purposes of illustrating the error correction step, we omit this step here.

and we next apply the unitary measurement channel corresponding to the syndrome operator $Z_2 Z_3$ using Tab. 1, which tells us that $\mathcal{O} \otimes \widetilde{\mathbb{1}}_{23} \to \mathcal{O} \otimes \widetilde{\mathbb{1}}_{23}$ while $\mathcal{O} \otimes \widetilde{Z}_{23} \to \mathcal{O} Z_2 Z_3 \otimes \widetilde{Z}_{23}$. The result is

$$\begin{aligned} Z_{\mathrm{L}}'' &= \mathsf{V}_{[23]}^\dagger \mathcal{R}^\dagger Z_{\mathrm{L}} \mathcal{R} \mathsf{V}_{[23]} \\ &= \frac{1}{2} [A Z_1 + B Z_2 + C Z_3 - D Z_1 Z_2 Z_3] \otimes \widetilde{\mathbb{1}}_{12} \otimes \widetilde{\mathbb{1}}_{23} \\ &+ \frac{1}{2} [D Z_1 + C Z_2 + B Z_3 - A Z_1 Z_2 Z_3] \otimes \widetilde{\mathbb{1}}_{12} \otimes \widetilde{Z}_{23} \\ &+ \frac{1}{2} [A Z_1 + B Z_2 - C Z_3 + D Z_1 Z_2 Z_3] \otimes \widetilde{Z}_{12} \otimes \widetilde{\mathbb{1}}_{23} \\ &+ \frac{1}{2} [D Z_1 + C Z_2 - B Z_3 + A Z_1 Z_2 Z_3] \otimes \widetilde{Z}_{12} \otimes \widetilde{Z}_{23} \,, \end{aligned} \quad (S31)$$

and applying the channel corresponding to measurement of the syndrome operator $Z_1 Z_2$ gives

$$\begin{aligned} Z_{\mathrm{L}}''' &= \mathsf{V}_{[12]}^\dagger \mathsf{V}_{[23]}^\dagger \mathcal{R}^\dagger Z_{\mathrm{L}} \mathcal{R} \mathsf{V}_{[23]} \mathsf{V}_{[12]} \\ &= \frac{1}{2} [A Z_1 + B Z_2 + C Z_3 - D Z_1 Z_2 Z_3] \otimes \widetilde{\mathbb{1}}_{12} \otimes \widetilde{\mathbb{1}}_{23} \\ &+ \frac{1}{2} [D Z_1 + C Z_2 + B Z_3 - A Z_1 Z_2 Z_3] \otimes \widetilde{\mathbb{1}}_{12} \otimes \widetilde{Z}_{23} \\ &+ \frac{1}{2} [B Z_1 + A Z_2 + D Z_3 - C Z_1 Z_2 Z_3] \otimes \widetilde{Z}_{12} \otimes \widetilde{\mathbb{1}}_{23} \\ &+ \frac{1}{2} [C Z_1 + D Z_2 + A Z_3 - B Z_1 Z_2 Z_3] \otimes \widetilde{Z}_{12} \otimes \widetilde{Z}_{23} \,, \end{aligned} \quad (S32)$$

and before applying the bit-flip error channel $\mathsf{V}_{[\mathrm{err}]}$, we deal with the Stinespring registers.

Note that the bit-flip error channel does not affect the Stinespring outcomes corresponding to the syndrome measurements, allowing us to consider the Stinespring content of (S32) first. We note that this operator must reproduce (S22b) acting on the initial state (S20) in the dilated Hilbert space (S24),

$$|\psi\rangle = (\alpha|000\rangle + \beta|111\rangle) \otimes |0\rangle_{12} \otimes |0\rangle_{23} \,, \quad (S33)$$

and regarding the Stinespring content of (S32), we note that all four terms act as the identity on the default Stinespring state $|0\rangle_{12} \otimes |0\rangle_{23}$. Hence, in the remainder, we simply assume that the Stinespring state is $|0\rangle_{12} \otimes |0\rangle_{23}$ and make no further reference thereto. The physical part of the Heisenberg-evolved logical operator $Z_\mathrm{L}$ is then given by

$$Z_{\mathrm{L}}''' = (Z_1 + Z_2 + Z_3 - Z_1 Z_2 Z_3)/2 \,, \quad (S34)$$

using the fact that $A + B + C + D = 1$.

The logical operator $Z_\mathrm{L}$ for a bit-flip error on site $i$ (where $i = 0$ indicates no bit-flip error), is given finally by conjugating $Z_{\mathrm{L}}'''$ with either the identity or $X_{1,2,3}$. We find

$$\begin{aligned} Z_{\mathrm{L}}^{(0)} &= Z_1 \left(\mathbb{1} + Z_1 Z_2 + Z_1 Z_3 - Z_2 Z_3\right)/2 & (S35a) \\ Z_{\mathrm{L}}^{(1)} &= Z_1 \left(Z_2 Z_3 + Z_1 Z_2 + Z_1 Z_3 - \mathbb{1}\right)/2 & (S35b) \\ Z_{\mathrm{L}}^{(2)} &= Z_1 \left(Z_2 Z_3 - Z_1 Z_2 + Z_1 Z_3 + \mathbb{1}\right)/2 & (S35c) \\ Z_{\mathrm{L}}^{(3)} &= Z_1 \left(Z_2 Z_3 + Z_1 Z_2 - Z_1 Z_3 + \mathbb{1}\right)/2 \,, & (S35d) \end{aligned}$$

for the four possible $X$ errors, and we see that $Z_{\mathrm{L}}^{(i)}(t)$ remains of the form $Z_1 f_z^{(i)} (\mathbb{1}, Z_1 Z_2, Z_2 Z_3)$[5], but with a modified function $f_z^{(i)}$ of the syndrome operators, which acts trivially on the initial state (S33).

Note that the $f$ terms in (S35) are always arranged so as to cancel the effect of $Z$ acting on the site $i$ where the error occurred, so that the operators (S35) always act as a genuine, logical $Z$, even after conjugating with the error operator (i.e., $Z_{\mathrm{L}}^{(i)}$ sends $|\psi(\alpha, \beta)\rangle \to |\psi(\alpha, -\beta)\rangle$, as the $f$ function acts as the identity. A similar procedure can be

———

[5] Each logical operator could equally well be cast in the form $Z_j f_z^{(i)} (\mathbb{1}, Z_1 Z_2, Z_2 Z_3)$ for any site $j = 1, 2, 3$.

used to verify that the error correction channel given by $\mathcal{R}$ (S26) maintains $X_\mathrm{L}$ in the form (S21a). Note that this toy model for QEC cannot correct for phase ($Z$) errors.

We note that the procedure described in this section for the three-qubit repetition code can be generalized—although the derivations are rather tedious—to describe genuine quantum error-correcting codes that are able to protect any qubit from an arbitrary error (bit-flip and sign errors). In those settings, the code is set up in such a way that, depending on the outcomes of the syndrome measurements, the desired logical operator's Heisenberg evolution is modified (as can be seen by evolving the logical operator backward in time in the Heisenberg picture, starting with the error correction step (S26), which generates Stinespring operator content) so as to commute with the error itself. If this does not occur, then a Stinespring $\widetilde{X}$ arises in the Heisenberg-evolved operator, which heralds the failure of quantum error correction, as off-diagonal Stinespring terms have weight zero for any particular measurement outcome. We explore such an error-correcting code in the following Section.

## 4. QUANTUM ERROR CORRECTING CODE: THE TRANSVERSE FIELD ISING MODEL

We now introduce the "transverse field Ising model" (TFIM) code—a teleportation protocol similar to the entanglement-swapping teleportation protocol (ESTP) of the main text. Like the ESTP, the TFIM protocol saturates the bound (S119) on the speed of quantum information in the presence of measurements, but has the advantage of being more realizable in extant experimental platforms. Further details of the TFIM code appear in [S6].

We first provide a brief crash course on Clifford circuits, then detail the gate set use by the TFIM code, before presenting the explicit circuit construction following [S6]. We then show that this code correctly teleports the logical qubit a distance $L = N - 1$, from the first site ($j = 1$) to the final site ($j = N$), by evolving the final $t = T$ logical operators in the Heisenberg picture and showing that they reproduce the $t = 0$ logic operators acting on the initial state (in the computational basis). Finally, we discuss the measurement-free case ($M = \mathcal{M} = 0$).

### 4.1. Clifford circuits

For a spin chain comprising $N$ qubits, the operator space $\mathrm{End}(\mathcal{H}_\mathrm{ph})$ is spanned by elements of the Pauli group,

$$\mathbf{P}_N \,=\, \left\{ \mathrm{e}^{\mathrm{i}\pi m/2}\, \sigma_1^{n_1} \otimes \cdots \otimes \sigma_N^{n_N} \,\Big|\, m, n_j \in \{0, 1, 2, 3\} \right\}, \qquad (\mathrm{S}36)$$

where $\sigma_j^n \in \{\mathbb{1}_j, X_j, Y_j, Z_j\}$ denotes the $n$th Pauli matrix acting on site $j$. The Clifford group $C_N$ (on $N$ qubits) consists of the set of physical unitary operators $V \in \mathrm{Aut}(\mathcal{H}_\mathrm{ph}) = \mathsf{U}(2^N)$ that *normalize* the Pauli group,

$$\mathbf{C}_N \,=\, \left\{ V \in \mathsf{U}\left(2^N\right) \,\big|\, V O V^\dagger \in \mathbf{P}_N \;\forall\, O \in \mathbf{P}_N \right\}, \qquad (\mathrm{S}37)$$

so that teh single-qubit Clifford group is generated by the phase gate $\mathbf{S} = \sqrt{Z}$ and the Hadamard gate,

$$H \,=\, \frac{1}{\sqrt{2}}\, (X + Z) \,=\, \frac{1}{\sqrt{2}} \begin{bmatrix} 1 & 1 \\ 1 & -1 \end{bmatrix} \begin{matrix} |0\rangle \\ |1\rangle \end{matrix} \;\to\; -\boxed{\mathbf{H}}- \,, \qquad (\mathrm{S}38)$$

represented above both as a matrix in the computational ($Z$) basis and in standard gate notation [S18]. The Hadamard gate rotates between the $X$ and $Z$ bases, with $H X H = Z$ (and vice versa).

The two-qubit Clifford group is generated by the single-qubit Clifford generators along with the two-qubit controlled NOT (CNOT) gate, which applies $X$ to the "target" qubit $j$ if the "control" qubit $i$ is in the state $|1\rangle$,

$$\mathrm{CNOT} \,=\, \begin{bmatrix} 1 & 0 & 0 & 0 \\ 0 & 1 & 0 & 0 \\ 0 & 0 & 0 & 1 \\ 0 & 0 & 1 & 0 \end{bmatrix} \begin{matrix} |00\rangle \\ |01\rangle \\ |10\rangle \\ |11\rangle \end{matrix} \;\to\; \begin{matrix} \bullet \\ \oplus \end{matrix} \,, \qquad (\mathrm{S}39)$$

again as a matrix in the computational basis and standard gate notation [S18], where the control qubit $i$ corresponds to the upper line above, and the target qubit $j$ to the lower line.

Finally, it is useful to define the two-qubit swap (SWAP) gate, which appears in the ESTP,

$$\text{SWAP} = \begin{bmatrix} 1 & 0 & 0 & 0 \\ 0 & 0 & 1 & 0 \\ 0 & 1 & 0 & 0 \\ 0 & 0 & 0 & 1 \end{bmatrix} \begin{matrix} |00\rangle \\ |01\rangle \\ |10\rangle \\ |11\rangle \end{matrix} \rightarrow \;\;\raisebox{-2pt}{\text{⨯}}\!\!\!\!\!\raisebox{-10pt}{\text{⨯}}\;\;, \tag{S40}$$

again in matrix form in the computational basis and one of several gate notations. The SWAP gate exchanges the states of the qubits on which it acts—i.e., $\text{SWAP}|ab\rangle = |ba\rangle$. Acting on qubits labelled $i$ and $j$, the SWAP gate can be written in terms of Pauli operators as $\text{SWAP} = \left(\mathbb{1} + X_i X_j + Y_i Y_j + Z_i Z_j\right)/2$. In practice, one can realize SWAP gates using various sequences of simpler single- and two-qubit gates.

For an arbitrary state $|\Psi\rangle \in \mathcal{H}$, we next define the associated *stabilizer* group $\text{Stab}(|\Psi\rangle)$ as the subset of unitary operators $S \in \text{Aut}(\mathcal{H}_{\text{ph}})$ including the identity operator $\mathbb{1}_{\text{ph}}$ that act trivially on $|\Psi\rangle$, i.e.,

$$\text{Stab}(|\Psi\rangle) \equiv \{ S \in \text{End}(\mathcal{H}) \,|\, S|\Psi\rangle = |\Psi\rangle \}, \tag{S41}$$

which forms a group under matrix multiplication. In the cases of interest, the stabilizer elements $S$ are Pauli strings.

Now consider a $1D$ spin chain with $N$ sites, where we wish to transfer a quantum state $|\psi\rangle$ from the first site ($j = 1$) to the last site ($j = N$). We initialize the $N$-qubit state in the $Z$-basis product state given by

$$|\Psi_0\rangle \,=\, |\psi\rangle \otimes |00\ldots 00\rangle \,=\, (\alpha|0\rangle_1 + \beta|1\rangle_1) \otimes |00\ldots 00\rangle, \tag{S42}$$

where $\alpha, \beta \in \mathbb{C}$ with $|\alpha|^2 + |\beta|^2 = 1$, and $|0\rangle$ and $|1\rangle$ correspond to the spin-1/2 states $|\uparrow\rangle$ and $|\downarrow\rangle$ in the $z$ basis.

Using a combination of local Clifford gates and measurements, the goal is to identify the appropriate *dilated* quantum channel $\mathcal{T}$ [S1–S3, S6] that produces the desired final quantum state,

$$\mathcal{T} \,:\, |\Psi_0\rangle \mapsto |\Phi\rangle_{1,\ldots,N-1} \otimes (\alpha|0\rangle_N + \beta|1\rangle_N), \tag{S43}$$

where $\mathcal{T}$ contains both Clifford and measurement unitaries and $|\Phi\rangle_{1,\ldots,N-1}$ denotes a many-body state on sites $j < N$.

The *logical operators* acting on the localized logical state $|\psi\rangle = \alpha|0\rangle + \beta|1\rangle$ are given by $X_\text{L} = X_1$ and $Z_\text{L} = Z_1$, such that $X_\text{L}|\psi(\alpha,\beta)\rangle = |\psi(\beta,\alpha)\rangle$ and $Z_\text{L}|\psi(\alpha,\beta)\rangle = |\psi(\alpha,-\beta)\rangle$. The initial stabilizer group $\mathbf{S}_0 = \text{Stab}(|\Psi_0\rangle)$ is generated by the commuting local operators $Z_2,\ldots,Z_N$; $Z_1$ is excluded, meaning that the logical operators commute with every element of $\mathbf{S}_0$. Many equivalent pairs of logical operators ($X_\text{L}$ and $Z_\text{L}$) can be identified for the many-body state $|\Psi_0\rangle$ (S42): Multiplying the local logical operators $X_1$ and $Z_1$ by *any* element of $S \in \text{Stab}(|\Psi_0\rangle)$,

$$\sigma_\text{L}^i \to \sigma_\text{L}^i S \,=\, \sigma_1^i S, \quad \forall S \in \text{Stab}(|\Psi_0\rangle), \tag{S44}$$

produces another valid set of logical operators, which act on the initial state (S42) as

$$\sigma_\text{L}^i \,|\Psi_0\rangle \,=\, \sigma_1^i \, S \,|\Psi_0\rangle \,=\, \sigma_1^i \,|\Psi_0\rangle, \tag{S45}$$

meaning that any logical operator of the form (S44) reproduces the action $\sigma_\text{L}^i = \sigma_1^i$ on the state $|\Psi_0\rangle$.

Crucially, in a Clifford circuit, all operators either commute or anticommute with one another; these [anti]commutation relations are preserved under unitary time evolution, i.e.,

$$[A, B] = 0 \implies \left[ \mathcal{W} A \mathcal{W}^\dagger, \mathcal{W} B \mathcal{W}^\dagger \right] = 0, \tag{S46}$$

for the commutator, and likewise for the *anti*commutator $\{A, B\}$. Thus, unitary dynamics preserves the facts that $(i)$ the initial logical operator $X_\text{L} = X_1$ commutes with all elements of the initial stabilizer group $\mathbf{S}_0$ and $(ii)$ the operator $X_\text{L} = X_1$ anticommutes with $Y_\text{L} = Y_1$ and $Z_\text{L} = Z_1$.

The action of any operator $O$ on the initial state $|\Psi_0\rangle$ (S42) is the same as that of the reverse time-evolved operator $O(-t)$ on the time-evolved state $|\Psi(t)\rangle$:

$$\mathcal{W} O|\Psi_0\rangle = (\mathcal{W} O \mathcal{W}^\dagger)\mathcal{W}|\Psi_0\rangle \,=\, \mathcal{W} O \mathcal{W}^\dagger |\Psi(t)\rangle = O(-t)|\Psi(t)\rangle, \tag{S47}$$

and hence, if $O$ stabilizes the initial state (i.e., $O|\Psi_0\rangle = |\Psi_0\rangle$) then $O(-t) = \mathcal{W} O \mathcal{W}^\dagger$ stabilizes the final state (i.e., $O(-t)|\Psi(t)\rangle = |\Psi(t)\rangle$). Thus, the stabilizer group at time $t$ is given straightforwardly in terms of the initial stabilizer group at time $t = 0$. Importantly, the simultaneous $+1$ eigenspace of the time-evolved stabilizer group,

$$\mathbf{S}_t \equiv \{ \mathcal{W} O \mathcal{W}^\dagger \,|\, \forall O \in \text{Stab}(|\Psi_0\rangle) \} \tag{S48}$$

defines the "codespace" at time $t$. We refer to these backward-time-evolved stabilizer elements $O(-t) \in \mathbf{S}_t$ as "check" operators, as they allow one to "check" whether or not the current state $|\Psi(t)\rangle$ at time $t$ is in the codespace. In the context of quantum error correction, local errors effectively rotate the state $|\Psi\rangle$ out of the codespace, and we then measure the check operators and use the resulting "error syndromes" to determine how to reverse the effect of any local errors to return $|\Psi\rangle$ to the codespace and restore the existence of logical operators $X_\text{L}$ and $Z_\text{L}$.

### 4.2. Entangling Clifford gates

Before specifying the TFIM code in full detail, we first review the Clifford gates comprising the time-evolution circuit $\mathcal{W}$ of the TFIM code. Note that these gates differ from those of the entanglement swapping teleportation protocol (ESTP) presented in the main text. While the TFIM's Clifford gates are arguably less intuitive than those of the ESTP, the TFIM protocol also saturates the bound on quantum teleportation while being *easier* to implement experimentally in Rydberg atom arrays [S6]. The primary two-qubit Clifford gate used for unitary encoding is given by

$$\mathcal{U}_i \equiv \frac{1}{\sqrt{2}} \left( Z_i + X_i X_{i+1} \right) = -\frac{\mathrm{i}}{\sqrt{2}} \left( \gamma_i \xi_i + \xi_i \gamma_{i+1} \right), \tag{S49}$$

where $\gamma$ and $\xi$ are Majorana fermion modes satisfying $\{\gamma_i, \gamma_j\} = \{\xi_i, \xi_j\} = 2\delta_{ij}$ and $\{\gamma_i, \xi_j\} = 0$. The two Majorana modes are related to the Pauli operators according to

$$\gamma_i = X_i \bigotimes_{j=1}^{i-1} Z_j, \qquad \xi_i = Y_i \bigotimes_{j=1}^{i-1} Z_j, \tag{S50}$$

and in terms of Majorana representation (S50), the action of the gate $\mathcal{U}_i$ (S49) is given by

$$\gamma_i \xleftrightarrow{\mathcal{U}_i} \gamma_{i+1}, \quad \xi_i \xrightarrow{\mathcal{U}_i} -\xi_i, \quad \xi_{i+1} \xrightarrow{\mathcal{U}_i} \xi_{i+1}, \tag{S51}$$

meaning that $\mathcal{U}_i$ shifts the $\gamma$ mode by one site while leaving $\xi$ unchanged (up to phase).

We introduce a second type of TFIM gate for the purpose of translating the $\xi$ mode,

$$\mathcal{V}_i = \frac{1}{\sqrt{2}} \left( Z_i + Y_i Y_{i+1} \right) = -\frac{\mathrm{i}}{\sqrt{2}} \left( \gamma_i \xi_i - \gamma_i \xi_{i+1} \right), \tag{S52}$$

whose action on the Majorana modes is given by

$$\gamma_i \xrightarrow{\mathcal{V}_i} -\gamma_i, \quad \gamma_{i+1} \xrightarrow{\mathcal{V}_i} \gamma_{i+1}, \quad \xi_i \xleftrightarrow{\mathcal{V}_i} \xi_{i+1}. \tag{S53}$$

The state transfer protocol involves the application of both $\mathcal{U}$ and $\mathcal{V}$ gates, in addition to local measurements of $Z_j$. Note that all of these operations commute with the global $\mathbb{Z}_2$ Ising symmetry,

$$\mathcal{G} \equiv \bigotimes_{j=1}^{N} Z_j = \mathrm{i}^{-N} \bigotimes_{j=1}^{N} \gamma_j \xi_j, \tag{S54}$$

which is an invariant of the circuit dynamics generated by $\mathcal{U}$ and $\mathcal{V}$, and has connections to recent discussions on symmetry-protected topological phases (SPTs) and error correction [S19–S21].

### 4.3. Circuit protocol

Having defined the requisite entangling Clifford gates, we now discuss the explicit details of the TFIM code. The protocol is depicted diagrammatically in Fig. S1 for $N = 25$ qubits, $M = 3$ local measurement *regions*, and circuit depth $T = 5$. Note that the total number of measurement *outcomes* is $\mathcal{M} = 6 = 2M$; the relation $\mathcal{M} \geq 2M$ is generic to teleportation [S22]. The time $T$ is taken to be the minimum circuit depth (i.e., number of circuit layers) required to realize all multi-site unitary gates (i.e., all $\mathcal{U}$ (S49) and $\mathcal{V}$ (S52) gates), parallelizing where possible.

Starting from the initial state (S42) with the logical qubit at site $j = 1$, we apply the entangling circuit $\mathcal{W}$, which encodes the logical qubit amongst all $N$ qubits. In Fig. S1, this corresponds to the "bilayer cone" of orange ($\mathcal{U}$) and green ($\mathcal{V}$) Clifford gates. We then perform local operations and classical communications (LOCC)—i.e., the dilated unitary measurement channel M followed by an error-correction unitary $\mathcal{R}$ that correctly realizes the logical state $|\psi\rangle$ on the final site. The measurements correspond to single-site Pauli $Z$ operators on the $B$ (red) and $C$ (blue) qubits, and are indicated in the figure by gray boxes with pointer dials; the classical information about these measurement outcomes is instantaneously communicated (via the dashed line in the figure) to determine $\mathcal{R}$.

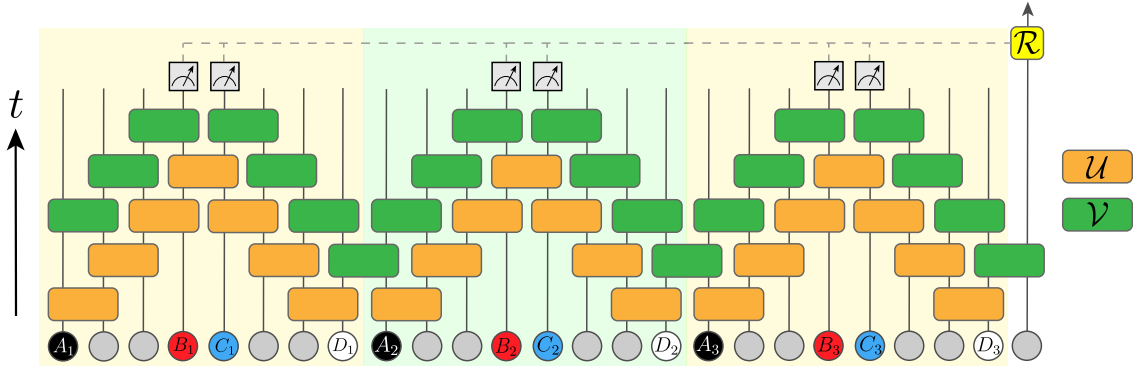

FIG. S1. Circuit depiction of the TFIM state-transfer protocol for $N = 25$ qubits, $M = 3$ measurement regions, and unitary circuit depth $T = 5$. The unitary channel $\mathcal{W}$ corresponds to the bilayered conic pattern of $\mathcal{U}$ (orange) and $\mathcal{V}$ (green) gates; the shaded circles below denote the qubits. The background shading distinguishes repeated "regions" corresponding to the qubits $A_s$ through $D_s$. The square boxes with pointer dials denote local $Z$-basis measurements. Following the unitary channel $\mathcal{W}$ and measurement channel $\mathsf{M}$, the local error-correcting unitary $\mathcal{R}$ (applied to the final site $j = N$) is determined by the measurement outcomes, which are transmitted via instantaneous classical communication (indicated by the dashed lines). Following the application of $\mathcal{R}$, the logical state $|\psi\rangle$ (originally on site $j = 1$) is transferred to site $j = N$.

Details of the TFIM protocol depend on the desired depth $T$ (of $\mathcal{W}$) and number of local measurement regions $M$. As in Fig. S1, the $1D$ qubit chain is divvied into $M$ repeating regions of $\ell = 2(T-1)$ qubits plus the final site[6]. The TFIM teleportation distance $L = N - 1 = 2M(T-1)$ saturates the main bound (S119) up to $\mathsf{O}(1)$ offsets.

The initial state is given in the computational basis by (S42), and can be written in terms of density matrices as

$$\rho_0 \;=\; |\psi\rangle\langle\psi| \otimes \prod_{j=2}^{N} \frac{\mathbb{1} + Z_j}{2} \;=\; |\psi\rangle\langle\psi| \otimes \frac{1}{2^{N-1}} \sum_{S \in \mathbf{S}_0} S \,, \tag{S55}$$

where $|\psi\rangle = \alpha |0\rangle + \beta |1\rangle$ (S42) is the initial logical qubit (on the first site $j = 1$), $\mathbf{S}_0 = \mathrm{Stab}(|\Psi_0\rangle) \subset \mathbf{C}_N$ is the stabilizer group (S41) for the state $|\Psi_0\rangle$ (S42), and the check operators $S \in \mathbf{S}_0$ only act nontrivially on sites $j \in [2, N]$[7].

Labelling the $\ell$-qubit regions from left to right as $s \in \{1, \ldots, M\}$, we construct $\mathcal{W}$ within each region $s$ by staggering $\mathcal{U}$ and then $\mathcal{V}$ gates in a "bilayered cone" as depicted in Fig. S1. All nonoverlapping gates are applied concurrently, and the circuit depth $T$ is the minimum number of layers required to implement all two-site Clifford gates. Importantly, the $\mathcal{U}$ and $\mathcal{V}$ gates are naturally realized in Rydberg atom arrays [S6].

Following the entangling circuit $\mathcal{W}$, one measures local $Z_j$ operators on a subset of qubits within each region and records the outcomes. To elucidate where measurements occur, we define the following four sets of sites:

$$A \;=\; \bigcup_{s=1}^{M} A_s \;=\; \bigcup_{s=1}^{M} \{\ell(s-1) + 1\}\,, \tag{S56a}$$

$$B \;=\; \bigcup_{s=1}^{M} B_s \;=\; \bigcup_{s=1}^{M} \{\ell(s-1) + T - 1\}\,, \tag{S56b}$$

$$C \;=\; \bigcup_{s=1}^{M} C_s \;=\; \bigcup_{s=1}^{M} \{\ell(s-1) + T\}\,, \tag{S56c}$$

$$D \;=\; \bigcup_{s=1}^{M} D_s \;=\; \bigcup_{s=1}^{M} \{\ell s\}\,, \tag{S56d}$$

which correspond respectively to the black ($A$), red ($B$), blue ($C$) and white ($D$) sites in Fig. S1. For $\Omega \in \{A, B, C, D\}$,

---

[6] The ESTP of the main text also features $M$ repeating regions of length $\ell = 2(T-1)$, but can accommodate an additional $T-1$ sites to the left of all repeating regions (including the initial site $j = 1$) and another two sites to the right of all repeating regions (including the final site $j = N = L + 1$). The total teleportation distance for the ESTP is $L = (2M+1)(T-1) + 1$.

[7] The check operators must commute with the logical operators, and therefore cannot act on the first site.

| $\mathcal{R}$ | Measurement Outcomes | |
|---|---|---|
| | $\nu\, m_B$ | $\nu\, m_C$ |
| $\mathbb{1}_N$ | 1 | 1 |
| $X_N$ | 1 | -1 |
| $Y_N$ | -1 | 1 |
| $Z_N$ | -1 | -1 |

TABLE 2. "Lookup" table giving the final-site error type (which is always the identity or a Pauli operator) based on particular products of the measurement outcomes, where $m_B$ and $m_C$ denote the products of $Z$-measurement outcomes on the site sets $B$ and $C$, respectively (S56), and $\nu = (-1)^{M+1}$ (S65). The state $|\psi\rangle$ is transferred up to a rotation error $\mathcal{R}$; because this error is involutory (i.e., $\mathcal{R}^2 = \mathbb{1}$), applying $\mathcal{R}$ again to site $j = N$ "undoes" the error, leaving $|\psi\rangle$.

we also define the associated $\mathcal{G}$-like operators (S54),

$$Z_\Omega = \prod_{j \in \Omega} Z_j. \tag{S57}$$

The measurement channel $\mathsf{M}$ that follows $\mathcal{W}$ involves measuring $Z_{B_s}$ and $Z_{C_s}$ for $1 \leq s \leq M$ (i.e., measuring $Z$ on all of the red and blue sites, as depicted in Fig. S1). The measurement outcomes are communicated instantaneously (via classical signal) to determine the error correction unitary $\mathcal{R}$ according to Tab. 2, which is applied to the final site $j = N$ to ensures that the state $|\psi\rangle$ recovers. Essentially, after applying $\mathsf{M} \circ \mathcal{W}$, the many-body state factorizes over $j = N$ and all other sites, with the final site in the state $\mathcal{R}|\psi\rangle$, for $\mathcal{R} \in \{\mathbb{1}, X_N, Y_N, Z_N\}$. Because $\mathcal{R}$ is a Pauli operator, applying it once again on site $j = N$ produces the desired state on that site: $\mathcal{R}^2|\psi\rangle = |\psi\rangle$.

It will prove convenient to introduce several nonlocal operators to help demonstrate how the TFIM protocol teleports quantum information. In the remainder of this Section, we use primes (e.g., $X'_i$) to denote operators that have been "dressed" by a product of $Z_j$ operators that are compatible with successful state transfer. We define

$$X'_N = X_N \times (-1)^{M+1} Z_B \tag{S58a}$$
$$Y'_N = Y_N \times (-1)^{M+1} Z_C, \tag{S58b}$$
$$\tag{S58c}$$

where $Z_B$ and $Z_C$ are defined in (S57). Evolving these operators in the Heisenberg picture, we find

$$\mathcal{W}^\dagger X'_N \mathcal{W} = X_1 \times \prod_{j \notin (D \cup \{1\})} Z_j \tag{S59a}$$
$$\mathcal{W}^\dagger Y'_N \mathcal{W} = Y_1 \times \prod_{j \notin (A \cup \{N\})} Z_j, \tag{S59b}$$

and the Heisenberg evolution of $Y'_N$ is depicted in Fig. S2. The operator $Z'_N$ follows from $Z = -\mathrm{i}XY$:

$$Z'_N = -\mathrm{i}\, X'_N Y'_N = Z_N \times Z_B Z_C, \tag{S60}$$

and by multiplying with (S59), we find

$$\mathcal{W}^\dagger Z'_N \mathcal{W} = Z_1 \times \prod_{j \notin (A \cup D \cup \{1,N\})} Z_j. \tag{S61}$$

Loosely speaking, measuring $Z$ on the $B$ and $C$ sites replaces $Z_B$ and $Z_C$ in a given expression by their measurement outcomes $m_{B,C}$[8]; this can be seen rigorously in Sec. 4.4 using the Stinespring formulation.

The signs of the resulting $X_N$ and $Y_N$ local operators indicate which error-correcting unitary $\mathcal{R}$ should be applied to the final site ($j = N$). For example, if $X_N$ and $Y_N$ both acquire an overall factor of $-1$ under the circuit $\mathsf{M} \circ \mathcal{W}$, then conjugation with $Z_N$ in the Heisenberg picture (i.e., due to the final unitary error-correction step $\mathcal{R} = Z_N$) will correct these sign errors. The general rules for the determination of $\mathcal{R}$ are given in Tab. 2.

---

[8] The outcome $m_B = \pm 1$ of measuring $Z_B$ is the product over the outcomes $m_j = \pm 1$ for each $j \in B$: $m_B = \prod_{j \in B} m_j$.

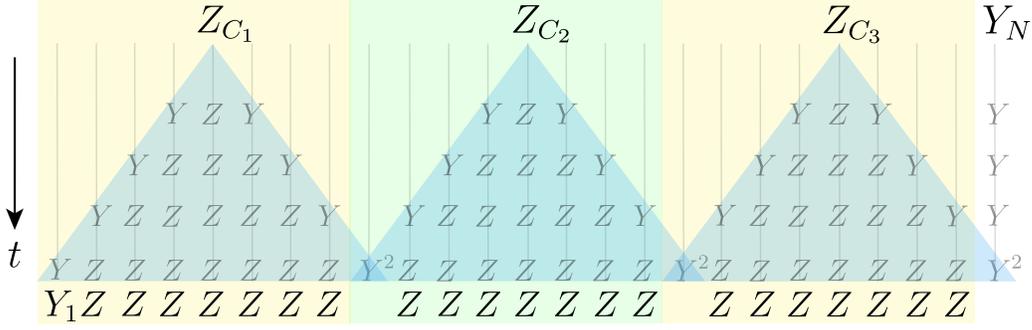

FIG. S2. The TFIM protocol is depicted in the Heisenberg picture for the $Y'_N$ logical operator (Stinespring qubits hidden). The local $Z$ operators grow with the Lieb-Robinson velocity $v$ of the circuit as depicted by the edges of the shaded cones. The check operators during the intermediate steps are shown inside their respective "light cones".

We conclude this discussion of the TFIM teleportation protocol by computing the maximum spatial distance $L$ over which this protocol can transfer a single, logical qubit for fixed depth $T$. The length of each measurement region is $\ell = 2(T-1)$, and so the total transfer distance in terms of $T, M$ and $L = N - 1$ is simply

$$L = \ell M = 2M(T-1), \tag{S62}$$

where the restriction $M \geq 1$ follows from the fact that *at least* one region must exist to apply the circuit.

Note that (S62) asymptotically saturates the bound (S119) of the main text, with $T_0 = -1$ and $M_0 = -1/2$. The fact that both of these $\mathsf{O}(1)$ offsets are negative implies the existence of a *slightly* more efficient protocol. In particular, the ESTP with measurements of $X_j X_{j+1}$ and $Z_j Z_{j+1}$ in each of the $M$ regions satisfies (S119) with $M_0 = T_0 = 0$. However, the ESTP and TFIM protocol both saturate the bound (S119) in the asymptotic limit of interest $L, T \gg 1$.

We also have the constraint that $M < N = L + 1$, since measuring *every* qubit would destroy all of the quantum information prepared in the initial state. However, $M$ counts the number of local *regions* in which measurements are made; in the TFIM protocol, each measurement acts on two neighboring qubits, so we have $M = \mathcal{M}/2$, where $\mathcal{M}$ is the number of measurements. Since we cannot measure the final site $f$ to which the qubit is teleported, we have $N - 1 = L \geq \mathcal{M} \geq 2M$, meaning that $M \leq L/2$. In fact, a generic quantum circuit that teleports $Q$ qubits using two-qubit unitary gates, projective measurements, and outcome-dependent feedback obeys $\mathcal{M} \geq 2M$ [S22] and $\mathcal{M} \leq N - Q$, so that $M \leq (N - Q)/2$ in the most general case (see also Sec. 8).

We comment that the circuit depth $T$ for the TFIM state-transfer protocol is smaller by a factor of roughly $1/M$ than that of the nearest-neighbor hopping protocol discussed in Sec. 4.5 (with depth $T \sim L$). In particular, if $M \sim L/2$, then a depth as small as $T = \mathsf{O}(1)$ is possible. Such a constant-depth circuit is experimentally appealing, since it could potentially be realized entirely within a given system's coherence time. In the limit of maximal $M \sim L/2$, we find that the minimum circuit depth is $T = 3$, which is also the minimum depth for the TFIM state-transfer protocol in general.

### 4.4. Recovering the logical operators: The Heisenberg picture

In the dilated Hilbert space, the measurement and error-correction channels ($\mathsf{M}$ and $\mathcal{R}$, respectively) are both *unitary*. Suppose that the operator $O_N$ on the final site $j = N$ is one of the logical operators *after* the TFIM protocol $\mathcal{T} = \mathcal{R} \circ \mathsf{M} \circ \mathcal{W}$ has concluded. We now show that this logical operator—evolved backward in time for time $T$ under the full circuit $\mathcal{T}$ in the Heisenberg picture—reproduces one of the initial logical operators for the initial state $|\Psi_0\rangle$ (S42). Note that $|\Psi_0\rangle \in \mathcal{H}_{\mathrm{dil}}$ is a state in the *dilated* Hilbert space, which includes $\mathcal{M} = 2M$ "outcome" qubits, each initialized in the state $|0\rangle_{\mathrm{ss}}$ (S14). All logical operators act trivially (as the identity) on the outcome registers in $\mathcal{H}_{\mathrm{ss}}$.

Essentially, $O_N$ is a valid logical operator on the final state (S43) if its Heisenberg-Stinespring evolution obeys

$$O_N(T) = \mathcal{W}^\dagger \mathsf{M}^\dagger \mathcal{R}^\dagger O_N \mathcal{R} \mathsf{M} \mathcal{W} = O_1 \otimes Z_\Gamma \otimes \widetilde{Z}_\Omega, \tag{S63}$$

where $\Gamma$ and $\Omega$ respectively denote arbitrary subsets of the physical and outcome qubits (though $\Gamma$ cannot contain site $j = 1$). The initial state's stabilizer group (S41) contains $Z_\Gamma$ and $\widetilde{Z}_\Omega$ (since all qubits are prepared in the state $|0\rangle$ with $Z|0\rangle = |0\rangle$). The only nontrivial operator in (S63) is the logical operator $O_1 \in \{X_L, Y_L, Z_L\} = \{X_1, Y_1, Z_1\}$.

We now consider the logical operators for the final state (S43), $X_L, Y_L = X_N, Y_N$ (with $Z_N = -\mathrm{i}\, X_N Y_N$). The error-correcting unitaries $\mathcal{R}_x$ and $\mathcal{R}_y$ ensure that $X_N$ and $Y_N$ (respectively) satisfy (S63), and that $O_1 = X_1$ in (S63)

for $X_\mathrm{L}(T)$ and $O_1 = Y_1$ for $Y_\mathrm{L}(T)$. The full error-correction unitary is given by the product $\mathcal{R} = \mathcal{R}_x \mathcal{R}_y$. Note that the error-correction unitaries may be applied in either order, with equivalent results up to a physically meaningless sign.

We first consider the putative logical operator $X_N$; the relevant error-correction unitary $\mathcal{R}_x$ is given by

$$\mathcal{R}_x = \frac{1}{2}\left(\widetilde{\mathbb{1}} + \nu \widetilde{Z}_B\right) + \frac{1}{2}\left(\widetilde{\mathbb{1}} - \nu \widetilde{Z}_B\right) Y_N, \tag{S64}$$

where $\widetilde{Z}_B = \prod_{j \in B} \widetilde{Z}_j$ is the product of $\widetilde{Z}$ operators acting on the Stinespring outcome registers corresponding to the measurements of all $B$ sites, and the "measurement parity" $\nu$ is defined as

$$\nu \equiv (-1)^{M+1} \in \{-1, +1\}, \tag{S65}$$

and if the product of the "$B$" outcomes is $+\nu$, then a $Y$ (possibly $Z = -\mathrm{i}XY$) error occurred, and $\mathcal{R}_x$ applies $Y_N$; if the product is $-\nu$, then no $Y$ (nor $Z$) error occurred, and $\mathcal{R}_x$ acts trivially, as dictated by Tab. 2. Analogously, the parity of the outcomes of all measurements of $C$ qubits determines whether an $X$ error occurred. If both $X$ and $Y$ errors are present, we say there is a $Z$ error (but there is no need to check for a $Z$ error separately)[9].

The channel $\mathcal{R}_x$ does not modify the logical operator $Y_N$ (i.e., $\mathcal{R}_x^\dagger Y_N \mathcal{R}_x = Y_N$). Applying $\mathcal{R}_x$ to $X_N$ gives

$$\begin{aligned}\mathcal{R}_x^\dagger X_N \mathcal{R}_x &= \frac{1}{4}\left\{\left(\widetilde{\mathbb{1}} + \nu \widetilde{Z}_B\right) + \left(\widetilde{\mathbb{1}} - \nu \widetilde{Z}_B\right) Y_N\right\} X_N \left\{\left(\widetilde{\mathbb{1}} + \nu \widetilde{Z}_B\right) + \left(\widetilde{\mathbb{1}} - \nu \widetilde{Z}_B\right) Y_N\right\} \\ &= \frac{1}{4} X_N \left\{\left(\widetilde{\mathbb{1}} + \nu \widetilde{Z}_B\right) - \left(\widetilde{\mathbb{1}} - \nu \widetilde{Z}_B\right) Y_N\right\} \left\{\left(\widetilde{\mathbb{1}} + \nu \widetilde{Z}_B\right) + \left(\widetilde{\mathbb{1}} - \nu \widetilde{Z}_B\right) Y_N\right\} \\ &= \frac{1}{2} X_N \left\{\left(\widetilde{\mathbb{1}} + \nu \widetilde{Z}_B\right) - \left(\widetilde{\mathbb{1}} - \nu \widetilde{Z}_B\right)\right\} \\ &= \nu X_N \otimes \widetilde{Z}_B, \end{aligned} \tag{S66}$$

and we next consider the measurement channel $\mathsf{M}$, noting that only the measurements of $B$ sites act nontrivially on (S66), since none of the measurements in $\mathsf{M}$ act on the final site ($j = N$), but $\mathsf{M}_B$ acts nontrivially on $\widetilde{Z}_B = \prod_{j \in B} \widetilde{Z}_j$.

Following (S5), the unitary $\mathsf{M}_B$ is given by the product over individual measurement unitaries,

$$\mathsf{M}_B = \prod_{j \in B} \left\{\frac{1}{2}(\mathbb{1} + Z_j)\widetilde{\mathbb{1}}_j + \frac{1}{2}(\mathbb{1} - Z_j)\widetilde{X}_j\right\}, \tag{S67}$$

and applying one of the $\mathsf{M}_B$ channels—corresponding to $b \in B$—to (S66) gives

$$\begin{aligned}\mathsf{M}_b^\dagger \mathcal{R}_x^\dagger X_N \mathcal{R}_x \mathsf{M}_b &= \left\{\frac{1}{2}(\mathbb{1} + Z_b) + \frac{1}{2}(\mathbb{1} - Z_b)\widetilde{X}_b\right\} \nu X_N \widetilde{Z}_B \left\{\frac{1}{2}(\mathbb{1} + Z_b) + \frac{1}{2}(\mathbb{1} - Z_b)\widetilde{X}_b\right\} \\ &= \nu X_N \widetilde{Z}_B \left\{\frac{1}{2}(\mathbb{1} + Z_b) - \frac{1}{2}(\mathbb{1} - Z_b)\widetilde{X}_b\right\} \left\{\frac{1}{2}(\mathbb{1} + Z_b) + \frac{1}{2}(\mathbb{1} - Z_b)\widetilde{X}_b\right\} \\ &= \nu X_N \widetilde{Z}_B \left\{\frac{1}{2}(\mathbb{1} + Z_b) - \frac{1}{2}(\mathbb{1} - Z_b)\right\} \\ &= \nu X_N Z_b \widetilde{Z}_B, \end{aligned} \tag{S68}$$

and applying the channels for the remaining $b$ sites gives

$$\mathsf{M}_B^\dagger \mathcal{R}_x^\dagger X_N \mathcal{R}_x \mathsf{M}_B = \nu X_N Z_B \widetilde{Z}_B, \tag{S69}$$

and finally, evolving via the entangling circuit $\mathcal{W}$ (depicted in Fig. S1) gives

$$\mathcal{W}^\dagger \mathsf{M}_B^\dagger \mathcal{R}_x^\dagger X_N \mathcal{R}_x \mathsf{M}_B \mathcal{W} = \left(X_1 \times \prod_{j \notin (D \cup \{1\})} Z_j\right) \otimes \widetilde{Z}_B, \tag{S70}$$

which indeed satisfies (S63), as $\widetilde{Z}_B$ acts trivially on the initial state (with all Stinespring registers in the state $|0\rangle$).

---

[9] Note that we also could have chosen the generating logical operators to be $X_\mathrm{L}$ and $Z_\mathrm{L}$, in which case one instead checks for $X$- and $Z$-type Pauli errors. A $Y$-type error simply corresponds to both $X$ and $Z$ errors.

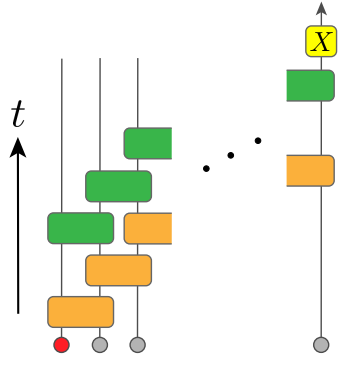

FIG. S3. The special quantum circuit construction for $M = 0$ (no measurements). The leftmost (red) site denotes the initial logical qubit to be transferred to the rightmost site. This protocol satisfies $L = T - 2$.

The analogous calculation for the $Y_N$ operator follows similarly: Using (S58b), we define

$$\mathcal{R}_y = \frac{1}{2}\left(\widetilde{\mathbb{1}} + \nu\widetilde{Z}_C\right) + \frac{1}{2}\left(\widetilde{\mathbb{1}} - \nu\widetilde{Z}_C\right)X_N, \tag{S71}$$

$$\tag{S72}$$

where, as with the $X$ case, we note that $\mathcal{R}_y^\dagger X_N \mathcal{R}_y = X_N$, so this channel does not affect the previous calculation. We likewise define the measurement channel for $C$ sites,

$$\mathsf{M}_C = \prod_{j \in C}\left\{\frac{1}{2}\left(\mathbb{1} + Z_j\right)\widetilde{\mathbb{1}}_j + \frac{1}{2}\left(\mathbb{1} - Z_j\right)\widetilde{X}_j\right\}, \tag{S73}$$

and evolving $Y_N$ under the full circuit $\mathcal{T}$ gives

$$\mathcal{W}^\dagger \mathsf{M}_C^\dagger \mathcal{R}_y^\dagger Y_N \mathcal{R}_y \mathsf{M}_C \mathcal{W} = \left(Y_1 \times \prod_{j \notin (A \cup \{N\})} Z_j\right) \otimes \widetilde{Z}_C, \tag{S74}$$

which is also of the form (S63), and again, $\widetilde{Z}_C$ acts trivially on the initial state.

Thus, the TFIM protocol (depicted in Fig. S1) transfers a logical state a distance $L = N - 1$ from site $j = 1$ to site $j = N$ with fidelity one upon implementing error correction using Tab. 2, as captured by the logical operators.

### 4.5. Special case: No measurements

Here we consider the special case in which no measurements are made ($M = 0$). The corresponding circuit in this case is equivalent to nearest-neighbor hopping (consecutive SWAP gates) up to an $\mathsf{O}(1)$ difference in circuit depth $T$, and is depicted in Fig. S3. We choose the initial logical operators to be

$$X_\mathrm{L} = X_1 \times \prod_{j=2}^{N} Z_k \tag{S75a}$$

$$Y_\mathrm{L} = Y_1 \times \prod_{j=2}^{N} Z_k, \tag{S75b}$$

with $Z_\mathrm{L} = -\mathrm{i}X_\mathrm{L}Y_\mathrm{L}$. After evolving under the time-evolution circuit $\mathcal{W}$ but before applying the local, error-correction unitary $\mathcal{R}$ on site $j = N$, the logical operators take the form

$$\mathcal{W}X_\mathrm{L}\mathcal{W}^\dagger = X_N \tag{S76a}$$

$$\mathcal{W}Y_\mathrm{L}\mathcal{W}^\dagger = -Y_N, \tag{S76b}$$

meaning that the initial state $|\psi\rangle$ on site $j = 1$ has been transferred to site $j = N$ with a bit-flip ($X$) error, which is corrected by applying $\mathcal{R} = X_N$. The total circuit depth—excluding the final error-correction unitary $X_N$—is $T = N + 1 = L + 2$, which is comparable to $T = L$ for nearest-neighbor hopping. We require $T \geq 3$, and the corresponding bound is $L \leq T - 2$, so that $T_0 = -2$ and $M_0 = 0$ in (S119).

# 5. EQUIVALENCE OF USEFUL QUANTUM TASKS

The TFIM code of Sec. 4 is a Clifford teleportation protocol that uses a modest number of measurements to enhance the distance over which information can be transferred. It is similar in spirit to the entanglement-swapping teleportation protocol (ESTP) of the main text. In both cases, we find that *using* the outcomes of measurements in $M$ regions speeds up quantum teleportation by a factor of $\sim 2M$ compared to unitary gates alone. A natural question is whether this factor $2M$ represents *the best possible speedup* to teleportation (and other useful quantum tasks)?

The remainder of the Supplementary Material provides an affirmative answer to the foregoing question by proving speed limits on generic quantum information dynamics in the presence of measurements. Essentially, using the Stinespring formalism [S1, S3] detailed in Sec. 2, we extend Lieb-Robinson bounds [S7] to quantum dynamics with measurements to find a general bound on the completion of useful quantum tasks. As a direct result of our speed bounds, one should not seek out protocols that are parametrically more efficient than those we have presented (e.g., the TFIM protocol of Sec. 4 and ESTP of the main text), as they *do not exist.*

We begin with two simple facts—Propositions 1 and 2—as a prelude to subsequent derivations. Working in the dilated Hilbert space $\mathcal{H}_{\mathrm{dil}}$, both time evolution and projective measurements are captured by the generic *unitary* channel $\mathcal{W} \in \mathrm{End}(\mathcal{H}_{\mathrm{dil}})$. We first show in Proposition 1 that, in order to teleport one qubit of information from site $i$ to site $f$, operators on $f$ evolved backwards in time in the Heisenberg-Stinespring picture must grow to have nontrivial support on site $i$ (i.e., the time-evolved operator must act nontrivially on the $i$th qubit). This is a "baby version" of the proof for the general speed limit in the presence of measurements: We primarily consider *operators* rather than states, as notions of locality are most precisely stated by determining the region upon which an operator acts nontrivially. We next show in Proposition 2 that any protocol that produces a Bell pair on sites $i$ and $f$ can also achieve teleportation from site $i$ to site $f$ (and vice versa) with only $\mathsf{O}(1)$ corrections to $M$ and $T$ and an extra qubit. This can be seen, e.g., by considering the standard teleportation protocol on three qubits discussed in the main text.

## 5.1. Connection between state transfer and operator growth

We now present and prove Proposition 1, directly relating teleportation and the growth of Pauli operators.

**Proposition 1.** *Suppose that $\mathcal{W}$ transfers an arbitrary single-qubit state $|\psi\rangle$ from site $i$ to $f$, i.e.,*

$$\mathcal{W}|\Psi_i(\alpha, \beta)\rangle = |\Psi_f(\alpha, \beta)\rangle, \tag{S77}$$

*where the above states are given by*

$$|\Psi_i(\alpha, \beta)\rangle = (\alpha|0\rangle_i + \beta|1\rangle_i) \otimes |\Phi_i\rangle_{\{i\}^{\mathrm{c}}}, \quad |\Psi_f(\alpha, \beta)\rangle = |\Phi_f\rangle_{\{f\}^{\mathrm{c}}} \otimes (\alpha|0\rangle_f + \beta|1\rangle_f), \quad \text{and } |\alpha|^2 + |\beta|^2 = 1, \tag{S78}$$

*where $\{j\}^{\mathrm{c}}$ denotes all qubits in $\mathcal{H}_{\mathrm{dil}}$ excluding $j$ (i.e., the complement of the set $\{j\}$), and both $|\Phi_i\rangle$ and $|\Phi_f\rangle$ are arbitrary many-body states of all qubits except $i$ and $f$, respectively. Then we must have*

$$\| [\mathcal{W}^\dagger X_f \mathcal{W}, Z_i] \| = 2, \tag{S79}$$

*where the* operator norm $\|\mathcal{O}\|$ *is defined as*

$$\|\mathcal{O}\| = \max_{\lambda \in \mathrm{eigs}} |\lambda|, \tag{S80}$$

*i.e., the largest-magnitude eigenvalue of $\mathcal{O}$.*

*Proof.* Applying the commutator (S79) to the initial state (S78) involves two parts:

$$\mathcal{W}^\dagger X_f \mathcal{W} Z_i |\Psi_i(\alpha, \beta)\rangle = \mathcal{W}^\dagger X_f \mathcal{W} |\Psi_i(\alpha, -\beta)\rangle = \mathcal{W}^\dagger X_f |\Psi_f(\alpha, -\beta)\rangle = \mathcal{W}^\dagger |\Psi_f(-\beta, \alpha)\rangle = |\Psi_i(-\beta, \alpha)\rangle$$

$$Z_i \mathcal{W}^\dagger X_f \mathcal{W} |\Psi_i(\alpha, \beta)\rangle = Z_i \mathcal{W}^\dagger X_f |\Psi_f(\alpha, \beta)\rangle = Z_i \mathcal{W}^\dagger |\Psi_f(\beta, \alpha)\rangle = Z_i |\Psi_i(\beta, \alpha)\rangle = -|\Psi_i(-\beta, \alpha)\rangle, \tag{S81a}$$

and subtracting the second line from the first gives the commutator

$$[\mathcal{W}^\dagger X_f \mathcal{W}, Z_i] |\Psi_i(\alpha, \beta)\rangle = 2|\Psi_i(-\beta, \alpha)\rangle \quad \Longrightarrow \quad \| [\mathcal{W}^\dagger X_f \mathcal{W}, Z_i] \| \geq 2, \tag{S82}$$

since the commutator acts on the initial state as $2X_i$, indicating that its maximal eigenvalue is at least two.

On the other hand, we have also that

$$\| [\mathcal{W}^\dagger X_f \mathcal{W}, Z_i] \| \leq 2 \|\mathcal{W}^\dagger X_f \mathcal{W}\| \|Z_i\| = 2 \quad \Longrightarrow \quad \| [\mathcal{W}^\dagger X_f \mathcal{W}, Z_i] \| \leq 2, \tag{S83}$$

which follows from the triangle inequality for the commutator norm,

$$\|[A, B]\| = \|AB - BA\| \leq \|AB\| + \|BA\| \leq 2\|A\|\|B\|, \tag{S84}$$

so that the only possibility is $\|\left[\mathcal{W}^\dagger X_f \mathcal{W}, Z_i\right]\| = 2$. $\square$

As a consequence of Proposition 1, the speed of state transfer is upper bounded by the speed with which $X_f$ "grows" to site $i$, at least in the sense of the Lieb-Robinson light cone [S23]. We note that, with an appropriately tailored initial state, state transfer can be achieved in $\mathsf{O}(1)$ time using $\mathsf{O}(1)$ measurements, independent of the distance $L$ between the task qubits $i$ and $f$. For example, if the initial state $|\psi_i\rangle$ involves a Bell pair between the sites $j$ and $f$ (where the site $j$ neighbors the site $i$), then a collective measurement on sites $j$ and $i$—followed by a unitary operation on site $f$ conditioned on the measurement outcome—achieves the standard teleportation protocol described in the main text. This also relates to the equivalence of certain quantum tasks, as we show in Proposition 2.

Thus, to derive *meaningful* bounds limiting the speed of information transfer in the presence of measurements, we must restrict our consideration to particular classes of experimentally feasible initial states. We first consider in Sec. 6 the standard case of an initial product state in the computational basis. Without loss of generality, we take this state to be $|\Psi(0)\rangle = |\mathbf{0}\rangle$, which is $|0\rangle = |\uparrow_z\rangle$ on all sites (in the $Z$ basis). In Sec. 6.8 we generalize the resulting bounds to short-range-entangled initial states via Corollary 9.

### 5.2. Quantum teleportation vs. Bell-pair creation

We now demonstrate that any state-transfer protocol is equivalent to a protocol that generates entanglement (in and of itself, a "useful quantum task"). To be specific, if a protocol can transfer an arbitrary state from site $i$ to site $f$, then up to $\mathsf{O}(1)$ corrections to the number of qubits $N$ (i.e., adding a qubit to the system), the time $T$ of Hamiltonian time evolution, and the number of measurement regions $M$ (or number of measurements $\mathcal{M}$), the same protocol also generates a Bell pair between $i$ and $f$ (and vice versa). In quantum Shannon theory [S24], this means that a teleportation channel is equivalent as a resource to one bit of entanglement, as long as classical communication is "free" (i.e., does not consume additional resources as quantified by $N$, $T$, and $M$). Classical communication is assumed to be free throughout—in the nonrelativistic limit of interest, classical signals (via photons) are *instantaneous*.

**Proposition 2.** *(a) Suppose that the quantum channel $\mathcal{W}$ in the dilated Hilbert space $\mathcal{H}_{\rm dil}$ is a quantum teleportation protocol satisfying (S77), where the initial state $|\Psi_i\rangle$ is a Kronecker product of the state $|\psi\rangle$ on site $i$ and the state $|0\rangle$ on all other sites. Then there exists another channel $\mathcal{W}' = \mathcal{W}_2(\mathcal{W} \otimes \mathbb{1}_R)\mathcal{W}_1$ that acts on the combination of the original system and an ancillary reference qubit (labelled "R") added in the neighborhood of qubit $i$, such that the state $\mathcal{W}'|\mathbf{0}\rangle$ contains a Bell pair shared by the qubits $i$ and $f$ (where $|\mathbf{0}\rangle$ denotes the state $|0\rangle$ on all sites). Moreover, the new channels $\mathcal{W}_1$ and $\mathcal{W}_2$ involve at most $\mathsf{O}(1)$ local unitaries and measurements.*

*(b) Conversely, suppose that $\mathcal{W}'$ produces a Bell pair on sites $i$ and $f$ from the initial state $|\mathbf{0}\rangle$. Then there exists another channel $\mathcal{W} = \mathcal{W}_2(\mathcal{W}' \otimes \mathbb{1}_R)\mathcal{W}_1$ that acts on the system and the ancilla qubit $R$ (in the neighborhood of $i$) such that $\mathcal{W}$ satisfies (S77) acting on the initial state $|\Psi_i\rangle = |\psi_i(\alpha,\beta)\rangle \otimes |\mathbf{0}\rangle_{\{i\}^c}$, where $\{i\}^c$ denotes all sites other than $i$, and the new channels $\mathcal{W}_1$ and $\mathcal{W}_2$ consist of at most $\mathsf{O}(1)$ local unitaries and measurements.*

*Proof.* We construct $\mathcal{W}_1$ and $\mathcal{W}_2$ explicitly for the two cases (a) and (b) using the unitary SWAP gate

$$\text{SWAP}\, |\phi\rangle_1 \otimes |\varphi\rangle_2 = |\varphi\rangle_1 \otimes |\phi\rangle_2, \tag{S85}$$

where 1 and 2 label the two qubits and $|\phi\rangle$ and $|\varphi\rangle$ are arbitrary states.

(a) The unitary $\mathcal{W}_1$ acts on converts the initial state of $i$ and $R$ (e.g., $|00\rangle$) to the Bell state $|\text{Bell}\rangle = (|00\rangle + |11\rangle)/\sqrt{2}$. After applying $\mathcal{W}$ to the original system $\{R\}^c$, (S77) implies that $R$ and $f$ form a Bell state $|\text{Bell}\rangle$. Then $\mathcal{W}_2$ is simply the SWAP gate (S85) on $\{i, R\}$; after exchanging $i$ and $R$, $i$ and $f$ are in the state $|\text{Bell}\rangle$.

(b) The unitary $\mathcal{W}_1$ acts on sites $i$ and $R$ as a SWAP gate (S85): $\mathcal{W}_1|\psi\rangle_i \otimes |0\rangle_R = |0\rangle_i \otimes |\psi\rangle_R$, with $|\psi\rangle = \alpha|0\rangle + \beta|1\rangle$. Then, after applying $\mathcal{W}'$, sites $i$ and $f$ form a Bell pair in the state $|\text{Bell}\rangle$. Finally, $\mathcal{W}_2$ is then the standard teleportation protocol on $\{R, i, f\}$ [S25] that transfers the state on $R$ to $f$ [S25], as detailed in the main text. $\square$

Crucially, Proposition 2 implies that any bound derived for the generation of a resource state with maximal entanglement between a pair of qubits also bounds quantum teleportation (state transfer) and the generation of other entangled states—up to small $\mathsf{O}(1)$ alterations, which we consider in detail in Secs. 6.8 and 7. In Sec. 6, we derive the main bound (S119) for the preparation of a Bell pair shared by qubits labelled $i$ and $f$ under generic protocols comprising continuous-time Hamiltonian evolution and measurements. This bound is stated in Theorem 5, and also applies to state transfer with minimal modification.

## 6. BOUNDS ON CONTINUOUS TIME DYNAMICS ASSISTED BY MEASUREMENTS

We now prove an extension of the Lieb-Robinson Theorem [S7] to local quantum dynamics with measurements, outcome-dependent operations, and instantaneous classical communication. This bound also extends to *arbitrary* local quantum channels—i.e., completely-positive trace-preserving (CPTP) maps, as we explain in Sec. 4.7 of the main text. The main speed limit we derive is stated in Theorem 5: Starting from a product state $|\mathbf{0}\rangle$, a protocol cannot output a Bell state shared by qubits $i$ and $f$ *unless* it satisfies $2(M+1)\,v\,T \geq L$, where $M$ counts the number of measurement *regions*, $T$ is the total time spent applying local unitary dynamics, and $L$ is the distance between the task qubits $i$ and $f$. The task of teleportation then obeys the same bound with $\mathsf{O}(1)$ modifications according to Proposition 2 above. In Sec. 7 we extend these bounds to the generation of other useful entangled (and/or correlated) resource states. In Sec. 8, we derive an analogous bound for the preparation of $Q$ Bell states (or teleportation of $Q$ logical qubits).

In a nutshell, the idea of Theorem 5 is that a Lieb-Robinson bound on the dynamics in the *dilated* Hilbert space ensures that the final state is $\epsilon$ close (in trace distance) to a state that has no entanglement between $i$ and $f$ (S112), as long as $2(M+1)\,v\,T \lesssim (1-\epsilon)L$, where $0 < \epsilon < 1$, up to asymptotically unimportant corrections.

We also stress that the techniques we use to prove the trace-distance condition (S112) and main bound (S119) are markedly different from those used to prove standard Lieb-Robinson bounds under time evolution or Lindblad dynamics alone. The latter proofs instead bound the norm of commutators between operators $A_x(Y)$ and $B_y(0)$ where $L = d(x,y)$. Additionally, the derivation of (S112) and (S119) does not involve using the unitary representation of generic CPTP maps to establish a standard Lieb-Robinson bound for the *dilated* Hilbert space $\mathcal{H}_{\text{dil}}$. Instead, we treat measurements (and other channels) on separate footing than time evolution to capture the instantaneous communication of outcomes and the crucial ingredient of outcome-dependent *feedback*.

Proving the bound (S119) with the full generality we desire requires numerous technical steps, which we divide among multiple subsections. Before stating Theorem 5 in Sec. 6.4, we rigorously define the model(s) we consider—and particularly, the formalization of measurement channels—with a useful result stated in Lemma 4. We also discuss crucial details of standard Lieb-Robinson bounds and separability of bipartite states. We then prove Theorem 5 in several steps. We first sketch the strategy of the proof, then prove a simple limiting case assuming that the result of Lemma 7 holds. We then prove Lemma 7 and generalize the results. Finally, we extend the speed limit of Theorem 5 to more general (i.e., short-range entangled) initial states in Sec. 6.8 via Corollary 9.

### 6.1. Model setup

We begin by precisely defining the class of protocols to be considered, along with some notation that will, eventually, be helpful. We first define an undirected graph $G = (V, E)$ where $V$ is the set of vertices and $E$ is the set of edges; each edge $X \in E$ also defines a *set* $X = \{v_1, v_2\}$, corresponding to the two vertices joined by the edge $X$. We assume that $G$ has bounded degree $\leq K$—i.e., that no vertex connects to more than $K$ edges.

For ease of presentation, we restrict our consideration to many-body systems wherein a qubit is assigned to each vertex of the graph, such that the physical Hilbert space is given by $\mathcal{H}_{\text{ph}} = (\mathbb{C}^2)^{\otimes |V|}$, where $|V|$ is the total number of vertices (qubits). We assume throughout the discussion to follow that the Hamiltonian $H$ takes the form[10]

$$H(t) = \sum_{X \in E} H_X(t), \tag{S86}$$

where $H_X(t) \in \text{End}(\mathbb{C}_u^2 \otimes \mathbb{C}_v^2)$ acts nontrivially only on the vertices $u$ and $v$ connected by edge $X = \{u, v\}$. We suppose that, at all times $t$ and for all edges $X$, the strength of the local terms is bounded by

$$\|H_X(t)\| \leq h, \tag{S87}$$

where $0 < h \in \mathbb{R}$ upper bounds the operator norm of the local terms comprising $H(t)$.

The Hamiltonian $H(t)$ generates operator dynamics via the Heisenberg equation of motion,

$$\partial_t A = \mathrm{i}\, \text{ad}_{H(t)} A = \mathrm{i}\,[\,H(t)\,,\,A\,]\,, \tag{S88}$$

where the dynamics generated by $H(t)$ are local: Information cannot propagate faster than the Lieb-Robinson velocity,

$$v \leq 2\mathrm{e}(K-1)h\,, \tag{S89}$$

---

[10] It is potentially straightforward to extend our results to systems with power-law interactions following [S23, S26, S27]. However, an alternative definition of locality is required; hence, we relegate the treatment of such long-range interacting systems to future work.

where e is Euler's number, $K$ is the maximal degree of $G$, and $h$ is the maximum local energy scale (S87). This is summarized by the following result, which we quote from the literature [S7, S28–S30].

**Theorem 3** (Lieb-Robinson Theorem). *Let $X, Y \subset V$ be subsets of the vertices of the graph $G$, and suppose that the operators $A_X$ and $B_Y$ act nontrivially only within $X$ and $Y$, respectively. If $r = d(X, Y)$ gives the distance between the two sets (along the shortest possible connecting path, fixed by $G$ itself), and $\|A_X\| = \|B_Y\| = 1$, then for some $0 < c_{\text{LR}} < \infty$ (where $c_{\text{LR}}$ depends only on $|X|$) we have that*

$$\|[A_X(t), B_Y]\| \leq c_{\text{LR}} \left(\frac{v\,t}{r}\right)^r. \tag{S90}$$

We refer the reader to [S7, S28–S30] for proofs of the Lieb Robinson Theorem. As a comment, one need only consider the case where $|X| = 2$, so that $c_{\text{LR}}$ is a fixed $\mathsf{O}(1)$ constant. This is most naturally justified in the formalism of [S31], where the commutator in (S90) is expanded as a sum of self-avoiding paths originating in $X$ and terminating in $Y$. The prefactor $c_{\text{LR}}$ is then chosen as an upper bound on the number of self-avoiding paths that start from $X$, no matter where they end, which removes any dependence on $Y$ from the bound (S90). Hence, the constant $c_{\text{LR}}$ in Theorem 3 is fixed as part of the proof itself, and crucially, is always some $\mathsf{O}(1)$ number for the Hamiltonians of interest.

We also note that the nature of the microscopic degrees of freedom and details of the Hamiltonian may result in modifications to the bound (S90) compared to the original Theorem [S7]. However, the resulting bound is generically of the form (S90). Since we restrict our consideration to systems of qubits, any differences compared to the original Lieb-Robinson Theorem [S7] are related to details of the particular Hamiltonian $H$. In most cases, one recovers a bound of the form (S90), where different Hamiltonians lead to different Lieb-Robinson velocities $v$.

We therefore expect that our results extend to quantum protocols involving time evolution under *any* Hamiltonian that admits a bound of the form (S90), including those with $k$-local, exponentially decaying, or even power-law interactions, and those that act on other types of degrees of freedom (with the likely exception of bosons [S32]). However, we note that long-range interactions of the form $V(x, y) \sim \exp(-|x - y|)$ or $V(x, y) \sim |x - y|^{-\alpha}$ for some $\alpha > 0$ require an alternate proof strategy, and are beyond the scope of this work.

Crucially, the bound we derive immediately extends to a larger class of models. Replacing the qubits on each vertex with $d$-state qudits (for $d \geq 2$), fermions, or Majorana mode and/or replacing the two-body nearest-neighbor terms (S86) with $k$-body terms (which act on vertices confined to a region whose size does not scale with $L, M, N, T$) merely alters the constant $c_{\text{LR}}$ in (S90), which is asymptotically unimportant to the bound (S119). Hence, (S119) and related bounds constrain useful quantum tasks implemented on systems with arbitrary, finite-dimensional local Hilbert spaces and local interactions between finitely many degrees of freedom, as we discuss in Sec. 4.6 of the main text.

A graph $G$ is said to have spatial dimension $D$ if $D$ is the smallest integer such that the number of sites at a given distance $y$ from any vertex $v \in V$ is upper bounded by $c_D\, y^{D-1}$, for some $y$-independent constant $c_D$[11]:

$$\sum_{u \in V} \delta_{y, d(u,v)} \leq c_D\, y^{D-1}, \tag{S91}$$

where $\delta_{a,b}$ is the Kronecker delta (which is 1 if $a = b$ and 0 otherwise). For sufficiently large distances between two qubits, we bound the time it takes to generate a Bell pair using a velocity $v_E$ of the form

$$v_E = \begin{cases} v & \text{if } G \text{ has spatial dimension, } D < \infty \\ (K - 1)\, v & \text{otherwise} \end{cases}, \tag{S92}$$

meaning that $v_E$ is the Lieb-Robinson velocity $v$ finite-dimensional graph $G$, while $v_E$ is enhanced by $(K - 1)$ as $D \to \infty$ (where $K$ is the maximal degree of $G$). Precise statements of the validity of this result appear in Theorem 5.

### 6.2. Measurement trajectories

In this work, we consider quantum dynamics in the presence of measurements. By convention, we take the initial state (at time $t = 0$) to be $|0\rangle$ in the computational basis—i.e., all physical and Stinespring qubits are initialized in the $+1$ eigenstate of $Z$. Our goal is then to constrain how quickly one can generate a Bell pair from this initial state, using an arbitrarily clever combination of projective measurements and local Hamilonian time evolution.

---

[11] Note that $D$ is only meaningful in the limit of a large number of vertices $|V| \gg 1$, where $c_D$ is an $\mathsf{O}(1)$ number that does not scale with $|V|$. The definition (S91) allows for fractal dimension $D$. For example, a linear chain of equally spaced qubits has $D = 1$ because, for any fixed vertex $v \in V$, there is a constant number of vertices $x \in V$ with $d(x, v) = y$; other examples are similarly intuitive.

Details of how to use the Stinespring formalism to record and access measurement outcomes appear in Sec. 2. We also allow for modification of the Hamiltonian $H(t)$ at any time $t$ after a measurement occurs, based on the outcome of that measurement (also known as "adaptive protocols"). For simplicity, it is helpful to define the Hamiltonian protocol $H^{\bm{n}}(t)$ that occurred *post facto*, based on the actual measurement outcomes that obtained (which, for convenience, are stored in the Stinespring qubits, whose post-measurement state is denoted by the binary outcome vector or "string" $\bm{n}$, which encodes the measurement trajectory). The operator that projects onto the trajectory $\bm{n}$ is $P_{\bm{n}}$ (S15), with associated probability $p_{\bm{n}}$ (S16). The actual Hamiltonian can then be written in the form

$$H(t) \equiv \sum_{\bm{n}} P_{\bm{n}} H^{\bm{n}}(t), \tag{S93}$$

where $P_{\bm{n}}$ acts only on the Stinespring qubits (S15), and $H^{\bm{n}}(t)$ acts only on physical qubits. Note that we do not distinguish between subscripts and superscripts, which we interchange as needed for formatting purposes.

By causality, we must have the redundancy $H^{\bm{n}}(t) = H^{\bm{n}'}(t)$ for all times $t$ prior to which the two trajectories $\bm{n}$ and $\bm{n}'$ have identical outcomes (i.e., only after the trajectories fork can the corresponding Hamiltonians differ). Note that any result proven in this formalism holds for quantum circuits comprising both unitary evolution and measurements, whose unitary component can always be formulated as a time-dependent Hamiltonian of the form (S93). For simplicity, we restrict to Hamiltonians with nearest-neighbor interactions (S86) on the graph $G$ of interest.

Note that we may interrupt the unitary evolution generated by $H(t)$ (S93) to measure some local operator. For simplicity, we restrict these measurements to involutory operators $A$ (which obey $A^2 = \mathbb{1}$) so that each measurement has only two possible outcomes (generally $\pm 1$), and thus the Stinespring degrees of freedoms are all qubits (with $|0\rangle$ corresponding to the "default" measurement outcome $+1$, and $|1\rangle$ corresponding to the other measurement outcome $-1$). We note that these are the most common operators one might encounter in qubit systems (corresponding to strings of Pauli operators), although in principle, one could also measure projectors directly as part of a hybrid protocol. We refer the reader to Sec. 2 for details of the Stinespring formalism, and note that, in addition to the physical Hilbert space, the calculations to follow also include Stinespring registers to record measurement outcomes.

We consider protocols $H(t)$ of the form (S93) for $0 \leq t \leq T$. We construct the *measurement set* $\mathbb{M}^{\bm{n}}$ for the outcome trajectory $\bm{n}$ as the set of measurement *regions* $\{S_m^{\bm{n}}\}$ measured during the course of the protocol $\mathcal{W}$ along trajectory $\bm{n}$ (the trajectory dependence reflects the fact that the particular choice of measurement regions may depend on prior outcomes). In (S94) we define $M$ to be the number of measurement regions, which is no larger than the number of Stinespring qubits (equality holds if each measurement region contains a single, binary measurement).

The measurement regions are themselves sets of sites: The $m$th measurement region along outcome trajectory $\bm{n}$ is naïvely given by the set of sites $\widetilde{S}_m^{\bm{n}}$ (the tilde reflects that this object may be relabelled during the construction of the measurement set $\mathbb{M}^{\bm{n}}$, detailed below. In our most general proof in Theorem 5, we allow for subsequent choices of measurement regions $\widetilde{S}_m^{\bm{n}}$ (as well as the choice of observable $\mathcal{O}_{\bm{n},m}$ to measure and the time $t_{\bm{n},m}$ at which to measure) to be conditioned on the outcomes of prior measurements. Note that we relabel the measurement regions after completing the recursive construction of the measurement set $\mathbb{M}^{\bm{n}}$, which proceeds as follows:

1. We initialize the measurement set as $\mathbb{M}^{\bm{n}} = \{\widetilde{S}_1^{\bm{n}}\}$ for the first measurement.

2. For the $m$th measurement, we update $\mathbb{M}^{\bm{n}}$ as follows:

    (a) If $\widetilde{S}_m^{\bm{n}} \in \mathbb{M}^{\bm{n}}$, then $\mathbb{M}_{\bm{n}}$ is not updated, because this set of qubits has already been counted as a measurement region in $\mathbb{M}_{\bm{n}}$.

    (b) Otherwise, if $\widetilde{S}_m^{\bm{n}} \notin \mathbb{M}^{\bm{n}}$, then we add $\widetilde{S}_m^{\bm{n}}$ to $\mathbb{M}^{\bm{n}}$ (i.e., $\mathbb{M}^{\bm{n}} \to \mathbb{M}^{\bm{n}} \cup \{\widetilde{S}_m^{\bm{n}}\}$).

3. Once the above prescription has been repeated for all measurements $m$ we next modify the measurement set $\mathbb{M}^{\bm{n}}$ as follows:

    (a) If there is no measurement region $\widetilde{S}_m^{\bm{n}} \in \mathbb{M}^{\bm{n}}$ that includes the *initial* qubit $i$ (i.e. $\nexists \widetilde{S}_m^{\bm{n}} \in \mathbb{M}^{\bm{n}}$ s.t. $i \in \widetilde{S}_m^{\bm{n}}$), then we add the set $S_{-1}^{\bm{n}} = \{i\}$ to $\mathbb{M}^{\bm{n}}$ (i.e., $\mathbb{M}^{\bm{n}} \to \mathbb{M}^{\bm{n}} \cup \{\{i\}\}$). If there is a set $\widetilde{S}_m^{\bm{n}}$ that contains the qubit $i$, then we rename this set "$S_{-1}^{\bm{n}}$".

    (b) If there is no measurement region $\widetilde{S}_m^{\bm{n}} \in \mathbb{M}^{\bm{n}}$ that includes the *final* qubit $f$ (i.e. $\nexists \widetilde{S}_m^{\bm{n}} \in \mathbb{M}^{\bm{n}}$ s.t. $f \in \widetilde{S}_m^{\bm{n}}$), then we add the set $S_0^{\bm{n}} = \{f\}$ to $\mathbb{M}^{\bm{n}}$ (i.e., $\mathbb{M}^{\bm{n}} \to \mathbb{M}^{\bm{n}} \cup \{\{f\}\}$). If there is a set $\widetilde{S}_m^{\bm{n}}$ that contains the qubit $f$, then we rename this set as "$S_0^{\bm{n}}$".

    (c) Finally, we relabel all remaining measurement regions in $\mathbb{M}^{\bm{n}}$ according to $\widetilde{S}_m^{\bm{n}} \to S_m^{\bm{n}}$. These regions appear in $\mathbb{M}^{\bm{n}}$ in order of increasing $m$, starting from $m = -1$ for the region containing site $i$. Lastly, if there is a missing value corresponding to $m = m_*$—due to conversion of a measurement region to the initial or final region (labelled $-1$ and $0$, respectively)—then we decrease the value of all $m > m_*$ by one.

The result is the measurement set $\mathbb{M}^{\boldsymbol{n}} = \{S_{-1}^{\boldsymbol{n}}, S_0^{\boldsymbol{n}}, S_1^{\boldsymbol{n}}, \cdots\}$, where $S_{-1}^{\boldsymbol{n}}$ always includes the site $i$ and $S_0^{\boldsymbol{n}}$ always includes the site $f$. We further define $M$ to be the number of measurement regions over *all trajectories*, given by

$$M + 2 \equiv \max_{\boldsymbol{n}} |\mathbb{M}^{\boldsymbol{n}}|, \tag{S94}$$

i.e., the greatest number of elements realized by a set $\mathbb{M}^{\boldsymbol{n}}$ (with $|\mathbb{M}^{\boldsymbol{n}}|$ elements) among all trajectories $\boldsymbol{n}$. We also define the maximum *range* of measurements (over all trajectories) as

$$l \equiv \max_{\boldsymbol{n}} \max_{S \in \mathbb{M}^{\boldsymbol{n}}} |S|, \tag{S95}$$

i.e., the greatest number of sites $|S|$ in any region $S$, over all measurement sets $\mathbb{M}^{\boldsymbol{n}}$ and trajectories $\boldsymbol{n}$.

We remark that there is some freedom in the definition of measurement regions. For example, in the ESTP considered in the main text and the TFIM protocol considered in Sec. 4, one can define the Bell pair measurements—always performed on two neighboring qubits—as two, single-site measurement regions, or a single, two-site measurement region. We choose the latter (a single, two-site region) because (*i*) our concern lies with notions of locality and (*ii*) because this implies saturation of the main bound of Theorem 5 (in the sense that the optimal protocol corresponds to $2v_E T \approx L/M$ instead of $v_E T \approx L/M$). Any difference in convention leads to $\mathsf{O}(1)$ modifications only.

In the Stinespring formalism, although one can view the Hamiltonian dynamics as having been generated by the "many-worlds" Hamiltonian $H(t) = \sum_{\boldsymbol{n}} P_{\boldsymbol{n}} H^{\boldsymbol{n}}(t)$—which acts on all possible measurement trajectories—it will prove more useful to consider the evolution generated by the Hamiltonians $H^{\boldsymbol{n}}(t)$ corresponding to *particular* outcome trajectories $\boldsymbol{n}$ in order to invoke standard Lieb-Robinson bounds. To recover our main results, we expand a generic operator $\mathcal{O}$—which initially acts nontrivially only on the physical qubits—according to

$$\mathcal{O} = \sum_{\boldsymbol{n}} \mathcal{O}_{\boldsymbol{n}}, \quad \text{where} \quad \mathcal{O}_{\boldsymbol{n}} \equiv \mathcal{O} \otimes P_{\boldsymbol{n}}, \tag{S96}$$

which follows from resolving the identity on the Stinespring qubits via

$$\sum_{\boldsymbol{n}} P_{\boldsymbol{n}} = \sum_{\boldsymbol{n}} |\boldsymbol{n}\rangle\langle\boldsymbol{n}| = \widetilde{\mathbb{1}}_{\text{ss}}. \tag{S97}$$

The evolution of $\mathcal{O}_{\boldsymbol{n}}$ (S96) in the Heisenberg picture is captured by

$$\mathcal{O}_{\boldsymbol{n}}(s) \equiv \mathcal{W}^\dagger(T; T-s) \, \mathcal{O}_{\boldsymbol{n}} \, \mathcal{W}(T; T-s), \tag{S98}$$

which only depends on the Hamiltonian (S93) and measurements in the trajectory $\boldsymbol{n}$, where $\mathcal{W}(T; T-s)$ is the unitary channel that evolves states in $\mathcal{H}_{\text{dil}}$ from time $t = T-s$ to time $t = T$. The dilated unitary $\mathcal{W}(T; T-s)$ captures *all* measurement trajectories. Note that we use the temporal variable $s$ for the backward time evolution of operators in the Heisenberg picture, and $t$ for forward time evolution of states in the Schrödinger picture).

**Lemma 4.** *The evolution of an observable $\mathcal{O}$, projected onto the measurement-outcome trajectory $\boldsymbol{n}$, satisfies the following two properties, captured by* (S99) *and* (S102):

1. *The time-evolved observable is defined by the exponential map*

$$\mathcal{O}_{\boldsymbol{n}}(s) = \widetilde{\mathbb{T}} \exp\left\{\int_{T-s}^T \mathrm{d}t' \mathcal{L}_{\boldsymbol{n}}(t')\right\} \mathcal{O}_{\boldsymbol{n}}, \tag{S99}$$

*where $\widetilde{\mathbb{T}}$ is the reverse-time-ordering operator (as operators are evolved "backwards" in time in the Heisenberg picture), and we introduced the superoperator Liouvillian $\mathcal{L}_{\boldsymbol{n}}(t)$[12], which contains Dirac-$\delta$ functions in $t$ as needed to account for measurements. The Liouvillian takes the general form*

$$\mathcal{L}_{\boldsymbol{n}}(t) = \mathrm{i}\,\mathrm{ad}_{H^{\boldsymbol{n}}(t)} + \sum_m \mathrm{i}\,\delta\left(t - t_{\boldsymbol{n},m}\right)\widehat{\mathsf{M}}_{\boldsymbol{n},m}, \tag{S100}$$

---

[12] A superoperator is an element of the Hilbert space $\mathrm{End}(\mathrm{End}(\mathcal{H}_{\text{dil}}))$—i.e., an operator on the space of operators on $\mathcal{H}_{\text{dil}}$.

where $\mathrm{ad}_A\, B \equiv [\,A,\,B\,]$ (S88) and $\widehat{\mathsf{M}}_{\bm{n},m}$ is the superoperator describing the mth Stinespring measurement channel along trajectory $\bm{n}$, which acts at "measurement time" $t_{\bm{n},m}$ and satisfies

$$\mathrm{e}^{\mathrm{i}\widehat{\mathsf{M}}_{\bm{n},m}}[A] = \mathsf{M}_{\bm{n},m}^{\dagger}\, A\, \mathsf{M}_{\bm{n},m}\,, \tag{S101}$$

where $\mathsf{M}_{\bm{n},m}$ acts unitarily *on* $\mathcal{H}_{\mathrm{dil}}$ *and represents the measurement of the mth observable along outcome trajectory* $\bm{n}$. *Crucially, the unitary* $\mathsf{M}_{\bm{n},m}$ *does not contain projectors* $P_{\bm{n}}$ *onto the trajectory* $\bm{n}$, *and is of the form* (S5). *The label* $\bm{n}$ *only reflects the fact that the operator being measured (and the corresponding region of support) may depend on prior outcomes in* $\bm{n}$. *When evolving operators backward in time in the Heisenberg picture, projection onto a measurement trajectory comes* first (S98), *allowing us to evolve* $\mathcal{O}_{\bm{n}}$ *directly under the combination of time-evolution and measurements and restrict to the trajectory* $\bm{n}$ *without further use of Stinespring projectors.*

2. *The time-evolved observable can be decomposed in the dilated Hilbert space according to*

$$\mathcal{O}_{\bm{n}}(s) = P_{\bm{n}_s} \otimes \mathcal{O}_{\bm{n}-\bm{n}_s}(s)_{\bm{n}}\,, \tag{S102}$$

*where* $\bm{n}_s$ *is the subset of* $\bm{n}$ *labelling the outcomes of measurements yet to be encountered in the Heisenberg evolution of the operator* $\mathcal{O}$ *after evolving by an amount* $s$ *along the measurement trajectory* $\bm{n}$. *The "trajectory"* $\bm{n}_s$ *reflects measurement outcomes that have already been recorded by time* $t = T - s$ *in the Schrödinger picture (since Heisenberg time* $s$ *corresponds to Schrödinger time* $t = T - s$*), which are the outcomes upon which the protocol* $H^{\bm{n}}(t)$ *can depend. The projector* $P_{\bm{n}_s}$ *acts on the Stinespring qubits in the subset* $\bm{n}_s \subset \bm{n}$ *that encode those previously recorded outcomes (in the Schrödinger picture). Similarly, we define* $P_{\bm{n}-\bm{n}_s}$ *to be the projector acting on the Stinespring qubits of outcomes after time* $t = T - s$, *such that* $P_{\bm{n}} = P_{\bm{n}_s} \otimes P_{\bm{n}-\bm{n}_s}$. *Then the operator* $\mathcal{O}_{\bm{n}-\bm{n}_s}(s)_{\bm{n}}$ *in* (S102) *corresponds to the Heisenberg evolution of* $\mathcal{O} \otimes P_{\bm{n}-\bm{n}_s}$ *under Liouvillian* $\mathcal{L}_{\bm{n}}(t)$, *and may act nontrivially on all later Stinespring qubits (not reflected in* $\bm{n}_s$*), along with all physical qubits.*

*Proof.* The basic idea of (S99) is that the instantaneous evolution of $\mathcal{O}_{\bm{n}}$ is generated by *either* $H^{\bm{n}}(t)$ acting only on the physical qubits, or by a measurement $\mathsf{M}_{\bm{n},m}$ that grows the Stinespring projector $P_{n_m}$ in $\mathcal{O}_{\bm{n}}$ onto the physical qubits by entangling the physical and Stinespring qubits. The basic idea of (S102) is that, at Heisenberg time $s$, the projectors in $P_{\bm{n}_s}$ have not yet been grown, because the corresponding measurements have not yet been implemented in the Heisenberg picture. We prove this result—that the time evolution of operators may be restricted to particular "trajectories" of measurement outcomes, captured by the projectors $P_{\bm{n}}$—recursively in the Heisenberg time $s$. Note that for any $\mathsf{M}_{\bm{n},m}$, there exists a corresponding superoperator $\widehat{\mathsf{M}}_{\bm{n},m}$ that can be exponentiated to realize (S101).

We now prove the result recursively, supposing that (S99) and (S102) hold at time $s$. Evolving for a small time step $s \to s^+$, there are two possibilities. The first possibility is that the evolution of $\mathcal{O}_{\bm{n}}(s)$ (S102) from $s \to s^+$ is generated solely by the continuous-time Hamiltonian (S93), which depends on the measurement trajectory $\bm{n}_s$. The time-evolution unitary corresponding to this small time step can be written as $\sum_{\bm{n}'_s} P_{\bm{n}'_s} \otimes U_{\bm{n}'_s}$, where $\bm{n}'_s$ denotes all trajectories so far at time $t = T - s$, as seen by states. Note that the channel cannot induce transitions between different Stinespring outcomes because the measurement-dependent Hamiltonian only depends on $\widetilde{Z}$ Pauli operators, all of which commute with $P_{\bm{n}_s}$. This does not change the structure in (S102):

$$\begin{aligned}\mathcal{O}_{\bm{n}}(s^+) &= \left(\sum_{\bm{n}'_s} P_{\bm{n}'_s} \otimes U_{\bm{n}'_s}^{\dagger}\right)\left(P_{\bm{n}_s} \otimes \mathcal{O}_{\bm{n}-\bm{n}_s}(s)_{\bm{n}}\right)\left(\sum_{\bm{n}'_s} P_{\bm{n}'_s} \otimes U_{\bm{n}'_s}\right)\\ &= P_{\bm{n}_s} \otimes \left(U_{\bm{n}_s}^{\dagger} \mathcal{O}_{\bm{n}-\bm{n}_s}(s)_{\bm{n}} U_{\bm{n}_s}\right) = U_{\bm{n}_s}^{\dagger} \mathcal{O}_{\bm{n}}(s)\, U_{\bm{n}_s}\,,\end{aligned} \tag{S103}$$

where we used the orthogonality condition $P_{\bm{n}_s} P_{\bm{n}'_s} = \delta_{\bm{n}_s,\bm{n}'_s} P_{\bm{n}_s}$. Furthermore, only $U_{\bm{n}_s}$ survives in (S103), so that time evolution is generated by $H^{\bm{n}}$ via the first term of (S100), which may be conditioned only on the outcomes in $\bm{n}_s$. Thus, although adaptive unitaries act on all outcome trajectories, (S103) implies that the evolution of operators restricted to the trajectory $\bm{n}_s$ (S102) only involves terms in the unitary corresponding to the trajectory $\bm{n}_s$. Hence, Hamiltonian evolution can be considered along particular trajectories without the use of additional projectors.

The second possibility is that the evolution of $\mathcal{O}_{\bm{n}}(s)$ (S102) from $s \to s^+$ involves performing a measurement in some trajectory $\bm{n}'_s$, which *may or may not be in* $\bm{n}_s$. Because measurements are instantaneous, they can be considered separately from both Hamiltonian evolution and other measurements[13]. Thus, without loss of generality, we assume

---

[13] This is a standard result enforced by the anti-time-ordering operator $\widetilde{\mathbb{T}}$, which requires that we Trotterize the Liouvillian into infinitesimal time steps in which *either* a single measurement or a unitary time-evolution update occurs. Additionally, two nominally simultaneous measurements at time $t = T - s$ can be split into one measurement at time $s^-$ (an infinitesimal step earlier) and another at time $s^+$ (an infinitesimal step later), by definition of the measurement set.

that the Hamiltonian part of (S100) can be neglected during the infinitesimal Heisenberg interval $s \to s^+$, which picks out a single, instantaneous measurement that acts only in the trajectory $\boldsymbol{n}'_s$, precisely at Schrödinger time $t = T - s$.

In the case that $\boldsymbol{n}'_s \neq \boldsymbol{n}_s$ at the time of measurement, we apply the dilated unitary measurement channel (S5),

$$\mathsf{M}_s \equiv P_{\boldsymbol{n}'_s} \otimes \mathsf{M}_{\boldsymbol{n}',m} + \left(1 - P_{\boldsymbol{n}'_s}\right) \otimes \mathbb{1}, \tag{S104}$$

where the second term reflects that nothing happens at time $s$ to trajectories other than $\boldsymbol{n}'_s$ (which undergo neither measurement nor Hamiltonian evolution). As a reminder, only $\boldsymbol{n}'_s$—which captures the outcomes of all measurements prior to (S104) in the Schrödinger picture—can affect the location of the measurement (S104), by causality. When applying the dilated unitary (S104) to $\mathcal{O}_{\boldsymbol{n}}(s)$, the projector $P_{\boldsymbol{n}_s}$ in (S102) picks the $(1 - P_{\boldsymbol{n}'_s}) \otimes \mathbb{1}$ part of $\mathsf{M}_s$ (which includes $P_{\boldsymbol{n}_s}$) similarly to (S103). In this case, the operator does not change: $\mathcal{O}_{\boldsymbol{n}}(s^+) = \mathcal{O}_{\boldsymbol{n}}(s)$.

Alternatively, suppose that the measurement happens exactly in trajectory $\boldsymbol{n}_s$ (with no prime). We label by "$a$" the Stinespring qubit that stores the outcome of this measurement at Heisenberg time $s$. Hence, just *after* the measurement at Schrödinger (forward) time $t = T - s$—and thus, just *before* the measurement at Heisenberg (backward) time $s$—the string $\boldsymbol{n}_s$ is given by $\boldsymbol{n}_s = (\boldsymbol{n}_{s^+}, n_a)$, where $n_a$ denotes the outcome of the measurement at $s$, stored in the Stinespring qubit $a$. The measurement unitary is then written as

$$\mathsf{M}_s = P_{\boldsymbol{n}_{s^+}} \otimes \mathsf{M}_{\boldsymbol{n},m} + \left(1 - P_{\boldsymbol{n}_{s^+}}\right) \otimes \mathbb{1}, \tag{S105}$$

and the projector in (S102) selects the first term,

$$\begin{aligned}\mathcal{O}_{\boldsymbol{n}}(s^+) &= \left(P_{\boldsymbol{n}_{s^+}} \otimes \mathsf{M}_{\boldsymbol{n},m}\right) \left(P_{\boldsymbol{n}_{s^+}} \otimes P_{n_a} \otimes \mathcal{O}_{\boldsymbol{n}-\boldsymbol{n}_s}(s)_{\boldsymbol{n}}\right) \left(P_{\boldsymbol{n}_{s^+}} \otimes \mathsf{M}_{\boldsymbol{n},m}\right) \\ &= P_{\boldsymbol{n}_{s^+}} \otimes \mathsf{M}_{\boldsymbol{n},m} \left(P_{n_a} \otimes \mathcal{O}_{\boldsymbol{n}}(s^-)\right) \mathsf{M}_{\boldsymbol{n},m} = \mathsf{M}_{\boldsymbol{n},m} \mathcal{O}_{\boldsymbol{n}}(s) \mathsf{M}_{\boldsymbol{n},m},\end{aligned} \tag{S106}$$

and thus, only the measurements in trajectory $\boldsymbol{n}$ contribute to the evolution, and they occur in at most $M$ measurement regions, $S_1^{\boldsymbol{n}}, \cdots, S_M^{\boldsymbol{n}}$. Furthermore, the direct product structure in (S102) continues at $s+$.

The above analysis implies that, if (S99) and (S102) hold at time $s$, they also hold at any subsequent time $s' > s$. Since both (S99) and (S102) hold initially at $s = 0$, they continue to hold under dynamical iteration of any adaptive protocol involving Hamiltonian time evolution and projective measurements, as outlined above. Therefore, in the Heisenberg picture, one can project operators onto trajectories $\boldsymbol{n}$ according to (S96), and evolve each trajectory independently without further explicit use of the trajectory projectors $P_n$. □

### 6.3. Separable states

Theorem 5 rules out the possibility of generating Bell pairs between qubits $i$ and $f$ separated by distances $L \gg MT$ by showing that the outcome-averaged reduced density matrix for $i$ and $f$ after some protocol $\mathcal{W}$,

$$\rho_{if} = \operatorname*{tr}_{\{i,f\}^c} \left[\mathcal{W} \ket{\mathbf{0}}\bra{\mathbf{0}} \mathcal{W}^\dagger\right], \tag{S107}$$

is close in trace distance to a *separable* state $\widetilde{\rho}_{if}$ that cannot possibly host a Bell pair. The definition of the $\widetilde{\rho}_{if}$ mirrors (S107), but with $\mathcal{W} \to \widetilde{\mathcal{W}}$. The residual density matrix for the sites $i$ and $f$ obtains after tracing out all other qubits in the full density matrix (where $\{i,f\}^c$ denotes the complementary set of all physical *and* Stinespring qubits *except* for the initial and final qubits $i$ and $f$). We stress that (S107) is averaged over all measurement outcomes (S19).

The *trace distance* between two operators is defined as

$$\|\rho_1 - \rho_2\|_1 \equiv \sum_\lambda |\lambda|, \quad \text{where } \{\lambda\} \text{ is the set of eigenvalues of } (\rho_1 - \rho_2), \tag{S108}$$

where the trace distance for proper density matrices is bounded by $\|\rho_1 - \rho_2\|_1 \leq 2$ (e.g., for orthogonal pure states).

Importantly, if two density matrices have trace distance $\|\rho_1 - \rho_2\|_1 \ll 1$, they effectively cannot be distinguished using any measurement protocol [S11, S33]. Thus, if the trace distance between two density matrices is small, they cannot have drastically different physical properties, as measuring some observable would allow for this distinction. Hence, one can check whether the state $\rho_{if}$ produced by the protocol $\mathcal{W}$ (S107) captures *useful entanglement* between qubits $i$ and $f$ by computing the trace distance (S108) to a *separable* state $\widetilde{\rho}_{if}$.

A bipartite state $\rho_{AB}$ supported on regions $A$ and $B$ is *separable* if and only if it can be written in the form

$$\rho_{AB} = \sum_k p_k |\psi_k\rangle\langle\psi_k|, \quad \text{where} \ |\psi_k\rangle = |\psi_k^{(a)}\rangle_A \otimes |\psi_k^{(b)}\rangle_B, \quad \sum_k p_k = 1, \ \text{and} \ p_k \geq 0 \ \forall \, k, \tag{S109}$$

which is equivalent to the statement that the state $\rho_{AB}$ has no entanglement between the regions $A$ and $B$. When $A$ and $B$ interact via *local operations and classical communication* (LOCC), separable states remain separable. Additionally, any separable state can be prepared from any initial state. A Bell state—which is the maximally entangled state of two qubits—is certainly not separable (as can be checked explicitly). Furthermore, any nonseparable state (such as the Bell pair) has nonvanishing trace distance to the set of separable states [S11, S33, S34], because the subset of separable states is closed in trace-norm topology in the set of all states [S34]. Given these implications, Theorem 5 rules out the possibility that $\rho_{if}$ corresponds to a Bell pair at sufficiently early times, and moreover, ensures that $\rho_{if}$ is nearly indistinguishable from an unentangled state $\widetilde{\rho}$ asymptotically in the regime $(M+1)T \lesssim L$.

### 6.4. Statement of the Theorem

We now state the main Theorem, which we prove in later subsections. After stating Theorem 5, we provide commentary on its implications and how to extract the main bound (S119) featured in the main text. We then sketch the proof strategy, which requires the results of Proposition 6 and Lemma 7.

**Theorem 5.** *Consider a quantum channel $\mathcal{W}$ applied to the initial product state $|\mathbf{0}\rangle$ in the dilated Hilbert space. Suppose that $\mathcal{W}$ involves (a) measurements in at most $M$ local regions (S94) with maximal measurement range $l$ (S95), and (b) Hamiltonian evolution (S86) applied for a time period $T$ with associated velocity $v_E$ (S92).*

1. *Suppose that the measurement locations are "prefixed," meaning that they do not depend on the measurement trajectory $\mathbf{n}$. Then, for any $\epsilon \in (0,1)$, suppose that the distance $L$ between qubits $i$ and $f$ obeys*

$$(M+1)(2 v_E T + l + 4) \leq (1-\epsilon)(L+1-l), \tag{S110}$$

*and we require that $T \gtrsim D$ if $G$ is a lattice with finite spatial dimension $D$. If we also have that*

$$vT \geq \max(e, D-3) \quad \text{and} \quad (D-1)\frac{\ln vT}{vT} \leq (1-\ln 2)\epsilon, \tag{S111}$$

*then there exists a separable state $\widetilde{\rho}_{if}$ with $\mathrm{tr}[\widetilde{\rho}_{if}] = 1$ whose proximity in trace distance (S108) to the final density matrix $\rho_{if}$ (S107) on qubits $i$ and $f$ generated by the true protocol $\mathcal{W}$ obeys*

$$\|\rho_{if} - \widetilde{\rho}_{if}\|_1 \leq \frac{c\,l}{\epsilon} 2^{v_E T + \frac{l}{2} - \frac{L+1-l}{2(M+1)}}, \tag{S112}$$

*where the constant $c > 0$ is determined by the structure of the graph $G$ according to (S138).*

2. *Suppose that the measurement locations are "adaptive," meaning that they may depend on the measurement trajectory $\mathbf{n}$ (i.e., the measurement protocol depends on past outcomes). For a finite-dimensional graph $G$, there exists a constant $c > 0$, an $\epsilon \in (0,1)$ and a separable state $\widetilde{\rho}_{if}$ with $\mathrm{tr}[\widetilde{\rho}_{if}] = 1$ whose proximity in trace distance (S108) to the final density matrix $\rho_{if}$ (S107) on qubits $i$ and $f$ generated by $\mathcal{W}$ obeys*

$$\|\rho_{if} - \widetilde{\rho}_{if}\|_1 \leq \frac{c}{\epsilon} L^{D-1} 2^{v_E T + \frac{l}{2} - \frac{(1-\epsilon)L+1-l}{2(M+1)}}, \tag{S113}$$

*where the constant $c > 0$ depends on details of the graph $G$ in a manner too tedious to calculate precisely, but does not scale with $L$, $M$, or $T$, and is thus asymptotically unimportant.*

We now comment on the interpretation of Theorem 5, explain how we extract the main bound of the main text (S119), and sketch the strategy for the Theorem's proof, which appears in Sec. 6.7 and involves the use of various Propositions and Lemmas that we state and prove along the way. Case (1) above is proven in Sec. 6.5.

We first focus on the case of prefixed measurement locations, since the general case (S113) is more complicated. First, we note that (S112) is similar in form to the main result (S90) of Theorem 3—the Lieb-Robinson Theorem for standard, unitary dynamics. There, (S90) states that the commutator between the time-evolved operator $A_X(t)$ and

the static operator $B_Y$ (where $X$ and $Y$ are separated by a distance $r$) is bounded by $(vt/r)^r$. This formula provides a useful prediction of the Lieb-Robinson velocity $v$ only in the asymptotic limit[14] $r \to \infty$ [S7, S28–S30].

Essentially, we define $\delta = vt - r$ in (S90) and take the $r \to \infty$ limit (note that $\delta$ remains fixed in the limit of interest, while $\delta \to -\infty$ in the trivial limit). The righthand side of (S90) takes the form

$$\lim_{r \to \infty} \left(\frac{vt}{r}\right)^r = \lim_{r \to \infty} \left(1 + \frac{\delta}{r}\right)^r \equiv e^{\delta} = e^{vt-r}, \tag{S114}$$

which upper bounds the commutator in (S90). In the trivial limit where $t \ll r/v$, we find that $\delta(r) \to -\infty$ as $r \to \infty$, so that the commutator vanishes. However, for $t \gtrsim r/v$, we keep $\delta(r)$ fixed as $r \to \infty$, so that (S114)—and thus, the commutator (S90)—may be asymptotically nonzero. Hence, it is most clear in the asymptotic limit $r \to \infty$ that quantum information in such local models travels with group velocity $v$.

We now consider the result (S112) of Theorem 5 in the asymptotic regime of interest, $L \to \infty$. With some simple manipulations and invoking the assumption (S110) of Theorem 5, (S112) can be rewritten for any fixed $M$ as

$$\|\rho_{if} - \widetilde{\rho}_{if}\|_1 \lesssim 2^{-\epsilon \frac{L+1-\ell}{2(M+1)}} \ll 1, \tag{S115}$$

where the rightmost inequality recovers as $L \to \infty$, with $\epsilon$ and $M$ fixed, nonzero constants compatible with (S110).

In other words, in the limit of interest—where the separation $L$ between $i$ and $f$ is large—Theorem 5 implies (S115), which guarantees that any protocol $\mathcal{W}$ that obeys the assumptions (S110) and (S111), cannot realize a Bell pair shared by the qubits $i$ and $f$. This conclusion follows from the proximity of the state to a separable state, which has no entanglement whatsoever between the qubits $i$ and $f$. Strictly speaking, (S115) applies to the *outcome-averaged* density matrices (S107) following the protocol $\mathcal{W}$ (or its nonentangling counterpart).

We now argue that Theorem 5 also precludes the possibility that the protocol $\mathcal{W}$ realizes a Bell pair on some nonvanishing fraction $\nu$ of all possible trajectories, while the outcome-averaged density matrix $\rho_{if}$ (S107) remains close to a separable state per (S115). As a simple concrete example, suppose that every outcome trajectory realizes one of the four possible Bell basis states $\{\,|\mathrm{Bell}_k\rangle\,|\,k=1,2,3,4\,\}$ on the qubits $i$ and $f$. We then partition the full the set $\{\bm{n}\}$ of outcome trajectories into into four subsets labelled $k=1,2,3,4$, corresponding to the Bell state that realizes on qubits $i$ and $f$. Suppose that the total probability to realize *some* trajectory in the subset $k$ is $p = 1/4$ for each trajectory $k$, meaning that all four Bell basis states realize on the qubits $i$ and $f$ with equal probabilities. Then, averaging over outcomes, the reduced density matrix is given by

$$\rho_{if} = \sum_{k=1}^{4} |\mathrm{Bell}_k\rangle\langle\mathrm{Bell}_k|_{if} = \frac{1}{4}\,\mathbb{1}_{if}, \tag{S116}$$

which is simply the maximally mixed state of the two qubits. We note that this state is indeed separable—and thus compatible with (S112)—and also precisely the scenario that recovers in the Standard Teleportation Protocol or ESTP of the main text, without using the outcomes of the measurements for error correction.

However, we next argue by contradiction that realizing Bell pairs in some finite subset of all trajectories with $\mathsf{O}(1)$ total probability is also impossible by Theorem 5. Returning to the simple example above, since we have access to all measurement outcomes in each trajectory, it should be possible to use these outcomes to determine which Bell state has realized on the qubits $i$ and $f$. With this knowledge, a local unitary acting on qubit $f$ alone can be used to rotate every Bell-basis state $|\mathrm{Bell}_k\rangle$ in every trajectory group $k$ into a particular Bell state, say $|\mathrm{Bell}_1\rangle$. Our knowledge of the time-dependent Hamiltonian along with the measurements made and their outcomes guarantees that this unitary can be constructed. We can then convert the separable state $\rho_{if}$ (S116) into the nonseparable Bell state $\rho_{if} = |\mathrm{Bell}_k\rangle\langle\mathrm{Bell}_k|_{if}$ using negligible resources (a single local operation). However, this violates Theorem 5, because the combined protocol (including this final error-correction step) outputs a unique Bell state using comparable resources, and thus violates (S112) for essentially the same values of $L, v, M, T$.

As a result, it must be the case that the final state $\rho_{if}$ contains little (if any) *useful* entanglement, just as the maximally mixed state (S116) contains no useful entanglement. In the case of the maximally mixed state $\rho \propto \mathbb{1}$ of $N$ qubits, this is most apparent from recognizing that the statistics of any Pauli measurement in this state are equivalent to having $N$ random classical bits (all outcomes are equally probable in this state). This is true despite the fact that this state can capture an ensemble of Bell states. Note that the realization of all four Bell basis states in the above example (and the equal probabilities of each subset trajectory, and thus each Bell basis state) is simply a convenient

---

[14] Attempting to extract the speed limit $v$ away from this $r \to \infty$ limit gives suboptimal predictions, due to complications associated with continuous time evolution involving tails of operators. This will hold for (S112) as well.

means to realize a separable outcome-averaged state that is transparently entangled in each trajectory subset. However, the counterexample holds regardless of the probabilities, and only relies on the fact that, if useful entanglement existed between the qubits $i$ and $f$ along an $\mathsf{O}(1)$ fraction of all measurement trajectories, one could use the knowledge of the protocol and measurement outcomes to perform local rotations (with negligible increase in $T$) to ensure that the outcome-averaged state were far from a separable state.

Hence, in the limit where $L$ is large and the assumptions (S110) and (S111) are satisfied, we are guaranteed that $\|\rho_{if} - \widetilde{\rho}_{if}\|_1 \ll 1$ (S115), meaning that the protocol $\mathcal{W}$ cannot possibly have generated a Bell pair shared by qubits $i$ and $f$—*either* on average or in any appreciable fraction of outcome trajectories. This follows directly from the proximity of the true state $\rho_{if}$ (S107) in trace distance to the separable state $\widetilde{\rho}_{if}$ [S11, S33, S34], combined with the arguments above for Lieb-Robinson bounds generally [S7, S28–S30], as well as those demonstrating that triviality of the outcome-averaged density matrix in Theorem 5 implies triviality of any finite fraction of measurement trajectories.

Crucially then, for $\rho_{if}$ (S107) to realize a Bell pair on sites $i$ and $f$ following the protocol $\mathcal{W}$, the assumption (S110) must be violated. In the asymptotic regime of interest, this implies that

$$L \lesssim 2\,(M+1)\,v\,T\;, \tag{S117}$$

and now, for the sake of argument, we briefly assume that there exists a choice of $\epsilon$ such that (S110) is satisfied *and* (S112) gives $\|\rho_{if} - \widetilde{\rho}_{if}\|_1 \sim 1$ as $L, T \to \infty$. In this limit, the righthand side of (S112) can be rewritten as

$$\|\rho_{if} - \widetilde{\rho}_{if}\|_1 \leq \widetilde{c}\, 2^{v_E T - \frac{L+1-l}{2(M+1)}}\;, \tag{S118}$$

where the prefactor $\widetilde{c}$ is finite, and therefore asymptotically unimportant[15]. Ignoring $\mathsf{O}(1)$ corrections that are independent of $L, T$, (S118) shows that $\rho_{if}$ cannot realize a Bell pair between sites $i$ and $f$ unless (S117) is obeyed, in violation of (S110). This holds on average and for any finite-measure subset of the outcome trajectories.

We therefore take (S117) to be the asymptotically correct bound for creating a Bell pair. Essentially, we replace "$\lesssim$" with "$\leq$" and provide for $\mathsf{O}(1)$ corrections to (S117), corresponding to the terms in (S110) that are independent of $L, T$, and $M$. The (possibly loose) bound (S117) is made generic by allowing for these asymptotically unimportant corrections, which we can absorb without loss of generality into $T$ and $M$ according to

$$L \leq (2\,M + M_0)\,v\,(T + T_0)\;, \tag{S119}$$

so that, in the absence of measurements, one recovers the usual Lieb-Robinson bound $L \leq M_0\,v\,(T + T_0)$, where, e.g., $M_0 = 1$ captures protocols that prepare a $1D$ state from left to right, while $M_0 = 2$ captures protocols that prepare a $1D$ state starting from the center of the chain. These $\mathsf{O}(1)$ terms may have particular physical meaning when we extend Theorem 5 to short-range-entangled initial states in Sec. 6.8, and to the preparation of other useful entangled states in Sec. 7. Note that, in the most general case of continuous-time protocols with adaptive measurement locations, $T_0$ may be $\mathsf{o}(L)$, rather than $\mathsf{O}(1)$, to account for a contribution to $v\,T_0$ of $(D-1)\log_2 L$.

We also note that (S119) gives an *upper* bound on $L$ for a particular $v$, $T$, and $M$. However, the existence of the ESTP of the main text, with $L = (2M+1)(T-1)$ (with $v = 1$), and the TFIM of Sec. 4, with $L = 2M(T-1)$, provide a *lower* bound on $L$ (namely, because we know that it is possible to teleport a quantum state *at least this distance* for a given $M$, $T$ using these protocols). The existence of protocols that saturate (S119) for a given task establishes that (S119) is *optimal* with respect to that task—i.e., no meaningful improvements (e.g., that scale in $L$, $T$, or $M$) can be made. The bound (S119) is optimal with respect to teleportation, and preparing Bell and GHZ (S166) states.

We conclude this discussion by stating the strategy for proving Theorem 5. We first consider a simplifying limit in which ($i$) the locations of measurements are determined from the outset and not altered by later outcomes and ($ii$) the measured sites can be partitioned into two sets, labelled $A$ and $B$, which obey $d(A, B) \geq L/(M+1)$, $i \in A$, and $f \in B$. After proving this special case, we then show that the condition ($ii$) must always be satisfied (see Lemma 7), before finally relaxing the first assumption ($i$)—and thus proving the Theorem generally—in Sec. 6.7.

The strategy for proving the minimalist bound in Theorem 5—when both assumptions ($i$) and ($ii$) hold—is as follows. Given the "true" hybrid protocol $\mathcal{W}$ that includes both time evolution and measurements[16], we consider a "modified" hybrid protocol $\widetilde{\mathcal{W}}$, which provably cannot generate any entanglement between the regions $A$ and $B$. To do this, we identify a cut $C$ of the system that—loosely speaking—bipartitions the entire graph into sites that are closer to region $A$, versus those closer to region $B$. We then define some unitary channel $\widetilde{\mathcal{W}}$, which is generated by the same measurement-based protocol as the true channel $\mathcal{W}$, but where we use the time-dependent Hamiltonian $\widetilde{H} \equiv H - H_C$,

---

[15] The prefactor $\widetilde{c}$ depends on small parameters in Theorem 5 (e.g., $c$, $l$, $\epsilon$, etc.), but is independent of the asymptotically large parameters $L, T, M$). The constant $\widetilde{c}$ may be as small as two, or as large as $\mathsf{O}(\epsilon^{-1})$.

[16] Recall that measurements are represented unitarily on the dilated Hilbert space, so $\mathcal{W}$ is unitary.

where terms $H_C$ consists of the terms in $H$ (S93) that act on both sides of the cut $C$. Hence, $\widetilde{H}$ cannot generate entanglement across the cut $C$, and cannot entangle the qubits $i \in A$ and $f \in B$ on opposite sides of $C$.

As we prove, acting on initially unentangled states, $\widetilde{\mathcal{W}}$ only generates states that are separable with respect to the regions $A$ and $B$. Thus, a Bell pair shared by $i$ and $f$ cannot be dynamically generated using $\widetilde{\mathcal{W}}$ if the sites $i$ and $f$ lie on different sides of the cut $C$. Moreover, we show using the Lieb-Robinson Theorem that the difference between $\widetilde{\mathcal{W}}|\mathbf{0}\rangle$ and $\mathcal{W}|\mathbf{0}\rangle$, restricted to the sites, $i$ and $f$, is exceptionally small at early times. Thus, $\mathcal{W}$ can only generate useful entanglement between $A$ and $B$ after sufficiently late times, bounded by (S110).

In Sec. 6.7, we extend the proof for prefixed measurements—and the corresponding bound on trace distance (S112)—to the more subtle case of protocols with *adaptive* measurement locations. The analogy to (S112) in the adaptive case is (S113), which involves the extra prefactor of $L^{D-1}$ compared to (S112), which we are confident is not physical, but an artifact of the proof strategy[17]. The corresponding bound (S119) contains an extra $(D-1)\log L$ correction to $T$. However, we note that the tail bounded in (S112) and (S113) are exactly zero for quantum circuits with strictly linear light cones—as long as (S110) is satisfied—so the $L^{D-1}$ factor is absent for quantum circuits. The factor also trivially vanishes for $D = 1$ and prefixed measurement locations.

### 6.5. Approximate early-time separability of $\rho_{if}$ with prefixed measurement locations

**Proposition 6.** *Suppose that the measurement locations do not depend on prior outcomes—i.e., $\mathbb{M}^n \to \mathbb{M} = \{S_i, S_f, S_1, \ldots, S_M\}$, $\forall \mathbf{n}$. Given a channel $\mathcal{W}$ generated by* (S100), *we define the reduced (or "residual") density matrix for the sites $\{i, f\}$* (S107); *we likewise define the* separable *reduced density matrix on $\{i, f\}$ as*

$$\widetilde{\rho}_{if} = \operatorname*{tr}_{\{i,f\}^c}\left[\widetilde{\mathcal{W}}|\mathbf{0}\rangle\langle\mathbf{0}|\widetilde{\mathcal{W}}^\dagger\right], \tag{S120}$$

*where the channel $\widetilde{\mathcal{W}}$—compared to $\mathcal{W}$—is instead generated by $\widetilde{H} \equiv H - H_C$ (i.e., the terms $H_C$ in the Hamiltonian* (S93) *acting on qubits neighboring the cut $C$ of the graph of physical qubits have been removed).*

*In this case,* (S110) *and* (S111) *imply* (S112) *for $\|\rho_{if} - \widetilde{\rho}_{if}\|_1$, with the density matrices given by* (S107) *and* (S120). *The bound* (S112) *implies approximate separability of $\rho_{if}$* (S107) *at early times $T$ compatible with* (S110) *and* (S111).

*Proof.* We choose a cut $C$ of the graph $G$ of the *physical* qubits; we also use "$C$" to denote the set of all physical qubits that have a nearest neighbor on the opposite side of the cut (i.e., $C$ denotes the cut itself, along with all qubits that border the cut on either side). Cutting the system also bipartitions the measurement set ($\mathbb{M} \to \mathbb{M}_A \cup \mathbb{M}_B$). The elements of $\mathbb{M}_A$ and $\mathbb{M}_B$ belong to opposite sides of the cut $C$, which they do not abut (i.e., the measurement sets in $\mathbb{M}_{A/B}$ are restricted to the region $A/B$, with no measurements on $C$ qubits). Note that we take $A = \cup_{S \in \mathbb{M}_A} S$ and $B = \cup_{S \in \mathbb{M}_B} S$ to correspond to the disjoint sets of *measured* sites on either side of the cut $C$. Additionally, the sites $i$ and $f$ lie on opposite sides of $C$ (i.e., $S_i \in \mathbb{M}_A, S_f \in \mathbb{M}_B$).

Throughout this proof, we assume that such a cut exists and is sufficiently far away from all measurement sets $S \in \mathbb{M}$, with the precise statement and its proof given in Lemma 7. We then choose $\widetilde{\rho}_{if}$ to be (S120), where the channel $\widetilde{\mathcal{W}}$ differs from $\mathcal{W}$ only in that the terms in the Hamiltonian that act on qubits in $C$ (i.e., across the cut) have been removed. See Fig. S5(a) for a cartoon depiction. Then, under $\widetilde{\mathcal{W}}$, the two regions delimited by the cut (i.e., $A$ and $B$) interact with each other only through sharing *classical* knowledge of the measurement outcomes on each side, but not through global *quantum* operations. In other words, physical qubits of the two sides, viewed as two parties, undergo LOCC. As a result, $\widetilde{\mathcal{W}}$ cannot generate entanglement between these two regions, and the final state $\widetilde{\rho}_{if}$ after applying the channel $\widetilde{\mathcal{W}}$ remains separable, as desired (the initial state is a [separable] product state).

Next, to recover the bound (S112), we consider some generic operator $\mathcal{O}$ supported only in $\{i, f\}$, to probe the trace distance between the two density matrices for the sites $i$ and $f$ (corresponding to $\mathcal{W}$ versus $\widetilde{\mathcal{W}}$):

$$\|\rho_{if} - \widetilde{\rho}_{if}\|_1 = \max_{\|\mathcal{O}\| \leq 1} \operatorname{tr}\left[\mathcal{O}(\rho_{if} - \widetilde{\rho}_{if})\right] = \max_{\|\mathcal{O}\| \leq 1} \left\langle \mathbf{0} \middle| \mathcal{W}^\dagger \mathcal{O} \mathcal{W} - \widetilde{\mathcal{W}}^\dagger \mathcal{O} \widetilde{\mathcal{W}} \middle| \mathbf{0} \right\rangle$$

$$\leq \max_{\|\mathcal{O}\| \leq 1} \|\mathcal{O}(T) - \widetilde{\mathcal{O}}(T)\| = \max_{\|\mathcal{O}\| \leq 1} \left\| \sum_{\mathbf{n}} \mathcal{O}_{\mathbf{n}}(T) - \widetilde{\mathcal{O}}_{\mathbf{n}}(T) \right\|, \tag{S121}$$

---

[17] Even with the $L^{D-1}$ prefactor, the bound $T(M+1) \gtrsim L$ rigorously holds for $T \gtrsim \ln L$ (where the $\log L$ correction is asymptotically unimportant). Note that the $L^{D-1}$ prefactor is absent in $D = 1$, for prefixed measurement locations, and for discrete time evolution (generated by a quantum circuit). We also believe that this factor is merely an artifact of the proof strategy (and is *not* physical). We are confident that there exists a proof strategy in which this factor does not appear (e.g., we expect the prefactor does not appear if we make the reasonable assumption that each measurement region is measured at most $\mathsf{O}(T)$ times, even for infinite dimensional graphs). In more general cases, it remains unclear how to remove this factor; we relegate such a refinement to future work.

which maximizes the (spectral) operator norm (S80) of the difference $\mathcal{O}(T) - \widetilde{\mathcal{O}}(T)$ for observables $\mathcal{O}$ with maximum-magnitude $\lambda \leq 1$ (so that $\|\mathcal{O}\| \leq 1$). Note that in the first line of (S121) we switched to the Heisenberg picture (in which one evolves operators instead of $\rho$). In the second line we denoted, e.g., $\mathcal{O}(T) = \mathcal{W}^\dagger \mathcal{O} \mathcal{W}$. Equivalently, the operators are decomposed onto measurement trajectories as defined in (S96), (S98), together with

$$\widetilde{\mathcal{O}}_{\boldsymbol{n}}(s) \equiv \widetilde{\mathcal{W}}^\dagger(T; T-s) \, \mathcal{O}_{\boldsymbol{n}} \, \widetilde{\mathcal{W}}(T; T-s) \,, \tag{S122}$$

as provided by Lemma 4, which will prove more convenient.

Since differences between $\widetilde{\mathcal{W}}$ and $\mathcal{W}$ only arise due to the terms $H_C$ in the Hamiltonian $H$ that act across the cut $C$, in order for $\mathcal{O}_{\boldsymbol{n}}(T)$ to differ from $\widetilde{\mathcal{O}}_{\boldsymbol{n}}(T)$, the former must first grow its support to the cut $C$ so that the cut Hamiltonian $H_C$ acts nontrivially to distinguish the two. Although measurements can grow $\mathcal{O}_{\boldsymbol{n}}(T)$ nonlocally on the graph $G$, this growth is restricted to the qubit sets $AB \equiv A \cup B$ where measurements occur, and only dynamics generated by $H^{\boldsymbol{n}}$ can grow the operator from $AB$ to qubits farther from the cut $C$. These local dynamics are constrained by Lieb-Robinson bounds (S90), even if the measurements lead to nonlocal operator growth *within $AB$*. We implement this using the (adjoint) Liouvillian representation (S99) for evolving operators $\widetilde{\mathcal{O}}_{\boldsymbol{n}}(s)$, with the modification

$$\widetilde{\mathcal{L}}_{\boldsymbol{n}}(t) = \mathcal{L}_{\boldsymbol{n}}(t) - \mathrm{i} \cdot \mathrm{ad}_{H_C^{\boldsymbol{n}}(t)} \,, \tag{S123}$$

where the only difference is the absence of evolution due to terms that appear in the cut Hamiltonian $H_C^{\boldsymbol{n}}$.

We next employ the Duhamel identity, finding

$$\mathcal{O}_{\boldsymbol{n}}(T) = \widetilde{\mathcal{O}}_{\boldsymbol{n}}(T) + \mathrm{i} \int_0^T \mathrm{d}s \, \widetilde{\mathbb{T}} \exp\left\{\int_0^{T-s} \mathrm{d}t' \, \widetilde{\mathcal{L}}(t')\right\} \mathrm{ad}_{H_C^{\boldsymbol{n}}(T-s)} \, \mathcal{O}_{\boldsymbol{n}}(s) \,, \tag{S124}$$

and next, from the triangle inequality, we conclude that

$$\begin{aligned}
\left\| \sum_{\boldsymbol{n}} \mathcal{O}_{\boldsymbol{n}}(T) - \widetilde{\mathcal{O}}_{\boldsymbol{n}}(T) \right\| &\leq \int_0^T \mathrm{d}s \, \left\| \sum_{\boldsymbol{n}} \left[ H_C^{\boldsymbol{n}}(T-s), \, \mathcal{O}_{\boldsymbol{n}}(s) \right] \right\| \\
&= \int_0^T \mathrm{d}s \, \left\| \sum_{\boldsymbol{n}_s} \sum_{\boldsymbol{n} \supset \boldsymbol{n}_s} \left[ H_C^{\boldsymbol{n}_s}(T-s), \, P_{\boldsymbol{n}_s} \mathcal{O}_{\boldsymbol{n}}(s) P_{\boldsymbol{n}_s} \right] \right\| \\
&= \int_0^T \mathrm{d}s \, \left\| \sum_{\boldsymbol{n}_s} P_{\boldsymbol{n}_s} \cdot \left[ H_C^{\boldsymbol{n}_s}(T-s), \, \sum_{\boldsymbol{n} \supset \boldsymbol{n}_s} \mathcal{O}_{\boldsymbol{n}}(s) \right] \cdot P_{\boldsymbol{n}_s} \right\| \\
&\leq \int_0^T \mathrm{d}s \, \max_{\boldsymbol{n}_s} \left\| \left[ H_C^{\boldsymbol{n}_s}(T-s), \, \mathcal{O}_{\boldsymbol{n}_s}(s) \right] \right\| \\
&\leq \max_{\boldsymbol{n}_s} \int_0^T \mathrm{d}s \, \sum_{X \subset C} \left\| \left[ H_X^{\boldsymbol{n}}(T-s), \, \mathcal{O}_{\boldsymbol{n}_s}(s) \right] \right\| \,,
\end{aligned} \tag{S125}$$

where, in the second line, we used $\mathcal{O}_{\boldsymbol{n}}(s) = P_{\boldsymbol{n}_s} \mathcal{O}_{\boldsymbol{n}}(s) P_{\boldsymbol{n}_s}$ (S102), and $H(T-s)$ only depends on the previous measurement outcomes $\boldsymbol{n}_s$. See Fig. S4(a) for a sketch of how the outcome trajectory $\boldsymbol{n}$ is partitioned into the previous and later outcomes $\boldsymbol{n}_s$ and $\boldsymbol{n} - \boldsymbol{n}_s$, respectively. Hence, we can move the sum over $\boldsymbol{n}$ with the same $\boldsymbol{n}_s$ into the commutator in the third line, and extract $P_{\boldsymbol{n}_s}$. The last line of (S125) recovers from the fact that

$$\mathcal{O}_{\boldsymbol{n}_s}(s) = \sum_{\boldsymbol{n} \supset \boldsymbol{n}_s} \mathcal{O}_{\boldsymbol{n}}(s) = \mathcal{W}^\dagger(T; T-s) \left( \mathcal{O} \otimes P_{\boldsymbol{n}_s} \right) \mathcal{W}(T; T-s) \,, \tag{S126}$$

along with orthogonality of $P_{\boldsymbol{n}_s}$. If the projectors $P_1, P_2$ are orthogonal, then for any two operators $\mathcal{O}_1, \mathcal{O}_2$,

$$\|P_1 \mathcal{O}_1 P_1 + P_2 \mathcal{O}_2 P_2\| = \max_\psi \|(P_1 \mathcal{O}_1 P_1 + P_2 \mathcal{O}_2 P_2)|\psi\rangle\| \leq \max_\psi \sqrt{\|\mathcal{O}_1 P_1 |\psi\rangle\|^2 + \|\mathcal{O}_2 P_2 |\psi\rangle\|^2}$$
$$\leq \max_\psi \sqrt{\|\mathcal{O}_1\|^2 \|P_1 |\psi\rangle\|^2 + \|\mathcal{O}_2\|^2 \|P_2 |\psi\rangle\|^2} \leq \max(\|\mathcal{O}_1\|, \|\mathcal{O}_2\|) \,, \tag{S127}$$

where there exists a choice of $|\psi\rangle$ with normalization such that $\|P_1 |\psi\rangle\|^2 + \|P_2 |\psi\rangle\|^2 \leq \||\psi\rangle\|^2 = 1$. Finally, in (S125), $X$ denotes edges in the set $C$ and we note that an edge can be viewed as a set containing the two vertices connected by that edge—here, the edge $X$ connects two qubits on either side of the cut $C$.

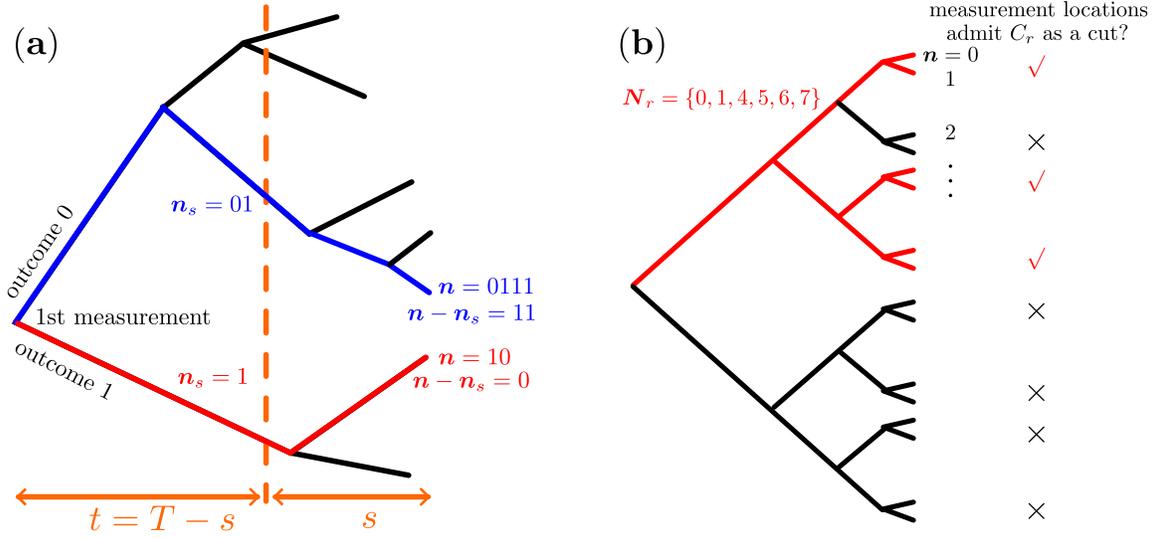

FIG. S4. (a) For a given backwards time $s$, any measurement trajectory $\bm{n}$, as a binary string, can be divided to two substrings $\bm{n}_s$ and $\bm{n} - \bm{n}_s$, where $\bm{n}_s$ denotes the measurement outcomes before time $t = T - s$. The Hamiltonian $H^{\bm{n}}(T-s) = H^{\bm{n}_s}(T-s)$ can only depend on the previous outcomes due to causality. The figure shows examples for two $\bm{n}$s (blue and red). Note that as indicated by the figure, measurements in different trajectories do not necessarily happen at the same time. (b) For adaptive measurement locations considered in Section 6.7, each trajectory $\bm{n}$ admits a number of cuts $C_r$ that are far $\gtrsim L/(2M)$ from the measurements. Thus for a given $C_r$, we ask each trajectory if it admits the cut: if so then the trajectory is included in the set $\bm{N}_r$. Note that, as indicated by the figure, two trajectories that only differ in the last digit share the same measurement locations, so they should agree on whether or not $C$ is a valid cut as specified in, e.g., Proposition 6.

The operator $\mathcal{O}_{\bm{n}_s}(s)$ is evolved on an enlarged graph $G_{\text{dil}}$ (corresponding to qubits in $\mathcal{H}_{\text{dil}}$), which includes both the physical qubits (the original vertex set $V = V_{\text{ph}}$) and the Stinespring qubits (the outcome vertex set $V_{\text{ss}}$). All measured observables act on qubits within the set $AB = A \cup B \subset V_{\text{ph}}$, meaning that the vertices in $AB$ are connected by *dilated* edges (elements of $E_{\text{dil}}$) to $V_{\text{ss}}$, their corresponding Stinespring qubits (which store the measurements' outcomes). Note that the edges $E_{\text{dil}}$ in the *enlarged* graph $G_{\text{dil}}$ are generated by terms in the dilated Hamiltonian that act on both the physical qubits in $AB$ and the Stinespring qubits in $V_{\text{ss}}$. Because the Stinespring qubits are all close to one another (separated by a handful of edges), all qubits in $AB$ are also close on the *enlarged* graph $G_{\text{dil}}$.

Therefore, under operator growth, $\mathcal{O}_{\bm{n}_s}(s)$ can reach all sites in $AB \cup V_{\text{ss}}$ in a short time $s$ due to measurements. On the other hand, the dynamics outside $AB \cup V_{\text{ss}}$ inherit entirely from the physical Hamiltonian $H^{\bm{n}}$ (S93)—which acts only in the *physical* part of the graph $G$—and hence the dynamics outside $AB$ remains local (see Fig. S5(a)). As a result, the Lieb-Robinson bound (Theorem 3) for the physical graph also applies to the enlarged graph:

$$\| [ H_X^{\bm{n}}(T-s), \mathcal{O}_{\bm{n}_s}(s) ] \| \leq c_{\text{LR}}\, h \left( \frac{v\, s}{d(X, AB)} \right)^{d(X, AB)}, \tag{S128}$$

where we have used $\|\mathcal{O}_{\bm{n}_s}(s)\| = \|\mathcal{O} \otimes P_{\bm{n}_s}\| = \|\mathcal{O}\| \leq 1$. Now, (S125) becomes

$$\left\| \sum_{\bm{n}} \mathcal{O}_{\bm{n}}(T) - \widetilde{\mathcal{O}}_{\bm{n}}(T) \right\| \leq \sum_{X \subset C} \frac{c_{\text{LR}}\, h\, T}{1 + d(X, AB)} \left( \frac{v\, T}{d(X, AB)} \right)^{d(X, AB)}$$

$$\leq \frac{c_{\text{LR}}\, h}{v} \sum_{X \subset C} \left( \frac{v\, T}{d(X, AB)} \right)^{d(X, AB)+1},$$

and now, using the fact that $AB = \cup_{m=-1}^{M} S_m$ we find

$$\leq \frac{c_{\text{LR}}\, h}{v} \sum_{X \subset C} \sum_{m=-1}^{M} \left( \frac{v\, T}{d(X, S_m)} \right)^{d(X, S_m)+1},$$

and changing the order of summation gives

$$\leq \frac{c_{\text{LR}} h}{v} \sum_{m=-1}^{M} \sum_{\substack{X \subset C \\ d(X,S_m) \geq d(C,S_m)}} \left(\frac{v\,T}{d(X,S_m)}\right)^{d(X,S_m)+1},$$

and defining $N_X(y)$ as the maximum number of edges $X \in E$ a distance $y$ from any connected measurement set $S_m$ of size $l$ (S95) [S7, S29] leads to the simplified expression

$$\left\| \sum_{\boldsymbol{n}} \mathcal{O}_{\boldsymbol{n}}(T) - \widetilde{\mathcal{O}}_{\boldsymbol{n}}(T) \right\| \leq \frac{c_{\text{LR}} h}{v} \sum_{m=-1}^{M} \sum_{y=d(C,S_m)}^{\infty} N_X(y) \left(\frac{v\,T}{y}\right)^{y+1}, \tag{S129}$$

and now, by simple counting we have

$$N_X(y) \leq \begin{cases} l\,K\,c_D\,y^{D-1} & \text{if } G \text{ has spatial dimension } D \\ l\,K\,(K-1)^y & \text{otherwise} \end{cases}, \tag{S130}$$

where $K$ is the maximal degree of the physical graph $G$ and $c_D$ is the constant defined in (S91). We now sum over $y$ in (S129) for these two cases separately. First, if $G$ has finite spatial dimension $D$, we have

$$\begin{aligned}
\sum_{y=y_0}^{\infty} l\,K\,c_D\,y^{D-1} \left(\frac{v\,T}{y}\right)^{y+1} &= l\,K\,c_D\,(v\,T)^{D-1} \sum_{y=y_0}^{\infty} \left(\frac{v\,T}{y}\right)^{y-D+2} \\
&\leq l\,K\,c_D\,(v\,T)^{D-1} \left(\frac{v\,T}{y_0}\right)^{y_0-D+2} \frac{1}{1-(1-\epsilon)} \\
&\leq \frac{l\,K\,c_D}{\epsilon} e^{(D-1)\ln y_0} \left(\frac{v\,T}{y_0}\right)^{y_0} \\
&\leq \frac{l\,K\,c_D}{\epsilon} e^{(1-\ln 2)(y_0-v\,T)} e^{v\,T-y_0} = \frac{l\,K\,c_D}{\epsilon} 2^{v\,T-y_0},
\end{aligned} \tag{S131}$$

where, in the second line, we have assumed that

$$y_0 = d(C,S_m) \geq \frac{v\,T}{1-\epsilon}, \tag{S132}$$

where $v_E = v$ (S92) and $0 < \epsilon < 1$ (S110), so that the summand decays faster than a geometric sequence,

$$\left(\frac{v\,T}{(y+1)}\right)^{(y+1)-D+2} \Big/ \left(\frac{v\,T}{y}\right)^{y-D+2} = \frac{v\,T}{y}\left(1+y^{-1}\right)^{-(y-D+3)} \leq \frac{v\,T}{y} \leq 1-\epsilon, \tag{S133}$$

where we used (S111) above. The above relation implies that

$$\sum_{y=y_0}^{\infty} \left(\frac{v\,T}{y}\right)^{y-D+2} = \sum_{n=0}^{\infty} \left(\frac{v\,T}{y_0+n}\right)^{y_0+n-D+2} \leq \left(\frac{v\,T}{y_0}\right)^{y_0-D+2} \sum_{n=0}^{\infty} (1-\epsilon)^n = \frac{1}{\epsilon} \left(\frac{v\,T}{y_0}\right)^{y_0-D+2}, \tag{S134}$$

and we note that in the third line of (S131) we used the identities

$$\left(\frac{a}{b}\right)^b \leq e^{a-b} \quad \text{and} \quad (D-1)\frac{\ln y_0}{y_0-v\,T} \leq (D-1)\frac{\ln y_0}{y_0}\frac{1}{1-(1-\epsilon)} \leq \frac{D-1}{\epsilon}\frac{\ln v\,T}{v\,T} \leq 1-\ln 2, \tag{S135}$$

which further invokes (S132) and (S111). Second, for a general graph $G$ with no finite spatial dimension $D$, we have

$$\sum_{y=y_0}^{\infty} l\,K\,(K-1)^y \left(\frac{v\,T}{y}\right)^{y+1} = \frac{l\,K}{K-1} \left(\frac{(K-1)\,v\,T}{y_0}\right)^{y_0+1} \frac{1}{1-(1-\epsilon)} \leq \frac{l\,K}{\epsilon} e^{v_E\,T-y_0}, \tag{S136}$$

where we assume (S132) with $v_E = (K-1)\,v$ (S92), and used the identities (S132-S135).

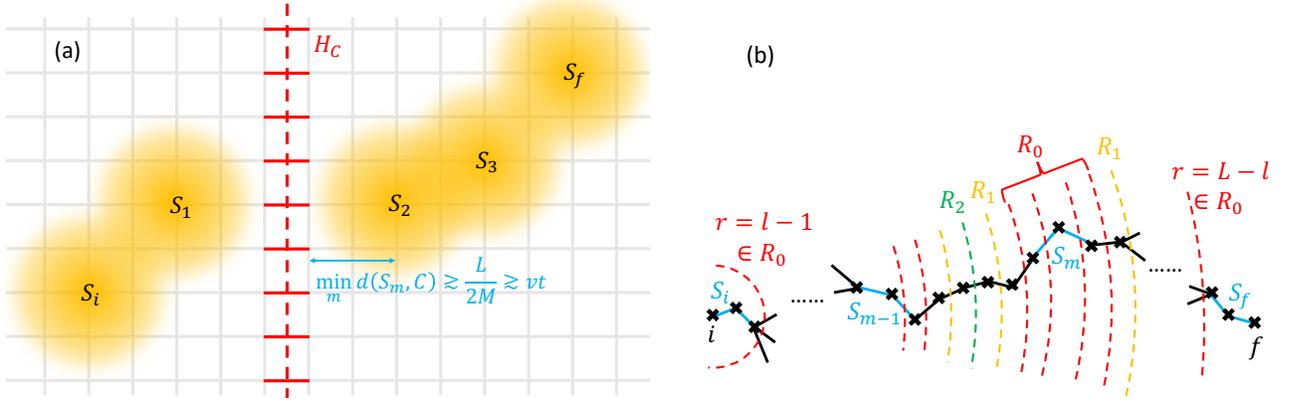

FIG. S5. (a) Schematic depiction of the Lieb-Robinson bound (S90) in a 2D lattice of physical qubits, with measurement range $l = 1$: $\mathcal{O}_{\bm{n}}(t)$ reaches the cut $C$ after $vT \approx L/2M$, even though the Stinespring qubits (not shown) facilitate rapid growth of the operator size within the physical vertex sets $A = S_i \cup S_1$ and $B = S_f \cup S_2 \cup S_3$. (b) Schematic depiction of the cuts $C_r$ and the sets $R_q$ in Lemma 7 and its proof, with range $l = 3$. Only the relevant vertices and edges of the graph $G$ are depicted. The desired cut $C_r$ corresponds to some $r \in R_q$ with $q \approx L/2M$.

Now, summarizing the above two cases that appear in (S130), the assumption (S132) yields

$$\left\| \sum_{\bm{n}} \mathcal{O}_{\bm{n}}(T) - \widetilde{\mathcal{O}}_{\bm{n}}(T) \right\| \leq \frac{c\,l}{8\epsilon} \sum_{m=-1}^{M} 2^{v_E T - d(C, S_m)}, \tag{S137}$$

where the constant $c$, above, is given by

$$c \equiv \frac{8\, c_{\mathrm{LR}}\, h\, K}{v} \times \begin{cases} c_D & \text{if } G \text{ has finite spatial dimension } D \\ 1 & \text{otherwise} \end{cases}, \tag{S138}$$

which only depends on the graph structure, since $v \propto h$ (S89), and where $c_D$ is defined in (S91).

Thus, in the case where the locations of measurements are not conditioned on prior outcomes, assuming (S110) and (S111)—and further assuming (S132), which we verify in Lemma 7 with choice $\epsilon' L = M + 1$, leading to (S140b) for the sum over $m$—we recover (S112) as required by connecting (S137) to the trace distance via (S121). □

### 6.6. Proof that the graph can be partitioned effectively

We now present and prove Lemma 7, which verifies the assumption (S132) invoked in Proposition 6,

$$d(C, S_m) > \frac{v_E T + \frac{l}{2} + 2}{1 - \epsilon} - \frac{l}{2} - 2 > \frac{v_E T}{1 - \epsilon}, \tag{S139}$$

by combining the Lemma's result (S140a) with the choice $\epsilon' L = M + 1$ and the assumption (S110) of Theorem 5. This proves that there always exists a cut (or "paritition") $C$ of the graph $G$ that is sufficiently far from all measurement regions. In fact, there are roughly $\epsilon L$ such desirable cuts $C$. Although any choice of cut $C$ is sufficient to prove Prop. 6 (where the measurement locations are "prefixed"), all are required in the more general *adaptive* case.

**Lemma 7.** *Consider a particular outcome trajectory $\bm{n}$ with corresponding measurement set $\mathbb{M} = \{S_{-1}, S_0, S_1, \cdots, S_M\}$ (S94), where we dispense with the $\bm{n}$ superscripts for convenience. The protocol can be fully adaptive: The locations, times, and nature of future measurements may all be conditioned upon prior outcomes. For each choice of $r \in \{l-1, l, \cdots, L-l\}$, we define the set of vertices within a distance $r$ of the initial site $i$ as $V_A(r) = \{u \in V \mid d(u, i) \leq r\}$, and we define its complement as $V_B(r) = V \setminus V_A(r)$ (i.e., all other qubits in the vertex set $V$). Then, the $r$-dependent cut $C_r$ corresponds to all edges connecting $V_A(r)$ and $V_B(r)$, with the regions $S_{-1} = S_i$ and $S_0 = S_f$ on opposite sides of $C_r$.*

*For any $0 < \epsilon' < 1$, there always exist $\epsilon' L$ choices of $r \in \{l-1, l, \cdots, L-l\}$, such that the cut $C_r$ of the graph $G$ separating the sets $V_{A/B}(r)$ satisfies*

$$\min_m d(C_r, S_m) > \frac{(1-\epsilon')L + 1 - l}{2(M+1)} - \frac{l+3}{2} \tag{S140a}$$

$$\sum_{m=-1}^{M} 2^{-d(C_r, S_m)} < \frac{M+1}{\epsilon' L} 2^{\frac{l+5}{2} - \frac{(1-\epsilon')L+1-l}{2(M+1)}}. \tag{S140b}$$

*Proof.* We first define the family of sets

$$R_q \equiv \left\{ r \in \mathbb{Z} \,\middle|\, l - 1 \leq r \leq L - l \quad \text{and} \quad \min_m d(C_r, S_m) = q \right\}, \tag{S141}$$

for any nonnegative integer $q$ (see Fig. S5(b) for a cartoon depiction). Essentially, $R_q$ is the set of choices of distances $r$ such that the distance between the corresponding cut $C_r$ (for each choice of $r$) and the closest measurement region to the $r$-dependent cut $C_r$ is exactly $q$. We want to show that the choices of $r$ satisfying (S140) are contained in some sets $R_q$ with $q \gtrsim L/M$. First, the size of each set $R_q$ (S141) is bounded by

$$|R_0| \leq M(l+1) + 2, \quad |R_q| \leq 2M + 2 \quad \forall q \geq 1, \tag{S142}$$

the reason being that each $S_m$ (for $1 \leq m \leq M$) is within distance $q = 0$ of at most $l + 1$ cuts $C_r$, and within distance $q \geq 1$ of at most two cuts $C_r$, while $S_i$ and $S_f$ can be within a distance $q \geq 0$ of at most *one* cut $C_r$. Since there are $L - 2(l-1)$ allowed values of $r$ in total, we have

$$\sum_{q > q_0} |R_q| \geq L - 2(l-1) - (M(l+1) + 2) - q_0 (2M+2) = L - l + 1 - (M+1)(l + 1 + 2q_0) \quad \forall q_0 \in \mathbb{Z}_+. \tag{S143}$$

Since we want the union of all $R_q$ with $q > q_0$ to contain at least $\epsilon' L$ elements of $r$, we must have

$$L - l + 1 - (M+1)(l + 1 + 2q_0) \geq \epsilon' L, \tag{S144}$$

and to satisfy the condition above, we choose

$$q_0 = \left\lfloor \frac{(1-\epsilon')L + 1 - l}{2(M+1)} - \frac{l+1}{2} \right\rfloor \in \left( \frac{(1-\epsilon')L + 1 - l}{2(M+1)} - \frac{l+3}{2}, \frac{(1-\epsilon')L + 1 - l}{2(M+1)} - \frac{l+1}{2} \right], \tag{S145}$$

where the floor function $\lfloor x \rfloor$ denotes the greatest integer less than or equal to $x$.

Summing the lefthand side of (S140b) over all values of $r$ in all sets $R_q$ (S141) with $q > q_0$ gives

$$\sum_{q > q_0} \sum_{r \in R_q} \sum_{m=-1}^{M} 2^{-d(C_r, S_m)} \leq \sum_{m=-1}^{M} \sum_{r : d(C_r, S_m) > q_0} 2^{-d(C_r, S_m)} \leq \sum_{m=-1}^{0} \sum_{\delta = q_0 + 1}^{\infty} 2^{-\delta} + \sum_{m=1}^{M} \sum_{\delta = q_0 + 1}^{\infty} 2^{1-\delta}$$

$$= (2M + 2) \frac{2^{-1-q_0}}{1 - 2^{-1}} = (M+1) 2^{1-q_0}, \tag{S146}$$

where we used the fact that $S_{m \geq 1}$ ($S_{m \leq 0}$) can be within distance $\delta \geq 1$ of at most two (one) cuts. For the $q_0$ determined in (S145), there are at least $\epsilon' L$ different $r$s that contribute to (S146); thus, at least $\epsilon' L$ choices of $r$ satisfy $r > q_0$, which yields (S140a), and the inequality

$$\sum_{m=-1}^{M} 2^{-d(C_r, S_m)} \leq \frac{M+1}{\epsilon' L} 2^{1-q_0}, \tag{S147}$$

which reduces to (S140b) upon replacing $q_0$ according to (S145). □

### 6.7. Extension to adaptive measurement locations

*Proof of Theorem 5.* We now consider the general case in which the locations and nature of the Hamiltonian $H(t)$ (S86) and measurements may be conditioned on the outcomes of prior measurements. Owing to these complications,

it will prove convenient to begin by considering the various trajectories $\boldsymbol{n}$ individually. For each trajectory $\boldsymbol{n}$, the measurement set $\mathbb{M}^{\boldsymbol{n}}$ is fixed, and according to Lemma 7 we have $\epsilon' L$ number of choices of cut $C_r$ that are far from the measurement locations in the sense of (S140). Alternatively, for each $r \in \{l-1, l, \cdots, L-l\}$, there is a set $\boldsymbol{N}_r$ of trajectories $\boldsymbol{n}$ that admits $C_r$ as a faraway cut (see Fig. S4(b) for an example). While $\boldsymbol{N}_r$ may vanish for some choices of $r$, Lemma 7 guarantees that the union of these sets for all $r$ covers all trajectories $\epsilon' L$ times:

$$\frac{1}{\epsilon' L} \sum_{r=l-1}^{L-l} P_{\boldsymbol{N}_r} = \sum_{\boldsymbol{n}} P_{\boldsymbol{n}} = 1, \quad \text{where} \quad P_{\boldsymbol{N}_r} = \sum_{\boldsymbol{n} \in \boldsymbol{N}_r} P_{\boldsymbol{n}_r}, \tag{S148}$$

and now, in analogy to the channel $\widetilde{\mathcal{W}}$ in Proposition 6, we define a modified channels $\widetilde{\mathcal{W}}_r$ for each $r$, which are identical to $\mathcal{W}$ except that the Hamiltonian terms across the cuts $C_r$ are removed. In particular, measurement locations in some trajectories of $\widetilde{\mathcal{W}}_r$ can be close to or even overlap with the cut $C_r$. This turns out to be fine when we later bound operator growth to the cut $C_r$, since we only care about the trajectories $\boldsymbol{n} \in \boldsymbol{N}_r$, for which the measurement locations are far from the cut. To be precise, for each $r$, we project the final state generated by $\widetilde{\mathcal{W}}_r$ onto $\boldsymbol{N}_r$:

$$\widetilde{\rho}'_{if,r} = \sum_{\boldsymbol{n} \in \boldsymbol{N}_r} \operatorname*{tr}_{\{i,f\}^c} \left[ P_{\boldsymbol{n}} \widetilde{\mathcal{W}}_r |\boldsymbol{0}\rangle\langle \boldsymbol{0}| \widetilde{\mathcal{W}}_r^\dagger P_{\boldsymbol{n}} \right] = \operatorname*{tr}_{\{i,f\}^c} \left[ P_{\boldsymbol{N}_r} \widetilde{\mathcal{W}}_r |\boldsymbol{0}\rangle\langle \boldsymbol{0}| \widetilde{\mathcal{W}}_r^\dagger P_{\boldsymbol{N}_r} \right], \tag{S149}$$

where the trace over the complementary set $\{i,f\}^c$ includes all physical and Stinespring qubits *except* $i$ and $f$. Since the initial state $|\boldsymbol{0}\rangle$ is separable, and $\widetilde{\mathcal{W}}_r$ involves LOCC with respect to the two sides of cut $C_r$ when restricted to trajectories in $\boldsymbol{N}_r$, the density matrix $\widetilde{\rho}'_{if,r}$ must be separable. Note that $\widetilde{\rho}'_{if,r}$ is not normalized; loosely speaking, $\operatorname{tr} \widetilde{\rho}'_{if,r} \approx p_{\boldsymbol{N}_r} = \langle \boldsymbol{0}| P_{\boldsymbol{N}_r}(T) |\boldsymbol{0}\rangle$ is the probability when the terms in $H_C$ are included in $H$.

Averaging $\widetilde{\rho}'_{if,r}$ over all $r$ leads to a generalization of the reference state (S120) to the case in which the locations of measurements *can* be conditioned on prior measurement outcomes,

$$\widetilde{\rho}_{if} = \frac{\widetilde{\rho}'_{if}}{\operatorname{tr}[\widetilde{\rho}'_{if}]}, \quad \text{where} \quad \widetilde{\rho}'_{if} = \frac{1}{\epsilon' L} \sum_{r=l-1}^{L-l} \widetilde{\rho}'_{if,r}, \tag{S150}$$

and since $\widetilde{\rho}_{if}$ is a convex sum of separable density matrices, it is itself separable. The $\epsilon' L$ denominator comes from the fact that each trajectory $\boldsymbol{n}$ contributes to $\epsilon' L$ number of $r$s. However, even with this denominator, the trace of $\widetilde{\rho}'_{if}$ is no longer guaranteed to be one, so we need to divide by a normalization factor in (S150).

We first bound the trace distance between $\rho_{if}$ and the unnormalized state $\widetilde{\rho}'_{if}$, and then show that $\operatorname{tr}[\widetilde{\rho}'_{if}] \sim 1$ nevertheless, so that $\|\rho_{if} - \widetilde{\rho}_{if}\|_1$ is also bounded. Following (S121), let $\mathcal{O}$ be any operator supported only in $\{i,f\}$,

$$\Delta \equiv \|\rho_{if} - \widetilde{\rho}'_{if}\|_1 = \max_{\|\mathcal{O}\|\leq 1} \operatorname{tr}\left[\mathcal{O}(\rho_{if} - \widetilde{\rho}'_{if})\right] = \max_{\|\mathcal{O}\|\leq 1} \frac{1}{\epsilon' L} \sum_{r=l-1}^{L-l} \operatorname{tr}\left[\mathcal{O}_{\boldsymbol{N}_r}(\rho_{if} - \widetilde{\rho}'_{if,r})\right]$$

$$= \max_{\|\mathcal{O}\|\leq 1} \frac{1}{\epsilon' L} \sum_{r=l-1}^{L-l} \langle \boldsymbol{0}| \mathcal{W}^\dagger \mathcal{O}_{\boldsymbol{N}_r} \mathcal{W} - \widetilde{\mathcal{W}}_r^\dagger \mathcal{O}_{\boldsymbol{N}_r} \widetilde{\mathcal{W}}_r |\boldsymbol{0}\rangle$$

$$\leq \frac{1}{\epsilon'} \max_{\|\mathcal{O}\|\leq 1} \max_r \left\| \mathcal{O}_{\boldsymbol{N}_r}(T) - \widetilde{\mathcal{O}}_{\boldsymbol{N}_r}(T) \right\| = \frac{1}{\epsilon'} \max_{\|\mathcal{O}\|\leq 1} \max_r \left\| \sum_{\boldsymbol{n}\in\boldsymbol{N}_r} \mathcal{O}_{\boldsymbol{n}}(T) - \widetilde{\mathcal{O}}_{\boldsymbol{n}}(T) \right\|, \tag{S151}$$

where in the first line we have used (S148). Hereafter, the $r$ dependence in $\widetilde{\mathcal{O}}_{\boldsymbol{n}}(T) = \widetilde{\mathcal{W}}_r^\dagger \widetilde{\mathcal{O}}_{\boldsymbol{n}} \widetilde{\mathcal{W}}_r$ is implicit.

Let us compare (S121) and (S151), where the former recovers the case in which the measurement set $\mathbb{M}^{\boldsymbol{n}}$ is fixed for all protocols. First, although in the general case we need to consider many cuts (and the corresponding reference protocols $\widetilde{\mathcal{W}}_r$) instead of just one, for bounding purpose it still suffices to focus on one cut $C_r$, as indicated by the $\max_r$ in (S151). The price to pay is an extra prefactor $1/\epsilon'$ that is mild. Second, (S151) luckily only involves a fraction of trajectories, since other trajectories outside of $\boldsymbol{N}_r$ are not guaranteed to have measurement locations far from the cut $C_r$. Thus, the proof below follows closely that of Proposition 6.

In particular, for any Heisenberg time $s$, a trajectory $\boldsymbol{n} \in \boldsymbol{N}_r$ has substring $\boldsymbol{n}_s$ recording the early measurement outcomes that have not been grown by the Heisenberg evolution in backwards time. The sum over trajectories is again split up into the "previous" and "later" outcomes, subject to

$$\sum_{\boldsymbol{n}\in\boldsymbol{N}_r} = \sum_{\boldsymbol{n}_s} \sum_{\boldsymbol{n}\in\boldsymbol{N}_r:\boldsymbol{n}\supset\boldsymbol{n}_s}, \tag{S152}$$

where it may be the case that, for some $\boldsymbol{n}_s$, the second sum vanishes. Then (S125) is easily modified to

$$\left\| \sum_{\boldsymbol{n} \in \boldsymbol{N}_r} \mathcal{O}_{\boldsymbol{n}}(T) - \widetilde{\mathcal{O}}_{\boldsymbol{n}}(T) \right\| \leq \max_{\boldsymbol{n}_s} \int_0^T ds \sum_{X \subset C_r} \left\| \left[ H_X^{\boldsymbol{n}}(T-s), \sum_{\boldsymbol{n} \in \boldsymbol{N}_r : \boldsymbol{n} \supset \boldsymbol{n}_s} \mathcal{O}_{\boldsymbol{n}}(s) \right] \right\|, \tag{S153}$$

and furthermore, the operator $\sum_{\boldsymbol{n} \in \boldsymbol{N}_r : \boldsymbol{n} \supset \boldsymbol{n}_s} \mathcal{O}_{\boldsymbol{n}}(s)$ has norm

$$\left\| \sum_{\boldsymbol{n} \in \boldsymbol{N}_r : \boldsymbol{n} \supset \boldsymbol{n}_s} \mathcal{O}_{\boldsymbol{n}}(s) \right\| = \left\| \mathcal{O} \otimes \sum_{\boldsymbol{n} \in \boldsymbol{N}_r : \boldsymbol{n} \supset \boldsymbol{n}_s} P_{\boldsymbol{n}} \right\| = 1, \tag{S154}$$

and is grown from measurement locations that are all a distance $\gtrsim L/(2M)$ (S140a) away from the cut $C_r$.

Applying the Lieb-Robinson bound (S128) for operator growth between set $X \subset C_r$ and $AB$—which now includes all lattice sites a distance (S140a) from the cut $C_r$—we recover the relation

$$\left\| \sum_{\boldsymbol{n} \in \boldsymbol{N}_r} \mathcal{O}_{\boldsymbol{n}}(T) - \widetilde{\mathcal{O}}_{\boldsymbol{n}}(T) \right\| \leq \frac{c}{2} L^{D-1} 2^{vT - \frac{(1-\epsilon')L + 1 - l}{2(M+1)} + \frac{l}{2}}, \tag{S155}$$

where the constant $c$ is proportional to $c_{\text{LR}} h K c_D / v$ up to a numerical constant that depends on asymptotically unimportant details of the graph $G$ as in (S138), and the derivation above applies in the case where the (physical) graph $G$ is finite dimensional, so that $\sum_{X \subset C_r} \lesssim |C_r| \lesssim L^{D-1}$ (or else the size of the cut $|C_r|$ may diverge).

Note that there is no $l/\epsilon$ factor compared to (S137), because we do not need to sum over individual measurement regions here. Nevertheless, there is an extra $1/\epsilon'$ in (S151) compared to (S121), so that choosing $\epsilon' = \epsilon$ leads to

$$\| \rho_{if} - \widetilde{\rho}'_{if} \|_1 = \Delta \leq \frac{c}{2\epsilon} L^{D-1} 2^{vT + \frac{l}{2} - \frac{(1-\epsilon)L + 1 - l}{2(M+1)}}, \tag{S156}$$

and we then have that $\operatorname{tr}[\widetilde{\rho}'_{if}]$ is close to unity by (S151):

$$| \operatorname{tr}[\widetilde{\rho}'_{if}] - 1 | = | \operatorname{tr}[\widetilde{\rho}'_{if}] - \operatorname{tr}[\rho_{if}] | \leq \| \rho_{if} - \widetilde{\rho}'_{if} \|_1 = \Delta, \tag{S157}$$

so that finally, combining (S156) with (S157) leads to

$$\begin{aligned} \| \rho_{if} - \widetilde{\rho}_{if} \|_1 &\leq \| \rho_{if} - \widetilde{\rho}'_{if} \|_1 + \| \widetilde{\rho}'_{if} - \widetilde{\rho}_{if} \|_1 \leq \Delta + | \operatorname{tr}[\widetilde{\rho}'_{if}] - 1 | \\ &\leq 2\Delta \leq \frac{c}{\epsilon} L^{D-1} 2^{vT + \frac{l}{2} - \frac{(1-\epsilon)L + 1 - l}{2(M+1)}}, \end{aligned} \tag{S158}$$

which becomes (S113). However, the constant $c$ for adaptive measurement protocols must be computed separately from the analogous $c$ for prefixed measurements; however, we forego calculating the adaptive constant $c$ as its derivation is tedious, it remains asymptotically unimportant, and due to the presence of this prefactor $L^{D-1}$, which we suspect to be an unphysical artifact of the proof strategy. $\square$

### 6.8. Extension to initial states with short-range entanglement

Thus far, we have considered only initial product states such as $|\boldsymbol{0}\rangle\langle\boldsymbol{0}|$. We now extend Theorem 5 and the resulting bound (S117) to a class of "short-range-entangled" (SRE) initial states $\rho_0$, which can be prepared according to

$$\rho_0 = \mathcal{W}_0 \rho' \mathcal{W}_0^\dagger, \tag{S159}$$

where the physical density matrix $\rho'$ is separable with respect to some cut $C$ (e.g., any product state such as $|\boldsymbol{0}\rangle\langle\boldsymbol{0}|$; see Sec. 6.6) and the protocol $\mathcal{W}_0$ is compatible with the requirements of Sec. 6.1, obeys Theorem 5), and involves measurements in $M'_0$ regions and Hamiltonian evolution for total duration $T'_0$, where both $M'_0$ and $T'_0$ are $\mathsf{O}(1)$.

To simplify the coming discussion, we make several observations. First, note that $\rho_0$ (S159) is the input for a protocol $\mathcal{W}$ that achieves some quantum task acting on $\rho_0$. Second, suppose that task has an associated "task distance" $L = d(i, f)$, where $i$ and $f$ label the "task qubits." Then we note that properties of the initial state $\rho_0$ (S159) some distance $r \gg L$ from sites $i$ and $f$ are not relevant to the bound (S117) on $\mathcal{W}$. Without loss of generality, we assume

that the qubits $i$ and $f$ lie roughly at two opposing "ends" of the graph $G$—so that $L$ is comparable to the size of $G$. In order to recover meaningful bounds for SRE initial states $\rho_0$ (S159), we require that the task distance $L$ and the resources $M'_0$ and $T'_0$ (of the protocol $\mathcal{W}_0$ that creates $\rho_0$ from $\rho'$) obey $M'_0, T'_0 \ll L$. This precludes, e.g., initial states $\rho_0$ (S159) that already contain Bell pairs on $i, f$. For convenience, we initially assume that the protocol $\mathcal{W}_0$ is optimal with respect to (S117), so that $M'_0$ and $T'_0$ reflect the *minimum* required to prepare $\rho_0$ (S159) from $\rho'$. It is straightforward to relax this assumption to generic protocols, and we make no such assumption in Corollary 9.

Now, suppose that some protocol $\mathcal{W}$ generates a Bell pair on the qubits $i$ and $f$ starting from the SRE initial state $\rho_0$ (S159). By construction, the *combined* protocol $\mathcal{W}' = \mathcal{W}\mathcal{W}_0$ generates a Bell pair on $i$ and $f$ starting from the *unentangled* (or separable) initial state $\rho'$. Consequently, $\mathcal{W}'$ obeys Theorem 5 and the main bound (S117) without modification—i.e., $L' \leq 2\,(M'+1)\,v\,T'$. From this bound on $\mathcal{W}' = \mathcal{W}\mathcal{W}_0$, we can infer a bound on the protocol of interest $\mathcal{W}$, which requires considering several scenarios involving $\rho_0$ and $\mathcal{W}_0$.

For the SRE initial state $\rho_0$ (S159) to be *advantageous* (compared to the separable initial state $\rho'$), it must provide for a reduction in resource requirements $M$ and $T$ for $\mathcal{W}$ compared to the protocol $\mathcal{W}'$ that acts on the initial state $\rho'$. In other words, the SRE state $\rho_0$ (S159) is advantageous if it is realized at some point along the protocol $\mathcal{W}'$ applied to the separable density matrix $\rho'$. By assumption, the protocol $\mathcal{W}_0$ in (S159) is optimal, so if the state $\rho_0$ (S159) realizes along $\mathcal{W}'$, then we have $M' = M + M'_0$ and $T' = T + T'_0$. Because $\mathcal{W}$ (with resources $M, T$) achieves the same useful quantum task as $\mathcal{W}'$ (with resources $M', T'$), and because $\mathcal{W}_0$ is optimal (meaning a more efficient protocol does not exist), the bound (S117) on $\mathcal{W}'$ immediately implies a bound for the protocol $\mathcal{W}$ given by

$$L \,=\, L' \,\leq\, 2\,(M'+1)\,v\,T' \,=\, 2\,(M+M'_0+1)\,v\,(T+T'_0)\,, \tag{S160}$$

where $L = d(i,f) = L'$ is the task distance for the protocols $\mathcal{W}$ and $\mathcal{W}'$ (which achieve the same "useful task"), $M', T'$ are the resources used by the combined protocol $\mathcal{W}' = \mathcal{W}\mathcal{W}_0$ acting on the separable (or product) initial state $\rho'$, and $M, T$ are the resources used by the protocol $\mathcal{W}$ that acts on the SRE initial state $\rho_0$ (S159).

Importantly, we note that (S160) is saturated when $(i)$) the protocol $\mathcal{W}'$ saturates the bound on $L' = L$, $(ii)$ the SRE state $\rho_0$ (S159) is *maximally advantageous* to the protocol $\mathcal{W}$ compared to $\rho'$, meaning that $\rho_0$ (S159) is realized from the state $\rho'$ along the protocol $\mathcal{W}'$, and $(iii)$ $M'_0, T'_0$ reflect the *minimum resources* required to prepare $\rho_0$ (S159) from the separable (or product) state $\rho'$ (i.e., $\mathcal{W}_0$ is optimal). If any of these conditions fail, then the bound (S160) still holds, but is not saturated. This occurs when one or more of $\mathcal{W}, \mathcal{W}'$, and $\mathcal{W}_0$ are suboptimal, or if the state $\rho_0$ (S159) is *not* realized along the protocol $\mathcal{W}'$ acting on the separable (or product) state $\rho'$. For example, suppose that the state $\rho_0$ (S159) does *not* realize at some point along the protocol $\mathcal{W}'$, but instead, the "nearby" state

$$\rho_1 \,=\, \mathcal{W}_1\,\rho'\,\mathcal{W}_1^\dagger \,\equiv\, \mathcal{W}_{10}\,\rho_0\,\mathcal{W}_{10}^\dagger \,=\, \mathcal{W}_{10}\,\mathcal{W}_0\,\rho'\,\mathcal{W}_0^\dagger\,\mathcal{W}_{10}^\dagger, \tag{S161}$$

realizes along $\mathcal{W}'$. We are guaranteed that there exists a choice of $\mathcal{W}_{10}$ such that $M_{10} \leq M'_0$ and $T_{10} \leq T'_0$—in the worst case, we choose $\mathcal{W}_{10} = \mathcal{W}_0^{-1}$ to "undo" the unhelpful entangled state $\rho_0$ (S159). The desired final state (e.g., with a Bell pair between qubits $i$ and $f$) is then given by

$$\rho(T) \,=\, \mathcal{W}_*\,\rho_1\,\mathcal{W}_*^\dagger \,=\, \mathcal{W}_*\,\mathcal{W}_1\,\rho'\,\mathcal{W}_1^\dagger\,\mathcal{W}_*^\dagger \,\equiv\, \mathcal{W}_*\,\mathcal{W}_{10}\,\rho_0\,\mathcal{W}_{10}^\dagger\,\mathcal{W}_*^\dagger \,=\, \mathcal{W}\,\rho_0\,\mathcal{W}^\dagger\,, \tag{S162}$$

where $\mathcal{W}' = \mathcal{W}_*\,\mathcal{W}_1 = \mathcal{W}_*\,\mathcal{W}_{10}\,\mathcal{W}_0 = \mathcal{W}\mathcal{W}_0$. In particular, the protocol $\mathcal{W}' = \mathcal{W}_*\,\mathcal{W}_1$ is efficient and obeys (S117), meaning that $M' = M_* + M'_1$ and $T' = T_* + T'_1$. The true protocol $\mathcal{W} = \mathcal{W}_*\,\mathcal{W}_{10}$ satisfies

$$M \,=\, M_* + M_{10} \,=\, M' + M_{10} - M'_1 \tag{S163a}$$
$$T \,=\, T_* + T_{10} \,=\, T' + T_{10} - T'_1\,, \tag{S163b}$$

for its resource requirements. Since the protocol $\mathcal{W}'$ obeys (S117), the protocol $\mathcal{W}$ obeys

$$L \,\leq\, 2\,(M + M'_1 - M_{10} + 1)\,v\,(T + T'_1 - T_{10})\,, \tag{S164}$$

which realizes (S160) for the *most favorable* initial states $\rho_0$ (S159), where $\mathcal{W}_{10} = \mathbb{1}$ and $\mathcal{W}_1 = \mathcal{W}_0$ so that $M_{10} = T_{10} = 0$ and $M'_1, T'_1 \to M'_0, T'_0$. For the *least favorable* initial states $\rho_0$ (S159), the closest state $\rho_1$ to $\rho_0$ (S159) realized along $\mathcal{W}'$ is simply the separable reference state $\rho'$. In this case, we have $\mathcal{W}_1 = \mathbb{1}$ and $\mathcal{W}_{10} = \mathcal{W}_0^{-1}$, and $\mathcal{W}$ obeys

$$L = 2\,(M - M'_0 + 1)\,v\,(T - T'_0)\,, \tag{S165}$$

meaning that the initial state $\rho_0$ was counterproductive to the task $\mathcal{W}$ (i.e., implementing $\mathcal{W}'$ directly is *more* efficient). In general, if a protocol obeys the main bound (S119) with negative values of $M'_0, T'_0$, it means that a more favorable initial state exists or a more efficient protocol exists. Finally, intermediately favorable choices of $\rho_0$ obey bounds between (S160) and (S165), generically captured by (S164). However, we stress that the protocols $\mathcal{W}_0$ and $\mathcal{W}' = \mathcal{W}\mathcal{W}_0$

need not saturate the bound (S119), nor does the SRE state $\rho_0$ (S159) have to be advantageous compared to the separable (or product) state $\rho'$ from which it is prepared. This just implies that $\mathcal{W}$ and $\rho_0$ are suboptimal.

We infer a bound on useful quantum tasks $\mathcal{W}$ acting on SRE initial states $\rho_0$ (S159) from the bound that holds for the most useful choices of $\rho_0$ and most efficient protocols $\mathcal{W}_0$ and $\mathcal{W}'$, corresponding to the bound (S160). We now formalize the foregoing arguments to extend Theorem 5 and the associated bound (S117) to general classes of entangled initial states $\rho_0$, which we specify in Definition 8. The extension of Theorem 5 appears in Corollary 9.

**Definition 8.** *Supposing that the physical qubits of a system correspond to the vertex set $V$ of some graph $G$,*

1. *A density matrix $\rho$ has "entanglement range $\xi$" if, for any tripartition $V = V_L \cup V_C \cup V_R$ of the physical qubits (which live on the vertices $V$ of the graph $G$) such that $d(V_L, V_R) \geq \xi + 1$, the reduced density matrix $\rho_{L,R} = \mathrm{tr}_C [\rho]$ (where the qubits in $V_C$ are traced out) is separable with respect to the two sides $V_L$ and $V_R$.*

2. *A density matrix $\rho$ is "short-range entangled of type $(M'_0, T'_0, \xi)$" if $\rho$ can be prepared from an initial state $\rho'$ with entanglement range $\xi \ll \mathrm{diam}(V)$ via some local protocol $\mathcal{W}_0$ (compatible with Sec. 6.1) utilizing the outcomes of local measurements in $M'_0$ regions and Hamiltonian dynamics for total time duration $T'_0$.*

In the literature, SRE states are typically defined as being preparable from a *product* state $\rho'$ using a unitary circuit $\mathcal{W}_0$ with finite depth $T'_0$. Note that (S159) captures such states, and also extends this definition to allow for *separable*[18] states $\rho'$ and *finite-resource* protocols $\mathcal{W}_0$ where the depth $T'_0$ and number of measurement regions $M'_0$ are both finite. In Def. 8, we identify an extended class of SRE states that can be prepared using finite-resource protocols $\mathcal{W}_0$ acting on states $\rho'$ with entanglement range $\xi > 0$, which are separable with respect to some cut $C$ of $V$ of thickness $\xi$ or greater.

In Corollary 9, we extend the main bound (S117) of Theorem 5 to useful quantum tasks $\mathcal{W}$ acting on arbitrary initial states $\rho_0$ that are short-range-entangled of type $(M'_0, T'_0, \xi)$, provided that $M'_0, T'_0, \xi \ll L$ (with $L$ the task distance). Of particular interest are the various entangled resource states known in the literature: Owing to their widespread use and knowledge of efficient preparation strategies, one might ask whether such states can be used to speed up useful quantum tasks that generate entanglement (e.g., teleportation and Bell-pair distillation).

Consider, e.g., the cat-like Greenberger-Horne-Zeilinger (GHZ) state [S8, S35–S37],

$$|\mathrm{GHZ}\rangle = \frac{1}{\sqrt{2}} \left( \bigotimes_{u \in V} |0\rangle_u + \bigotimes_{u \in V} |1\rangle_u \right) = \frac{1}{\sqrt{2}} \left( |00\ldots 00\rangle + |11\ldots 11\rangle \right), \tag{S166}$$

which becomes separable upon measuring *any* bulk qubit. Hence, the GHZ state (S166) has entanglement range $\xi = 1$. In other words, *any* bipartition that also traces out a bulk qubit converts the GHZ state (S166) to a separable state (for comparison, $\xi = 0$ reflects a product state, which becomes separable under any bipartition *without* tracing out any qubits). While the GHZ state is technically SRE of type $(0, 0, 1)$, this perspective is not useful.

In Theorem 11 we prove that any protocol $\mathcal{W}_0$ that prepares the GHZ state $|\mathrm{GHZ}\rangle$ on $N \sim L$ qubits must satisfy (S117). We caution, however, that Corollary 9 only extends the bound (S117) of Theorem 5 in the context of quantum tasks that generate entanglement between qubits $i$ and $f$ (in other words, useful quantum tasks that are bounded by Theorem 5). This includes quantum error correction (QEC), teleportation, Bell pair distillation, and other tasks. However, the preparation of the GHZ state (S166) and several other resource states requires a modified bound, which we state in Theorem 11 of Sec. 7. The latter bounds apply only to initial states $\rho_0$ (S159) that can be prepared from a product state $\rho'$. Importantly, starting from a product state, the GHZ-preparation protocol $\mathcal{W}_0$ obeys $2(M'_0 + 1) v T'_0 \geq L \sim N$ (see [S36, S37] and Sec. 7.2 for explicit protocols). Thus, preparing $|\mathrm{GHZ}\rangle$ on all physical qubits is no easier than generating a Bell pair on $\{i, f\}$ (from an initial product state).

The question is then: Instead of starting from a product state, is it qualitatively easier to generate a Bell pair on the qubits $i$ and $f$ starting from some entangled initial state that already encodes "nonlocal" entanglement, such as $|\mathrm{GHZ}\rangle$ (S166)? The answer is *no*, due to the peculiar nature of multiparty entanglement in such states. We first consider the GHZ state (S166) before discussing generic short-range-entangled initial states of type $(M'_0, T'_0, \xi)$.

Although $|\mathrm{GHZ}\rangle$ (S166) is nonlocally entangled, it becomes separable as soon as any single qubit is measured. As a result, the proof of Theorem 5 for an initial GHZ state is modified as follows. Ignoring the trajectory dependence of measurement locations for simplicity, recall that the proof of Theorem 5 in Sec. 6.7 involves partitioning all physical qubits into two sets $V = V_L \cup V_R$, corresponding to the left and right sides of some cut $C$. The cut $C$ avoids all measurement regions in $\mathbb{M}^n$, and we assume without loss of generality that $i \in V_L$ and $f \in V_R$. In the case of the GHZ initial state (S166), the cut $C$ must contain internal qubits (since some qubits must be traced out to realize a

---

[18] Here, "separability" is defined with respect to a cut $C$ of the type described in Sec. 6.6.

separable state, since $\xi = 1$ for GHZ compared to $\xi = 0$ for a product state). Accordingly, $V = V_L \cup V_C \cup V_R$, where the two sides $L/R$ are connected indirectly only through $V_C$ (i.e., $d(V_L, V_R) \geq 2$). To construct the reference protocol $\widetilde{\mathcal{W}}$, we delete all links in the Hamiltonian $H$ (S86) between $V_C$ and $V_{L/R}$, so that $\widetilde{\mathcal{W}}$ again represents a LOCC channel for $V_L$ and $V_R$ that does not couple the two parties under unitary time evolution directly. Moreover, the initial density matrix on $V_L \cup V_R$ is separable, even though the global state is $|\text{GHZ}\rangle$. This separability is the only property of the initial state used in the original proof, so Theorem 5 generalizes to $|\text{GHZ}\rangle$ immediately, with the only difference being the appearance of constant factors (e.g., $\xi$) due to the "thickness" of the cut $V_C$ (see Corollary 9 below).

The same arguments straightforwardly generalize Theorem 5 and the main bound (S119) to entangled initial states with similar properties to the GHZ state (S166). In fact, the exact same proof given in Sec. 6.7 applies to initial states with entanglement range $\xi$ (see the first part of Definition 8) with one minor adjustment. Essentially, a product state has entanglement range $\xi = 0$, and so the cut $C$ used in the proof of Theorem 5 had no "thickness"—it merely bipartitions the set of physical vertices $V$. For an initial state with entanglement range $\xi > 0$, we simply require that the cut $C$ have thickness $\xi$ or greater (so that $\xi$ qubits between the sides $V_{L,R}$ are traced out, ensuring that the state of all remaining qubits is separable per Definition 8). We can still choose an optimal cut a distance $r \sim L/2M$ (S140a) from all measurement regions and complete the proof as we did for initial product states, provided that $\xi \ll L$. This ensures that any corrections to $r = \min_m d(C, S_m)$ are $\mathsf{O}(1)$, and therefore negligible in the asmyptotic limit $L, T \gg 1$ from which the bound is extracted, as discussed in Sec. 6.4.

It is also straightforward to extend Theorem 5 and the main bound (S119) from initial states with entanglement range $\xi$ to those that are short-range entangled of type $(M_0', T_0', \xi)$. Recall that a protocol $\mathcal{W}$ that acts on a state $\rho_0 = \mathcal{W}_0 \rho' \mathcal{W}_0^{-1}$ (S159) obeys a bound (S160), which we infer from a bound on the protocol $\mathcal{W}' = \mathcal{W}\mathcal{W}_0$ that acts on the initial state $\rho'$. The arguments above imply that, if $\rho'$ is has entanglement range $\xi \ll L$, then $\mathcal{W}'$ obeys (S117) without modification. Because any state $\rho_0$ that is SRE of type $(M_0', T_0', \xi)$ is preparable using finite resources $M_0', T_0'$ from the entangled state $\rho'$, the protocol $\mathcal{W}$ that acts on $\rho_0$ obeys (S160), i.e.,

$$L \leq 2\,(M + M_0' + 1)\,v\,(T + T_0')\,, \tag{S167}$$

which follows directly from Theorem 5, the bound (S117), and the foregoing derivations, all of which are summarized in Corollary 9. The bound (S167) applies directly to the preparation of Bell states shared by the qubits $i$ and $f$ with $d(i, f) = L$, to quantum teleportation over distance $L$, and to quantum error correcting codes with code distance $\mathsf{O}(L)$, up to $\mathsf{O}(1)$ corrections. In Sec. 7 we show that an analogous same bound applies to other useful quantum tasks.

**Corollary 9.** *Consider a dilated quantum channel $\mathcal{W}$ comprising (a) $M$ total measurement regions (S94), with maximum measurement range $l$ (S95), and (b) local Hamiltonian dynamics (S86) for an interval of duration $T$, with associated velocity $v_E$ (S92). Suppose that the protocol $\mathcal{W}$ realizes a Bell pair on the physical qubits $i$ and $f$ when applied to a physical initial state $\rho_0$ that is short-range entangled of type $(M_0', T_0', \xi)$, where all Stinespring qubits are initialized in the state $|\mathbf{0}\rangle$. Then, for any $\epsilon \in (0, 1)$, suppose that the distance $L$ between the qubits obeys*

$$(M + M_0' + 1)\,[2\,v_E\,(T + T_0') + l + 4 + \xi] \leq (1 - \epsilon)\,(L + 1 - l)\,, \tag{S168}$$

*and we require that $T + T_0' \gtrsim D$ if $G$ is a lattice with finite spatial dimension $D$. If we also have*

$$v\,(T + T_0') \geq \max\,(\mathrm{e}, D - 3)\,, \quad \text{and} \quad (D - 1)\,\frac{\ln\,[v\,(T + T_0')]}{v\,(T + T_0')} \leq (1 - \ln 2)\,\epsilon\,, \tag{S169}$$

*then there exists a separable state $\widetilde{\rho}_{if}$ (with $\mathrm{tr}\left[\widetilde{\rho}_{if}\right] = 1$) that is close in trace distance to the final density matrix $\rho_{if}$ of qubits $i$ and $f$ (S107), such that*

$$\|\rho_{if} - \widetilde{\rho}_{if}\|_1 \leq \frac{1}{\epsilon}\,2^{v_E\,(T + T_0') + \frac{l}{2}} \times \begin{cases} c\,l\,2^{-\frac{L + 1 - l}{2(M + M_0' + 1)}} & \text{prefixed measurement locations} \\ c\,L^{D-1}\,2^{-\frac{(1-\epsilon)L + 1 - l}{2(M + M_0' + 1)}} & \text{adaptive measurement locations} \end{cases}. \tag{S170}$$

*where the constants $c > 0$ depends only on the structure of the graph $G$[19].*

The interpretation of Corollary 9 mimics that of Theorem 5; the result of the Corollary is the bound (S167), which resembles the main bound (S119) for initial product states. Importantly, the interpretation of the quantities $M_0'$ and

---

[19] Here and below, we combine the two cases for compactness, although strictly speaking the case of adaptive measurement locations does not rely on conditions (S168) and (S169), as stated in Theorem 5, and the two constants may $c$ differ by $\mathsf{O}(1)$ amounts.

$T'_0$ in (S167) relates to the protocol $\mathcal{W}_0$ that prepares the SRE initial state $\rho_0$ (i.e., using $M_0$ local measurement regions and Hamiltonian evolution for duration $T_0$) from an initial state $\rho'$ with entanglement range $\xi \ll L$. In (S119), we also use the same variables to capture small, $\mathsf{O}(1)$ corrections that are unimportant in the asymptotic limit $L, T \gg 1$.

We conclude with two comments on the long-standing problem of classifying entanglement in many-body states. First, the short-rangedness of entanglement in GHZ-like states is well known in the literature on topological order [S38]. While the common definition of SRE states is similar to that of Definition 8, it is closer to (S159) for a product state $\rho'$. In particular, the literature defines SRE states as the set of states with nonzero entanglement entropy for some bipartition that are convertible to a product state using a local, invertible, finite-depth transformation that is not necessarily unitary (i.e., if such a finite-depth protocol can create the state from an initial product state, it is said to be short-range entangled). However, states with entanglement range $\xi$ (e.g., the GHZ state) have short-range entanglement by construction. It would be interesting to investigate the relation between our definitions of SRE states and states with entanglement range $\xi \ll N$, and the definition of SRE states in, e.g., [S38]. Second, it is common to consider whether the entanglement entropy of *pure* many-body states obeys an "area law" (for some bipartition of the graph): If the state's bipartite entanglement entropy scales with the size of the smaller of the two regions, it is said to be "volume law"; if the entanglement entropy scales with the size of the *boundary* between the two regions, it is said to be "area law". For example, both $|\text{GHZ}\rangle$, and the W state [S10],

$$|W\rangle \;=\; \frac{1}{\sqrt{|V|}} \sum_{u \in V} |1\rangle_u \otimes |\mathbf{0}\rangle_{V \setminus \{u\}}, \tag{S171}$$

have area-law entanglement, despite being difficult to prepare (at least by any known protocol with $L \lesssim (M'_0+1)\,v\,T'_0$). Although we proved $|\text{GHZ}\rangle$ is qualitatively no better than product states for generating Bell pairs (or performing state transfer), it is not immediately clear from our proof whether the condition $L \lesssim (M+1)\,v\,T$ also applies to $|W\rangle$ (S171), since it is not short-range entangled of a small range $\xi$. We discuss the W state further in Sec. 7.3.

### 6.9. Summary of the bound

Consider a physical system with $N$ degrees of freedom assigned to the vertices $V$ of some graph $G$. Suppose that $G$ has finite spatial dimension $D$[20] and the Hilbert space $\mathcal{H}_v$ on each vertex $v \in V$ has bounded dimension (i.e., the vertices host qubits, $d$-state qudits, fermions, or Majorana modes). Suppose that a protocol $\mathcal{W}$ applied to the initial state $\rho_0$ achieves a useful quantum task involving the sites $i$ and $f$ (so that the "task distance" is $L = d(i,f)$), using a combination of time evolution, local quantum channels, and measurements in $M$ local regions.

Proposition 2 establishes that teleporting a state $|\psi\rangle$ from $i$ to $f$ is equivalent to creating a Bell pair between $i$ and $f$, up to $\mathsf{O}(1)$ corrections to $M, N, T$. Note that Proposition 10 (of Sec. 7) extends the resulting bound to the preparation of resource states with finite entanglement and/or correlations between the qubits $i$ and $f$.

In particular, suppose that time evolution is generated by some time-dependent local Hamiltonian $H(t) = \sum_j H_j(t)$ (S86), where each local term $H_j$ may be conditioned on the outcomes of any prior measurements, has maximal eigenvalue $h$ (S87) for any time $t$, and acts only on the $k_j = |\Omega_j|$ vertices $v \in \Omega_j$ with $k_j, \text{diam}(\Omega_j) \ll L, |V|$. Denote by $T$ the total duration of Hamiltonian evolution; in the case where $H(t)$ generates a local quantum circuit, $T$ is the total circuit depth (i.e., the minimum number of layers of multi-qubit gates, paralellizing where possible). In either case, $H(t)$ obeys the standard Lieb-Robinson Theorem [S7] with Lieb-Robinson velocity $v$[21].

Additionally, we allow every aspect of the measurement protocol associated with $\mathcal{W}$ at a given time $t$ to be conditioned upon the outcomes of all prior measurements. The set $\mathbb{M}$ of measurement regions is constructed as described in Sec. 6.2; the number of measurement regions $M$ is determined from $\mathbb{M}$ according to (S94), and we further define the maximum *range* $l$ of measurements via (S95). The measurements are assumed to be instantaneous, but may be projective (strong), weak, or realize "generalized" measurements. Note that the outcomes of all measurements can be communicated instantaneously, and we allow for arbitrary local quantum channels in $\mathcal{W}$.

Suppose that the protocol $\mathcal{W}$ acts on an initial state $\rho_0$ that is short-range entangled of type $(M'_0, T'_0, \xi)$, as specified in Definition 8. In other words, the SRE initial state $\rho_0$ can be written in the form

$$\rho_0 \;=\; \mathcal{W}_0 \, \rho' \, \mathcal{W}_0^\dagger, \tag{S172}$$

where $\mathcal{W}_0$ itself achieves a useful quantum task (as it generates entanglement) and obeys precisely the same constraints articulated above for $\mathcal{W}$. In particular, $\mathcal{W}_0$ involves measurements in $M'_0$ local regions and Hamiltonian evolution for

---

[20] If the spatial dimension $D$ of the graph $G$ is infinite (a zero-dimensional system), we merely multiply the Lieb-Robinson velocity (S89) by a factor of $(K-1)$, where $K$ is the maximal degree of $G$, as dictated by (S92).

[21] For example, $v \leq 2\,\mathrm{e}\,(K-1)\,h$ (S89) for a nearest-neighbor Hamiltonian $H$, where $K$ is the maximal degree of the graph $G$.

total time $T_0'$ (or is a quantum circuit with depth $T_0'$). Importantly, the state $\rho'$ in (S172) has entanglement range $\xi \ll L$—i.e., it becomes separable upon tracing out the qubits internal to any cut $C$ of thickness $\xi$ or greater that separates the qubits $i$ and $f$, as described in Definition 8 (this includes product states, e.g., $|0\rangle\langle 0|$).

If the generic conditions above hold, Corollary 9 applies directly to the protocol $\mathcal{W}$ acting on the SRE initial state $\rho_0$ (S172). In the case of *prefixed measurement locations*, for any $\epsilon \in (0,1)$, if the task distance $L$ obeys

$$(1-\epsilon)\,(L+1-l) \geq (M + M_0' + 1)\,[2\,v_E\,(T+T_0') + l + 4 + \xi]\,, \tag{S173}$$

and we restrict to times $v\,(T+T_0') \geq \max(\mathrm{e}, D-3)$, then the reduced density matrix $\rho_{if}$ for the task qubits $i$ and $f$ generated by the protocol $\mathcal{W}$ is arbitrarily close in trace distance (S112) to a separable density matrix $\widetilde{\rho}_{if}$:

$$\|\rho_{if} - \widetilde{\rho}_{if}\|_1 \leq \frac{c\,l}{4\epsilon} \exp\left\{-\frac{\ln 2}{2}\left[\xi + \epsilon\,\frac{L+1-l}{M+M_0'+1}\right]\right\} \ll 1\,, \tag{S174}$$

which is arbitrarily close to zero in the asymptotic limit $L, T \gg 1$ (for fixed $M$).

Essentially, the state $\rho$ prepared by the protocol $\mathcal{W}$ is arbitrarily close in trace distance to a separable state (S174) if the assumption (S173) holds. In other words, if $M, T$ of the protocol $\mathcal{W}$ obey (S173) for some $\epsilon$, then the protocol $\mathcal{W}$ cannot have generated entanglement or correlations (nor transferred quantum information) between the qubits $i$ and $f$. This means that $\mathcal{W}$ can only achieve a useful quantum task on the qubits $i$ and $f$ (separated by distance $L = d(i,f)$) if the assumption (S173) *fails*. In the large $L,T$ limit from which bounds are extracted, this corresponds to

$$L < 2\,(M + M_0' + 1)\,v\,(T+T_0') + \mathsf{O}(1)\,, \tag{S175}$$

where the correction on the right is $\mathsf{O}(1)$ in that it does not scale with $L, M, T$. Note that other asymptotically unimportant details may be modify the bound for finite $L, T$, in which case we define the quantities

$$T_0 \equiv T_0' + \delta T \quad \text{and} \quad M_0 \equiv 2\,(M_0' + \delta M + 1)\,, \tag{S176}$$

where $\delta T, \delta M \ll L, T$ are both $\mathsf{O}(1)$, and we recover the most general bound

$$L \leq (2\,M + M_0)\,v\,(T+T_0)\,, \tag{S177}$$

which is the first equation of the main text, and our main result. Note that the standard Lieb-Robinson bound $L \leq (2)\,v\,T$ recovers in the measurement-free limit $M \to 0$, and constrains the preparation of a Bell pair on sites $i$ and $f$ starting from a point midway between the two sites. However, in the absence of measurements, unitary state transfer obeys the bound $L \leq v\,T$. While both "staircase" (i.e., left-to-right) and "light-cone" (middle-out) circuits can prepare states on $N$ qubits, e.g., only staircase circuits can perform state transfer. These distinctions are captured by the prefactor $M_0$, which for the "light-cone" protocols relevant to measurements defaults to $M_0 = 2$; however, for "staircase" protocols, the fact that $L \leq vT$ is captured by taking $M_0 = 1$. In this sense, the redefinition (S176) facilitates the recovery of standard $M = 0$ bounds, while also more generally capturing the small corrections that may be present at finite size, but are asymptotically unimportant. In the remainder of the Supplement, we replace $T_0'$ and $M_0'$ with $T_0$ and $M_0$ for notational convenience. Matching the bound (S177) of the main text follows from (S176).

## 7. BOUNDS ON PREPARING OTHER USEFUL ENTANGLED RESOURCE STATES

In Sec. 6.8 we presented Corollary 9, which extends the main bound of Theorem 5 to protocols $\mathcal{W}$ that prepare a Bell pair on qubits a distance $L$ apart, starting from short-range-entangled initial states. We consider initial states with "entanglement range $\xi$" and those that are "short-range entangled of type $(M_0, T_0, \xi)$" (see Definition 8). States of the former type are directly compatible with Theorem 5, provided that the cut used in the proof is chosen to have thickness $\xi$ or greater. States of the latter type can be prepared from states of the former type using a finite-resource protocol $\mathcal{W}_0$ that uses the outcomes of local measurements in $M_0$ regions and involves local Hamiltonian time evolution for duration $T_0$. The resource requirements ($M_0$ and $T_0$) for preparing such states appears in the main bound of Corollary 9. For example, the GHZ state (S166) has entanglement range $\xi = 1$ and is short-range entangled of type $(0,0,1)$.

We now derive bounds constraining the preparation of various useful correlated and nonlocally entangled resource states of experimental interest, such as the GHZ state (S166), which cannot be optimally bounded using Theorem 5 directly. Instead of formulating the bound in terms of entanglement between the task qubits $i$ and $f$, we instead derive a bound in terms of correlations between qubits $i$ and $f$. The resulting bounds constrain the resource requirements $(M, T)$ for preparing a number of correlated resource states, and in some cases imply bounds on converting between

various types of useful entangled states. Since the states of interest involve more than two qubits, it will prove useful to generalize the "task distance" $L$ for generic resource states via

$$L \equiv \max_{u,v \in V} d(u,v), \tag{S178}$$

i.e., $L$ is the maximum separation between any two vertices $u$ and $v$ of the physical graph $G$, which constrains the preparation of resource states with finite entanglement and/or correlations between well-separated qubits.

### 7.1. Speed of preparing GHZ-like states

Here we generalize Theorem 5 to constrain the preparation of the Greenberger-Horne-Zeilinger (GHZ) state [S8, S35] defined in (S166). The bound also extends to other entangled states with similar properties. Before considering the GHZ state (S166) itself, we first observe that, if a protocol $\mathcal{W}$ generates a Bell state shared by the qubits $i$ and $f$ in some measurement trajectory $\boldsymbol{n}$, then the qubits $i$ and $f$ are *maximally correlated*. To quantify this correlation, we define the probability-weighted correlations of sites $\{i, f\}$ for the outcome trajectory $\boldsymbol{n}$ as

$$\begin{aligned}
\operatorname{Cor}(i, f)_{\boldsymbol{n}} &\equiv \max_{\|\mathcal{O}_i\|, \|\mathcal{O}_f\| \leq 1} \left\{ \langle \mathcal{O}_i \mathcal{O}_f P_{\boldsymbol{n}} \rangle - \frac{1}{p_{\boldsymbol{n}}} \langle \mathcal{O}_i P_{\boldsymbol{n}} \rangle \langle \mathcal{O}_f P_{\boldsymbol{n}} \rangle \right\} \\
&= \max_{\|\mathcal{O}_i\|, \|\mathcal{O}_f\| \leq 1} \left\{ \langle \mathcal{O}_{i,\boldsymbol{n}} \mathcal{O}_{f,\boldsymbol{n}} \rangle - \frac{1}{p_{\boldsymbol{n}}} \langle \mathcal{O}_{i,\boldsymbol{n}} \rangle \langle \mathcal{O}_{f,\boldsymbol{n}} \rangle \right\},
\end{aligned} \tag{S179}$$

where, as discussed in Sec. 2, expectation values of the form $\langle A \rangle$ are evaluated in the *dilated* Hilbert space with respect to the *final* density matrix according to $\langle A \rangle = \operatorname{tr}\left[ A \mathcal{W} \varrho \mathcal{W}^\dagger \right]$ (S17), where the trace runs over $\mathcal{H}_{\text{dil}}$.

In (S179) above, $\mathcal{O}_i$ and $\mathcal{O}_f$ are Hermitian operators that act nontrivially only on sites $i$ and $f$, respectively. The outcome-projected operators $\mathcal{O}_{i,\boldsymbol{n}}$ and $\mathcal{O}_{f,\boldsymbol{n}}$ are defined according to (S96), and $p_{\boldsymbol{n}} = \langle P_{\boldsymbol{n}} \rangle$ is the probability of recovering the sequence of measurement outcomes $\boldsymbol{n}$ (S16). To connect the correlation function (S179) to GHZ-like states, we now state and prove Proposition 10.

**Proposition 10.** *Suppose that the protocol $\mathcal{W}$ acting on the dilated Hilbert space $\mathcal{H}_{\text{dil}}$ satisfies*

$$\mathcal{W}|\mathbf{0}\rangle = \sum_{\boldsymbol{n}} \sqrt{p_{\boldsymbol{n}}} \left( \frac{1}{\sqrt{2}} |00\rangle_{if} \otimes |\Psi_{\boldsymbol{n}}\rangle_{V \setminus \{i,f\}} + \frac{1}{\sqrt{2}} |11\rangle_{if} \otimes |\Phi_{\boldsymbol{n}}\rangle_{V \setminus \{i,f\}} \right) \otimes |\boldsymbol{n}\rangle_{\text{ss}}, \tag{S180}$$

*where $|\Phi_{\boldsymbol{n}}\rangle$ and $|\Psi_{\boldsymbol{n}}\rangle$ are arbitrary many-body states of all physical qubits in $V$ except for the task qubits $i$ and $f$. This implies that $\operatorname{Cor}(i,f)_{\boldsymbol{n}} = p_{\boldsymbol{n}}$ (S179) for all outcome trajectories $\boldsymbol{n}$.*

*Proof.* We first make the choice $\mathcal{O}_{i/f} = Z_{i/f}$ and evaluate (S179) explicitly in the state (S180), finding

$$\langle Z_i Z_f P_{\boldsymbol{n}} \rangle = \langle P_{\boldsymbol{n}} \rangle = p_{\boldsymbol{n}} \tag{S181a}$$
$$\langle Z_i P_{\boldsymbol{n}} \rangle = \langle Z_f P_{\boldsymbol{n}} \rangle = 0, \tag{S181b}$$

and given that $\operatorname{Cor}(i,f)_{\boldsymbol{n}}$ is defined as the maximum over unit-norm operators (S179), we determine that

$$\operatorname{Cor}(i,f)_{\boldsymbol{n}} \geq p_{\boldsymbol{n}}. \tag{S182}$$

Now, it simply remains to prove that $\operatorname{Cor}(i,f)_{\boldsymbol{n}} \leq p_{\boldsymbol{n}}$ for any state. To see this, let

$$A = \mathcal{O}_{i,\boldsymbol{n}} - \langle \mathcal{O}_{i,\boldsymbol{n}} \rangle \quad \text{and} \quad B = \mathcal{O}_{f,\boldsymbol{n}} - \langle \mathcal{O}_{f,\boldsymbol{n}} \rangle, \tag{S183}$$

and, because $[A, B] = 0$, we have

$$0 \leq \langle (A - \zeta B)^2 \rangle = \langle A^2 \rangle + \zeta^2 \langle B^2 \rangle - 2\zeta \langle AB \rangle, \tag{S184}$$

for any $\zeta$. Minimizing over $\zeta$ and noting that the disconnected piece of (S179) vanishes for the choice (S183), we have

$$\operatorname{Cor}(i,f)_{\boldsymbol{n}}^2 = \max_{\mathcal{O}_i, \mathcal{O}_f} \langle AB \rangle^2 \leq \max_{\mathcal{O}_i, \mathcal{O}_f} \langle A^2 \rangle \langle B^2 \rangle \leq \max_{\mathcal{O}_i, \mathcal{O}_f} \langle \mathcal{O}_{i,\boldsymbol{n}}^2 \rangle \langle \mathcal{O}_{f,\boldsymbol{n}}^2 \rangle \leq p_{\boldsymbol{n}}^2, \tag{S185}$$

which proves the claim for the state (S180). □

More generally, we define the outcome-averaged correlation function for qubits $i$ and $f$,

$$\overline{\mathrm{Cor}}(i,f) \equiv \max_{\|\mathcal{O}_i\|, \|\mathcal{O}_f\| \le 1} \sum_{\boldsymbol{n}} \left\{ \langle \mathcal{O}_i \, \mathcal{O}_f \, P_{\boldsymbol{n}} \rangle - \frac{1}{p_{\boldsymbol{n}}} \langle \mathcal{O}_i \, P_{\boldsymbol{n}} \rangle \langle \mathcal{O}_f \, P_{\boldsymbol{n}} \rangle \right\}, \tag{S186}$$

where (S186) can be nonzero in states with *multipartite* entanglement (namely, if $i$ and $f$ do not share a Bell pair. Importantly, (S186) does *not* capture classical correlations due to measurements, which do not necessarily reflect useful entanglement. An important example with $\overline{\mathrm{Cor}}(i,f) = 1$ is the GHZ state, where $|\Psi_{\boldsymbol{n}}\rangle = |0\ldots 0\rangle$ and $|\Phi_{\boldsymbol{n}}\rangle = |1\ldots 1\rangle$ (on the physical qubits). We now prove that generating any such state also requires $2\,v\,(M+1)\,T \ge L$ by showing that obtaining a state of the form (S180) is bounded by the same Lieb-Robinson arguments invoked in Theorem 5.

As a remark, one can also consider bounding the quantity $\sum_{\boldsymbol{n}} \mathrm{Cor}(i,f)_{\boldsymbol{n}} \in [\overline{\mathrm{Cor}}(i,f), 1]$, where the maximization is over operators $\mathcal{O}_i(\boldsymbol{n}), \mathcal{O}_f(\boldsymbol{n})$ that may depend on the trajectory, in contrast to (S186). The result will be the same as bounding $\overline{\mathrm{Cor}}(i,f)$ up to an $\mathsf{O}(1)$ correction to $T$, because of the following. For any protocol that achieves $\sum_{\boldsymbol{n}} \mathrm{Cor}(i,f)_{\boldsymbol{n}} \sim 1$, it can be followed by local rotations on $i$ and $f$ that depend on the trajectory $\boldsymbol{n}$, such that the optimal operators $\mathcal{O}_i(\boldsymbol{n}), \mathcal{O}_f(\boldsymbol{n})$ are rotated to ones that do *not* depend on $\boldsymbol{n}$ anymore. This procedure only adds $\mathsf{O}(1)$ to the total time $T$, and makes $\overline{\mathrm{Cor}}(i,f) \sim 1$ large. Therefore it suffices to bound $\overline{\mathrm{Cor}}(i,f)$.

Since the final state is generated by applying the protocol $\mathcal{W}$ to the initial state $|\boldsymbol{0}\rangle$, (S186) is equivalent to

$$\overline{\mathrm{Cor}}(i,f) = \max_{\|\mathcal{O}_i\|, \|\mathcal{O}_f\| \le 1} \sum_{\boldsymbol{n}} \langle \mathcal{O}_{i,\boldsymbol{n}}(T)\, \mathcal{O}_{f,\boldsymbol{n}}(T) \rangle_0 - \frac{1}{p_{\boldsymbol{n}}} \langle \mathcal{O}_{i,\boldsymbol{n}}(T) \rangle_0 \langle \mathcal{O}_{f,\boldsymbol{n}}(T) \rangle_0, \tag{S187}$$

where the "0" subscript in the expectation values $\langle A \rangle_0$ reflects that they are evaluated in the *initial* state $|\boldsymbol{0}\rangle$. We now prove that the GHZ state (S166) of the full physical system cannot be prepared any more efficiently than a Bell pair on the qubits $i$ and $f$ at the system's edges using an adaptive, measurement-assisted protocol $\mathcal{W}$.

**Theorem 11.** *Consider a quantum channel $\mathcal{W}$ involving (a) local measurements in $M$ total regions (S94), with maximum measurement range $l$ (S95), and (b) local Hamiltonian dynamics (S86) over a time interval of total duration $T$, with associated velocity $v_E$ (S92). Suppose that we apply this channel to a pure, short-range-entangled initial state $\rho_0$ of type $(M_0, T_0, 0)$—i.e., $\rho_0$ can be prepared from some state $\rho'$ with entanglement range $\xi = 0$ (a product state, e.g., $|\boldsymbol{0}\rangle$) using some protocol $\mathcal{W}_0$ involving local measurements in $M_0$ regions and local Hamiltonian evolution for duration $T_0$. Following (S178), we identify the qubits $i$ and $f$ with two opposing edges of the graph $G$ of physical qubits, and denote by $L$ their separation. For any $\epsilon \in (0, 1)$, suppose that (S168) and (S169) hold. Then the final state obeys*

$$\overline{\mathrm{Cor}}(i,f) \le \frac{1}{\epsilon} 2^{v_E(T+T_0) + \frac{l}{2} + 2} \times \begin{cases} c\, l\, 2^{-\frac{L+1-l}{2(M+M_0+1)}} & \text{prefixed measurement locations} \\ c\, L^{D-1}\, 2^{-\frac{(1-\epsilon)(L+1-l)}{2(M+M_0+1)}} & \text{adaptive measurement locations} \end{cases}, \tag{S188}$$

*where the constants $c > 0$ are the same constants that appear in (S112) and (S113) of Theorem 5, respectively.*

*Proof.* We assume for now that the initial state is $|\boldsymbol{0}\rangle$, since the generalization to the case with $M_0, T_0 > 0$ is straightforward using the arguments developed in Sec. 6.8 and Corollary 9. The strategy below is quite similar to those employed in previous proofs, and we first consider the case of prefixed measurement locations.

We utilize the same cut $C_r$ that was identified in Sec. 6.5, and build the same reference protocol $\widetilde{\mathcal{W}}$ by removing the cut Hamiltonian. However, here we investigate a different quantity—i.e., $\overline{\mathrm{Cor}}(i,f)$ instead of $\|\rho_{if} - \widetilde{\rho}_{if}\|_1$. In the reference dynamics, the physical state remains a pure state that is a direct product of the states on opposite sides of the cut $C_r$. Consequently, correlations between the sites $i$ and $f$ vanish in the reference trajectory:

$$\langle \widetilde{\mathcal{O}}_{i,\boldsymbol{n}}(T)\, \widetilde{\mathcal{O}}_{f,\boldsymbol{n}}(T) \rangle_0 - \frac{1}{\widetilde{p}_{\boldsymbol{n}}} \langle \widetilde{\mathcal{O}}_{i,\boldsymbol{n}}(T) \rangle \langle \widetilde{\mathcal{O}}_{f,\boldsymbol{n}}(T) \rangle_0 = 0, \quad \forall\, \mathcal{O}_i, \mathcal{O}_f. \tag{S189}$$

As a result, the same correlation function for the true channel $\mathcal{W}$ is deformed by:

$$\overline{\mathrm{Cor}}(i,f) = \sum_{\boldsymbol{n}} \langle \mathcal{O}_{i,\boldsymbol{n}}(T) \mathcal{O}_{f,\boldsymbol{n}}(T) \rangle_0 - \frac{1}{p_{\boldsymbol{n}}} \langle \mathcal{O}_{i,\boldsymbol{n}}(T) \rangle \langle \mathcal{O}_{f,\boldsymbol{n}}(T) \rangle_0 - \langle \widetilde{\mathcal{O}}_{i,\boldsymbol{n}}(T) \widetilde{\mathcal{O}}_{f,\boldsymbol{n}}(T) \rangle_0 + \frac{1}{\widetilde{p}_{\boldsymbol{n}}} \langle \widetilde{\mathcal{O}}_{i,\boldsymbol{n}}(T) \rangle \langle \widetilde{\mathcal{O}}_{f,\boldsymbol{n}}(T) \rangle_0 ,$$

using the fact that the correlator for the reference trajectory vanishes (S189). Manipulating this expression gives

$$\overline{\mathrm{Cor}}(i,f) = \sum_{\boldsymbol{n}} \langle \mathcal{O}_{i,\boldsymbol{n}}(T)\mathcal{O}_{f,\boldsymbol{n}}(T) - \widetilde{\mathcal{O}}_{i,\boldsymbol{n}}(T)\widetilde{\mathcal{O}}_{f,\boldsymbol{n}}(T)\rangle_0 - \frac{1}{p_{\boldsymbol{n}}} \langle \mathcal{O}_{i,\boldsymbol{n}}(T) - \widetilde{\mathcal{O}}_{i,\boldsymbol{n}}(T)\rangle_0 \langle \mathcal{O}_{f,\boldsymbol{n}}(T)\rangle_0$$

$$- \frac{1}{\widetilde{p}_{\boldsymbol{n}}} \langle \widetilde{\mathcal{O}}_{i,\boldsymbol{n}}(T)\rangle_0 \langle \mathcal{O}_{f,\boldsymbol{n}}(T) - \widetilde{\mathcal{O}}_{f,\boldsymbol{n}}(T)\rangle_0 + \left(\frac{1}{\widetilde{p}_{\boldsymbol{n}}} - \frac{1}{p_{\boldsymbol{n}}}\right) \langle \widetilde{\mathcal{O}}_{i,\boldsymbol{n}}(T)\rangle_0 \langle \mathcal{O}_{f,\boldsymbol{n}}(T)\rangle_0$$

$$\leq \max_{\|\mathcal{O}\|\leq 1} \left\|\sum_{\boldsymbol{n}} \mathcal{O}_{\boldsymbol{n}}(T) - \widetilde{\mathcal{O}}_{\boldsymbol{n}}(T)\right\| + \left\|\sum_{\boldsymbol{n}} a_{\boldsymbol{n}} \left[\mathcal{O}_{i,\boldsymbol{n}}(T) - \widetilde{\mathcal{O}}_{i,\boldsymbol{n}}(T)\right]\right\|$$

$$+ \left\|\sum_{\boldsymbol{n}} b_{\boldsymbol{n}} \left[\mathcal{O}_{f,\boldsymbol{n}}(T) - \widetilde{\mathcal{O}}_{f,\boldsymbol{n}}(T)\right]\right\| + \left\|\sum_{\boldsymbol{n}} c_{\boldsymbol{n}} \left[P_{\boldsymbol{n}}(T) - \widetilde{P}_{\boldsymbol{n}}(T)\right]\right\|$$

$$\leq 4 \max_{|a_{\boldsymbol{n}}|\leq 1} \max_{\|\mathcal{O}\|\leq 1} \left\|\sum_{\boldsymbol{n}} a_{\boldsymbol{n}} \left[\mathcal{O}_{\boldsymbol{n}}(T) - \widetilde{\mathcal{O}}_{\boldsymbol{n}}(T)\right]\right\|, \tag{S190}$$

where, in the third line, we defined

$$a_{\boldsymbol{n}} \equiv p_{\boldsymbol{n}}^{-1} \langle \mathcal{O}_{f,\boldsymbol{n}}(T)\rangle_0, \tag{S191}$$

which satisfies $|a_{\boldsymbol{n}}| \leq 1$, with $b_{\boldsymbol{n}}$ and $c_{\boldsymbol{n}}$ defined (and bounded) similarly for their corresponding terms. Thus, the four terms above are of the same form, and are combined into a single term in the last line of (S190) by further maximizing over the set of coefficients (e.g., $\{a_{\boldsymbol{n}}\}$). The insertion of a trajectory-dependent coefficient $a_{\boldsymbol{n}}$ is the only difference (beyond the extra prefactor 4) compared to (S121) in the previous proof. Following (S125), we now have

$$\left\|\sum_{\boldsymbol{n}} a_{\boldsymbol{n}} \left[\mathcal{O}_{\boldsymbol{n}}(T) - \widetilde{\mathcal{O}}_{\boldsymbol{n}}(T)\right]\right\| \leq \int_0^T \mathrm{d}s \max_{\boldsymbol{n}_s} \left\|\left[H_C^{\boldsymbol{n}_s}(T-s), \sum_{\boldsymbol{n}\supset\boldsymbol{n}_s} a_{\boldsymbol{n}} \mathcal{O}_{\boldsymbol{n}}(s)\right]\right\|, \tag{S192}$$

which does not have $\mathcal{O}_{\boldsymbol{n}_s}(s)$ in the commutator like (S125). Nevertheless, we only used two properties of $\mathcal{O}_{\boldsymbol{n}_s}(s)$ before. First, it is grown from the measurement locations $AB$ by the Hamiltonian, so that we can apply Lieb-Robinson bound (S128). Second, the prefactor in the Lieb-Robinson bound involves $\|\mathcal{O}_{\boldsymbol{n}_s}\| \leq 1$. These two conditions also hold for $\sum_{\boldsymbol{n}\supset\boldsymbol{n}_s} a_{\boldsymbol{n}} \mathcal{O}_{\boldsymbol{n}}$ here. The first one is trivial, while the second comes from

$$\left\|\sum_{\boldsymbol{n}\supset\boldsymbol{n}_s} a_{\boldsymbol{n}} \mathcal{O}_{\boldsymbol{n}}\right\| = \left\|\mathcal{O} \otimes \sum_{\boldsymbol{n}\supset\boldsymbol{n}_s} a_{\boldsymbol{n}} P_{\boldsymbol{n}}\right\| = \|\mathcal{O}\| \left\|\sum_{\boldsymbol{n}\supset\boldsymbol{n}_s} a_{\boldsymbol{n}} P_{\boldsymbol{n}}\right\| \leq 1, \tag{S193}$$

where we used $|a_{\boldsymbol{n}}| \leq 1$ and the orthogonality of projectors. Hence, the bound (S137) holds upon inserting $a_{\boldsymbol{n}}$, giving

$$\overline{\mathrm{Cor}}(i,f) \leq 4 \frac{cl}{\epsilon} 2^{v_E T + \frac{l}{2} - \frac{L+1-l}{2(M+1)}} \tag{S194}$$

where the constant $c$ is given by (S138), as with the analogous expression (S112) in Theorem 5.

Finally, we consider adaptive measurement locations. Recall that each trajectory $\boldsymbol{n}$ admits $\epsilon' L$ good cuts $C_r$ that are far $\sim L/(2M)$ from any measurement location. Then, for each cut $C_r$, we define a reference protocol $\widetilde{W}_r$, and there exists a set $\boldsymbol{N}_r$ of trajectories that admit that cut $C_r$. Similar to (S150), now we separate operators (rather than density matrices) according to contributions from different $C_r$, e.g.,

$$\mathcal{O}_i(T) = \frac{1}{\epsilon' L} \sum_r \sum_{\boldsymbol{n}\in\boldsymbol{N}_r} \mathcal{O}_{i,\boldsymbol{n}}(T), \quad \widetilde{\mathcal{O}}_i(T) = \frac{1}{\epsilon' L} \sum_r \sum_{\boldsymbol{n}\in\boldsymbol{N}_r} \widetilde{\mathcal{O}}_{i,\boldsymbol{n}}(T), \tag{S195}$$

and following the procedure in (S190), we recover

$$\overline{\mathrm{Cor}}(i,f) \leq \frac{4}{\epsilon' L} \sum_r \max_{|a_{\boldsymbol{n}}|\leq 1} \max_{\|\mathcal{O}\|\leq 1} \left\|\sum_{\boldsymbol{n}\in\boldsymbol{N}_r} a_{\boldsymbol{n}} \left[\mathcal{O}_{\boldsymbol{n}}(T) - \widetilde{\mathcal{O}}_{\boldsymbol{n}}(T)\right]\right\|, \tag{S196}$$

and it is then a straightforward exercise to combine the derivations of Sec. 6.7 with the above argument that inserting $a_{\boldsymbol{n}}$ does not change the bound. The result is just four times the right hand side of (S113), which becomes (S188). □

We make two remarks on the initial state used in Theorem 11. First, it is straightforward to generalize to mixed states that can be prepared using a $(M_0, T_0)$ protocol from a classical ensemble of pure product states—the only difference is that one must consider each pure state individually for $\overline{\text{Cor}}(i,f)$ to be small, then average over the classical ensemble at the end. Second, unlike Theorem 5, Theorem 11 does *not* generalize to initial states with entanglement range $\xi$ (like the GHZ state with $\xi = 1$), which may contain nonlocal multipartite entanglement[22]: After all, it is easy to prepare a GHZ from itself! Following the interpretation of (S112) in Theorem 5, we find that (S188) gives rise to a bound of the form (S119) for preparing a GHZ state (S166) on $N = L + 1$ sites. In other words, Theorem 11 implies an identical bound to (S177) as recovers from Theorem 5 and Corollary 9; the only difference is the restriction to SRE initial states $\rho_0$ (S159) of type $(M_0, T_0, \xi = 0)$, which can be prepared from a product state $\rho'$ using a minimum of $M_0$ measurement regions and evolution for time $T_0$. Preparing the GHZ state on $N$ qubits therefore obeys

$$N \lesssim 2\,(M + M_0 + 1)\,v\,(T + T_0)\,, \tag{S197}$$

which is identical to (S177) up to redefinition of certain $\mathsf{O}(1)$ quantities, and the caveat that the initial state $\rho_0$ be preparable using some finite-resource protocol $\mathcal{W}_0$ from a *product* state $\rho'$.

### 7.2. Optimal protocol for preparing the GHZ state

We now motivate the bound (S188) using an explicit protocol that generates the GHZ state (S166) on a $1D$ chain of $N$ qubits in the initial state $|\mathbf{0}\rangle$ (on all physical and Stinespring qubits) using two-local unitary Clifford gates, local measurements, and classical communications. The corresponding protocol $\mathcal{W}$ is detailed below and depicted diagrammatically in Fig. S6. We conclude with a brief sketch of how to generalize this protocol to dimensions $D > 1$.

The GHZ protocol we present involves generating local GHZ states (S166) in local regions, and subsequently patching all regions together to form a single GHZ state. It will prove convenient to define the GHZ state $|\text{GHZ}\rangle_Y$ for some set of vertices $Y \subset V$. In $1D$, we take $Y = [i, f]$ be the interval $Y = \{i, i+1, \ldots, f-1, f\} \subset V$, i.e.,

$$|\text{GHZ}\rangle_Y \;=\; \frac{1}{\sqrt{2}}\left(\bigotimes_{n \in Y}|0\rangle_n + \bigotimes_{n \in Y}|1\rangle_n\right) \;=\; \frac{1}{\sqrt{2}}\left(|00\ldots00\rangle_Y + |11\ldots11\rangle_Y\right)\,, \tag{S198}$$

which is a stabilizer state [S40] whose stabilizer group (S41) for the case $Y = [i, f]$ in $1D$ is given by

$$\mathbf{S}_Y \;=\; \mathbf{S}_{i,f} \;=\; \text{Stab}\left(|\text{GHZ}\rangle_{[i,f]}\right) \;=\; \left\{Z_i Z_{i+1},\, Z_{i+1} Z_{i+2},\, \ldots,\, Z_{f-1} Z_f,\, \bar{X}_{i,f} \equiv X_i X_{i+1} \cdots X_{f-1} X_f \right\}\,, \tag{S199}$$

which can be straightforwardly extended to noncontiguous regions and higher spatial dimensions $D > 1$.

In general, the $\bar{X}_Y$ stabilizer generator for some set $Y \subset V$ of physical qubits (with $\ell = |Y|$) is simply the product of $X_j$ over all sites $j \in Y$. The $\ell - 1$ other stabilizer generators require specifying an ordering of the qubits $j \in Y$ (where we label the first qubit as $i$, the last qubit as $f$, and all others as $n_1, \ldots, n_{\ell-2}$), so that the other stabilizers are $Z_i Z_{n_1}, Z_{n_1} Z_{n_2}, \ldots, Z_{n_{\ell-2}} Z_f$ (i.e., every $Z_j$ for $j \in Y$ appears in exactly two stabilizer generators, with the exception of the endpoint operators $Z_i$ and $Z_f$, which appear only once). The $\bar{X}$ operator always acts as $X_j$ on all qubits $j \in Y$, while the $Z_j Z_{j'}$ operators can always be multiplied together to produce arbitrary two-site $Z_k Z_{k'}$ operators.

We denote by $|\text{GHZ}\rangle$ the GHZ state (S166) on *all* physical qubits in $V$; correspondingly, $\mathbf{S}_{\text{GHZ}}$ denotes the group that stabilizes $|\text{GHZ}\rangle$ and $\bar{X}$ denotes the $X$-like stabilizer for the full GHZ state. Importantly, because $|\text{GHZ}\rangle_{[i,j]}$ is a stabilizer state [S40], it can be generated using a Clifford circuit from the initial state $|\mathbf{0}\rangle$ on $Y = [i, j]$. Moreover, the state is fully specified at a given time $t$ by its stabilizer group at time $t$. The stabilizer group for a GHZ state on $|Y|$ qubits has $|Y|$ independent generators, which form a complete set of mutually-commuting observables, which fully specify the GHZ state. Thus, we can keep track of the local GHZ states by tracking their stabilizer generators, and a GHZ state recovers on *all* physical sites in $V$ when the generators reproduce (S199) on the full system. Additionally, the evolution of the stabilizer generators is easily tracked in the context of Clifford circuits. We now discuss the key ingredients of the GHZ preparation protocol, corresponding to two distinct "stages".

The first stage of GHZ state preparation involves unitarily expanding the weight of numerous local GHZ states (S198) using CNOT gates (S39). Consider a $1D$ qubit chain in which the local state $|\text{GHZ}\rangle_{i,j} \otimes |0\rangle_{j+1}$ is realized on the qubits $i$ through $j+1$ (where qubit $j+1$ is not yet part of the GHZ cluster $i \ldots j$). We then produce the state $|\text{GHZ}\rangle_{i,j+1}$ by applying a CNOT gate (S39) to qubits $j$ and $j+1$ with $j$ the "control" qubit (although any qubit

---

[22] There are even multipartite entangled states with $\xi = 0$; see [S39] for an example.

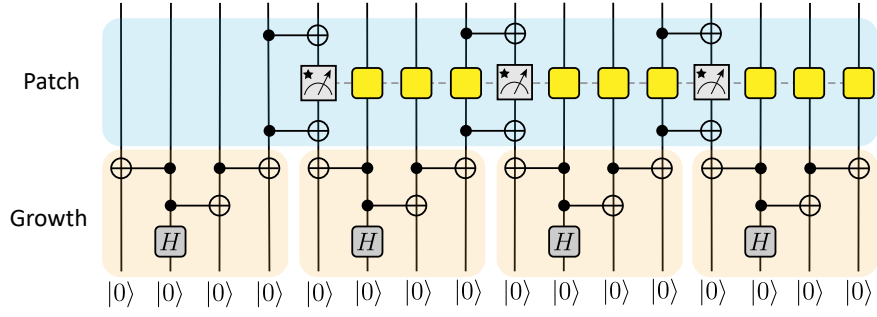

FIG. S6. Circuit diagram for the protocol $\mathcal{W}$ that generates the GHZ state (S166) on a $1D$ chain of $N = 16$ qubits from the state $|\mathbf{0}\rangle$ using $T = 4$ and $M = 3$. During the first "growth" stage (the four orange-shaded regions), *local* GHZ states are seeded and grown within their respective Lieb-Robinson light cones. During the second "patch" stage (the blue-shaded region), the individual GHZ states are "patched" into a single GHZ state of all $N$ qubits. The boxes with starred pointer dials denote a local $Z$ measurement followed by a reset of the measured qubit to the $|0\rangle$ state. The yellow boxes denote error-correction channels, which act as $X$ if the outcome of the measurement in the same patch was 1, and as $\mathbb{1}$ otherwise.

$k \in [i,j]$ will do) and $j + 1$ the "target" qubit. Clearly, this joins the state $|0\rangle_{j+1}$ with the $|0\rangle_i \otimes \cdots \otimes |0\rangle_j$ part of the GHZ state (S198), and joins the state $X_{j+1}|0\rangle_{j+1} = |1\rangle_{j+1}$ with the $|1\rangle_i \otimes \cdots \otimes |1\rangle_j$ part of the GHZ state (as $X_{j+1}$ is applied only if the control qubit $j$ in the cluster $i \ldots j$ is in the state $|1\rangle_j$).

Prior to applying the CNOT gate (S39), the state of the qubits $i \ldots j + 1$ is stabilized by all stabilizers of the GHZ state $|\text{GHZ}\rangle_{i,j}$, along with $Z_{j+1}$ for the additional qubit at $j+1$. The final state after applying the CNOT gate is the GHZ state on qubits $i, \ldots, j+1$ (S198); the $\ell + 1$ stabilizer generators are given by (S199). This can be seen directly by evolving the initial stabilizer generators in time, which gives

$$\bar{X}_{i,j} \longrightarrow \bar{X}_{i,j+1} \equiv X_i \cdots X_j X_{j+1}, \quad Z_{j+1} \longrightarrow Z_j Z_{j+1}, \quad \text{and} \quad Z_k Z_{k+1} \longrightarrow Z_k Z_{k+1} \ \forall k \in [i, j-1], \qquad \text{(S200)}$$

under the "growth" update. Combining (S200) with the original stabilizer generators (S199) for the $\ell$-site cluster $i \ldots j$ gives the stabilizer group for the $(\ell + 1)$-site cluster $i \ldots j + 1$ (S199) (i.e., the $\ell + 1$ stabilizer generators take the form $\{Z_i Z_{i+1}, Z_{i+1} Z_{i+2}, \ldots, Z_{j-1} Z_j, Z_j Z_{j+1}\} \cup \{\bar{X}_{i,j+1} = X_i X_{i+1} \cdots X_j X_{j+1}\}$). For a $1D$ qubit chain, the cluster $i \ldots j$ supporting the local GHZ state $|\text{GHZ}\rangle_{i,j}$ (S198) is generally *bounded*, which allows for the cluster to be grown by two sites in each time step of the "growth" stage of the Clifford circuit. Applying CNOT gates (S39) to both boundaries of the cluster—with sites $i$ and $j$ the target qubits and sites $i - 1$ and $j + 1$ the control qubits—grows the GHZ cluster by two qubits each time step (one on each side of the cluster).

The second stage involves "patching" two local GHZ states $|\text{GHZ}\rangle_{i,j}$ and $|\text{GHZ}\rangle_{j+1,k}$ (S198) on neighboring clusters $i \ldots j$ and $j+1 \ldots k$ into a single GHZ state on the combined cluster $i \ldots k$. Starting from the state $|\text{GHZ}\rangle_{i,j} \otimes |\text{GHZ}\rangle_{j+1,k}$ for the two clusters, we first apply a CNOT gate (S39) to qubits $j$ and $j + 1$—with $j$ the control qubit. We then measure $Z_{j+1}$ and apply an error-correction unitary $\mathcal{R}$ and another CNOT gate to produce the state $|\text{GHZ}\rangle_{i,k}$. Following the CNOT gate, the stabilizer generators are transformed according to

$$\bar{X}_{i,j} \to \bar{X}_{i,j+1} = X_i X_{i+1} \cdots X_j X_{j+1}, \quad Z_{j+1} Z_{j+2} \to Z_j Z_{j+1} Z_{j+2}, \qquad \text{(S201)}$$

with all other stabilizer generators (i.e., $\bar{X}_{j+1,k}$ and $Z_n Z_{n+1}$ for $n \in \{i, i+1, \ldots, j-1, j, j+2, j+3, \ldots, k-1, k\}$) unchanged. Measurement of $Z_{j+1}$ is accomplished by the dilated unitary operator

$$\mathsf{M}_{j+1} \equiv \frac{1}{2}\left(\mathbb{1} + Z_{j+1}\right) \widetilde{\mathbb{1}}_{\text{ss},j+1} + \frac{1}{2}\left(\mathbb{1} - Z_{j+1}\right) \widetilde{X}_{\text{ss},j+1}, \qquad \text{(S202)}$$

where the Stinespring qubit labelled $j + 1$ is initialized in the state $|0\rangle$: If the measurement outcome is "0", we apply $\widetilde{\mathbb{1}}_{\text{ss},j+1}$; if the outcome is "1", we apply $\widetilde{X}_{\text{ss},j+1}$. The stabilizer generators that do not act on qubit $j + 1$ are unaffected by the measurement. Both of the $\bar{X}$-like generators $\bar{X}_{i,j+1}$ and $\bar{X}_{j+1,k}$ are annihilated for either measurement outcome and on average; however, their product $\bar{X}_{i,k\backslash j+1} = \bar{X}_{i,j+1} \bar{X}_{j+1,k}$ is unaffected by the measurement. The operator $Z_j Z_{j+1} Z_{j+2}$ acts as $Z_j |0\rangle\langle 0|_{j+1} Z_{j+2}$ or $-Z_j |1\rangle\langle 1|_{j+1} Z_{j+2}$, for measurement outcomes 0 and 1, respectively.

Following this measurement, several operations are necessary to ensure that the final state is $|\text{GHZ}\rangle_{i,k}$ (S198). These operations are collected in the error-correction unitary

$$\mathcal{R}_{j+1} \equiv \widetilde{P}^{(0)}_{\text{ss},j+1} \otimes \mathbb{1} + \widetilde{P}^{(1)}_{\text{ss},j+1} \otimes \prod_{n=j+1}^{k} X_n, \qquad \text{(S203)}$$

so that the $Z_j Z_{j+2}$-like stabilizer for either outcome is simply $Z_j Z_{j+2}$.

Acting on the initial state $|\mathbf{0}\rangle$, if outcome "0" recovers we have

$$Z_j P_{j+1}^{(0)} Z_{j+2} \to Z_j Z_{j+2} \to Z_j, \tag{S204}$$

while if outcome "1" recovers, we have

$$- Z_j P_{j+1}^{(1)} Z_{j+2} \to -Z_j \left( X_{j+1} P_{j+1}^{(1)} X_{j+1} \right) \left( X_{j+2} Z_{j+2} X_{j+2} \right) = Z_j P_{j+1}^{(0)} Z_{j+2} \to Z_j Z_{j+2}, \tag{S205}$$

as expected. Under the combination of measurement (S202) and error correction (S203), we have

$$Z_n Z_{n+1} \to Z_n Z_{n+1} P_{j+1}^{(0)} \widetilde{\mathbb{1}}_{j+1} = Z_n Z_{n+1} \tag{S206a}$$

$$Z_{j+1} Z_{j+2} \to Z_j P_{j+1}^{(0)} Z_{j+2} \widetilde{\mathbb{1}}_{j+1} = Z_j Z_{j+2} \tag{S206b}$$

$$\bar{X}'_{i,k} \to X_i \cdots X_j P_{j+1}^{(0)} \widetilde{\mathbb{1}}_{j+1} X_{j+2} \cdots X_k = \bar{X}'_{i,k}, \tag{S206c}$$

and we note that an additional stabilizer generator is required, since we lost one of the two $\bar{X}$-type generators when we applied (S202). The missing stabilizer generator is given simply by $Z_{j+1}$.

Hence, the state at this point is a GHZ state (S198) on the qubits $i,\ldots,j,j+2,\ldots,k = [i,k] \setminus \{j+1\}$. The associated stabilizer generators are simply $Z_n Z_{n+1}$ (for $n \in \{i, i+1, \ldots, k-1\} \setminus \{j\}$), $Z_j Z_{j+2}$, $\bar{X}_{\{i,\ldots,k\}\setminus\{j+1\}}$, and $Z_{j+1}$ (for the removed site $j+1$). Finally, applying a CNOT gate (S39) to the qubits $j$ and $j+1$ (with $j$ the control qubit and $j+1$ the target qubit), we again "grow" the GHZ state to absorb the missing site $j+1$ as described above (S200). Finally, the GHZ state is realized on all qubits $i,\ldots,k$, and the stabilizers are given by (S199). This "patching" subroutine thus requires $T = 2$ layers of Clifford gates and $M = 1$ measurement regions.

An optimal protocol for generating the GHZ state (S166) is then specified by the following prescription:

1. Initialize $M+1$ local two-site $|\text{GHZ}\rangle$ states (S198) in equally spaced regions of $\ell = 2(T'+1)$ qubits. The two-site $|\text{GHZ}\rangle$ is equivalent to the Bell state $|\text{Bell}\rangle = 2^{-1/2}(|00\rangle + |11\rangle)$, and can be produce using Bell encoding (a Hadamard gate on the left qubit $n$ followed by a CNOT gate, with $n$ the control and $n+1$ the target qubits).

2. Apply the "growth" stage to expand each GHZ state by two qubits per time step. There are $T'$ growth steps in total, after which each GHZ state involves all $\ell = 2T'+2$ qubits in the region in which it was spawned.

3. Apply the "patch" stage to each neighboring pair of $\ell$-site GHZ states in any order (parallelizing where possible). After the final CNOT gate, the GHZ state (S166) obtains on the full system. When applying multiple patches in parallel, we may find that the string of $X_n$ operators prescribed in (S203) is cancelled in certain patches but not others, depending where the outcome "1" is recorded upon measuring $Z_n$.

The 1D protocol is depicted for $N = 16$, $T = 4$, $M = 3$ in Fig. S6. The three steps enumerated in the recipe above involve a total depth of $T = T' + 2$ layers of nearest-neighbor Clifford gates. We note that it is possible to apply two CNOT gates in *each* layer of the growth stage, so that the first layer of CNOT gates creates three-qubit GHZ states in each region, so that $\ell(T) = 2(T'+2) = 2T$. The task distance obeys (S177) with

$$L = N - 1 = 2(M+1)T - 1, \tag{S207}$$

for $M \geq 0$ and $T \geq 3$. This saturates the bound (S177), indicating that the protocol is optimal. This also implies that the bound (S177) as derived from Theorem 11 is also optimal with respect to preparing the GHZ state (S166).

Finally, we briefly sketch the protocol for preparing the GHZ state (S166) on graphs $G$ with spatial dimension $D > 1$, referring to the recipe enumerated above. For convenience, we restrict to the scenario where the vertex set $V$ realizes a [hyper]cubic lattice[23]. Rather than line segments, the system is then partitioned into $M+1$ $D$-dimensional "blocks" (e.g., containing $\ell^D \sim (2 + 2T')^D$ qubits apiece). In order to patch these regions together, we must identify an ordering of the patches (the need for this ordering is most clear when considering the stabilizer generators).

Starting from the initial state $|\mathbf{0}\rangle$ on all qubits, the first step is to seed single-site GHZ states in the center of each $D$-dimensional block using a Hadamard gate $H_j$ (where $H_j|0\rangle = (|0\rangle + |1\rangle)/\sqrt{2}$). We then apply CNOT gates in parallel, with the seed site $j$ the control qubit and all sites that neighbor $j$ the target qubits (note that these

---

[23] We use the term "hypercube" to refer to generalizations of the square and cube to dimensions $D > 3$; the hypercube is distinguished by being equal in spatial extent along all $D$ Cartesian axes.

operations commute). In $1D$, this incorporates two new qubits on either side; in $2D$ this incorporates four new qubits in a + shape surrounding vertex $j$; in $D \geq 3$ this incorporates $2 \cdot D$ new qubits. On the $2D$ square lattice, this creates diamond-shape patterns of $\ell(t) = 2t^2 - 2t + 1$ qubits that participate in a local GHZ state. Note that smaller regions may be required at the system's corners and edges to "fill in" the remaining qubits. Generally speaking, this growth protocol is most efficient, and generalizes straightforwardly to higher dimension.

All local blocks are expanded in size until their interfaces meet, at which point the graph $G$ is tiled with local GHZ states (S166). The ordering assigned to the local regions dictates the application of CNOTs and measurements in the "patch" stage (in parallel), which is largely unaffected by the extension to $D > 1$. The main complication is the string of $X_n$ operators in (S203), which we treat in analogy to Jordan-Wigner strings that arise in mapping bosonic qubit operators to Dirac fermion operators. Generally speaking, a classical computer implements this "decoding" stage, and will determine the appropriate choice of $\mathcal{R}$ (S203). Generally speaking, there are two (likely equivalent) options for extending (S203) to higher dimensions: The first option is that we modify the $X_n$ string in (S203) to be contained to the block to which the qubit $j+1$ belongs, so that large strings on other blocks are unnecessary; the second option is that we specify an ordering of sites in the $D$ lattice so that a $1D$ chain is uniquely specified—the Jordan-Wigner-like string in (S203) then snakes its way along this $1D$ chain. In practice, it is only necessary to introduce $X_n$ operators between regions in which the outcome "1" obtains upon measuring $Z_{j+1}$. Essentially, the measurements of $Z_n$ operators are "syndrome" measurements that uniquely specify where $X_{n'}$ operators need be applied.

Finally, after a GHZ state (S198) has been realized on each local block (after $T'$ layers of "growth"), we apply the "patch" stage (in parallel). The patch stage of the protocol is largely unaffected by the extension to $D > 1$. The main complication is the string of $X_n$ operators in (S203), which we treat in analogy to Jordan-Wigner strings that arise in mapping bosonic qubit operators to Dirac fermion operators. Generally speaking, a classical computer implements this "decoding" stage, and will determine the appropriate choice of $\mathcal{R}$ (S203). Generally speaking, there are two (likely equivalent) options for extending (S203) to higher dimensions: The first option is that we modify the $X_n$ string in (S203) to be contained to the block to which the qubit $j+1$ belongs, so that large strings on other blocks are unnecessary; the second option is that we specify an ordering of sites in the $D$ lattice so that a $1D$ chain is uniquely specified—the Jordan-Wigner-like string in (S203) then snakes its way along this $1D$ chain. In practice, it is only necessary to introduce $X_n$ operators between regions in which the outcome "1" obtains upon measuring $Z_{j+1}$. Essentially, the measurements of $Z_n$ operators are "syndrome" measurements that uniquely specify where $X_{n'}$ operators need be applied. This protocol continues to saturate the bound (S177) in higher dimensions.

### 7.3. Speed of preparing Dicke and W states

The GHZ state (S166) considered in Sec. 7.1 is invariant under permutations of the individual spins. That is to say, the GHZ state is invariant under arbitrary one-to-one relabelings of the qubits. Such permutation-symmetric states constitute the *Dicke manifold* of a system of physical qubits assigned to the vertices $V$ of the graph $G$; the Dicke manifold hosts other highly entangled states, as well. Of particular importance are the "Dicke states" [S9]

$$|D_k\rangle = \binom{|V|}{k}^{-1/2} \sum_{\substack{X \subset V \\ |X| = k}} |\mathbf{1}\rangle_X \otimes |\mathbf{0}\rangle_{V \setminus X} \quad \text{for} \quad 1 \leq k < |V|, \tag{S208}$$

which have been shown to be useful resource states for various tasks related to quantum information, sensing, and metrology [S41–S44]. The $k = 1$ Dicke state $|D_1\rangle$ is also known as the "W state" $|W_V\rangle$ (S171) for the vertex set $V$,

$$|W_V\rangle \equiv \frac{1}{\sqrt{|V|}} \left( \sum_{u \in V} X_u \right) |\mathbf{0}\rangle = |D_1\rangle, \tag{S209}$$

where $|\mathbf{0}\rangle$ initializes all qubits in the vertex set $V$ in the state $|0\rangle$. The W state on $V$ (S209) is an equal-weight superposition over all states in which a single qubit is in the state $|1\rangle$, with all other qubits in the state $|0\rangle$.

The time $T$ required to prepare entangled states (S208) in the Dicke manifold is bounded by Theorem 11, which we now restate in this particular context. For simplicity, we assume that the spins live on the vertices $V$ of a $D$-dimensional graph $G$, with $|V| = C_D L^D$ for some $\mathsf{O}(1)$ constant $C_D$, with $L$ the maximal distance between vertices (S178).

**Corollary 12.** *Consider a dilated quantum channel $\mathcal{W}$ involving (a) $M$ total measurement regions (S94) with measurement range $l$ (S95) and (b) Hamiltonian dynamics (S86) for total duration $T$, with associated velocity $v_E$ (S92). The channel $\mathcal{W}$ produces some final state $|\Psi\rangle$ acting on a short-range-entangled initial state $\rho_0 = \mathcal{W}_0 \rho' \mathcal{W}_0^\dagger$ (S159), which itself is prepared from a product state $\rho'$ using measurements in $M_0$ regions and Hamiltonian evolution*

for duration $T_0$. Then, for any $\epsilon \in (0, 1)$, there exists an $\mathsf{O}(1)$ constant $C_\epsilon$ such that, for sufficiently large $L$ (S178), the final state $|\Psi\rangle$ cannot be the Dicke state $|D_k\rangle$ (S208) for any $k \in \{1, 2, \ldots, |V| - 1\}$, unless

$$2\left(M + M_0 + 1\right)\left(v_E\left(T + T_0\right) + (3D - 1)\log_2 L + C_\epsilon\right) \geq (1 - \epsilon) L, \tag{S210}$$

where the factor $(3D - 1)$ in front of the log term is replaced by $2D$ for *prefixed* measurement locations. Note that the $\mathsf{O}(1)$ constant $C_\epsilon$ does not scale with $L$, $T$, or $M$, but is otherwise arbitrary.

*Proof.* We begin by identifying the qubits $i$ and $f$ as the most-separated pair of vertices in $V$, such that $L = d(i, f)$ (this choice need not be unique). We then note that the correlation $\mathrm{Cor}(i, f)$ (S179) between qubits $i$ and $f$ is nonnegligible (i.e., finite) when evaluated in the Dicke state $|D_k\rangle$. This allows us to apply Theorem 11 for the Dicke state $|D_k\rangle$. Following the proof of Proposition 10, we choose $\mathcal{O}_i = Z_i$, $\mathcal{O}_f = Z_f$, and then compute

$$\begin{aligned}\langle Z_i \rangle = \langle Z_f \rangle &= \binom{|V|}{k}^{-1}\left[\binom{|V| - 1}{k} - \binom{|V| - 1}{k - 1}\right] = 1 - \frac{2k}{|V|} \\ \langle Z_i Z_f \rangle &= \binom{|V|}{k}^{-1}\left[\binom{|V| - 2}{k} + \binom{|V| - 2}{k - 2} - 2\binom{|V| - 2}{k - 1}\right] = 1 - \frac{4k\left(|V| - k\right)}{|V|\left(|V| - 1\right)},\end{aligned} \tag{S211}$$

so that the correlator $\mathrm{Cor}(i, f)$ (S179) obeys

$$\mathrm{Cor}(i, f) \geq |\langle Z_i Z_f \rangle - \langle Z_i \rangle\langle Z_f \rangle| = \frac{4k\left(|V| - k\right)}{|V|^2\left(|V| - 1\right)} \geq \frac{4}{|V|^2} = \left(\frac{2}{C_D L^D}\right)^2. \tag{S212}$$

However, we know that the final state $|\Psi\rangle$ must obey (S188). Therefore, $|\Psi\rangle$ is a Dicke state $|D_k\rangle$ (S208) only if

$$4\frac{c}{\epsilon} L^{D-1} 2^{v_E (T + T_0) + \frac{l}{2} - \frac{(1-\epsilon)L + 1 - l}{2(M + M_0 + 1)}} \geq \left(\frac{2}{C_D L^D}\right)^2, \tag{S213}$$

for adaptive measurement locations, which reduces to (S210) for sufficiently large $L$. The case of prefixed measurement locations follows similarly: One simply replaces the left hand side of (S213) by the first line of (S188), so that the prefactor of $\log L$ in (S210) becomes $2D$ just from the right hand side of (S213). □

The bound (S210) also constrains channels $\mathcal{W}$ that *approximately* produce a Dicke state $|D_k\rangle$ (S208), as long as the final state $|\Psi\rangle$ is sufficiently close to $|D_k\rangle$ that $\mathrm{Cor}(I, F)$ is finite. In the absence of measurements ($M = 0$), (S210) reduces to the usual Lieb-Robinson bound (S90) for generating correlations [S45]; in general, we expect that (S210) is optimal up to factors of $\log_2 L$, which may be an artifact of the proof strategy.

In the particular case of the W state (S171)—and for Dicke states with finite $k$—it is indeed possible to remove the $\log_2 L$ term in (S210) for *prefixed* measurement locations, in which case the inequality

$$3\left(M + M_0 + 1\right)\left[v_E\left(T + T_0\right) + C_\epsilon\right] \geq L, \tag{S214}$$

replaces the fully generic result (S210). This requires a slight generalization of Theorem 11 in which we bound $\overline{\mathrm{Cor}}(I, F)$ for two *regions* $I$ and $F$ instead of two *sites*, because $\overline{\mathrm{Cor}}(i, f)$ in a Dicke state is algebraically small in $d(i, f)$ (S212). However, this generalization follows the strategy for $I = \{i\}$ and $F = \{f\}$ straightforwardly. The resulting bound is the same as (S188) up to a numerical prefactor, as long as $I$ and $F$ do not face each other with large plane-like surfaces. For example, it is fine if the sets $I$ and $F$ correspond to two conical regions whose apexes (pointy parts) face each other, but it is problematic if the bases of the two cones face one another. This condition ensures that the $l$ factor in (S130) does not blow up with the size of the regions like $\ell \sim |I|$, since the initial and final sites $i$ and $f$ (or sets of sites)) are included as measurement regions via (S94).

In $D$ spatial dimensions (with $D$ finite), one can always choose two such regions $I$ and $F$, each containing $c_W |V|$ qubits (where $1 > c_W > 0$ is a constant), such that the distance $d(I, F)$ between the regions $I$ and $F$ is larger than $2L/3$. Then, as long as the correlation function $\mathrm{Cor}(I, F)$ between $I$ and $F$ in the target state is finite, (S214) follows immediately from the generalization of (S188) described above. The correlation between two parties $A, B$ with $|A| = |B| = c_W N \equiv c_W |V|$ is indeed finite—i.e., $\Omega(1)$—for the W state (S171), since

$$|W_V\rangle = \sqrt{c_W}\left(|W_A\rangle \otimes |\mathbf{0}\rangle_B + |\mathbf{0}\rangle_A \otimes |W_B\rangle\right) \otimes |\mathbf{0}\rangle_C + \sqrt{1 - 2c_W}|\mathbf{0}\rangle_{AB} \otimes |W_C\rangle, \tag{S215}$$

where $C = (A \cup B)^c$, so that $\mathcal{O}_A = |\mathbf{0}\rangle_A\langle\mathbf{0}| - |W_A\rangle\langle W_A|$ (and likewise for $\mathcal{O}_B$) leads to $\mathrm{Cor}(A, B) \geq (1 - 2c_W)^2 - (1 - 4c_W) = 4c_W^2$. It is an open question whether all Dicke states (with $k > 1$) have this property of finite correlations; any Dicke state with finite correlations also obeys the asymptotically tighter bound (S214).

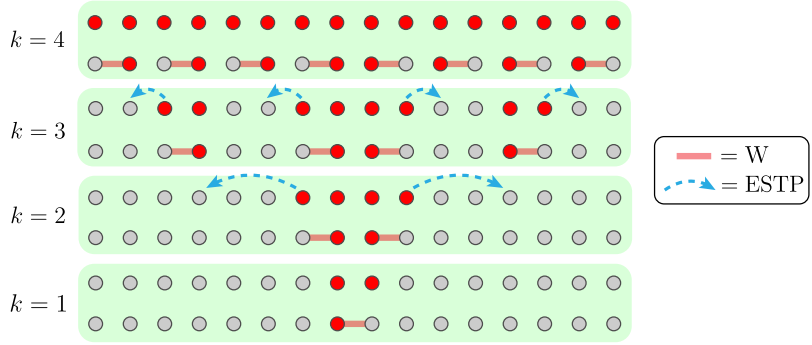

FIG. S7. A protocol that generates a W state is depicted for a $1D$ chain of $N = 16$ qubits. The gray circles denote qubits still in the initial $|0\rangle$ state; the red circles denote the set of qubits $A \subset V$ that realize a collective W state $|W_A\rangle$ (S209). That state is equivalent to the $k = 1$ Dicke state $|D_1\rangle$ (S208) on $A$.

We now motivate the bound (S214)—and the expectation of optimality up to corrections of $\mathsf{o}(L)$[24]—by presenting a simple, tree-like, W-state-generating protocol [S46] in $1D$ using nearest-neighbor gates.

### 7.4. Logarithmic-depth protocol for preparing the W state

We now present a protocol for preparing the W state $|W\rangle$ (S209) on $N$ qubits in depth $\mathsf{O}(\log N)$ using measurements. The W state $|W\rangle$ is equivalent to the $k = 1$ Dicke state $|D_1\rangle$ (S208). In particular, we generate the W state on a $1D$ chain of $N = 2^n$ qubits (where $n \in \mathbb{N}$ is some positive integer) starting from the product state $|\mathbf{0}\rangle$ using the protocol $\mathcal{W}$ involving nearest-neighbor non-Clifford gates and, optionally, measurements and feedback. The protocol $\mathcal{W}$ admits tradeoffs between unitary depth $T$ and the number of measurements $\mathcal{M}$ (and measurement regions $M$).

If we minimize measurements (so that $M = \mathcal{M} = 0$), we find that $T = N/2 + \log_2 N - 1$ (S218)—note that $T = \mathsf{O}(N)$ saturates the measurement-free Lieb-Robinson bound (S90). However, since the protocol prepares the W state (S209) starting from the center of the chain, one would expect $T = N/2$ to be optimal; in this sense, the additional depth $T \sim \log_2 N$ seems to capture an "extra" overhead involved in preparing the W state.

If we instead minimize depth $T$, the same protocol $\mathcal{W}$ prepares the W state (S209) on $N$ sites from the product state $|\mathbf{0}\rangle$ using total depth $T = 3 \log_2 N - 3$, a total of $\mathcal{M} = (N \log_2 N - 4N + 8)/2$ independent measurement *outcomes*, and $M = N/2 - 2 \log_2 N + 2$ distinct measurement *regions*. While the depth $T = \mathsf{O}(\log N)$ is faster than the linear-depth protocols known to the literature [S37, S47], the combined resources $2(M+1)vT = \mathsf{O}(N \log_2 N)$ are suboptimal relative the corresponding bounds (S208) and (S214). We conjecture that the protocol described herein—rather than the bounds (S208) and (S214)—are suboptimal with respect to preparing the W state. We also note that, in contrast to the GHZ state, which can be prepared using measurements in depth $T = \mathsf{O}(1)$, it appears that the W state always requires depth $T \gtrsim \log_2 N$ to prepare, even using measurements and outcome-dependent feedback. We relegate a proof of this conjecture—and the conjecture that the bounds (S208) and (S214) are optimal—to future work [S12].

We now consider the protocol $\mathcal{W}$ that generates $|W\rangle$ (S171) on a $1D$ chain of $N = 2^n$ qubits from the product state $|\mathbf{0}\rangle$. We comment that the W state is equivalent to the state $|1\rangle_{k=0} \otimes |\mathbf{0}\rangle_{k>0}$ with the $k = 0$ Fourier mode excited and all other Fourier modes in the 0 state—i.e., the quantum Fourier transform (QFT) [S11, S47] of the state $|1\rangle \otimes |\mathbf{0}\rangle$. The QFT protocols in the literature are purely unitary with depth $T \sim \mathsf{O}(N \log_2 N)$ [S47]. By comparison, the $M = 0$ limit of our protocol for $N = 2^n$ qubits requires depth $T = 2^{n-1} + n - 1 = (N + 2 \log_2 N - 2)/2 < N$, meaning that $T = \mathsf{o}(N)$ saturates the measurement-free Lieb-Robinson bound (S90) [S7] in Theorem 3.

In contrast to teleportation and the preparation of GHZ states, we have yet to identify a protocol that uses measurements to realize a *multiplicative* speedup (e.g., $v \to 2(M+1)v$). This may be related to the fact that the W state is not a stabilizer state and cannot be generated via Clifford gates alone [S12]. Additionally, because measurements destroy the W state (S171), there is no obvious hybrid protocol to replace (S216) or facilitate the patching of two neighboring W states into a single W state, as in the GHZ protocol presented in Sec. 7.2 (see Fig. S6).

The following non-Clifford unitary gate will prove useful in the construction of the W state (S171). The two-qubit

---

[24] Note that any power of $\log_2 y$ vanishes faster than $y$ as $y \to 0$, and therefore $(\log_2 y)^n \in \mathsf{o}(y)$ is subleading compared to $\mathsf{O}(y)$ terms.

W gate $W_{i,j}$ acts on the ordered pair of qubits $(i,j)$ as

$$W_{i,j} \equiv \frac{1}{\sqrt{2}} \begin{bmatrix} \sqrt{2} & 0 & 0 & 0 \\ 0 & 0 & 1 & -1 \\ 0 & 0 & 1 & 1 \\ 0 & \sqrt{2} & 0 & 0 \end{bmatrix} \begin{matrix} |0_i 0_j\rangle \\ |0_i 1_j\rangle \\ |1_i 0_j\rangle \\ |1_i 1_j\rangle \end{matrix} \rightarrow \quad , \tag{S216}$$

which consists of a controlled Hadamard (CH) gate followed by a CNOT gate. The first gate (CH) applies a single-qubit Hadamard gate (S38) to the target qubit $j$ if the control qubit $i$ (on the left) is in the state $|1\rangle$; the CNOT gate (S39) applies $X_i$ if the control qubit $j$ is in the state $|1\rangle$. The operator W (S216) is *not* a Clifford operation (see Sec. 4.1), and generically maps Pauli strings to *superpositions* of Pauli strings.

The protocol that generates the W state on $V$ (S209) uses the following recursive "subroutine" to extend the W state $|W_A\rangle$ on the set $A \subset V$ to the W state $|W_{A \cup B}\rangle$ on the set $A \cup B \subset V$,

$$|W_{A \cup B}\rangle = \left(\prod_{a=1}^{|A|} W_{a,b(a)}\right) |W_A\rangle_A \otimes |\mathbf{0}\rangle_B, \tag{S217}$$

provided that $|A| = |B|$, where $a$ indexes qubits in $A$, and $b(a)$ uniquely identifies the qubit $b(a) \in B$ that accompanies the qubit $a \in A$. This pairing is arbitrary—the only requirement is that each qubit appears only once in (S217).

Since the qubit pairs $a, b \in A, B$ do not overlap, the nonlocal gates $W_{i,j}$ (S216) can be applied in parallel. However, as detailed in Sec. 6.1, all of our continuous-time bounds presume nearest-neighbor interactions only (S86). In general, representing the subroutine (S217)—which grows the W state (S209) on $A \subset V$ to a W state on $A \cup B \subset V$—using nearest-neighbor gates only likely requires combining the gates $W_{i,j}$ (S216) with a state transfer protocol (e.g., the TFIM code presented in Sec. 4 or the ESTP protocol considered in the main text).

The W-state-preparation protocol involves $n$ "rounds" labelled $k$, as depicted in Fig. S7. Starting with $k = 1$, we seed the initial two-site W state (S209), which is equivalent to one of the four Bell states. In each subsequent step $2 \leq k \leq n$, we apply W gates (S216) as in (S217): The $2^{k-1}$ qubits already incorporated into the W state at the start of round $k$ are the "$a$" sites in (S217); their nearest neighbors (in the direction of the closest end of the chain) are the corresponding "$b(a)$" sites in (S217). This step is depicted in Fig. S7 via the red horizontal lines connecting the incorporated $a$ qubits in red to their corresponding $b(a)$ qubits still in gray. In the same round (for $2 \leq k \leq n-1$), we then move the newly incorporated $b(a)$ qubits away from the chain's center such that all $2^k$ incorporated qubits are equally spaced by a distance $L_k = 2^{n-k} - 1$ (corresponding to the dashed blue lines in Fig. S7).

Importantly, the W gates (S216) are all applied to nearest neighbors, and it is only during the separation of the qubits that a measurement-induced speedup is possible. Essentially, one can apply the ESTP of the main text—or the TFIM code of Sec. 4—to speed up the qubit teleportation in steps $2 \leq k \leq n-3$. In steps $k = n$ and $k = 1$, no teleportation is required; in step $k = n-1$, we instead use a single layer of SWAP gates. However, meaningful speedups can be achieved for all other rounds $2 \leq k \leq n-2$ using measurements.

While various tradeoffs between $M$ and $T$ are possible, we consider the two scenarios that minimize these respective quantities. The case with minimum $M = 0$ (and maximum depth $T$) eschews the ESTP for teleportation, and instead uses $L_k = 2^{n-k} - 1$ layers of SWAP gates for $2 \leq k \leq n-1$. The case with minimum $T$ (and maximum $M$), on the other hand, applies the ESTP for all teleportation steps $2 \leq k \leq n-2$, using two-qubit Bell measurements (of $X_j X_{j+1}$ and $Z_j Z_{j+1}$) instead of computational basis measurements (of $Z_j$ and $Z_{j+1}$). In each of the $2^{k-1}$ teleportation "regions," the protocol uses a single layer of two-qubit gates to form $(L_k - 1)/2 = 2^{n-k-1} - 1$ consecutive Bell pairs, each of which is measured in the Bell basis. Importantly, only the outermost two teleportation regions utilize new measurement regions, so each round $k \in [2, n-2]$ has depth $T_k = 3$, $M_k = L_k - 1 = 2^{n-k} - 2$ new measurement regions, and $\mathcal{M}_k = 2^{k-1} M_k = 2^{n-1} - 2^k$ total measurement *outcomes*.

We now provide a detailed explanation of the protocol $\mathcal{W}$ that prepares the W state (S171) on a $1D$ chain of $N = 2^n$ qubits starting from the state $|\mathbf{0}\rangle$. The protocol uses $n = \log_2(N)$ steps as depicted in Fig. S7. The qubits in gray are in the state $|0\rangle$, while the qubits in red have been incorporated into the W state. The first step $k = 1$ is special: We first seed a two-site W state (S209) on the central two qubits $j, j+1$ (where $j = 2^{n-1}$) using a modified Bell encoding channel that first applies $H_j \otimes X_{j+1}$—where $H$ is the Hadamard gate (S38)—followed by a CNOT gate (with $j$ the control qubit and $j+1$ the target qubit), with total cost $T = 1$. We next iterate the following subroutine for $2 \leq k \leq n$:

1. For round $k \in [2, n]$, the current state realizes a W state (S209) on the set $V_{k-1}$ (the $|V_{k-1}| = 2^{k-1}$ red qubits in Fig. S7); the other $2^{n+1-k}$ qubits remain in the state $|0\rangle$ (the gray qubits in Fig. S7). For each qubit $v \in V_{k-1}$, we apply the W gate W (S216) pairwise to the qubit $v$ and the nearest-neighboring qubit that is closer to the boundary of the system (farther from the central qubits). The central cluster of two qubits involved in the W

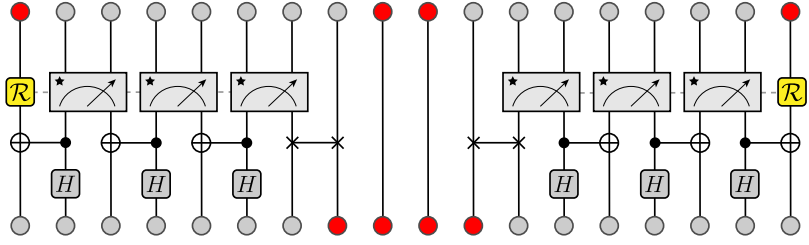

FIG. S8. Circuit diagram showing the second stage of round $k = n - 4$ of the W-state-preparation protocol depicted in Fig. S7 (gray circles denote qubits in the state $|0\rangle$, while red circles correspond to qubits in the W state). For round $k = n - 4$, the $2^{n-5}$ newly incorporated qubits (in red, including the outer two red qubits above) are teleported a distance $L_k = 2^{n-k} - 1 = 7$ sites using the ESTP of the main text with the smallest possible depth $T = 1$. This is possible by performing two-site measurements in the Bell basis (i.e., we measure of $X_j X_{j+1}$ and $Z_j Z_{j+1}$ instead of $Z_j$ and $Z_{j+1}$). The prescription for determining the error-correction channel $\mathcal{R}$ is identical. In the teleportation step $k \in [2, n-2]$, the $2^{k-1}$ newly incorporated qubits live at sites $i_s$ and must be moved by $L_k = 2^{n-k} - 1$ away from the center to site $f_s$. For convenience, we consider the right half of the chain. Teleportation is achieved by applying a SWAP gate (S40) to sites $i_s, i_s + 1$ and seeding $2^{n-k-1} - 1$ Bell pairs on the $2^{n-k} - 2$ qubits in the interval $[i_s + 2, f_s]$. We then perform Bell-basis measurements on neighboring qubits $i_s + 1, i_s + 2$ through $f_s - 2, f_s - 1$, and the outcomes determine the QEC channel $\mathcal{R} \in \{\mathbb{1}, X_{f_s}, Y_{f_s}, Z_{f_s}\}$, which concludes the teleportation.

state, e.g., always becomes a four-qubit cluster. All other qubits in $V_{k-1}$ adjoin the less central of their two neighboring qubits into the W state, realizing a W state on $V_k$ (S209). This step requires circuit depth $T = 2$.

2. For steps $k \in [2, n-1]$, following the W gate (S216), the $2^{k-1}$ qubits newly incorporated into $V_k$ (and thus, the W state) must be teleported a distance $L_k = 2^{n-k} - 1$ away from the central qubits (in each local "block"), so that all qubits that participate in the W state are equally spaced in each half of the chain (by distance $L_k$) going into step $k + 1$ (while the central pair remain fixed). Teleportation is not required in steps $k = 1$ and $k = n$, and there are two limiting scenarios to consider for all other rounds $k \in [2, n-1]$:

    (a) The first scenario is purely unitary (i.e., $M = 0$). The teleportation of the newly incorporated qubits a distance $L_k = 2^{n-k} - 1$ is achieved using $T_k = L_k$ layers of SWAP gates (S40). This scenario is optimal with respect to the usual Lieb-Robinson bound (S90) of Theorem 3 [S7] (and always optimal in round $k = n - 1$).

    (b) The second scenario uses the entanglement-swapping teleportation protocol (ESTP) of the main text to teleport qubits. For contrast, we minimize the depth $T$ at the cost of measuring many qubits. For reference, the teleportation of the newly generated qubits at the chain's center (sites $2^{n-1} - 1$ and $2^{n-1} + 2$) using the ESTP is depicted in Fig. S8 for step $k = n - 4$ so that $L_k = 7$. For convenience, consider the qubit at site $j = 2^{n-1} + 2$ (the rightmost red qubit at the bottom of Fig. S8), which must be teleported a distance $L_k = 2^{n-k} - 1$. This is achieved by applying a SWAP gate to sites $j, j + 1$ and seeding $2^{n-k-1} - 1$ Bell pairs on sites $j + 2$ through $j + 2^{n-k} - 1$ via Bell encoding $\mathcal{B}_{j,p} = \text{CNOT}(j + 2p \to j + 2p + 1) H_{j+2p}$ for $1 \leq p \leq 2^{n-k-1} - 1$ (where $p$ labels the Bell pairs). We then perform Bell measurements, so that $Z_A$ is replaced by the measurements of $X_{j+2p-1} X_{j+2p}$ and $Z_B$ is replaced by the measurements of $Z_{j+2p-1} Z_{j+2p}$ for $1 \leq p \leq 2^{n-k-1} - 1$ (compared to the main text). The error-correction channel $\mathcal{R}$ is determined exactly as in the main text; the application of $\mathcal{R}$ completes the teleportation.

After completing the final round $n = \log_2 N$, a W state is realized on all $N = 2^n$ qubits (S171), as shown in Fig. S7 for $N = 16$ ($n = 4$). We now consider the resource requirements for the two scenarios described above.

First, consider the measurement-free case ($M = 0$). Round $k = 1$ costs $T = 1$ while round $k = n$ costs $T = 2$. All other rounds require $T = 2$ layers for the "growth" portion and $T = L_k = 2^{n-k} - 1$ layers of SWAP gates to move the newly incorporated qubits. The total circuit depth is given by

$$T_{\rm W} = 1 + \sum_{k=2}^{n-1} \left(2 + 2^{n-k} - 1\right) + 2 = 2^{n-1} + n - 1 = \frac{1}{2}\left(N + 2\log_2 N - 2\right) < N, \tag{S218}$$

which is $\mathsf{O}(N)$, saturating the standard Lieb-Robinson bound $T \leq N$ (S90) from Theorem 3 [S7]. In the asymptotic limit $N \to \infty$, we have $T/N \to 1/2$, saturating the standard "light-cone" Lieb-Robinson bound (S90).

Alternatively, consider the scenario in which we minimize the circuit depth $T$ (S218), at the cost of measuring many qubits. This is achieved by using the ESTP for the teleportation stage in each round $k \in [2, n-2]$. In rounds $k = 1, n$ there is no teleportation, while in round $k = n - 1$, the ESTP is replaced by a single layer of SWAP gates (S40). Hence,

round $k = 1$ costs $T = 1$, round $n$ costs $T = 2$, round $n - 1$ costs $T = 3$, and all other rounds require depth $T \geq 3$, which depends on how many measurements are used in the parallel ESTP protocols.

We now work out the total resource costs in the case where $T$ is minimized, so that $T = 3$ for all rounds $k \in [2, n-1]$, and rounds $k \in [2, n-2]$ all involve measurements. The total contribution to the depth $T$ is captured by (S219a). Now consider the number of measurements $\mathcal{M}$ and number of measurement regions $M$, noting that there is no contribution from the rounds $k = 1, n-1, n$. We note that all $2^{k-1}$ teleportation protocols for step $k$ can be applied in parallel, with total depth $T = 1$. In round $k \in [2, n-2]$, there are $2^{k-1}$ "teleportation regions" following the doubling of the W state, corresponding to $2^{k-1}$ newly "activated" qubits that must be teleported a distance $L_k = 2^{n-k} - 1$. This is achieved in depth $T = 1$ by creating $M'_k = (L_k - 1)/2 = 2^{n-k-1} - 1$ Bell pairs in each teleportation region (along with a SWAP gate, as depicted in Fig. S8). Since there are $2^{k-1}$ teleportation regions, each with $M'_k$ Bell pairs, each of which is measured twice, the total number of *outcomes* in round $k$ is $\mathcal{M}_k = 2 \times 2^{k-1} \times (2^{n-k-1} - 1) = 2^{n-1} - 2^k$. However, all measurements of Bell pairs are made in measurements used in the prior round, excepting the outermost two regions in each round. Hence, the contribution to the number of measurement *regions* is $M_k = 2\, M'_k = 2^{n-k} - 2$. The cumulative resources required in the case of minimum depth are

$$T_{\rm W} = 1 + \sum_{k=2}^{n-1} 3 + 2 = 3n - 3 = 3\log_2 N - 3 \tag{S219a}$$

$$M_{\rm W} = \sum_{k=2}^{n-2} \left(2^{n-k} - 2\right) = 2^{n-1} - 2n + 2 = \frac{1}{2}\left(N - 4\log_2 N + 4\right) < N \tag{S219b}$$

$$\mathcal{M}_{\rm W} = \sum_{k=2}^{n-2} \left(2^{n-1} - 2^k\right) = n\, 2^{n-1} - 2^{n+1} + 4 = \frac{1}{2}\left(N \log_2 N - 4N + 8\right) < N \log_2 N, \tag{S219c}$$

where $T$, $M$, and $\mathcal{M}$ are respectively the total circuit depth, number of unique measurement regions, and number of measurement outcomes utilized in the protocol that prepares the W state (S171) on $N = 2^n$ qubits. Importantly, we note that the number of measurements utilized above is maximal: The number of outcomes is $\mathsf{O}(N \log N)$ (S219c), but the number of measurement *regions* (which enters a given bound) is roughly $\mathsf{O}(N)$, with $M_{\rm W} < N$ (S219b). Most striking is the total circuit depth $T_{\rm W} = \mathsf{O}(\log N)$ (S219a).

We comment that any tree-like all-to-all protocol can be implemented on $1D$ nearest-neighbor architectures with similar overhead using the routing scheme described above [S46]. Using a similar technique with CNOT gates, we can create $|{\rm GHZ}_N\rangle$ with $T = \mathsf{O}(\log N)$ and $M = \mathsf{O}(N)$. However, this tree-like GHZ protocol is not the most efficient use of resources: Recall Sec. 7.2 where we identified a protocol that prepares the GHZ state on $N$ qubits (S166) using $T = \mathsf{O}(1)$, $M = \mathsf{O}(N)$. Although we have not identified a similar measurement-assisted protocol for preparing the W state (S171) in a manner that saturates the bound (S210), we cannot rule out the existence of such a protocol. Identifying a protocol that saturates the bound (S210) would confirm the optimality of that bound. Alternatively, deriving a tighter bound than (S210) that *is* saturated by the aforementioned protocol could prove that the protocol above is optimal. Absent an obvious means to demonstrate either case, we defer such refinements to future work. However, if the resource costs are counted as $MT$, then minimizing the resource requirements is accomplished by a purely unitary protocol (i.e., $M = 0$), which saturates (S90) from Theorem 3 [S7].

### 7.5. Speed of preparing spin-squeezed states

Another class of multipartite entangled states corresponds to "spin-squeezed" states [S13]. We define the *collective* spin operators (analogous to total angular momentum),

$$J_\alpha = \frac{1}{2} \sum_{u \in V} \sigma_u^\alpha, \tag{S220}$$

where $\sigma_u^\alpha$ is the Pauli matrix corresponding to spin of qubit $u$ along the axis $\alpha \in \{x, y, z\}$. These collective spin operators (S220) obey the Heisenberg uncertainty relation

$$\Delta J_\alpha \, \Delta J_\beta \geq \frac{1}{2} |\langle [J_\alpha, J_\beta] \rangle| = \frac{1}{2} |\langle J_\gamma \rangle|, \tag{S221}$$

for the *variances* $(\Delta J_\alpha)^2 = \langle J_\alpha^2 \rangle - \langle J_\alpha \rangle^2$, where the labels $\alpha, \beta, \gamma$ are all distinct (i.e., $\{\alpha, \beta, \gamma\}$ is some permutation of $\{x, y, z\}$). For example, the state $|\mathbf{0}\rangle$—in which all spins are oriented along the positive $z$ axis in the state $|0\rangle = |\uparrow_z\rangle$—has equal variance along the perpendicular directions $\Delta J_x = \Delta J_y = \Delta J_\perp$, where $\Delta J_\perp$ saturates (S221).

We now consider a spin-squeezed state [S13]. Without loss of generality, we choose a coordinate frame such that $\langle \vec{J} \rangle = \langle J_z \rangle \equiv J$ (i.e., the expectation values along the perpendicular axes $x, y \perp z$ is zero). The Heisenberg relation (S221) then dictates that $\Delta J_x \Delta J_y \geq \langle J_z \rangle/2 = |J|/2$. Then, for example, to ensure that the variance $\Delta J_x$ in the $x$ direction is small, the variance $\Delta J_y$ in the $y$ direction must be *large*, to compensate. We quantify this tradeoff using the spin-squeezing parameter [S13],

$$\xi^2 \equiv |V| \min_{\alpha \perp \langle \vec{J} \rangle} \frac{(\Delta J_\alpha)^2}{\langle \vec{J} \rangle^2}, \tag{S222}$$

where $\alpha$ runs over coordinate axes perpendicular to $\langle \vec{J} \rangle$. If the expectation value $\langle \vec{J} \rangle$ of the collective spin components (S220) points along the $z$ axis, then $\alpha \in \{x, y\}$. We note that spin-squeezed states with $\xi^2 < 1$ are useful for quantum sensing [S48, S49] and probing entanglement [S50].

The preparation of any permutation-symmetric spin-squeezed state in which the spin-squeezing parameter (S222) obeys $1 - \xi^2 = \mathsf{O}(1)$ is bounded by a constraint of the form (S210) on the time and measurements needed. This can be verified using a similar strategy as was used in the proof of Corollary 12; the reason is that the correlation function $\mathrm{Cor}(i, f)$ (S179) between any two sites $i$ and $f$ is related to the squeezing parameter [S13] by

$$\mathrm{Cor}(i, f) \geq - \min_{\alpha \perp \langle \vec{J} \rangle} \left( \langle \sigma_i^\alpha \sigma_f^\alpha \rangle - \langle \sigma_i^\alpha \rangle \langle \sigma_f^\alpha \rangle \right) \geq \frac{1 - \xi^2}{|V| - 1}, \tag{S223}$$

and if the squeezing is parametrically strong (i.e., $\xi^2 = |V|^{-\nu}$ for some $0 < \nu \leq 1$), then a slight generalization of Theorem 11 bounds the preparation of such states, even if they are not permutation symmetric. The key is that strong squeezing, together with the uncertainty relation (S221), imply large two-qubit correlations, on average:

$$\sum_{u,v \in V} \mathrm{Cor}(u, v) \geq \sum_{u,v \in V} \langle \sigma_u^\beta \sigma_v^\beta \rangle - \langle \sigma_u^\beta \rangle \langle \sigma_v^\beta \rangle = 4 \left( \Delta J_\beta \right)^2$$

$$\geq \frac{\langle \vec{J} \rangle^2}{(\Delta J_\alpha)^2} = \frac{|V|}{\xi^2} = |V|^{1+\nu}, \tag{S224}$$

where $\alpha$ and $\beta$ are orthogonal to one another and to $\langle \vec{J} \rangle$, where $\alpha$ corresponds to the axis that minimizes the righthand side of (S222). We use (S224), in proving the following Corollary on the speed of preparing strongly spin-squeezed states, for which $\xi^2 = |V|^{-\nu}$ (S222) for some $0 < \nu \leq 1$.

**Corollary 13.** *Consider a quantum channel $\mathcal{W}$ involving (a) local measurements in $M$ total regions (S94) and (b) Hamiltonian dynamics (S86) for total temporal duration $T$. Suppose that the dilated channel $\mathcal{W}$ acts on $|V|$ qubits, which we identify with the vertices $v \in V$ of the $D$-dimensional graph $G$; and suppose that $\mathcal{W}$ produces some desired final state $|\Psi\rangle$, starting from the initial product state $|\mathbf{0}\rangle$. Then, there exists a positive constant $c' = \mathsf{O}(1)$ such that, if $|\Psi\rangle$ is a spin-squeezed state with $\xi^2 = |V|^{-\nu}$ for some $\nu \in (0, 1]$ (S222), and the relation*

$$T^D < c' |V|^\nu, \tag{S225}$$

*holds, then we have*

$$M T^D \geq \Omega \left( |V|^{\frac{1+\nu}{2}} \right), \tag{S226}$$

*where the "big $\Omega$" function $\Omega(y)$ denotes a quantity that is at least as large as its argument $y$.*

*Proof.* Instead of using Theorem 11 directly, we instead revisit the strategy invoked in its proof. That strategy relies on the notion of locality in the enlarged graph $G_{\mathrm{dil}}$ articulated below (S125) and depicted in Fig. S5(a). Suppose that some vertex $u \in V$ is far from all measurement regions (i.e., $d(u, S_m) \gg vT$, $\forall m$). Then the support of some local operator $\mathcal{O}_u$ (initially supported only on $u$), when evolved backwards by time $T$ in the Heisenberg-Stinespring picture, is effectively constrained to the ball $B_r(u) = \{ v \in V \mid d(u,v) \leq r \}$ of radius $r = \eta vT$ centered about the vertex $u$, where $\eta$ is constant that is independent of $|V|, T, M$, but large enough to guarantee (S227).

The support of the Heisenberg-evolved operator $\mathcal{O}_u(T)$ remains far from all measurement regions in $\mathbb{M}^n$, which effectively ensures that measurements cannot be used to teleport the operator out of $B_r(u)$. In the language of building up correlations, this implies that the qubit $u$ cannot become highly correlated with spins outside of the ball $B_r(u)$. By tracking the proof of Theorem 11 in this slightly modified context, one finds that

$$\sum_{v \notin B_r(u)} \mathrm{Cor}(u, v) \leq 1, \tag{S227}$$

for a sufficiently large $\eta$ independent of $|V|, T, M$. As a result, the site $u$ can only contribute

$$\sum_{v \in V} \mathrm{Cor}(u,v) \leq |B_r(u)| + 1 = \mathsf{O}(T^D)\,, \tag{S228}$$

to the total correlation on the lefthand side of (S224). We then define $S_{\mathrm{far}}$ as the set of vertices a distance of $r$ or greater from all measurement regions: $S_{\mathrm{far}} = \{\, v \in V \,|\, d(v, S_m) \geq r \,\forall\, m \in [1, M]\,\})$. Since there are at most $|V|$ such vertices in $S_{\mathrm{far}}$, their total contribution to (S224) is at most

$$2 \sum_{u \in S_{\mathrm{far}}} \sum_{v \in V} \mathrm{Cor}(u,v) = \mathsf{O}(|V|\,T^D) \leq \frac{1}{2}|V|^{1+\nu}\,, \tag{S229}$$

provided that (S225) holds, with a sufficiently small constant $c'$ that is independent of $|V|$. As a result, the terms included in the sum in (S224) that are not counted in (S229) should satisfy

$$\sum_{u,v \in S \setminus S_{\mathrm{far}}} \mathrm{Cor}(u,v) \geq \frac{1}{2}|V|^{1+\nu}\,. \tag{S230}$$

However, the set $S \setminus S_{\mathrm{far}}$ includes *at most* $\mathsf{O}(M T^D)$ vertices, so the lefthand side of (S230) can be replaced by $\mathsf{O}(M^2 T^{2D})$, from which (S226) follows. □

To clarify, we make the assumption (S225) in Corollary 13 because the alternative,

$$T^D \gtrsim |V|^\nu\,, \tag{S231}$$

would imply a trivial bound on $M$: essentially, $M \geq 0$ instead of (S226). In $D = 1$, a protocol that satisfies (S231) with $M = 0$ (no measurements) can be constructed as follows. We first divide the whole system into $N_c = |V|^{1-\nu}$ contiguous clusters $V_n$ each containing $N = |V|^\nu = |V_n|$ qubits. We then apply a unitary quantum circuit within each cluster to produce the spin-squeezed state $|\psi\rangle$ on the $N$ qubits in the cluster $V_n$, where $|\psi\rangle$ satisfies $\xi^2 \sim 1/N$ (S222). For concreteness, $|\psi\rangle$ can be chosen as a state in the Dicke manifold of the cluster $V_n$ [S51], in which case $|\psi\rangle$ can be prepared using a circuit with depth $T = \mathsf{O}(N)$ obeying (S231) [S52]. Finally, one can verify that the state $|\Psi\rangle = |\psi\rangle^{\otimes N_c}$ of the *whole* system $V$ has $|\langle \vec{J} \rangle| = J \sim |V|$ and $\xi^2 \sim |V|^{-\nu}$ (S222). Note that the recent work [S53] may provide a blueprint for generalizing the above protocol to $D > 1$.

### 7.6. Speed of preparing critical states

Another class of states one might seek to prepare more efficiently using measurements correspond to *critical* states. These states may correspond to a quantum critical point or a conformal field theory (CFT), and are characterized by power-law correlations $\overline{\mathrm{Cor}}(i, f) \sim d(i, f)^{-\alpha}$ for any two qubits $i, f \in V$ and some $\alpha > 0$. Bounding the preparation of such states follows almost trivially from Theorem 11, and is captured by Corollary 14 below.

**Corollary 14.** *Consider a dilated quantum channel $\mathcal{W}$ involving (a) $M$ total measurement regions (S94) with measurement range $l$ (S95) and (b) Hamiltonian dynamics (S86) for total duration $T$, with associated velocity $v_E$ (S92). The channel $\mathcal{W}$ produces some final state $|\Psi\rangle$ acting on a short-range-entangled initial state $\rho_0 = \mathcal{W}_0\, \rho'\, \mathcal{W}_0^\dagger$ (S159), which itself is prepared from a product state $\rho'$ using measurements in $M_0$ regions and Hamiltonian evolution for duration $T_0$. Then, for any $\epsilon \in (0, 1)$, there exists an $\mathsf{O}(1)$ constant $C_\epsilon$ such that, for sufficiently large $L$ (S178), the final state $|\Psi\rangle$ cannot realize power-law correlations $\overline{\mathrm{Cor}}(x, y) \propto |x - y|^{-\alpha}$ for some $\alpha > 0$ unless*

$$2\,(M + M_0 + 1)\,[v_E\,(T + T_0) + (\alpha + D - 1)\log_2 L + C_\epsilon] \geq (1 - \epsilon)\,L\,, \tag{S232}$$

*where the factor $(D - 1)$ is absent in the case of* prefixed *measurement locations.*

The proof of Corollary 14 follows the proof of Theorem 11 and Corollary 12, with the only modification that one multiplies a factor of $L^{-\alpha}$ to the righthand side of (S213), since the correlation function $\overline{\mathrm{Cor}}(i, f)$ between two qubits $i, f \in V$ with $d(i, f) = L$ is proportional to $L^{-\alpha}$, rather than an $\mathsf{O}(1)$ constant as for Dicke states (S208).

## 8. MULTI-QUBIT BOUNDS

Having examined speed limits for the preparation of a well-separated Bell pair (or equivalently, teleporting a single qubit of information), we now provide a partial generalization of Theorem 5 to the generation of $Q > 1$ Bell pairs (or equivalently—up to an $\mathsf{O}(Q)$ difference in resource requirements—teleporting $Q$ qubits). Obviously, one can generate the Bell pairs consecutively, and the total cost (quantified by $T$ and $M$) is simply the sum of the costs for the individual protocols. An alternative would be to generate the Bell pairs *concurrently* (i.e., "in parallel"), so that the *total* time $T$ for generating $Q$ Bell pairs is roughly the time $T$ required to produce a single Bell pair, with an additional $\mathsf{O}(Q)$ overhead. Moreover, for suitable geometries, the measurements in each of the parallel protocols can be performed using the same set of qubits, so that the total number of measurement regions $M$ is the same as for a single Bell pair. In this case, preparing $Q > 1$ Bell pairs would have roughly the same cost as preparing a single Bell pair ($Q = 1$).

However, even in the parallel protocol described above, the total number of measurement outcomes *utilized*—rather than the number of distinct measurement *regions*, reflected in $M$—is indeed $Q$ times larger than for a single Bell pair. We denote by $\mathcal{M}$ the number of measurement *outcomes* required to generate $Q$ Bell pairs with error correction, while $M$ remains the number of measurement *regions*. We then prove that a $Q$-fold enhancement of $\mathcal{M}$ is the minimum increase in the number of measurement outcomes required to create $Q$ Bell pairs (compared to a single Bell pair).

We set up the model exactly as in the beginning of Sec. 6, with two differences. First, as mentioned above, we use $\mathcal{M}$ to count the number of measurement outcomes (or equivalently, the number of Stinespring qubits). The measurement set $\mathbb{M}^{\boldsymbol{n}} = \{S_1^{\boldsymbol{n}}, \cdots, S_{\mathcal{M}}^{\boldsymbol{n}}\}$ may then contain identical elements that appear multiple times, meaning that the same set of $l$ qubits are measured at multiple points during the protocol. We assume that each measurement has only two outcomes, as we have before[25]. Second, given two sets of initial and final qubits $I = \{i_1, \ldots, i_Q\}$ and $F = \{f_1, \ldots, f_Q\}$, the desired final state following completion of the protocol is $\bigotimes_{q=1}^{Q} |\text{Bell}\rangle_{i_q, f_q}$, or equivalently, a maximally entangled pure state shared by $I$ and $F$. We set $L = d(I, F)$, and our goal is to show that any such protocol requires $Q L \lesssim (\mathcal{M} + 1) v T$. This is confirmed by the following Theorem.

**Theorem 15.** *Suppose that a channel $\mathcal{W}$ realizes a final state $\rho(T)$ with $Q \geq 1$ Bell pairs shared between pairs of qubits in the respective sets $I, F \subset V$ when applied to the product state $|\mathbf{0}\rangle$ in the dilated Hilbert space. Further suppose that $\mathcal{W}$ involves (i) performing $\mathcal{M}$ total measurements with maximum range $l$ and (2) evolution under local Hamiltonian $H$ with associated velocity $v$ for total time $T$. For any $\epsilon \in (0, 1)$ and any positive integer $Q' \leq Q$, suppose that*

$$(\lfloor \mathcal{M}/Q' \rfloor + 1)(2 v_E T + l + 4) \leq (1 - \epsilon) L, \tag{S233}$$

*where $\lfloor x \rfloor$ denotes the greatest integer that is no larger than $x$, and we take $T$ to be large compared to the spatial dimension $D$ if the graph is a finite-dimensional lattice, as constrained by* (S111).

*Then there exists a reduced density matrix $\sigma_{IF}$ with $\operatorname{tr}[\sigma_{IF}] = 1$ that is close in trace distance to the true final reduced density matrix $\rho_{IF} = \operatorname{tr}_{(I \cup F)^c}[\mathcal{W} |\mathbf{0}\rangle\langle\mathbf{0}| \mathcal{W}^\dagger]$ on $I$ and $F$ under $\mathcal{W}$, such that the fidelity obeys*

$$F(\rho_{IF}, \sigma_{IF}) \geq 2^{1-Q'}, \tag{S234}$$

*and where $\sigma_{IF}$ is close in trace distance to a separable state $\widetilde{\sigma}_{IF}$ with $\operatorname{tr}[\widetilde{\sigma}_{IF}] = 1$:*

$$\|\sigma_{IF} - \widetilde{\sigma}_{IF}\|_1 \leq \frac{Q'}{\epsilon} 2^{v_E T + \frac{l}{2}} \times \begin{cases} c\, l\, 2^{-L/2(\lfloor \mathcal{M}/Q' \rfloor + 1)}, & \text{prefixed measurement locations} \\ c\, L^{D-1} 2^{-(1-\epsilon)L/2(\lfloor \mathcal{M}/Q' \rfloor + 1)}, & \text{adaptive measurement locations} \end{cases}, \tag{S235}$$

*where the constants $c > 0$ are the same ones in* (S112) *and* (S113), *for the two cases respectively.*

Before proving Theorem 15, we first show that (S234) implies that $\rho_{IF}$ is far from any maximally entangled state on $I$ and $F$, given that $\sigma_{IF}$ is close to a separable state, as implied by (S233) and (S235). We have invoked the fidelity $F(\rho, \sigma)$ for two density matrices, defined by

$$F(\rho, \sigma) \equiv \|\sqrt{\rho}\sqrt{\sigma}\|_1^2 = F(\sigma, \rho), \tag{S236}$$

which reduces to $\langle \psi | \sigma | \psi \rangle$ when $\rho = |\psi\rangle\langle\psi|$ is a pure state. When the state $\psi$ is maximally entangled with respect to $I$ and $F$, its fidelity to any separable state is no larger than $2^{-Q}$, since in such cases, $\psi$ can be written as

$$|\psi\rangle = 2^{-Q/2} \left(|11\rangle + |22\rangle + \cdots + |2^Q, 2^Q\rangle\right), \tag{S237}$$

---

[25] A Bell measurement counts as a single region toward $M$, but as two measurement outcomes toward $\mathcal{M}$, because the outcomes of two single-qubit measurements are needed per Bell pair. This can easily be generalized beyond binary measurements.

using some suitable basis $\{|1\rangle, \ldots, |2^Q\rangle\}$ for each of the $Q$-qubit sets $I$ and $F$.

For any (pure) product state on $I$ and $F$ of the form,

$$|\phi\rangle = \left(\sum_{k=1}^{2^Q} a_k |k\rangle_I\right) \otimes \left(\sum_{k=1}^{2^Q} b_k |k\rangle_F\right), \tag{S238}$$

the fidelity of $|\phi\rangle$ with $|\psi\rangle$ is

$$F(|\psi\rangle, |\phi\rangle) = |\langle\psi|\phi\rangle|^2 = 2^{-Q} |\sum_{k=1}^{2^Q} a_k b_k|^2 \leq 2^{-Q} \left(\sum_{k=1}^{2^Q} |a_k|^2\right)\left(\sum_{k=1}^{2^Q} |b_k|^2\right) = 2^{-Q}, \tag{S239}$$

and since any separable state $\widetilde{\sigma}$ is a convex sum of pure product states, it follows that $F(|\psi\rangle, \widetilde{\sigma}) \leq 2^{-Q}$ because of the linearity of the fidelity function $F(|\psi\rangle, \sigma) = \langle\psi|\sigma|\psi\rangle$. As a result, $\rho_{IF}$ in (S234) cannot correspond to such a state $\psi$ until (S233) is violated. In fact, (S234) and (S250) together imply

$$F(\rho_{IF}, \widetilde{\sigma}_{IF}) \geq 2^{1-Q'} - \mathcal{O}\left(\sqrt{2}^{v_E T + \frac{l}{2} - \frac{L}{2(\lfloor \mathcal{M}/Q' \rfloor + 1)}}\right) > 2^{-Q}, \tag{S240}$$

from the relation between fidelity and trace distance [S11]. This means that $\rho_{IF}$ is also not a maximally entangled state if (S233) holds—this state effectively contains at most $Q' - 1$ Bell pairs.

*Proof.* As before, we first focus on prefixed measurement locations, and consider reference protocols that differ from the original $\mathcal{W}$ by the removal of Hamiltonian terms acting on both sides of the cut $C$ that partitions the graph $G$ of physical qubits. However, the cut $C$ is chosen in a slightly different manner than in previous proofs. Here we demand that all measurements $S_m$ and all sites in the sets $I$ and $F$ are of a distance $\gtrsim Q'L/2\mathcal{M}$ away from the cut $C$—excepting $Q' - 1$ measurements, which may be closer to $C$. Such a cut exists by Lemma 7: We first sort and relabel the measurement regions $\{S_m\}$ by their distance to $I$, so that

$$d(S_1, I) \leq d(S_2, I) \leq \ldots \leq d(S_{\mathcal{M}}, I), \tag{S241}$$

and we then sort the measurement regions into $Q'$ groups, with the $q$th group being $\{S_{mQ'+q} \mid m = 0, 1, \ldots\}$. There is a group that contains $\lfloor \mathcal{M}/Q' \rfloor$ measurements, and there is a cut $C_r$ that is a distance $\gtrsim Q'L/2\mathcal{M}$ to all measurements in this group, according to Lemma 7. Supposing that $d(S_{mQ'+q}, I) \leq r \leq d(S_{(m+1)Q'+q}, I)$, then (S241) ensures that all measurements in other measurement regions are farther than $Q'L/2\mathcal{M}$ from the cut, except for the $Q'-1$ measurements $S_{mQ'+q+1}, \ldots, S_{(m+1)Q'+q-1}$. Let $V_{\text{ss},C}$ denote the Stinespring qubits recording these $Q'-1$ measurements that are close to the cut, and let $V_{\text{ss},F}$ be the other farther Stinespring qubits. Lemma 7 implies that

$$\min_{m \in V_{\text{ss},F}} d(C, S_m) \quad \text{and} \quad d(C, I) \quad \text{and} \quad d(C, F) > \frac{L}{2(\lfloor \mathcal{M}/Q' \rfloor + 1)} - \frac{l}{2} - 2 \tag{S242a}$$

$$2^{-d(C,I)} + 2^{-d(C,F)} + \sum_{m \in V_{\text{ss},F}} 2^{-d(C,S_m)} < Q' 2^{3 + \frac{l}{2} - \frac{L}{2(\lfloor \mathcal{M}/Q' \rfloor + 1)}}, \tag{S242b}$$

where the latter relation follows from summing over (S140b) in Lemma 7 applied to the $Q'$ groups (but deleting the closest $Q' - 1$ sites from the sum).

We next introduce some notation before defining the states $\sigma_{IF}$ and $\widetilde{\sigma}_{IF}$. Let $\boldsymbol{n}_C$ ($\boldsymbol{n}_F$) denote any measurement outcome stored in $V_{\text{ss},C}$ ($V_{\text{ss},F}$), and $P_{\boldsymbol{n}_C}$ ($P_{\boldsymbol{n}_F}$) is the corresponding projector on Stinespring qubits. For each $\boldsymbol{n}_C$, define the protocol $\mathcal{W}_{\boldsymbol{n}_C}$ to be identical to $\mathcal{W}$, except that the Hamiltonian in any trajectory $(\boldsymbol{n}'_C, \boldsymbol{n}_F)$ is not $H^{(\boldsymbol{n}'_C, \boldsymbol{n}_F)}$, but instead $H^{(\boldsymbol{n}_C, \boldsymbol{n}_F)}$. In other words, the classical information stored in $V_{\text{ss},C}$ is not used; rather, the Hamiltonian is applied as though that information would be *fixed* to $\boldsymbol{n}_C$, even if the trajectory has $\boldsymbol{n}'_C \neq \boldsymbol{n}_C$. The measurement basis for each measurement also does not depend on $\boldsymbol{n}'_C$, but is predetermined by $\boldsymbol{n}_C$. Following Lemma 4, the channel $\mathcal{W}_{\boldsymbol{n}_C}$ acts identically to $\mathcal{W}$ on any initial operator containing the projector $P_{\boldsymbol{n}_C}$,

$$\mathcal{W}^\dagger \left(P_{\boldsymbol{n}_C} \otimes \mathcal{O}\right) \mathcal{W} = \mathcal{W}^\dagger_{\boldsymbol{n}_C} \left(P_{\boldsymbol{n}_C} \otimes \mathcal{O}\right) \mathcal{W}_{\boldsymbol{n}_C}, \tag{S243}$$

since the projector automatically selects the "correct" evolution in trajectories with $\boldsymbol{n}_C$. An example is then $P_{\boldsymbol{n}_C}(T) = \mathcal{W}^\dagger_{\boldsymbol{n}_C} P_{\boldsymbol{n}_C} \mathcal{W}_{\boldsymbol{n}_C}$. Furthermore, we define $\widetilde{\mathcal{W}}_{\boldsymbol{n}_C}$, which differs from $\mathcal{W}_{\boldsymbol{n}_C}$ only in that the cut Hamiltonian $H_C$ is deleted, so that $\widetilde{\mathcal{W}}_{\boldsymbol{n}_C}$ corresponds to a LOCC protocol with respect to the two sides of the cut $C$.

Now we define the state $\sigma_{IF}$ as an averaged final state produced by the protocols $\mathcal{W}_{\boldsymbol{n}_\mathrm{C}}$:

$$\sigma_{IF} = \sum_{\boldsymbol{n}_\mathrm{C}} p_{\boldsymbol{n}_\mathrm{C}} \operatorname*{tr}_{(I\cup F)^c} \left[ \mathcal{W}_{\boldsymbol{n}_\mathrm{C}} |\boldsymbol{0}\rangle\langle\boldsymbol{0}| \mathcal{W}_{\boldsymbol{n}_\mathrm{C}}^\dagger \right], \quad \text{where} \quad p_{\boldsymbol{n}_\mathrm{C}} = \langle P_{\boldsymbol{n}_\mathrm{C}}(T) \rangle_0, \tag{S244}$$

where $\langle A \rangle_0 = \langle \boldsymbol{0}|A|\boldsymbol{0}\rangle$ is the expectation value evaluated in the initial state $|\boldsymbol{0}\rangle$, and $\operatorname{tr}[\sigma_{IF}] = 1$ follows from the relation, $\sum_{\boldsymbol{n}_\mathrm{C}} p_{\boldsymbol{n}_\mathrm{C}} = 1$. We further define the "reference" (or "cut") version according to

$$\widetilde{\sigma}_{IF} = \sum_{\boldsymbol{n}_\mathrm{C}} p_{\boldsymbol{n}_\mathrm{C}} \operatorname*{tr}_{(I\cup F)^c} \left[ \widetilde{\mathcal{W}}_{\boldsymbol{n}_\mathrm{C}} |\boldsymbol{0}\rangle\langle\boldsymbol{0}| \widetilde{\mathcal{W}}_{\boldsymbol{n}_\mathrm{C}}^\dagger \right], \tag{S245}$$

where again we have $\operatorname{tr}[\widetilde{\sigma}_{IF}] = 1$. Having constructed the states, we first verify (S234) using the fact that

$$\rho_{IF} = \sum_{\boldsymbol{n}_\mathrm{C}} \operatorname*{tr}_{(I\cup F)^c} \left[ P_{\boldsymbol{n}_\mathrm{C}} \mathcal{W} |\boldsymbol{0}\rangle\langle\boldsymbol{0}| \mathcal{W}^\dagger P_{\boldsymbol{n}_\mathrm{C}} \right] = \sum_{\boldsymbol{n}_\mathrm{C}} p_{\boldsymbol{n}_\mathrm{C}} \operatorname*{tr}_{(I\cup F)^c} \left[ |\psi_{\boldsymbol{n}_\mathrm{C}}\rangle\langle\psi_{\boldsymbol{n}_\mathrm{C}}| \right], \tag{S246}$$

where we have defined

$$|\psi_{\boldsymbol{n}_\mathrm{C}}\rangle = p_{\boldsymbol{n}_\mathrm{C}}^{-1/2} P_{\boldsymbol{n}_\mathrm{C}} \mathcal{W}_{\boldsymbol{n}_\mathrm{C}} |\boldsymbol{0}\rangle, \tag{S247}$$

and the first equality in (S246) follows from idempotency and completeness of the projectors $P_{\boldsymbol{n}_\mathrm{C}}$, while the second equality in (S246) follows from (S243). Since the normalized state $|\psi_{\boldsymbol{n}_\mathrm{C}}\rangle$ (S247) comes from a projection $P_{\boldsymbol{n}_\mathrm{C}}$ onto the state $\mathcal{W}_{\boldsymbol{n}_\mathrm{C}} |\boldsymbol{0}\rangle$, $|\psi_{\boldsymbol{n}_\mathrm{C}}\rangle$ should inherit some properties of $\mathcal{W}_{\boldsymbol{n}_\mathrm{C}} |\boldsymbol{0}\rangle$. In particular, the entanglement between $I$ and $F$ cannot be drastically increased by the projection—roughly speaking $P_{\boldsymbol{n}_\mathrm{C}}$ can create at most $Q' - 1$ Bell pairs shared by the two parties ($I$ and $F$), on average (see [S54] for a related discussion). To be precise, the weight of $|\psi_{\boldsymbol{n}_\mathrm{C}}\rangle$ (S247) is related to its fidelity to the state $\mathcal{W}_{\boldsymbol{n}_\mathrm{C}} |\boldsymbol{0}\rangle$,

$$F(|\psi_{\boldsymbol{n}_\mathrm{C}}\rangle, \mathcal{W}_{\boldsymbol{n}_\mathrm{C}} |\boldsymbol{0}\rangle) = \left\langle \boldsymbol{0} \middle| P_{\boldsymbol{n}_\mathrm{C}}(T) \middle| \boldsymbol{0} \right\rangle^2 / p_{\boldsymbol{n}_\mathrm{C}} = p_{\boldsymbol{n}_\mathrm{C}}. \tag{S248}$$

Using the joint concavity property of the square root of the fidelity function $F(\cdot, \cdot)$ [S11] leads to (S234),

$$\sqrt{F(\rho_{IF}, \sigma_{IF})} \geq \sum_{\boldsymbol{n}_\mathrm{C}} p_{\boldsymbol{n}_\mathrm{C}} \sqrt{F\left( \operatorname*{tr}_{(I\cup F)^c} \left[ |\psi_{\boldsymbol{n}_\mathrm{C}}\rangle\langle\psi_{\boldsymbol{n}_\mathrm{C}}| \right], \operatorname*{tr}_{(I\cup F)^c} \left[ \mathcal{W}_{\boldsymbol{n}_\mathrm{C}} |\boldsymbol{0}\rangle\langle\boldsymbol{0}| \mathcal{W}_{\boldsymbol{n}_\mathrm{C}}^\dagger \right] \right)}$$
$$\geq \sum_{\boldsymbol{n}_\mathrm{C}} p_{\boldsymbol{n}_\mathrm{C}} \sqrt{F\left( |\psi_{\boldsymbol{n}_\mathrm{C}}\rangle, \mathcal{W}_{\boldsymbol{n}_\mathrm{C}} |\boldsymbol{0}\rangle \right)} = \sum_{\boldsymbol{n}_\mathrm{C}} p_{\boldsymbol{n}_\mathrm{C}}^{3/2} \geq \sqrt{2^{1-Q'}}, \tag{S249}$$

where we used the fact that the fidelity is monotonic when tracing out subsystems: $F(\rho_{AB}, \sigma_{AB}) \leq F(\rho_A, \sigma_A)$, and that there are only $2^{Q'-1}$ choices of $\boldsymbol{n}_\mathrm{C}$.

It remains to verify (S235). The idea is that, for each $\boldsymbol{n}_\mathrm{C}$, the protocol $\mathcal{W}_{\boldsymbol{n}_\mathrm{C}}$ does not use classical information regarding the outcomes of any measurements close to the cut $C$. When probing the trace distance using an arbitrary operator $\mathcal{O}$ supported only on $I \cup F$, one can show as in previous proofs that the $\boldsymbol{n}_\mathrm{C}$ terms in (S244) and (S245) are close to each other. As a result, the sums over all such terms are also close:

$$\|\sigma_{IF} - \widetilde{\sigma}_{IF}\|_1 \leq \sum_{\boldsymbol{n}_\mathrm{C}} p_{\boldsymbol{n}_\mathrm{C}} \left\| \operatorname*{tr}_{(I\cup F)^c} \left[ \mathcal{W}_{\boldsymbol{n}_\mathrm{C}} |\boldsymbol{0}\rangle\langle\boldsymbol{0}| \mathcal{W}_{\boldsymbol{n}_\mathrm{C}}^\dagger \right] - \operatorname*{tr}_{(I\cup F)^c} \left[ \widetilde{\mathcal{W}}_{\boldsymbol{n}_\mathrm{C}} |\boldsymbol{0}\rangle\langle\boldsymbol{0}| \widetilde{\mathcal{W}}_{\boldsymbol{n}_\mathrm{C}}^\dagger \right] \right\|_1$$
$$\leq \left( \sum_{\boldsymbol{n}_\mathrm{C}} p_{\boldsymbol{n}_\mathrm{C}} \right) \max_{\boldsymbol{n}_\mathrm{C}} \max_{\|\mathcal{O}\|\leq 1} \operatorname{tr}\left[ \mathcal{O} \left( \mathcal{W}_{\boldsymbol{n}_\mathrm{C}} |\boldsymbol{0}\rangle\langle\boldsymbol{0}| \mathcal{W}_{\boldsymbol{n}_\mathrm{C}}^\dagger - \widetilde{\mathcal{W}}_{\boldsymbol{n}_\mathrm{C}} |\boldsymbol{0}\rangle\langle\boldsymbol{0}| \widetilde{\mathcal{W}}_{\boldsymbol{n}_\mathrm{C}}^\dagger \right) \right]$$
$$= \max_{\boldsymbol{n}_\mathrm{C}} \max_{\|\mathcal{O}\|\leq 1} \left\langle \boldsymbol{0} \middle| \mathcal{W}_{\boldsymbol{n}_\mathrm{C}}^\dagger \mathcal{O} \mathcal{W}_{\boldsymbol{n}_\mathrm{C}} - \widetilde{\mathcal{W}}_{\boldsymbol{n}_\mathrm{C}}^\dagger \mathcal{O} \widetilde{\mathcal{W}}_{\boldsymbol{n}_\mathrm{C}} \middle| \boldsymbol{0} \right\rangle$$
$$\leq \max_{\boldsymbol{n}_\mathrm{C}} \max_{\|\mathcal{O}\|\leq 1} \| \mathcal{W}_{\boldsymbol{n}_\mathrm{C}}^\dagger \mathcal{O} \mathcal{W}_{\boldsymbol{n}_\mathrm{C}} - \widetilde{\mathcal{W}}_{\boldsymbol{n}_\mathrm{C}}^\dagger \mathcal{O} \widetilde{\mathcal{W}}_{\boldsymbol{n}_\mathrm{C}} \|. \tag{S250}$$

Moreover, we claim that the operator difference in the end of (S250) is bounded similarly to (S137), with $Q'-1$ measurements regions close to the cut excluded in the sum. The reason is that $\mathcal{W}^\dagger_{\bm{n}_\mathrm{C}} \mathcal{O} \mathcal{W}_{\bm{n}_\mathrm{C}}$ does not grow support on the Stinespring qubits in $V_{\mathrm{ss},C}$, since the protocol $\mathcal{W}_{\bm{n}_\mathrm{C}}$ does not utilize the classical information stored in $V_{\mathrm{ss},C}$ following measurements: The classical bit is *fixed* to be $\bm{n}_\mathrm{C}$ regardless of trajectory! Therefore the previous Lieb-Robinson analysis applies here, and we recover (S235) for prefixed measurement locations from (S242b).

To tackle adaptive measurement locations, we use the method introduced in the proof of Theorem 5 in Sec. 6.7, which considers all possible cuts $C_r$. For each $C_r$, there exists a set of trajectories $\bm{N}_r$ in which the support of *all* measured observables are a distance $q \gtrsim Q'L/(2\mathcal{M})$ to $C_r$, except for $Q'-1$ measurements (i.e., outcomes). We then define $\mathcal{W}_{r,\bm{n}_\mathrm{C}}$ in relation to $\mathcal{W}$ as a protocol that does not use the outcomes of those $Q'-1$ measurements near $C_r$, but instead prefixes them to $\bm{n}_\mathrm{C}$. As usual, the reference protocol $\widetilde{\mathcal{W}}_{r,\bm{n}_\mathrm{C}}$ further removes all terms $H_C$ in the Hamiltonian that act across $C_r$. The states are now

$$\rho_{IF} = \frac{1}{\epsilon' L} \sum_r \sum_{\bm{n}_\mathrm{C}} \operatorname*{tr}_{(I\cup F)^c} \left[ P_{r,\bm{n}_\mathrm{C}} \mathcal{W}_{r,\bm{n}_\mathrm{C}} |\bm{0}\rangle\langle\bm{0}| \mathcal{W}^\dagger_{r,\bm{n}_\mathrm{C}} P_{r,\bm{n}_\mathrm{C}} \right] \tag{S251a}$$

$$\sigma_{IF} = \frac{1}{\epsilon' L} \sum_r \sum_{\bm{n}_\mathrm{C}} p_{r,\bm{n}_\mathrm{C}} \operatorname*{tr}_{(I\cup F)^c} \left[ \mathcal{W}_{r,\bm{n}_\mathrm{C}} |\bm{0}\rangle\langle\bm{0}| \mathcal{W}^\dagger_{r,\bm{n}_\mathrm{C}} \right] \tag{S251b}$$

$$\widetilde{\sigma}_{IF} = \frac{1}{\epsilon' L} \sum_r \sum_{\bm{n}_\mathrm{C}} p_{r,\bm{n}_\mathrm{C}} \operatorname*{tr}_{(I\cup F)^c} \left[ \widetilde{\mathcal{W}}_{r,\bm{n}_\mathrm{C}} |\bm{0}\rangle\langle\bm{0}| \widetilde{\mathcal{W}}^\dagger_{r,\bm{n}_\mathrm{C}} \right], \tag{S251c}$$

where we have

$$p_{r,\bm{n}_\mathrm{C}} = \left\langle \bm{0} \middle| \mathcal{W}^\dagger_{r,\bm{n}_\mathrm{C}} P_{r,\bm{n}_\mathrm{C}} \mathcal{W}_{r,\bm{n}_\mathrm{C}} \middle| \bm{0} \right\rangle, \tag{S252}$$

and with these definitions, the proof for prefixed measurement locations extends to protocols with adaptive measurements following the Proof of Theorem 5 in Sec. 6.7, giving (S235). □

In the $Q=1$ case, optimal protocols such as the ESTP of the main text or the TFIM code of Sec. 4 obey

$$2(M+1)\,vT = 2(\mathcal{M}/2+1)vT = (\mathcal{M}+2)vT \approx L, \tag{S253}$$

because each two-site measurement region produces two single-site Pauli measurement *outcomes*. On the other hand, Theorem 15 suggests the possibility of accomplishing this $Q=1$ protocol with only half as many measurements:

$$2\,vT(\mathcal{M}+1) \approx L, \tag{S254}$$

which we suspect is not the case. We conjecture that one in fact needs *two* measurement outcomes for each of the $Q$ logical qubits, and for each repeating "measurement region" (loosely speaking, this factor of two comes from the need to correct both $X$ and $Z$ errors). For example, in a $Q$-qubit teleportation protocol in which all outcome-dependent recovery operations are elements of the dilated Clifford group, if there are $M$ regions of measurement (i.e., each logical qubit's operator light cone is reflected $M$ times), then all qubits may be teleported a distance $L \approx (2M+1)vT$, provided that there are $\mathcal{M} = 2QM$ total outcomes (i.e., $2Q$ per region), so that $L \approx (\mathcal{M}/Q+1)vT$, which matches (S253), but not (S254) [S22]. In general, we suspect that (S253) is optimal, though we leave this conjectured tightening of the multi-qubit bound for generic protocols to future work.

---



\end{document}